 \documentclass[oneside,12pt]{Classes/CUEDthesisPSnPDF}
\usepackage{enumerate}
\usepackage{amssymb}
\usepackage{amsmath}
\usepackage{graphicx,epsfig}
\usepackage{braket}

\usepackage{subfigure}
\usepackage{multirow}
\usepackage{listings}

\usepackage{enumerate}
\usepackage{comment}
\usepackage{mathrsfs}

\newcommand{\ee}{\end{equation}}
\newcommand{\be}{\begin{equation}}
\newcommand{\bea}{\begin{eqnarray}}
\newcommand{\eea}{\end{eqnarray}}

\newcommand{\ba}{\begin{aligned}}
\newcommand{\ea}{\end{aligned}}

\newcommand{\eu}{{\rm e}}
\newcommand{\ii}{{\rm i}}
\newcommand{\de}{{\displaystyle\rm\mathstrut d}}
\newcommand{\sn}{{\rm sn}}
\newcommand{\cn}{{\rm cn}}
\newcommand{\dn}{{\rm dn}}
\newcommand{\Tr}{{\rm Tr}}

\newcommand{\tx}{\tilde{x}}
\newcommand{\Ord}{{\mathcal O}}

\newcommand{\sgn}{{\rm sign}}

\def\XXint#1#2#3{{\setbox0=\hbox{$#1{#2#3}{\int}$}
     \vcenter{\hbox{$#2#3$}}\kern-.5\wd0}}

\ifpdf
    \pdfcatalog { /PageMode (/UseOutlines)
                  /OpenAction (fitbh)  }
\fi

\title{Quantum Correlations in\\ Field Theory and\\ Integrable Systems}

\ifpdf
  \collegeordept{\href{http://www.df.unibo.it}{Department of Physics}}
  \university{\href{http://www.unibo.it}{University of Bologna}}
	\author{\href{mailto:stefano.evangelisti@gmail.com}{Stefano Evangelisti}}
  \crest{\includegraphics[width=40mm]{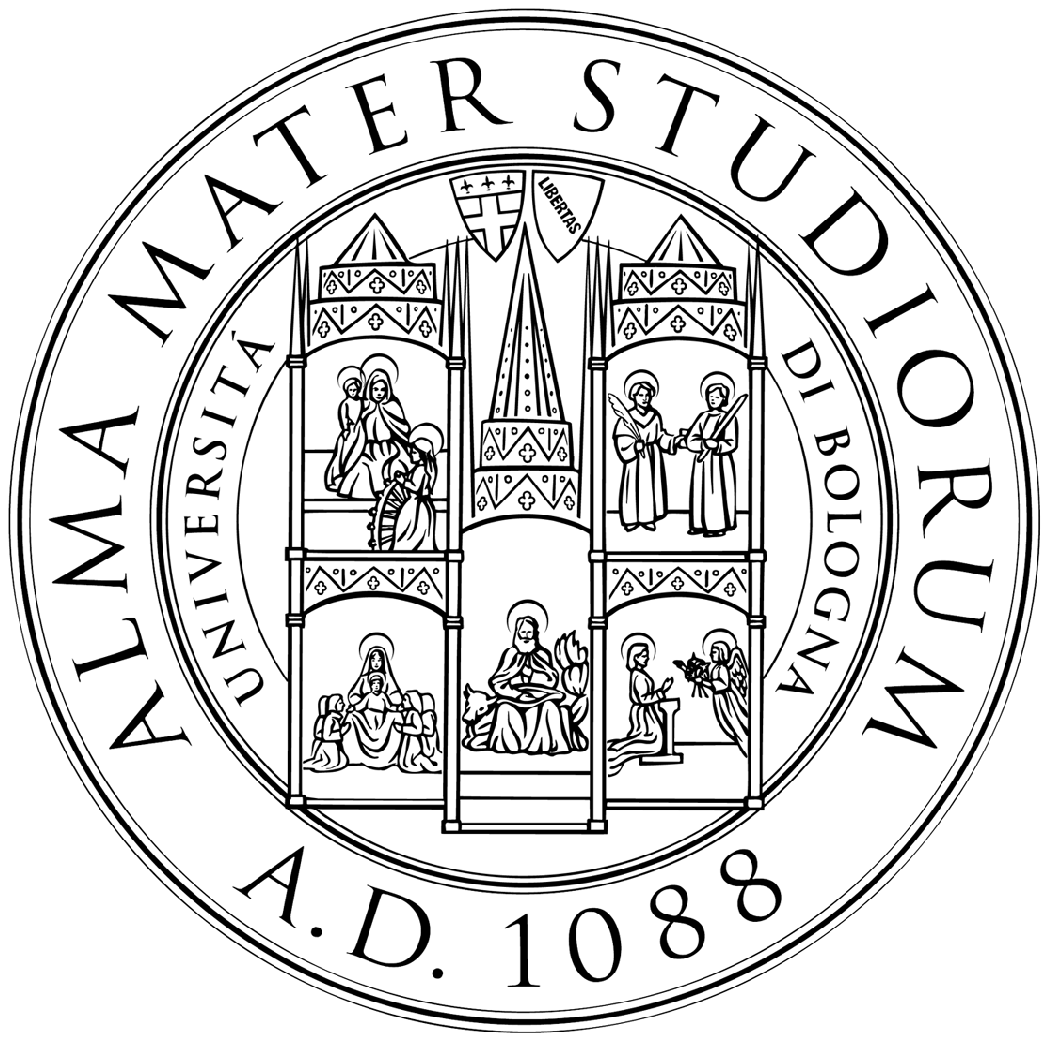}
\else
	\collegeordept{Department of Physics and Astronomy}
  \university{University of Bologna}
	\author{Candidate: Stefano Evangelisti}
	\supervisor{Supervisor: Prof. Francesco Ravanini}
	\cosupervisor{Co-Supervisor: Prof. Elisa Ercolessi}
	\externalsupervisor{External Supervisor: Prof. Fabian Essler (Oxford U.)}
	\cordinator{Ph.D. Coordinator: Prof. Fabio Ortolani}

  \crest{\includegraphics[bb = 0 0 280 336, width=40mm]{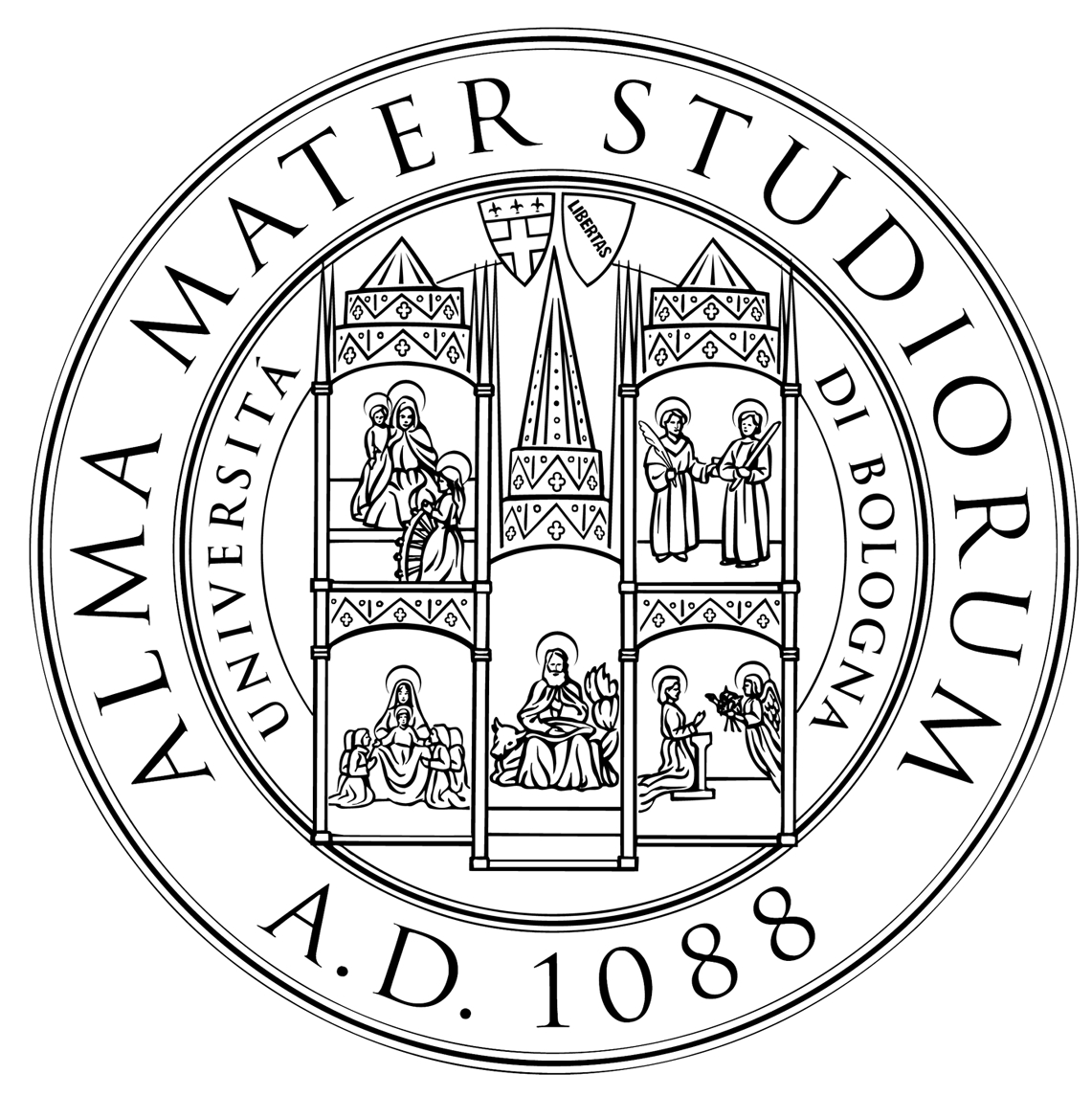}}
\fi
\degree{Doctor of Philosophy}
\degreedate{The 22nd of February 2013}

\usepackage{StyleFiles/watermark}

\begin{document}




\maketitle

\setcounter{secnumdepth}{3}
\setcounter{tocdepth}{3}

\frontmatter 
\pagenumbering{roman}

\begin{dedication} 

I would like to dedicate this thesis to my loving parents.\\
 Their support has always been priceless. 

\end{dedication}




\begin{acknowledgements}      

I am happy to thank Francesco Ravanini and Elisa Ercolessi for their supervision on my graduate projects and for their long-lasting support over the last four years,  and also Fabio Franchini for all the insights he gave me and the discussions we have had during our fruitful collaboration. It is also a pleasure to thank Fabian Essler - my supervisor during my year at Oxford and Worcester College - for his important contribution to my scientific career and for all the instructive meetings we have had in his office, where I learnt a lot, not only about Physics. In particular he really helped me forge a strong work ethic, which is amongst the most important things I have learnt during my graduate studies. 
I am also indebted to Maurizio Fagotti, who has been a fantastic collaborator and a great friend during my stay at Oxford and beyond. His help has been priceless and some of my scientific results would have never seen the light of day without the huge number of insights he gave me. \\
Said that, I now want to thank all the people from Worcester College, especially my wonderful housemates at Worcester House, plus a very good friend from the lab and Jesus College, Neil Robinson, with whom I shared many very interesting discussions. My year in UK has been unforgettable, and the merit is also yours, guys. A special thankyou also goes to the Oxford University Italian Society, the best hub for all italian students at Oxford (and not only italian). I have been proud of being your secretary, a memory that I will keep with me forever. \\
Finally I really really want to thank all my friends from Collegio Alma Mater, Residenza Bononia and Department of Physics, Bologna, Italy, where I spent part of my graduate years, and where I met some of my present best friends.

\end{acknowledgements}



\tableofcontents
\listoffigures
\addcontentsline{toc}{chapter}{Preface}
\chapter*{Preface}

This thesis is based upon the work I performed during my graduate studies and is basically made of two parts, corresponding to the two different projects I have conducted as a PhD student: The first one, on entanglement entropies, dates back to my Master thesis, where we started off by studying the Von Neumann entropy of the XYZ spin-1/2 chain. This project has been carried on mainly at the University of Bologna, under the guidance of Prof. F. Ravanini and Prof. E. Ercolessi, and with the fruitful collaboration of Dr. F. Franchini, who now works at the Massachusett Institute of Technology, USA; the second part is instead based on the project I have conducted
 at the University of Oxford on the out-of-equilibrium physics, with Prof. F.H.L. Essler (supervisor) and Dr. M. Fagotti (post-doc). At Oxford I spent one year of my doctorate as a graduate visiting student, member of Worcester College. 
\mainmatter 
\chapter{Introduction}
\ifpdf
    \graphicspath{{Introduction/IntroductionFigs/PNG/}{Introduction/IntroductionFigs/PDF/}{Introduction/IntroductionFigs/}}
\else
    \graphicspath{{Introduction/IntroductionFigs/EPS/}{Introduction/IntroductionFigs/}}
\fi

In this thesis we shall investigate some properties of integrable one-dimensional quantum systems. From a theoretical point of view one-dimensional models are particularly interesting because they are strongly interacting, since particles cannot avoid each other in their motion, and collisions can {\em never} be ignored. Furthermore, in one dimension also the role of the quantum statistics is somehow special, since it is impossible to exchange two particles without having them interacting. In addition to this, it has to be said that exaclty solvable models are of great importance in physics not just from a theoretical point of view, but also from the experimentalist's perspective. The reason is that in such cases theoretical and experimental results can be compared directly and without any ambiguity. 
Moreover they often generate new and non-trivial situations, which could not be found perturbatively. 
In this dissertation we shall focus on two important aspects of integrable one-dimensional models: Their entanglement properties at equilibrium and their dynamic correlators after a quantum quench. The phenomenon of entanglement is probably amongst the most fundamental properties of quantum systems, distinguishing the quantum from the classical world. Even if it was first discussed more than 75 years ago (\cite{EPR} and \cite{Schrod}) the interest in studying the entanglement, especially in many-body systems, is still growing. Entanglement manifests itself for instance when a measurement of an observable of a subsystem affects drastically and istantaneously the outcome of a measurement  of another part of the system, no matter how distant they are in space. What is extremely fascinating about this effect is the fact that such a correlation propagates at infinite speed. The interest in studying and understanding the features of entangled states  has received an huge boost with the advent of ``quantum information'' in the '90, where entanglement is considered a resource, because such entangled states are the basis for enhancing the efficiency of quantum computation protocols (\cite{Nielsen}). Starting within the context of quantum information a huge progress to quantify the entanglement has been made, which has then found important applications in the study of extended many-body systems. In this other framework the entanglement entropy - which is a popular way for quantifying the entanglement content of a quantum system, even if, not the only one - is considered an important indicator of quantum phase transitions, and its behaviour varying the system sizes, geometries and other physical quantities such as the mass-gap uncovers universal features characterizing the critical points (see the seminal paper \cite{Calabrese2}, and also \cite{Calabrese10}, \cite{Calabrese9}). The first part of this thesis will be therefore devoted to the study of the entanglement entropy in one-dimensional integrable systems, with a special focus on the XYZ spin-1/2 chain, which, in addition to being integrable is also an interacting model. We will derive its Renyi entropies in the thermodynamic limit (and also the Von Neumann as a special case), and its behaviour in different phases and for different values of the mass-gap will be analysed (chapter \ref{chap2} and \ref{chap3}). In the second part of this thesis we will study the dynamics of correlators after a quantum quench, which represent a powerful tool to measure how perturbations and signals propagate across a quantum chain (or a continuous field theory).   
From an experimental point of view the study of the out-of-equilibrium physics has been a difficult task for many years, since genuine quantum properties of a system cannot be preserved for a long time, because of decoherence and dissipation. Only in the last decade the physics of ultra-cold atomic gases overcame these problems, thanks to the fact that they are high-tunable systems, weakly coupled to the external enviroment so that quantum coherence can be preserved for larger times. Basically they behave as  ``quantum simulators'', where the interactions with external potentials may be modified dynamically. Moreover the experimental realization of low-dimensional systems has unreveiled the important role that dimensionality and conservation laws play in quantum non-equilibrium dynamic. 
These fascinating properties have recently been addressed in an experiment on the time-evolution of non-equilibrium Bose gases in one dimension, which is often refered to as the equivalent of the ``Newton's cradle'' (\cite{Weiss}). One of the most important open problems is represented by the characterization of a system that evolves out-of-equilibrium after a sudden change of one (or more) external parameter(s). This kind of time-evolution is commonly called quantum quench. Even if in theory the time-evolution of local observables could be computed from first principles, this is almost always an incredibly hard task, which could be  difficult to solve even numerically.
For these reasons exploiting the most advanced mathematical techniques or developing different approaches in order to takle this problem is very important to draw general conclusion about quantum quenches. For instance if local observables become stationary for long times after a quench (although the entire system will never attain equilibrium), the characterization of these steady states becomes a fundamental issue if we want to compute expectation values at late times without solving the too complicated non-equilibrium dynamics. In this regard some features of the system play a crucial role, such as integrability, which means the existence of an infinite number of independent conservation laws. The general belief is that non-integrable  systems reach a stationary state which can be described by means of a single parameter, an effective temperature. The state at late times is to all purposes equivalent to a thermal one with that temperature. For integrable systems a single parameter is no longer sufficient to describe the state at late times, and it is widely believed that the behaviour of local observables could be explained by the so called generalized Gibbs ensemble (GGE) (\cite{Rigol}). In this dissertation a special attention will be given to the semi-classical approach for computing quantum correlators, which will be applied to the transverse field Ising chain (TFIC) and the O(3) non-linear sigma model (chapter \ref{chap4}). In both cases we will derive the asymptotic behavior of the two-point correlator of the order parameter after a quench. Comparisons with other exact techniques and numerical results will be discussed. Finally we will show that, from a general point of view, if equal-time correlators of local observables at late times after a quench are described by a particular statistical ensemble, then also different-times correlators are described by the same ensemble (chapter \ref{chap_final}). In other words when a stationary state exists, static and dynamic properties of a system are characterized by the same ensemble, which we believe is a new and important result. \\
The structure of this thesis is the following: Chapters \ref{cap:2} and \ref{chap_ising} are reviews on the entanglement properties and the semi-classical method for the TFIC respectively, and their purpose is just to introduce the reader to the content of the subsequent chapters, which contain the original results of this dissertation. These chapters themselves are structured like articles, with an abstract (containing the references to the corresponding published papers) at the beginning and conclusions at the end. In particular in the third chapter we will study in detail the entanglement entropies of the XYZ spin-$1/2$ model, in the entirety of its phase diagram and in the thermodynamic limit. We will see how these entropies signal the lines of phase transition of the spin chain, and how points with the same value of the entropy lying in different phases are connected by a modular transformation in the parameter space. Yet we will discover that the entanglement entropies close to the ferromagnetic critical points of the model describe what we call an essential critical point. In practice in any neighbor of these points it is possible to find any possible (real) value of the entropy. Our analytical results will be also checked numerically by means of DMRG simulations. In the fourth chapter we will study analytically the corrections to the leading terms in the Renyi entropy of both the model on a lattice and the model in the continuous scaling limit, finding significant deviations from na$\ddot{i}$ve expectations.  In particular we will show that finite size and finite mass effects yield different contributions, and thus violate simple scaling arguments. \\
In the sixth chapter the focus will turn to the semi-classical approach to the non-equilibrium dynamics of the O(3) non-linear sigma model, where we will predict quench-dependent relaxation times and correlation lenghts. The same method will be also applied to the TFIC, where the semiclassics can be directly compared to other exact techniques, to unveil the limits of this method. Finally in the seventh chapter we will consider in general the problem of dynamical correlations after a quantum quench in integrable systems, where we will show that in the absence of long-range interactions in the final Hamiltonian the dynamics is determined by the same ensemble that describes  static correlators. For many integrable models we know that these correlators of local observables after a quantum quench relax to stationary values, which are described by a GGE. Therefore the same GGE then determines dynamical correlators and we will also see that the basic form of the fluctuation-dissipation theorem still holds.\\ 
Finally let us notice that, even if there is a logical and chronological flow from the first to the last chapter, each one has been kept voluntarily independent, and can be read singularly.




\chapter{Entanglement in integrable systems}
\label{cap:2}
\ifpdf
    \graphicspath{{Chapter1/Chapter1Figs/PNG/}{Chapter1/Chapter1Figs/PDF/}{Chapter1/Chapter1Figs/}}
\else
    \graphicspath{{Chapter1/Chapter1Figs/EPS/}{Chapter1/Chapter1Figs/}}
\fi

In this first chapter we will introduce the description of entanglement in many-body systems and review the main properties of the reduce density matrices and the entanglement entropies. Some analytical techniques for computing these quantities will be discussed, along with a brief introduction to the Density Matrix Renormalization Group (DMRG), which is the most efficient numerical algorithm to obtain the low energy physics of quantum many-body systems with high accuracy. Both analytical and numerical approaches will be used later in this thesis to describe the entanglement features of the XYZ spin-1/2 chain. \\
This chapter is based on the recent review article ``{\em Entanglement in solvable many-particle models}'', written by I. Peschel, arXiv:1109.0159 and  Braz. J. Phys. {\bf 42}, 267 (2012), even if the material used here has been adapted to the purpose. 

\section{The general framework}
\markboth{\MakeUppercase{\thechapter. Entanglement in integrable systems }}{\thechapter. Entanglement in integrable systems }
The notion of entanglement can be dated back to 1935 when it was introduced by E. Schr${\rm\ddot{o}}$dinger in a series of three articles (\cite{Schrod}). At the same time A. Einstein, B. Podolski and N. Rosen discussed the implication of their famous {\em gedankenexperiment}, intended to reveal what they believed to be inadequacies of quantum mechanics. Nowadays the same kind of experiments is usually formulated by considering a pair of spins (the original EPR paradox considered positions and momenta of quantum particles),  and typically this is the framework in which one encounters entanglement first. The concept of entanglement has to do with the fundamental features of quantum mechanics and the information content of wave functions. For a long time it was a topic discussed mainly in the quantum optics community and for physical systems with few degrees of freedom. However over the last three decades it has seen a revival thanks to inputs from many different areas, such as the theory of black holes, the analytic and numerical investigation of quantum spin chains and the field of quantum information. In the first part of this thesis we will be interested mainly on the entanglement entropies of spin chains, or more in general on entanglement in condensed matter physics. In this case one almost always deals with large systems and many degrees of freedom. When investigating the entanglement properties of a quantum system the protocol is usually the following: we start with the total system which has been prepared in a given quantum state $|\psi\rangle$, then we divide the system into two parts (in space or at the level of the Hilbert space)\footnote{Also the multi-partite entanglement can be investigated, but this topic will not be addressed in this thesis.}, and finally we ask ourselves how the two parts are coupled in $|\psi\rangle$. There is a general way to answer this question, namely one must bring the quantum state  $|\psi\rangle$ into a well-defined standard form, which will explicitly display the coupling between the two parts the system has been divided into. This technique is called {\em Schmidt decomposition}, we will descibe it in detail in a moment. In practice one uses mathematical objects which determine the properties of a subsystem, the so-called {\em reduced density matrices} (RDMs). As we will see in the follow they also contain information on the entanglement and for this reason they will be the basic {\em tool} throughout the first half of this thesis. \\
The quantum states we will consider are ground states of integrable models. These models are very important in physics, not just from a theoretical point of view but also from the experimentalist's perspective, because in such cases analytical and experimental results can be compared without any ambiguity. As in other contexts, they serve as point of orientation which allow to study the features of the problem and to develop a solid feeling on the overall picture. Moreover the entanglement properties of a system turn out to be crucial for the performance of a Density Matrix Renormalization Group simulation (DMRG), a numerical technique which is deeply used in numerical investigations in condensed matter physics.                                    

\section{The Schmidt decomposition}

Let us start by considering a quantum state $|\psi\rangle$ and divide the system into parts A and B (Alice and Bob). We can immediately expand $|\psi\rangle$ as:
\begin{equation}
\label{chap1_eq1}
|\psi^{A\cup B}\rangle = {\displaystyle\sum_{m,n}A_{m,n}|\psi_{m}^{A}\rangle|\psi_{n}^{B}\rangle},
\end{equation}
where $\{ \psi_{m}^{A} \}$ and $\{ \psi_{n}^{B} \}$ are orthonormal bases in the two Hilbert spaces $\mathcal{H}_{A}$ and $\mathcal{H}_{B}$. Note that the sum in (\ref{chap1_eq1}) is double and in general the matrix of coefficients $A_{m,n}$ is rectangular, since the dimensions of the two Hilbert spaces can differ. Nevertheless we can factorize it by using the singular value decomposition method (SVD). This is a widely used technique to decompose a matrix into several component matrices, exposing many of the useful and interesting properties of the original matrix. The decomposition of a matrix is often called {\em factorization}. In practice any rectangular matrix $A$ (real or complex) can be written as:
\begin{equation}
\label{chap1_eq2}
{\bf A}={\bf U}{\bf\Lambda} {\bf V}^{\dagger},
\end{equation}
where ${\bf U}$ is a $m\times m$ unitary matrix (${\bf U}{\bf U^{\dagger}}=\mathbb{I}$), ${\bf V}$ is a $n\times n$ unitary matrix (${\bf V}{\bf V^{\dagger}}=\mathbb{I}$), and $\Lambda$ is a $m\times n$ matrix whose off-diagonal entries are all 0's and whose diagonal elements satisfy:
\begin{equation}
\label{chap1_eq3}
\lambda_1\geq\lambda_2\geq\dots\lambda_{r}\geq 0.
\end{equation}
This allows us to re-write equation (\ref{chap1_eq1}) as:
\begin{equation}
\label{chap1_eq4}
|\psi^{A\cup B}\rangle = {\displaystyle\sum_{m,k,p,n}U_{m,k}\Lambda_{k,p}V_{p,n}^{\dagger}|\psi_{m}^{A}\rangle|\psi_{n}^{B}\rangle}.
\end{equation}
Now by making use of the fact that $\Lambda_{k,p}\equiv \lambda_{p}\delta_{k,p}$, and combining $|\psi_{m}^{A}\rangle$ with ${\bf U}$ and $|\psi_{n}^{B}\rangle$ with ${\bf V}$ we end up with the following representation for the state $|\psi^{A\cup B}\rangle$:
\begin{equation}
\label{chap1_eq5}
|\psi\rangle = {\displaystyle\sum_{k}\lambda_{k}|\psi_{k}^{A}\rangle|\psi_{k}^{B}\rangle},
\end{equation}
where $k=1,\dots,r$, and $r$ is defined in (\ref{chap1_eq3}) ($r \leq \min{(m,n)}$). This representation is called the {\emph Schmidt decomposition} (\cite{Schmidt}), and has the following properties:
\begin{enumerate}
\item  $|\psi\rangle$ is expressed as a single sum, and the number of coefficients is limited by the dimension of the smaller Hilbert space.
\item  The sum of the squared of the coefficients $\sum |\lambda_{k}|^{2}=1$ if the state $|\psi\rangle$ is normalized.
\item  The state  $|\psi\rangle$ is separable if and only if $\exists !\,\lambda_{i} \neq 0$, $i=1,\dots,r$.  
\item  As we shall see, the entanglement is encoded in the coefficients $\lambda_{k}$. 
\end{enumerate}
In particular it is worth noticing that when $\lambda_1=1$ and $\lambda_{k}=0$ for $k>1$ the quantum state $|\psi\rangle$ becomes a product state, therefore there is no entanglement in the system. On the contrary, when $\lambda_{k}=1/\sqrt{r}$ for all $k$ all terms hold equal weight, and this is the situation where the entaglement is maximal.
A few examples for the Schmidt decomposition in a system of two spins one-half are the following: {\bf (a)} $|\psi_1\rangle = |\uparrow\rangle|\downarrow\rangle$, this is trivially a product state, measuring the z-component of the left spin does not interfere with a later measure done on the right one. For this reason this state is not entangled. {\bf (b)} $|\psi_2\rangle = |\uparrow\rangle \left[a |\uparrow\rangle+b  |\downarrow\rangle  \right]$, again this is a product state. {\bf (c)} $|\psi_3\rangle = a|\uparrow\rangle|\uparrow\rangle+b|\downarrow\rangle|\downarrow\rangle$, this state, differently, is an entangled state. The result of a measure conducted on the left spin will immediately tell us what another observer will find out after measuring the right one. 

\section{Reduced density matrices}

The Schmidt structure discussed in the previous section can be also derived from the density matrices associated with the state $|\psi\rangle$, which is also the standard way to obtain it. Let us start by considering the total density matrix $\rho$ of the system:
\begin{equation}
\label{chap1_eq6}
\rho=|\psi\rangle\langle\psi|,
\end{equation}
and, for a chosen splitting, one can trace over the degrees of freedom of one subsystem. This would leave us with one of the following two density matrices:
\begin{equation}
\label{chap1_eq7}
\rho_{A}={\rm Tr_{B}}(\rho),\qquad\qquad \rho_{B}={\rm Tr_{A}}(\rho),
\end{equation}
These hermitian operators can be used to compute arbitrary expectation values of physical quantities in the subsystems. Assuming that the state $|\psi\rangle$ has the Schmidt form (\ref{chap1_eq5}), we can immediately re-write the density matrix $\rho$ as:
\begin{equation}
\label{chap1_eq8}
\rho = |\psi\rangle\langle\psi| = {\displaystyle\sum_{k,p}\lambda_{k}\lambda_{p}^{*}|\psi_{k}^{A}\rangle|\psi_{k}^{B}\rangle\langle\psi_{p}^{A}|\langle\psi_{p}^{B}|},
\end{equation}
Taking now the trace over the degrees of freedom of one of the two subsystem leaves us with the following formula for the reduced density matrices:
\begin{equation}
\label{chap1_eq9}
\rho_{\alpha=(A,B)} ={\displaystyle\sum_{k}|\lambda_{k}|^{2}|\psi_{k}^{\alpha}\rangle\langle\psi_{k}^{\alpha}|}.
\end{equation}
From this equation we immediately infer that $\rho_{A}$ and $\rho_{B}$ have the same non-zero eigenvalues, which are given by square of the Schmidt coefficients $w_{k}=|\lambda_{k}|^{2}$. Moreover their eigenfunctions are exactly the Schmidt functions $|\psi_{k}^{\alpha}\rangle$. Therefore the eigenvalue spectrum of the $\rho_{\alpha}$ gives directly the weights of the Schmidt decomposition, and a look at this spectrum shows the basic entanglement features of the quantum state, for a chosen bipartition.
For these reasons it is often referred to as {\em entanglement spectrum} (\cite{Haldane1}). \\
In general $\rho_{\alpha}$ describes a mixed state. A generic expectation value in subsystem $\alpha$ is given by:
\begin{equation}
\label{chap1_eq10}
\langle A_{\alpha}\rangle = {\displaystyle\sum_{k}|\lambda_{k}|^{2}\langle\psi_{k}^{\alpha}|A_{\alpha}|\psi_{k}^{\alpha}\rangle}.
\end{equation}
Since $\rho_{\alpha}$ is hermitian and has non-negative eigenvalues, one can always write:
\begin{equation}
\label{chap1_eq11}
\rho_{\alpha}={1 \over \ Z}{\rm e}^{-H_{\alpha}},
\end{equation}
where $Z$ is a normalization constant and the operator $H_{\alpha}$ is called {\em entanglement Hamiltonian}. We will encounter this form again when we will deal with the entanglement entropy of integrable models.\\
Usually one has the state $|\psi\rangle$ in the form (\ref{chap1_eq1}) and then
\begin{equation}
\label{chap1_eq12}
\rho=|\psi\rangle\langle\psi| = {\displaystyle\sum_{m,n,k,p}A_{m,n}A_{k,p}^{*}}|\psi_{m}^{A}\rangle|\psi_{n}^{B}\rangle\langle\psi_{k}^{A}|\langle\psi_{p}^{B}|,
\end{equation}
and taking the trace over $|\psi_{n}^{B}\rangle$ gives $n=p$ and we get:
\begin{equation}
\label{chap1_eq13}
\rho_{A} = {\displaystyle\sum_{m,m'}\sum_{n} A_{m,n}A_{n,m'}^{\dagger}}|\psi_{m}^{A}\rangle\langle\psi_{m'}^{A}|.
\end{equation}
From this expression we clearly see that $\rho_{A}$ contains the square hermitian matrix ${\bf A}{\bf A}^{\dagger}$ and similarly $\rho_{B}$ contains ${\bf A}^{\dagger}{\bf A}$. The form (\ref{chap1_eq9}) in then obtained by diagonalizing these matrices, and this is the general approach. 

\section{The DMRG algorithm}

In this subsection we briefly introduce a numerical procedure called {\em Density Matrix Renormalization Group} which will be used to test our analytical results for the entanglement entropy of the XYZ spin chain. This method, which was introduced by Steven White in $1992$ (\cite{White1} and \cite{White2}), makes a direct use of the Schmidt decomposition and the reduced density matrix's properties we introduced before.\\
Let us consider for simplicity a quantum spin one-half chain, with open ends. The DMRG algorithm, in its simplest version, would execute the following steps:
\begin{enumerate}[I]
\item {\bf Starting point}: \\
Start by taking a small system, let say of $5-10$ sites. In this framework compute the ground state {\em exactly} (by using a specific algorithm to do that). 
\item {\bf Schmidt decomposition}:\\
Divide the system into two halves, and calculate the corresponding RDM's. \\
Diagonalize them and obtain the Schmidt coefficients and Schmidt states.
\item {\bf Approximation}:
Keep only the $m$ states with the largest weight $w_{m}$. The number $m$ is normally given as an input to the program.
\item{\bf Enlargement}:\\
Insert two additional sites in the center. Form a new Hamiltonian in the basis of kept and additional states. Compute again the ground state.
\end{enumerate}
Go back to point (II) and repeat. For the numerical scheme to work well the form of the Schmidt spectra is crucial. In order to have good performances, a rapid drop of the eigenvalues $w_{n}$ is a necessary condition, such that only a small number of Schmidt states has to be retained. This condition is normally satisfied by non-critical chains. In some cases even a very small number of Schmdt states can give a fantastic accuracy for the ground state energy. It is therefore extremely important to understand the features of the RDM spectra and this motivates us to study them in the solvable cases, which will be one of the topics of this thesis. 

\section{Entanglement entropies}

The full RDM spectra give a very clear impression of the entanglement in a bipartite system, but nonetheless it would be desirable to have a simpler measure through one single number. Since the eigenvalues of the RDM's can be interpreted as probabilities, and can consider the Shannon entropy, as used in probability theory, to characterize the $w_{n}$. In this quantum scenario this gives the so called (von Neumann) entanglement entropy:
 \begin{equation}
\label{chap1_eq14}
{\cal S}_{\alpha}\equiv -{\rm Tr}(\rho_{\alpha}\ln\rho_{\alpha})=-{\displaystyle\sum_{n}w_{n}\ln(w_n)},
\end{equation} 
which is the most common entanglement measure for bipartitions. The subscript $\alpha$ here refers to the subsystem's indexes. The von Neumann entropy has the following properties:
\begin{enumerate}[(a)]
\item If $\rho$ represents a pure state then ${\cal S}(\rho_{\alpha})=0$.
\item ${\cal S}(\rho_{\alpha})={\cal S}(U^{\dagger}\rho_{\alpha}U)$ for any unitary operator $U$.
\item {\rm max} ${\cal S}(\rho_{\alpha})=\ln(D_{\alpha})$, where $D_{\alpha}$ is the dimension of the Hilbert space $\mathcal{H}_{\alpha}$.
\item If $\lambda_1+\lambda_2+\dots+\lambda_n=1$, where $\lambda_i\geq 0$ for all $i=1,\dots,n$, then\\
${\cal S}(\lambda_1\rho_1+\lambda_2\rho_2+\dots+\lambda_n\rho_n)\geq  \lambda_1 {\cal S}(\rho_1)+\lambda_2 {\cal S}(\rho_2)+\dots+ \lambda_n {\cal S}(\rho_n)$.
\item If a system can be splitted into two subsystems $A$ and $B$ we have:\\
${\cal S}(\rho_{AB})\leq {\cal S}(\rho_{A})+{\cal S}(\rho_{B})$,\\
and the equality holds when there is no correlation between the two subsystems, that means: $\rho_{AB}=\rho_{A}\otimes\rho_{B}$.
\end{enumerate}
From property (c) we are led to a simple interpretation of ${\cal S}$. Writing $S=\ln(M_{\rm eff})$ we can interpret ${\rm e}^{{\cal S}}$ as an effective number of states in the Schmidt decomposition. Another possible measure of the entanglement is the Renyi entropy, which is defined as (\cite{Renyi}):
\begin{equation}
\label{chap1_eq15}
{\cal S}_{n}\equiv {1 \over 1-n}\ln{\rm Tr}(\rho_{\alpha}^{n}),
\end{equation} 
where $n$ can also be non-integer (and complex as well). $S_n$ has similar properties as $S$ and the same extremal values $S_n=0$ and $S_n=\ln D_{\alpha}$. It is worth noticing that in the limit $n \to 1$ it reduces to the von Neumann entropy. 
Morover varying the parameter $n$ in (\ref{chap1_eq15}) gives us access to a lot of information on $\rho_{\alpha}$, including its full spectrum (\cite{Calabrese1} and \cite{Franchini1}). The important point is that both entropies measure a {\em mutual connection} and in general they are not proportional to the size of the system, as usual thermodynamics entropies are.

\section{Exactly solvable models}

\subsection{Free lattice models}
The simplest examples of exactly solvable models are the free lattice models. Let us consider models with a quadratic Hamiltonian in fermionic or bosonic operators, 
in their ground state. Examples belonging to this family are fermionic hopping models with conserved number of particles, whose Hamiltonian reads as:
\begin{equation}
\label{chap1_eq16}
H_{\rm hop} = -{1 \over 2}{\displaystyle\sum_{<i,j>}t_{i,j}c_{i}^{\dagger}c_{j}},
\end{equation} 
where the sum is over nearest neighbours. Another interesting example is the XY spin one-half chain, which has been proven to be equivalent to free fermions via the Jordan-Wigner transformation. The most general Hamiltonian for such a model can be written as:
\begin{equation}
\label{chap1_eq17}
H_{\rm XY} = {\displaystyle\sum_{i}\left[{1+\gamma \over 2}\sigma_{i}^{x}\sigma_{i+1}^{x}+  {1-\gamma \over 2}\sigma_{i}^{y}\sigma_{i+1}^{y}\right]}-h{\displaystyle\sum_{i}\sigma_{i}^{z}},
\end{equation} 
where $\sigma_{i}^{\alpha}$ are the Pauli matrices at site $i$ and $h$ is the magnetic field along the z-direction. For $\gamma = 0$ this reduces to the XX model, corresponding to (\ref{chap1_eq16}) with nearest-neighbour hopping and can also describe a model of hard-core bosons. For $\gamma \neq 0$ is exibits pair creation and annihilation terms. The case $\gamma =1$ needs special attention: in this case the XY model becomes the transverse field Ising chain (TFIC), which can be re-written as:
\begin{equation}
\label{chap1_eq18}
H_{\rm Ising} = -\lambda{\displaystyle\sum_{i}\sigma_{i}^{x}\sigma_{i+1}^{x}}-{\displaystyle\sum_{i}\sigma_{i}^{z}}.
\end{equation} 
Despite the simplicity of this model a huge amount of physics has been learned by analyzing the TFIC in and out-of-equilibrium.  
Moreover this model is a paradigm of quantum critical behaviour and quantum phase transitions (\cite{Sachdev1}): At zero temperature in the thermodynamic limit it exhibits ferromagnetic ($\lambda>1$) and paramegnetic ($\lambda<1$) phases, separeted by a quantum critical point at $\lambda_{c}=1$. \\
Notice that the solubility of the models in itself does not necessarily mean that the RDM's are easy to compute. In other integrable models, as for instance the XXZ spin chain, the formulae turn out to be very complicated, see (\cite{Sato1} and \cite{Sato2}). However free lattice models  have eigenstates with special properties which allow us to make a simple general statement: the reduced density matrices for the ground state can be written as:
\begin{equation}
\label{chap1_eq19}
\rho_{\alpha}={1 \over Z}{\rm e}^{-H_{\alpha}},\qquad\qquad H_{\alpha}={\displaystyle\sum_{i=1}^{L}\epsilon_{i}f_{i}^{\dagger}f_{i}},
\end{equation} 
where $L$ is the number of sites in subsystem $\alpha$ and the operatos $(f_{i}^{\dagger},f_{i})$ are fermionic or bosonic creation and annihilation operators for single particle states with eigenvalue $\epsilon_{i}$. These operators are related to the original ones in the subsystem by a canonical transformation. It is suggestive to notice that $\rho_{\alpha}$ has the form of a thermal density matrix with an effective Hamiltonian $H_{\alpha}$ which is of the same free-particle type as $H$, and also it is already in a diagonal form. The constant $Z$ ensures the correct normalization ${\rm tr}\rho_{\alpha}=1$.\\
This form of teh RDM is rather surprising since one finds a similar situation in a system in contact with a thermal bath. However no assumption about the relative sizes of the two coupled systems has been made here. More importantly, the operator $H_{\alpha}$ is {\em not} the Hamiltonian $H$ restricted to the subsystem $\alpha$. Hence (\ref{chap1_eq19}) is not a true Boltzmann formula. As we will see for the TFIC, $H_{\alpha}$ corresponds to an inhomogeneous system even if the subsystem it describes is homogeneous, and this holds true in general.\\
In order to explicitly obtain $\rho_{\alpha}$ one can basically follow three pathways: {\bf (1)} the first one consists in computing the RDM directly, tracing over the degrees of freedom outside the subsystem $\alpha$. This method can be used for instance to solve a system  of coupled quantum oscillators (\cite{Peschel1}).\\
{\bf (2)} The second method to calculate the RDM $\rho_{\alpha}$ reduces the whole problem to the computation of correlation functions, and it has been widely used for free-fermion systems. An example can be a model of electrons hopping on $N$ lattice sites in a state described by a Slater determinant. In such a state, all many-particle correlation functions factorize into products of one-particle functions:
\begin{equation}
\label{chap1_eq20}
\langle c_{i}^{\dagger}c_{j}^{\dagger}c_{k}c_{l}\rangle = \langle c_{i}^{\dagger}c_{k}\rangle\langle c_{j}^{\dagger}c_{l}\rangle-\langle c_{i}^{\dagger}c_{l}\rangle\langle c_{j}^{\dagger}c_{k}\rangle,
\end{equation}
where the ground state $|\dots\rangle$ is just the Fermi sea of the electrons.
If all the sites involved in these expectation values belong to the same subsystem, a calculation using the reduced density matrix must give the same result. This is guaranteed by Wick's theorem if $\rho_{\alpha}$ is the exponential of a free-fermion operator:
\begin{equation}
\label{chap1_eq21}
\rho_{\alpha}\propto \exp(-{\displaystyle\sum_{i,j=1}^{L}h_{i,j}c_{i}^{\dagger}c_{j}})
\end{equation}
 where $i$ and $j$ are sites within the subsystem. With the analytic form of $\rho_{\alpha}$ fixed, the hopping matrix ${\bf h}$ is then determined in a way that it gives the correct one-particle correlation functions $C_{i,j}=\langle c_{i}^{\dagger}c_{j}\rangle$. The two matrices are diagonalized by the same transformation and one finds that the following relation holds:
 \begin{equation}
\label{chap1_eq22}
{\bf h}=\ln[(1-{\bf C}/{\bf C})].
\end{equation}
The same formula also relates the eigenvalues of ${\bf h}$ and ${\bf C}$.\\
{\bf (3)} The third method that one can use to compute the RDM $\rho_{\alpha}$ exploits the relation between one-dimensional quantum systems and two-dimensional classical statistical models. Being this method the most relevant with regards to the results discussed in this thesis\footnote{This is the technique used in the next chapter to derive the entanglement entropy of the XYZ 1/2-spin chain.}, it will be discussed separately in the next section.  

\subsection{Integrable models and corner transfer matrices}

In one dimension one can exploit the relations between quantum chains and two-dimensional classical statistical models. As we have already said, although these quantum models are integrable and their ground state is know analytically, a direct calculation of $\rho$ is in general a difficul task. The difficulties arising  in a direct calculation ca be avoided by mapping the quantum chain onto two-dimensional classical spin systems. This connection was firstly pointed out by Nishino et al. (\cite{Nishino2}) and (\cite{Nishino1}), the density matrix of the quantum chain is the partition function of a two-dimensional strip with a cut perpendicular to it. In fact if the quantum Hamiltonian $H$ and the classical row-to-row transfer matrix $T$ commute $[H,T]=0$ the ground state of $H$ is also an eigenstate of $T$. Therefore the reduced density matrix of a subsystem $A$ of the chain ($\rho_{A}={\rm Tr}_{B}|\psi\rangle\langle\psi|$, with $B$ the complement of $A$) is the partition function of two half-infinite strips, one extending from $-\infty$ to $0$ and the other from $+\infty$ to $0$, with the spins in $B$ {\em identified} (see figure (\ref{fig:1})).  This procedure works for the ground state of a number of integrable quantum chains. For instance the TFIC can in this way be related to a two-dimensional Ising model on a square lattice which is rotated through $45$ with respect to the horizontal (\cite{Peschel2}). In the same way, the XY chain is connected to an Ising model on a triangular lattice \cite{Peschel3}. 
\begin{figure}[t]
   \begin{center}
   \includegraphics[width=8cm]{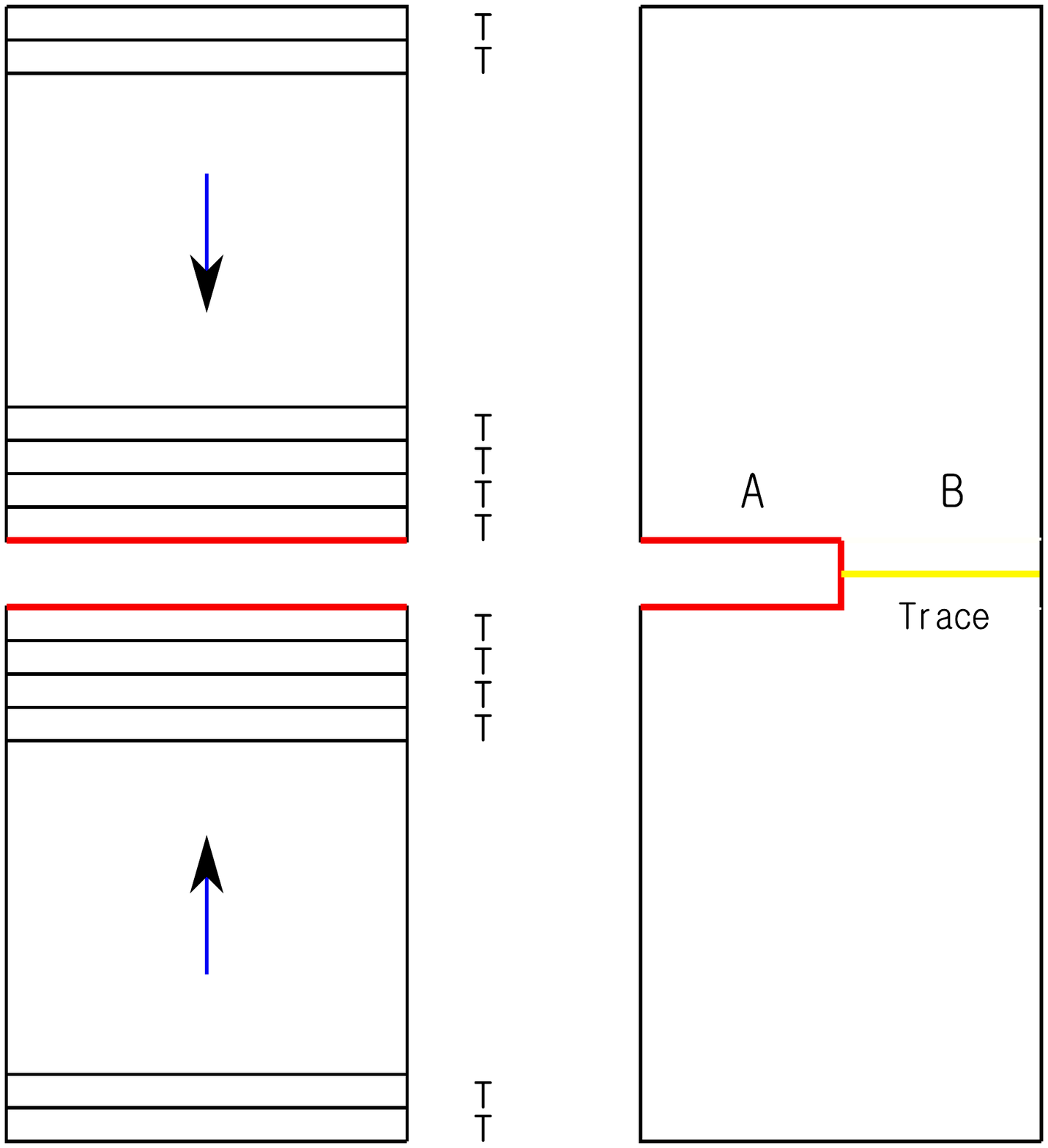}
   \caption{Density matrices for a quantum model as a two-dimensional partition function. Left: expression for $\rho$. Right: expression for $\rho_{A}$.}
\label{fig:1}
\end{center}
\end{figure}
Analogous corrispondences link the XXZ and XYZ and higher-spin chains to vertex models, as we shall see in detail in the next chapter. In order to use these relations, however, one needs to actually compute the classical partition function. This is possible thanks to the help of the corner transfer matrices (CTM's) which have been introduced by R. Baxter (\cite{Baxter}). These are partition functions of whole quadrants as shown in figure (\ref{fig:2}), or sextants when dealing with the triangular lattice. If we multiply these transfer matrices we can then obtain the reduced density matrix for half-chain as:
\begin{equation}
\label{chap1_eq23}
\rho_{\alpha}\sim ABCD.
\end{equation}
Since $\rho_{\alpha}$ is given by an infinite strip, one also needs infinite-size CTM's in this expression. Luckily it is exactly in this limit that the CTM's are known for several non-critical integrable models and have the form:
\begin{equation}
\label{chap1_eq23}
A = {\rm e}^{-u\,H_{\rm CTM}},
\end{equation}
where $u$ is a parameter measuring the anisotropy of the two-dimensional model. 
\begin{figure}[t]
   \begin{center}
   \includegraphics[width=6cm]{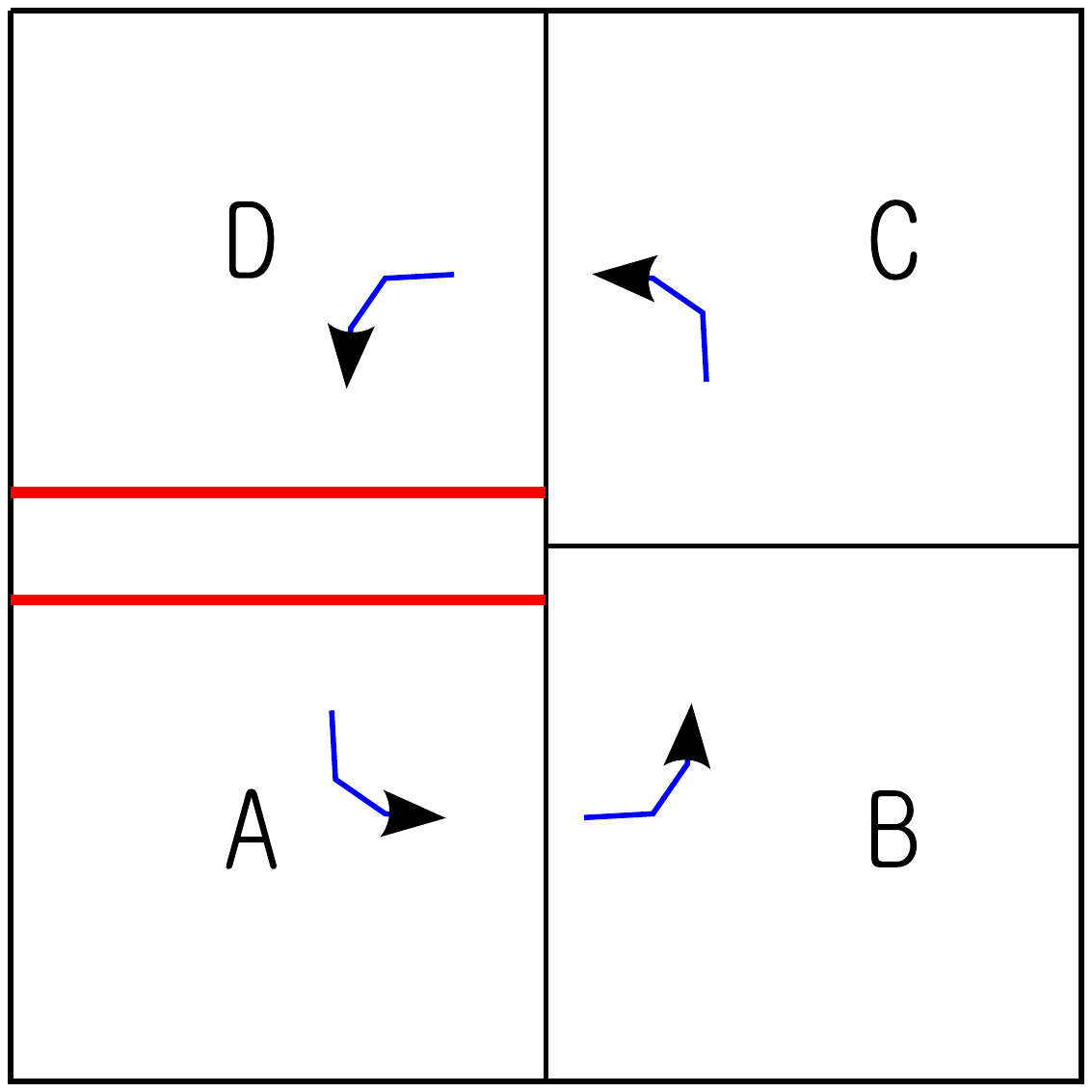}
   \caption{Two-dimensional system lattice built from four corner transfer matrices, $A$, $B$, $C$ and $D$. The arrows indicate the direction of transfer. }
\label{fig:2}
\end{center}
\end{figure}
This special form is a direct consequence of the star-triangle relations (the classical analogous of the Yang-Baxter relations) on which the integrability rests (\cite{Cardy1}). The effective Hamiltonian $H_{\rm CTM}$ can usually be diagonalized by means of fermionization. According to the scheme of derivation adopted here, formula (\ref{chap1_eq23}) applies to half of an infinite chain, but in practice the chain has only to be much longer than the correlation length. \\
The method outlined here is very general, however for integrable chains which satisfy suitable Yang-Baxter equations (\cite{Peschel2}) it is possible to write the exponent of equation (\ref{chap1_eq23}) as $u H_{\rm CTM}=\epsilon \hat{O}$, with $\epsilon$ the scale giving the distance between the energy levels, and $\hat{O}$ is an operator with integer eigenvalues. Using this property the entropy is given by:
\begin{equation}
\label{chap1_eq24}
{\cal S}=-{\rm Tr}\rho_{A}\ln\rho_{A}=-{\rm Tr}{\displaystyle\frac{\rho_{A}\ln\rho_{A}}{{\rm Tr}\rho_{A}}}+\ln{\rm Tr}\rho_{A}=-\epsilon{\displaystyle\frac{\partial\ln Z}{\partial\epsilon}}+\ln Z,
\end{equation}
where we defined $Z={\rm Tr}\rho_{A}={\rm Tr}{\rm e}^{-u\,H_{\rm CTM}}$. 
In the next section we will explicitly show the results for the entanglement entropies of the TFIC,  which represents the simplest possible applications of this analytical technique.

\subsection{The quantum Ising chain}

Let us consider the Hamiltonian defined in (\ref{chap1_eq18}), where the quantum transition is driven by the parameter $\lambda$. The classical equivalent of  (\ref{chap1_eq18}) is the two-dimensional Ising model. For $\lambda=0$ the ground state of the model is a quantum paramagnet with all the spins aligned with the magnetic field in the $x$ direction, and $\langle\sigma_{i}^{z}\rangle=0$. In the opposite limit $\lambda=\infty$ the contribution of the magnetic field to the interactions is negligible and the ground state is ferromagnetic with $\langle\sigma_{i}^{z}\rangle=\pm 1$. The second-order phase transition between these two phases happens at $\lambda=1$, and the critical exponent characterizing the divergence of the correlation length is $\nu = 1$, that is $\xi \simeq |\lambda -1|^{-1}$. 
The CTM's of the Ising model may be diagonalized in terms of fermionic operators, and the effectve Hamiltonian reads as (\cite{Peschel2}):
\begin{equation}
\label{chap1_eq25}
H_{\rm CTM}={\displaystyle\sum_{j=0}^{\infty}\epsilon_{j}f_{j}^{\dagger}f_{j}}.
\end{equation}
The energy levels are given by:
\begin{equation}
\label{chap1_eq26}
\epsilon_{j}=
\begin{cases}
(2j+1)\epsilon\qquad\qquad {\rm for}\, \lambda<1,\\ 
2j\epsilon\qquad\qquad\qquad\,\,\,{\rm for}\, \lambda>1,
\end{cases}
\end{equation}
with $\epsilon=\pi\frac{K(\sqrt{1-k^{2}})}{K(k)}$, where $K(k)$ is the complete elliptic integral of second kind and $k = {\rm min}[\lambda,\lambda^{-1}]$. For $\lambda < 1$ we obtain the following expression for the partition function:
\begin{equation}
\label{chap1_eq27}
Z = {\rm Tr}{\rm e}^{-u H_{\rm CTM}}= {\displaystyle\prod_{j=0}^{\infty}}\left[1+{\rm e}^{-\epsilon(2j+1)}\right],
\end{equation}
end the entropy from equation (\ref{chap1_eq24}) becomes:
\begin{equation}
\label{chap1_eq28}
{\cal S}=\epsilon{\displaystyle\sum_{j=0}^{\infty}\frac{2j+1}{1+{\rm e}^{(2j+1)\epsilon}}}+{\displaystyle\sum_{j=0}^{\infty}\ln(1+{\rm e}^{-(2j+1)\epsilon})}.
\end{equation}
The same computation in the ferromagnetic phase gives ($\lambda > 1$):
\begin{equation}
\label{chap1_eq29}
{\cal S}=\epsilon{\displaystyle\sum_{j=0}^{\infty}\frac{2j}{1+{\rm e}^{(2j)\epsilon}}}+{\displaystyle\sum_{j=0}^{\infty}\ln(1+{\rm e}^{-(2j)\epsilon})}.
\end{equation}
Let us notice that the limiting values of the entropy are $S(0) = 0$ and $S(\infty) = \ln 2$, in agreement with the
expectation that the pure ferromagnetic ground state ($\lambda = \infty$) owns two possible accessible
configurations with opposite magnetization (that is,  $S(\infty) = \ln 2$) whereas the pure quantum
paramagnetic ground state ($\lambda = 0$) has only one configuration available with all the spins
aligned in the direction of the magnetic field $x$ and therefore the resulting entropy is zero. 
Moreover the entropy has a devergence at the quantum critical point $\lambda =1$, which we want now to analyze further.
For $\lambda \to 1$ $\epsilon \to 0$ and we can approximate the sums by integrals:
\begin{equation}
\label{chap1_eq30}
{\cal S}\approx{\displaystyle\int_{0}^{\infty}dx \left(\frac{2x\epsilon}{1+{\rm e}^{2x\epsilon}}+\ln (1+{\rm e}^{-2x\epsilon})\right)}={\displaystyle\frac{\pi^{2}}{12\epsilon}},
\end{equation}
and by $\approx$ we mean in the critical region. If we now use $K(0)= \pi/2$ and $K(x)=-1/2\ln(1-x)+O[(1-x)^{0}]$ (\cite{Abramowitz}), 
\begin{equation}
\label{chap1_eq31}
{\cal S}\approx {\displaystyle\frac{\pi^{2}}{12\epsilon}} \sim -\frac{1}{12}\ln(1-k)=\frac{1}{12}\ln\xi + C,
\end{equation}
where in the last step we used $\xi \propto |1-k|^{-1}$. 
\begin{figure}[t]
 \begin{center}
   \includegraphics[width=8cm]{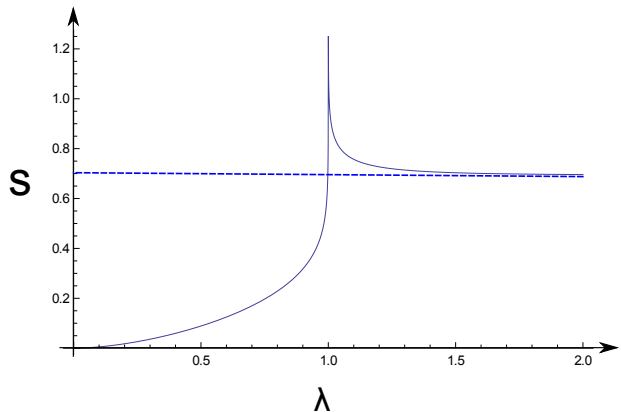}
   \caption{Entanglement entropy between two halves of the infinite TFIC as a function of $\lambda$.  }
\label{fig:1}
\end{center}
\end{figure}
This result agrees with the general result of Cardy and Calabrese (\cite{Calabrese2}), with central charge $c=1/2$ for the Ising model. The constant $C$ is not universal, and as we shall see, it contains a contribution from the boundary entropy of Affleck and Ludwig (\cite{Affleck}).
This limiting case shows a {\em logarithmic} critical behaviour. The effective number of states in the Schmidt decomposition, however, has a standard power-law behaviour:
\begin{equation}
\label{chap1_eq32}
M_{\rm eff}\sim \xi^{1/12},
\end{equation}
and in this regard the coefficient of the logarithm in (\ref{chap1_eq31}) can be interpreted as a critical exponent. The Renyi entropy for this model reads as:
\begin{equation}
\label{chap1_eq33}
{\cal S}_{n}=\frac{1}{24}(1+1/n)\ln\xi+C',
\end{equation}
where again $C'$ is a non-univeral constant. An unusual structure is seen if one looks at the next (subleading) corrections in the expansion. One finds that there are terms of the form $\xi^{-k/n}$ with $k=1,2,3,\dots$, where the power depends on the Renyi index $n$ which determines the number of windings in the path integral form of $\rho_{\alpha}^{n}$. We will discuss this interesting phenomenon in more details in the next chapter, where we shall use the same method developed here to derive the entanglement entropy of the XYZ one-half spin chain (the method is applicable to all integrable models satisfying a proper Yang-Baxter relation).

\ifpdf
  \pdfbookmark[2]{bookmark text is here}{And this is what I want bookmarked}
\fi



\chapter{Exact entanglement entropy of the XYZ model}
\label{chap2}
\ifpdf
    \graphicspath{{Chapter2/Chapter2Figs/PNG/}{Chapter2/Chapter2Figs/PDF/}{Chapter2/Chapter2Figs/}}
\else
    \graphicspath{{Chapter2/Chapter2Figs/EPS/}{Chapter2/Chapter2Figs/}}
\fi
In this chapter we will study the Renyi entropy of the one-dimensional XYZ spin-$1/2$ chain in the entirety of its phase diagram. The model has several quantum critical lines corresponding to rotated XXZ chains in their paramagnetic phase, and four tri-critical points where these phases join.
Moreover we will see how to parametrize the whole phase diagram by using elliptic functions, and in particular how points with the same values of the entropy lying in different phases are connected throught a modular transformation. This approach results to be completely equivalent to the Baxter reparametrization procedure.\\
This chapter is basically a review of the following papers: ``{\em Exact entanglement entropy of the XYZ model and its sine-Gordon limit}'', E. Ercolessi, S. Evangelisti and F. Ravanini, arXiv:0905.4000, Phys. Lett. A, {\bf 374} (2010); ``{\em Essential singularity in the Renyi entanglement entropy of the one-dimensional XYZ spin-1/2 chain }'', E. Ercolessi, S. Evangelisti, F. Franchini and F. Ravanini, arXiv:1008.3892, Phys. Rev. B, {\bf 83} (2011); ``{\em Modular invariance in the gapped XYZ spin chain}'', E. Ercolessi, S. Evangelisti, F. Franchini and F. Ravanini, arXiv:1301.6758.  

\section{The XYZ chain and its symmetries}
\label{sec:symmetries}
\markboth{\MakeUppercase{\thechapter. Exact entanglement entropy of the XYZ model }}{\thechapter. Exact entanglement entropy of the XYZ model }
The quantum spin-$\frac{1}{2}$ ferromagnetic XYZ chain can be described by the following Hamiltonian
\begin{equation}
   \hat{H}_{XYZ} =
   -{\displaystyle \sum_{n}} \left( J_{x}\sigma_{n}^{x} \sigma_{n+1}^{x}
   + J_{y}\sigma_{n}^{y} \sigma_{n+1}^{y} + J_{z}\sigma_{n}^{z} \sigma_{n+1}^{z} \right) \; ,
   \label{eq:XYZ1}
\end{equation}
where the $\sigma_{n}^{\alpha}$ ($\alpha=x,y,z$) are the Pauli matrices acting on the site $n$, the sum ranges over all sites $n$ of the chain and the constants $J_{x}$, $J_{y}$ and $J_{z}$ take into account the degree of anisotropy of the model.

\begin{figure}[t]
\begin{center}
	\includegraphics[width=17cm]{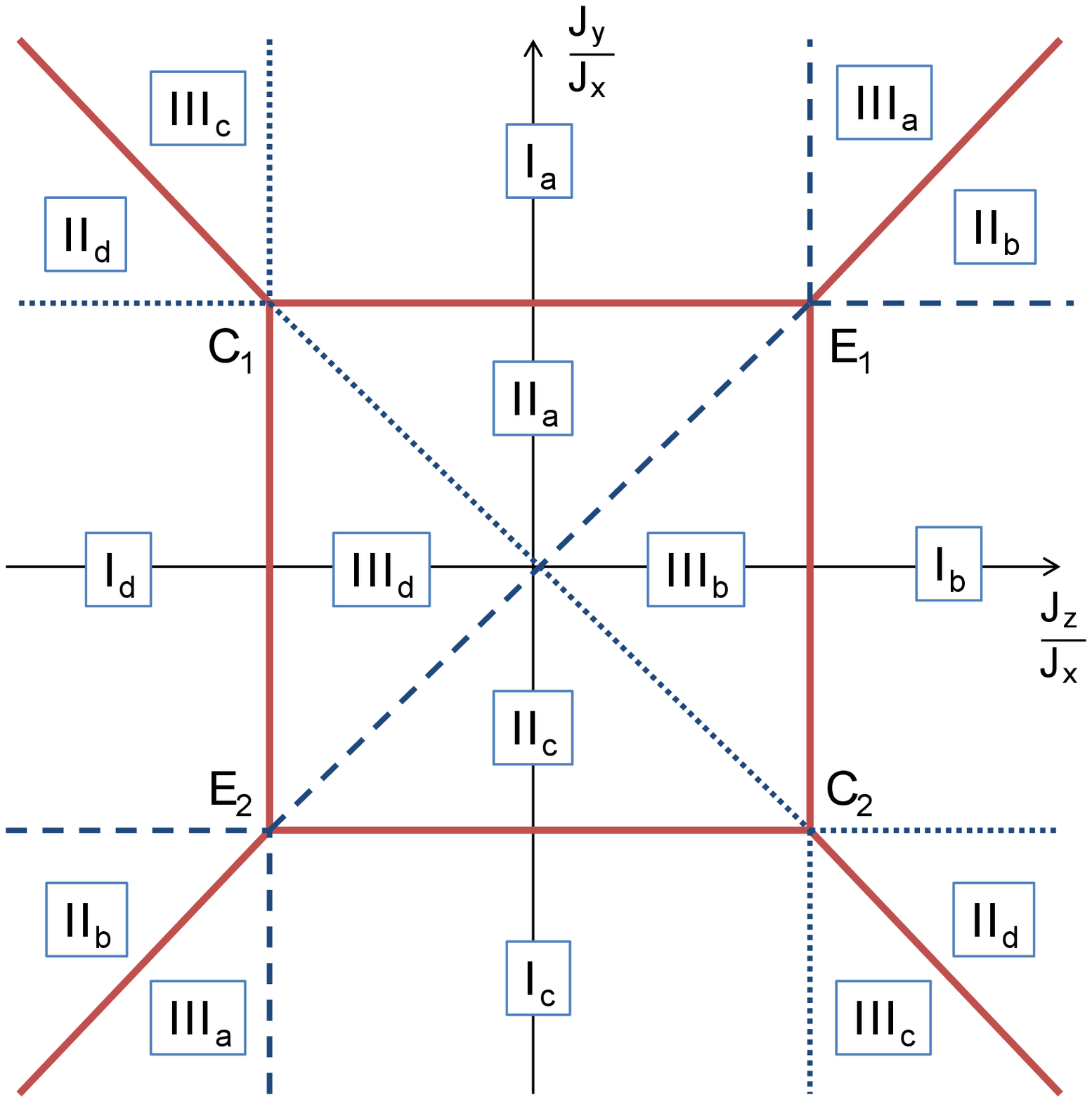}
   \caption{Phase Diagram of the XYZ model in the $\left( {J_y \over J_x} , {J_z \over J_x} \right)$ plane.}
\end{center}	
   \label{fig:phasediagram}
\end{figure}

In figure \ref{fig:phasediagram} we draw a cartoon of the phase diagram of the XYZ model in the $\left( {J_y \over J_x}, {J_z \over J_x} \right)$ plane. We divided it into 12 regions, whose role will become clear as we proceed, named ${\rm I}_{a,b,c}$, ${\rm II}_{a,b,c}$, ${\rm III}_{a,b,c}$ . Bold (red-online) continuous lines indicate the gapless phases. To understand better the model, let us first look at the $J_y=J_x$ line. Here we have the familiar XXZ model and we recognize the critical (paramagnetic) phase for 
$\left| {J_z \over J_x} \right|<1$ and the (anti-)ferromagnetic Ising phases for $\left| {J_z \over J_x} \right| >1$.
The same physics can be observed for $J_y= -J_x$. There, one can rotate every other spin by 180 degrees around the x-axis to recover a traditional XXZ model. Note, however, that $J_z$ changes sign under this transformation and therefore the ferromagnetic and anti-ferromagnetic Ising phases are reversed.
If we look at the lines $J_z = \pm J_x$, we are observing a XYX model, i.e. a rotated XXZ model. Thus the phases are the same as before.
Finally, along the diagonals $J_y = \pm J_z$ we have a XYY model of the form
\be
   \hat{H}_{XYY} =
   -{\displaystyle \sum_{n}} \left[ J_x \sigma_{n}^{x} \sigma_{n+1}^{x}
   + J_{y} \left( \sigma_{n}^{y} \sigma_{n+1}^{y} \pm \sigma_{n}^{z} \sigma_{n+1}^{z} \right) \right] \; .
   \label{HXYY}
\ee
Thus the paramagnetic phase is for $\left| {J_y \over J_x } \right| > 1 $ and the Ising phases for $\left| {J_y \over J_x} \right| <1$, with the plus sign ($J_y = J_z$) for Ising ferromagnet and the minus ($J_y = -J_z$) for the anti-ferromagnet.

In this phase diagram we find four {\it tri-critical points} at $\left( {J_y \over J_x} , {J_z \over J_x} \right) = (\pm 1, \pm 1)$.
As each paramagnetic phase can be described as a sine-Gordon theory with $\beta^{2} = 8 \pi$ at the anti-ferromagnetic isotropic point, flowing to a $\beta^{2} = 0$ theory at the ferromagnetic Heisenberg point, the critical nature of these tri-critical points is quite different. Assuming $J_x >0$, at $C_{1}= \left( {J_y \over J_x} , {J_z \over J_x} \right) = (1, - 1)$ and $C_{2} = \left( {J_y \over J_x} , {J_z \over J_x} \right) = (- 1, 1)$ we have two conformal points dividing three equivalent (rotated) sine-Gordon theories with $\beta^{2} = 8 \pi$. At $E_{1} = \left( {J_y \over J_x} , {J_z \over J_x} \right) = (1, 1)$ and $E_{2} = \left( {J_y \over J_x} , {J_z \over J_x} \right) = (- 1, - 1)$, we have two $\beta^{2} = 0$ points which are no longer conformal, since the low energy excitations have a quadratic spectrum there. The former points correspond to an antiferromagnetic Heisenberg chain at the BKT transition, whilst the latter correspond to an Heisenberg ferromagnet at its first order phase transition. The different nature of the tri-critical points has been highlighted from an entanglement entropy point of view in (\cite{Ercolessi2011}) .

We have seen that the model is invariant under a ${\pi \over 2}$ rotation ${\bf R}_\alpha = \prod_n \eu^{\ii {\pi \over 4} \sigma_n^\alpha}$ of every spin around one of the axes $\alpha=x,y,z$, followed by the interchange of the couplings $ 1/2 (\epsilon^{\alpha \beta \gamma})^{2} \tilde{\bf R}_{\beta,\gamma}$ on the perpendicular plane (for instance, $\tilde{\bf R}_{y,z} \,J_y = J_z$, $\tilde{\bf R}_{y,z} \, J_z = J_y$). Note that the composition of two rotations, say ${\bf R}_y \cdot {\bf R}_z$ is compensated by the exchange of the coupling in {\bf opposite} order: $\tilde{\bf R}_{x,y} \cdot \tilde{\bf R}_{x,z}$. Similarly, the inversion of every other spin in a plane ${\bf P}_\alpha = \prod_n \sigma_{2 n}^\alpha$ can be compensated by changing the signs of the couplings on that plane $1/2 (\epsilon^{\alpha \beta \gamma})^{2} \tilde{\bf P}_{\beta \gamma}$ (where, $\tilde{\bf P}_{y,z} \,J_y = -J_y$, $\tilde{\bf P}_{y,z} \, J_z = -J_z$). Three operations are sufficient to generate all these symmetries, for instance ${\bf R}_x$, ${\bf R}_y$ and ${\bf P}_x$ (and $\tilde{\bf R}_{y,z}$, $\tilde{\bf R}_{x,z}$ and $\tilde{\bf P}_{y,z}$).

These generators relate the different regions of the phase diagram: ${\bf R}_x {\rm I}_a = {\rm I}_b$, ${\bf R}_y {\rm I}_a = {\rm III}_a$, ${\bf P}_x {\rm I}_a = {\rm I}_c$ and so on. Note that the action of the $\tilde{\bf R}_{\alpha, \beta}$ is to exchange the coupling, and we have six different orderings of the three coupling constants. The action of $\tilde{\bf P}_{y,z}$ additionally doubles the possible phases given by a set of three couplings, giving the total of twelve regions of figure \ref{fig:phasediagram}.
We will show that these operators are equivalent to an extension of the modular group, acting in parameter space.

\section{Baxter's solution}

The Hamiltonian (\ref{eq:XYZ1}) commutes with the transfer matrices of the zero-field eight vertex model (\cite{Baxter}) and this means that they can be diagonalized together and they share the same eigenvectors. 
Indeed, as shown by Sutherland (\cite{Sutherland}), when the coupling constants of the XYZ model are related to the parameters $\Gamma$ and $\Delta$ of the eight vertex model\footnote{We adopt the conventions of \cite{Baxter} on the relation among $\Gamma,\Delta$ and the Boltzmann weights of the XYZ chain.} at zero external field by the relations
\begin{equation}
   J_{x}:J_{y}:J_{z}=1:\Gamma:\Delta \; ,
   \label{eq:XYZ4bis}
\end{equation}
the row-to-row transfer matrix ${\bf T}$ of the latter model commutes with the Hamiltonian $\hat{H}$ of the former.
In the principal regime of the eight vertex model it is customary (\cite{Baxter}) to parametrize the constants $\Gamma$ and $\Delta$ in terms of elliptic functions
\begin{equation}
   \Gamma = \frac{1+k\;\mbox{sn}^{2}(\ii \lambda; k)}{1-k\;\mbox{sn}^{2}(\ii \lambda;k)} \; ,
   \qquad\qquad
   \Delta = -\frac{\mbox{cn}(\ii \lambda;k)\;\mbox{dn}(\ii \lambda;k)}{1-k\;\mbox{sn}^{2}(\ii \lambda;k)} \; ,
   \label{eq:XYZ3bis}
\end{equation}
where $\mbox{sn}(z;k)$, $\mbox{cn}(z;k)$ and $\mbox{dn}(z;k)$ are Jacoby elliptic functions of parameter $k$, while $\lambda$ and $k$ are the argument and the parameter (respectively) whose natural regimes are 
\begin{equation}
   0 \le k \le 1\; , \qquad \qquad 0 \le \lambda \le K(k') \; ,
   \label{eq:XYZ4}
\end{equation}
$K(k')$ being the complete elliptic integral of the first kind (\ref{Kdef}) of argument $k'=\sqrt{1-k^{2}}$.

In light of (\ref{eq:XYZ4bis}), without loss of generality, we rescale the energy and set $J_{x}=1$. 
We recall that the parametrization (\ref{eq:XYZ3bis}) of the parameters is particularly suitable to describe an anti-ferroelectric phase of the eight vertex model, corresponding to $\Delta \le -1$, $|\Gamma| \le 1$. However, using the symmetries of the model and the freedom under the rearrangement of parameters, it can be used in all cases by re-defining the relations that hold between $\Delta$ and $\Gamma$ and the Boltzmann weights (for more details see (\cite{Baxter})). We can apply the same procedure to extract the physical parameters of the XYZ model in the whole of the phase diagram in a fairly compact way. In fact, when $\left|J_{y}\right|<1$ and $\left|J_{z}\right|<1$ we have:
\begin{equation}
   \label{st11}
   \left\{
   \begin{array}{r c l}
      \Gamma & = &  {\displaystyle\frac{ \left| J_z - J_y \right| - \left| J_z + J_y \right| }  {\left| J_z - J_y \right| + \left| J_z + J_y \right|}} \; , \\
      \Delta & = &    {\displaystyle-\frac{2}{\left| J_z - J_y \right| + \left| J_z + J_y \right|}} \, ,
   \end{array} \right.
\end{equation}
while for $\left|J_{y}\right|>1$ or $\left|J_{z}\right|>1$, we get:
\begin{equation}
   \label{st12}
   \left\{
   \begin{array}{r c l}
      \Gamma & = &  {\displaystyle\frac
      {\min\left[1,\left|\frac{\left|J_z - J_y\right| - \left|J_z + J_y\right|}{2}\right|\right]}
      {\max\left[1,\left|\frac{\left|J_z - J_y\right| - \left|J_z + J_y\right|}{2}\right|\right]}
      \cdot \mbox{sgn} \Big( \left|J_z -J_y\right| - \left|J_z + J_y\right| \Big)} \; ,\\
      \, \\
      \Delta & = &
      {\displaystyle-\frac{1}{2}\frac{\left|J_z - J_y\right| + \left|J_z + J_y\right|}
      {\max\left[1,\left|\frac{\left|J_z - J_y \right| - \left|J_z + J_ y \right|}{2}\right|\right]}} \; ,
   \end{array} \right.
\end{equation}
where $\Gamma$ and $\Delta$ are given by (\ref{eq:XYZ3bis}).
For instance, for $|J_y| \le 1$ and $J_z \le-1$ we recover $J_y = \Gamma$ and $J_z = \Delta$, while for $J_y \ge 1$ and $|J_z| \le 1$ we have $J_y = - \Delta$ and $J_z = -\Gamma$. In figure \ref{fig:phasediagram} we also divided the phase diagram in the different regions where a given parametrization applies.

While Baxter's procedure allows to access the full phase-diagram of the model, it introduces artificial discontinuities at the boundaries between the regions, as indicated by the {\it max} and {\it min} functions and the associated absolute values. In section (\ref{sect:newparameters}) we will show that, with a suitable analytical continuation of the elliptic parameters, we can extend the parametrization of a given region to the whole phase diagram in a continuous way, which numerically coincides with Baxter's prescription.

\section{Von Neumann and Renyi entropies}

We are now interested in studying the entanglement properties of the XYZ model in the thermodynamic limit. To this extend, we start with its ground state $\mid0\,\rangle\in\mathcal{H}$ and divide the system in two parts, which we take as the semi-infinite left and right chains, such that $\mathcal{H}=\mathcal{H}_{R}\otimes\mathcal{H}_{L}$ (here we prefer to use the notation $R$ and $L$ instead of $A$ and $B$ for the two subsystems). Then we introduce the density matrix of the whole system $\rho\equiv \mid0\,\rangle\langle\,0\mid$, and trace out one of the two half-chain:
\begin{equation}
   \rho_{R}={\rm Tr}_{\mathcal{H}_{L}}(\rho).
   \label{added}
\end{equation}

According to (\cite{Nishino1}, \cite{Nishino2} and \cite{Peschel1}), in the thermodynamic limit we can calculate the reduced
density matrix $\rho_{R}$ as product of the four Corner Transfer Matrices $A=C$
and $B=D$ of the eight vertex model at zero field:
\begin{equation}
   \rho_{R}(\bar{\sigma},\bar{\sigma}')
   = (ABCD)_{\bar{\sigma},\bar{\sigma}'}
   = (AB)_{\bar{\sigma},\bar{\sigma}'}^{2} \; ,
   \label{eq:XYZ4a}
\end{equation}
where $\bar{\sigma}$ and $\bar{\sigma}'$ are particular spin configurations of the semi-infinite chain.\\
Generally speaking, the above quantity does not satisfy the constraint ${\rm Tr}\rho_{R}=1$, so we must consider a normalized version $\rho_{R}'$.
Notice taking the trace of the reduced density matrix over the remaining half-line we get the normalization of $\rho_{R}$, which corresponds to the partition function of the eight vertex model:
\be
   \mathcal{Z} \equiv {\rm Tr}\rho_{R}.
   \label{Zdef}
\ee

For the zero field eight vertex model with fixed boundary conditions, Baxter constructed an explicit form of the Corner Transfer Matrices (CTM's) in the thermodynamic limit (\cite{Baxter}, \cite{Baxter1} and \cite{Baxter2}). In (\cite{Evangelisti}) it was shown that, using relation (\ref{eq:XYZ4a}), the reduced density operator can be written as
\begin{equation}
   \rho_{R}=\left(\begin{array}{cc}
        1 & 0\\
        0 & x^{2}\end{array}\right)
   \otimes\left(\begin{array}{cc}
        1 & 0\\
        0 & x^{4}\end{array}\right)
   \otimes\left(\begin{array}{cc}
        1 & 0\\
        0 & x^{6}\end{array}\right)
   \otimes \ldots \; ,
   \label{eq:XYZ6}
\end{equation}
where $x \equiv \exp[-\pi\lambda/2I(k)]$.
We notice that $\rho_{R}$ is a function of $\lambda$ and $k$ only through $x$.
Furthermore, (\ref{eq:XYZ6}) can be rewritten as
\begin{equation}
   \rho_{R}=(AB)^{2}=\eu^{-\epsilon\hat{O}} \; .
   \label{eq:XYZ7}
\end{equation}
$\hat{O}$ is an operator with integer eigenvalues ( which was defined before equation (\ref{chap1_eq24})) and also:
\begin{equation}
   \epsilon \equiv \pi \: {\lambda \over 2I(k)} \; .
   \label{eq:eps}
\end{equation}
The Von Neumann entropy ${\cal S}$ can be calculated easily according to (\ref{chap1_eq24}):
\begin{equation}
   {\cal S}
   = -\mbox{Tr}\rho_{R}'\ln\rho_{R}'
   = -\epsilon{\displaystyle \frac{\partial\ln\mathcal{Z}}{\partial\epsilon}}+\ln\mathcal{Z} \; ,
   \label{eq:XYZ8}
\end{equation}
where the partition function can be written as
\begin{equation}
   \mathcal{Z} = {\displaystyle \prod_{j=1}^{\infty}}(1+x^{2j})
   = {\displaystyle \prod_{j=1}^{\infty}}(1+\eu^{-\pi j \lambda/I(k)}) \; .
   \label{eq:XYZ9}
\end{equation}
Thus we obtain an exact analytic expression for the entanglement entropy of the XYZ model as
\begin{equation}
   {\cal S} = 2\epsilon{\displaystyle \sum_{j=1}^{\infty}}{\displaystyle \frac{j}{(1+\eu^{2j\epsilon})}}
   + {\displaystyle \sum_{j=1}^{\infty}}\ln(1+\eu^{-2j\epsilon}) \; ,
   \label{eq:XYZ10}
\end{equation}
which is valid for generic values of $\lambda$ and $k$. 
Another quantity of interest to quantify the entanglement is the Renyi entropy
\begin{equation}
   {\cal S}_{\alpha} \equiv {\displaystyle\frac{1}{1-\alpha}}\ln\mbox{Tr}[\rho_{R}^{\prime \alpha}] \; ,
   \label{eq:renyi1}
\end{equation}
where $\alpha$ is a parameter. 

The Renyi entropy can be seen as a generalized series expansion for the von Neumann entropy and it is directly related to the spectrum of the reduced density matrix, see, for instance, (\cite{Franchini1}). As we have already seen, in the limit $\alpha \to 1$ it reduces to the von Neumann entropy.

Using the factorized expression (\ref{eq:XYZ6}) for $\rho$, we can be easily write
\begin{equation}
   \rho_{R}^\alpha=\left(\begin{array}{cc}
        1 & 0\\
        0 & x^{2\alpha}\end{array}\right)
   \otimes\left(\begin{array}{cc}
        1 & 0\\
        0 & x^{4\alpha}\end{array}\right)
   \otimes\left(\begin{array}{cc}
        1 & 0\\
        0 & x^{6\alpha}\end{array}\right)
   \otimes \ldots \; .
   \label{eq:renyi2}
\end{equation}
With this result and noting that
\begin{equation}
   \label{eq:renyi3}
   \mbox{Tr}\rho_{R}^{\prime \alpha} =
   \mbox{Tr}\left ({\displaystyle\frac{\rho_{R}}{\mathcal{Z}}}\right)^{\alpha} =   {\displaystyle\frac{\mbox{Tr}\rho_{R}^{\alpha}}{\mathcal{Z}^{\alpha}} } =  {\displaystyle\frac{\prod_{j=1}^{\infty}\left(1+\eu^{-2j\alpha\epsilon} \right)}{\mathcal{Z}^{\alpha}} } \; ,
\end{equation}
it is now trivial to compute the Renyi entropy of the XYZ model:
\begin{eqnarray}
   {\cal S}_{\alpha}  & = &
   {\displaystyle\frac{1}{1-\alpha}} \left [ \ln {\displaystyle\frac{1}{\mathcal{Z}^{\alpha}}} + \sum_{j=1}^{\infty}\ln \left ( 1+\eu^{-2j\alpha\epsilon}  \right) \right]
   \nonumber \\
   & = & {\displaystyle\frac{\alpha}{\alpha-1}} \sum_{j=1}^{\infty}\ln\left ( 1+\eu^{-2j\epsilon} \right)
   + {\displaystyle\frac{1}{1-\alpha}}\sum_{j=1}^{\infty}\ln\left ( 1+\eu^{-2j\alpha\epsilon} \right) \; .
   \label{eq:renyi4bis}
\end{eqnarray}
Moreover, by introducting the generalized partition function:
\be
   \mathcal{Z}_{\alpha} 
   = {\displaystyle \prod_{j=1}^{\infty}}(1+\eu^{-2j\alpha\epsilon})\;.
   \label{eq:XYZ10bis}
\ee
we can re-write the Von Neumann entropy as:
\be
   {\cal S}=\left.\left(1-\frac{\partial}{\partial\alpha}\right) \ln\mathcal{Z}_{\alpha}\right|_{\alpha = 1}
   \label{eq:XYZ10tris}
\ee
and the Renyi entropy as:
\be
   {\cal S}_{\alpha}=\lim_{\beta\to 1}{\displaystyle\frac{\beta\ln\mathcal{Z}_{\alpha}-\alpha\ln\mathcal{Z}_{\beta}}{\beta-\alpha}}\;.
   \label{eq:XYZ10quatris}
\ee
These two last expressions are completely equivalent to (\ref{eq:XYZ10}) and (\ref{eq:renyi4}) respectively.
When $\epsilon\ll1$, i.e. in the scaling limit analogous to the one of (\cite{Weston}), formula (\ref{eq:XYZ10})
can be approximated by its Euler-Maclaurin asymptotic expansion which was introduced in (\ref{chap1_eq30}), yielding
\begin{eqnarray}
{\cal S} & = & {\displaystyle \int_{0}^{\infty}}\;{\rm d}x\;\left({\displaystyle \frac{x\epsilon}{1+e^{x\epsilon}}}+\ln(1+e^{-x\epsilon})\right)-{\displaystyle \frac{\ln2}{2}}+O(\epsilon)\nonumber \\
\, & = & {\displaystyle \frac{\pi^{2}}{6}}{\displaystyle \frac{1}{\epsilon}}-{\displaystyle \frac{\ln2}{2}}+O(\epsilon)
\label{eq:XYZ11}
\end{eqnarray}
This will be used in the next subsection where we will check our analytic results against some special known cases, namely the XXZ and
the XY chain.

\subsection{The XXZ and XY spin-$1/2$ chain limits}

Let us first consider the case $k=0$ (i.e. $\Gamma=1$) and the limit
$\lambda\to 0^{+}$, which corresponds to the spin $1/2$ XXZ. In this
limit the eight vertex model reduces to the six vertex model. Let
us note that formula (\ref{eq:XYZ11}) coincides exactly with the
one proposed by Weston (\cite{Weston}) which was obtained in the study
of more general spin $\kappa/2$. In this limit the approximation
$\epsilon\ll1$ is still valid and we can use the result of equation
(\ref{eq:XYZ11}). From relation (\ref{eq:XYZ3bis}), it follows that
\begin{equation}
\lambda={\displaystyle\sqrt{2}}\sqrt{-\Delta-1}+O\left((-1-\Delta)^{3/2}\right)\label{eq:XYZ12}
\end{equation}
 Thus equation (\ref{eq:XYZ11}) gives\begin{equation}
{\cal S}=\frac{\pi^{2}}{12\sqrt{2}}\,\frac{1}{\sqrt{-\Delta-1}}-\frac{\ln(2)}{2}+O\left((-1-\Delta)^{1/2}\right)\label{eq:XYZ13}\end{equation}
 which can be written in a simpler form if we recall that, when $\epsilon\to0$
(i.e. $\Delta\rightarrow-1^{-}$), the correlation length is given
by (\cite{Baxter}) 
\begin{equation}
\ln\frac{\xi}{a}={\displaystyle {\frac{\pi^{2}}{2\epsilon}}-2\ln(2)+O(e^{-\pi^{2}/\epsilon})}
\end{equation}
 where $a$ is the lattice spacing. Recalling that \begin{equation}
\epsilon={\displaystyle \sqrt{2}}\sqrt{-\Delta-1}+O\left((-1-\Delta)^{3/2}\right)\end{equation}
 the expression for the entanglement entropy becomes \begin{equation}
{\cal S}=\frac{1}{6}\ln\frac{\xi}{a}+U+O\left((-1-\Delta)^{1/2}\right)\end{equation}
 where $U=-\ln(2)/6$. This last expression confirms the expected
${\cal S}\sim c\ln(\xi)+U$ with $c=1$, which is exactly what
one should expect, the XXZ model along its critical line being a free
massless bosonic field theory with $c=1$.

\begin{figure}[t]
\begin{center}
\includegraphics[width=8cm]{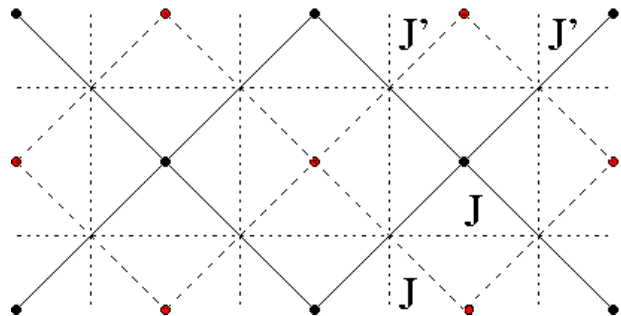} 
\par\end{center}
\caption{Decoupled anisotropic Ising lattices. Horizontal and vertical lines
belong to the original eight vertex model lattice, diagonal lines
belong to the dual Ising lattice.}
\label{fig:ising} 
\end{figure}

As a second check, we consider the case $\Gamma=0$, which corresponds
to the XY chain. It is convenient now to describe the corresponding
eight vertex model by using Ising-like variables which are located
on the dual lattice (\cite{Baxter}), thus obtaining an anisotropic
Ising lattice, rotated by $\pi/4$ with respect to the original one,
with interactions round the face with coupling constants $J,J'$,
as shown in figure \ref{fig:ising}. %

In our case the Ising lattice decouples into two single sublattices
with interactions among nearest neighbors. Now \begin{equation}
\Delta=\sinh(2\beta J)\sinh(2\beta J')\equiv k_{I}^{-1}\label{eq:XYZ17}\end{equation}
 so that, using the elliptic parametrization, one has\begin{equation}
\lambda=\frac{1}{2}I(k')\label{eq:XYZ18}\end{equation}
 Thus $\epsilon$ of equation (\ref{eq:eps}) becomes \begin{equation}
\epsilon=\frac{\pi I(k_{I}')}{2I(k_{I})}\label{eq:XYZ19}\end{equation}
 Let us approach the critical line of the anisotropic Ising model
from the ferromagnetic phase, i.e. let us assume that $k_{I}\to1^{-}$.
In this case it is straightforward to write \begin{equation}
\epsilon=-{\displaystyle \frac{\pi^{2}}{2\ln(1-k_{I})}}+O\left(\ln^{-2}(1-k_{I})\right)\label{eq:XYZ20}\end{equation}
 so that the entanglement entropy becomes\begin{equation}
{\cal S}=-\frac{1}{6}\ln(1-k_{I})+O\left(\ln^{-1}(1-k_{I})\right)\label{eq:XYZ21}\end{equation}
 Since $\xi^{-1}=(1-k_{I})+O\left((1-k_{I})^{2}\right)$, we can easily
conclude that\begin{equation}
{\cal S}=\frac{1}{6}\ln\frac{\xi}{a}+O\left(\ln^{-1}(1-k_{I})\right)\label{eq:XYZ22}\end{equation}
 where again the leading term confirms the general expected result
with $c=1$. This result is in agreement with what found in previous
works (\cite{Vidal}, \cite{Calabrese2}, \cite{Peschel3}, \cite{Korepin}, \cite{Korepin2}) by means of different approaches.

\subsection{The sine-Gordon limit}

In (\cite{Luther}, \cite{McCoy}) it has been proposed that a particular scaling
limit of the XYZ model yields the sine-Gordon theory. In this section
we will use this connection to compute the exact entanglement entropy between
two semi-infinite intervals of a 1+1 dimensional sine-Gordon model.
In his article, Luther (\cite{Luther}) showed that in the scaling
limit, where $a\to0$ while keeping the mass gap constant, the parameters
of the XYZ model and those of the sine-Gordon theory are connected
by the following relation (keeping $J_{x}=1$ from the beginning)\begin{equation}
M=8\pi\,\left({\displaystyle \frac{\sin\mu}{\mu}}\right)\,\left({\displaystyle \frac{l_{r}}{4}}\right)^{\pi/\mu}\label{eq:SG1}\end{equation}
 where the parameter $\mu$ is defined as\begin{equation}
\mu\equiv\pi\left(1-{\displaystyle \frac{\beta^{2}}{8\pi}}\right)=\arccos\left({\displaystyle -J_{z}}\right)\label{eq:SG2}\end{equation}
 Here $M$ is the sine-Gordon solitonic mass, and $l_{r}=l\, a^{-\mu/\pi}$,
with\begin{equation}
l^{2}={\displaystyle \frac{1-J_{y}^{2}}{1-J_{z}^{2}}}\label{eq:SG3}\end{equation}
 These relations tell us how the coupling constant $J_{z}$ is connected
to the parameter $\beta$ of sine-Gordon, and how $J_{y}$ scales
when we take the scaling limit $a\to0$. It is clear from equation
(\ref{eq:SG3}) that in this limit $J_{y}\to1^{-}$. In the following
we work in the repulsive regime $4\pi<\beta^{2}<8\pi$ (which corresponds
to $0<\mu<\pi/2$ and $-1<J_{z}<0$). In this regime the mass gap
of the theory is the soliton mass $M$. Taking this limit we use the
following parametrization of the XYZ coupling constants\begin{equation}
\Gamma={\displaystyle \frac{J_{z}}{J_{y}}}\qquad,\qquad\Delta={\displaystyle \frac{1}{J_{y}}}\label{eq:SG4}\end{equation}
 which amounts to a reparametrization of the Boltzmann weights of
XYZ suitable for the $|\Delta|\leq1$ disordered regime where we are
working now (see chapter 10 of (\cite{Baxter}) for details). As a consequence
of such reparametrization a minus sign appears in front of both equations
(\ref{eq:XYZ3bis}). Taking the sine-Gordon limit, $\lambda$ and
$k$ parametrizing $\Gamma$ and $\Delta$ must now satisfy the following
constraint\begin{equation}
\mbox{sn}^{2}(i\lambda)=-{\displaystyle \frac{\frac{J_{z}}{J_{y}}+1}{k-k\frac{J_{z}}{J_{y}}}}\label{eq:SG5}\end{equation}
 Considering the parametrization of $\Delta$ and using the properties
of the Jacobian elliptic functions we can write\begin{equation}
\Delta^{2}=\frac{\mbox{cn}^{2}(i\lambda)\mbox{dn}^{2}(i\lambda)}{(1-k\;\mbox{sn}^{2}(i\lambda))^{2}}={\displaystyle \frac{\left(k(1-\frac{J_{z}}{J_{y}})+\frac{J_{z}}{J_{y}}+1\right ) \left(k(1+\frac{J_{z}}{J_{y}})-\frac{J_{z}}{J_{y}}+1\right)}{4k}}\label{eq:SG6}\end{equation}
 Expanding around $k\to1^{-}$ and collecting $\Delta^{2}=1/J_{y}^{2}$
from both sides of the equation we find\begin{equation}
\Delta^{2}=1+\frac{1}{4}(1-J_{z}^{2})(k-1)^{2}+O(k-1)^{3}\label{eq:SG7}\end{equation}
 Using equation (\ref{eq:SG1}) we obtain\begin{equation}
l^{2}=l_{r}^{2}a^{2\mu/\pi}=4^{2-3\mu/\pi}\left({\displaystyle \frac{M\mu a}{\pi\sin\mu}}\right)^{2\mu/\pi}\label{eq:SG8}\end{equation}
 where $\mu$ is completely fixed by choosing a particular value of
$J_{z}$. Now using the definition (\ref{eq:SG3}) and (\ref{eq:SG4})
we find\begin{equation}
\Delta^{2}=1+(1-J_{z}^{2})4^{2-3\mu/\pi}\left({\displaystyle \frac{M\mu a}{\pi\sin\mu}}\right)^{2\mu/\pi}+O(a^{4\mu/\pi})\label{eq:SG9}\end{equation}
 which is valid when $a\to0$. Comparing equation (\ref{eq:SG7})
with (\ref{eq:SG9}) we can identify in which way $k$ scales to $1^{-}$\begin{equation}
k=1-2^{3(1-\mu/\pi)}\left({\displaystyle \frac{M\mu a}{\pi\sin\mu}}\right)^{\mu/\pi}+O(a^{2\mu/\pi})\label{eq:SG10}\end{equation}
 Remembering the constraint (\ref{eq:SG5}) and using the previous
expression for $k$ we have\begin{equation}
\mbox{sn}^{2}(i\lambda)={\displaystyle \frac{-J_{z}-1}{1-J_{z}}+O(a^{\mu/\pi})}\label{eq:SG11}\end{equation}
 When $k\to1$ the elliptic function sn reduces to an hyperbolic tangent,
thus we obtain\begin{equation}
\tan^{2}\lambda={\displaystyle \frac{1+J_{z}}{1-J_{z}}+O(a^{\mu/\pi})\quad\longrightarrow\quad\lambda=\arctan\sqrt{{\displaystyle \frac{1+J_{z}}{1-J_{z}}}}+O(a^{\mu/\pi})}\label{eq:SG12}\end{equation}
 Now we can evaluate the expression (\ref{eq:XYZ10}) in this limit.
Using the following asymptotic behaviour of the elliptic integral
$I(x)$\begin{equation}
I(x)\approx-\frac{1}{2}\ln(1-x)+\frac{3}{2}\ln2+O(1-x),\qquad x\approx1^{-}\label{eq:SG13}\end{equation}
 along with the approximation (\ref{eq:XYZ11}), we can write the
exact entanglement entropy of a bipartite XYZ model in the sine-Gordon
limit\begin{equation}
{\cal S}_{sG}=-{\displaystyle \frac{\pi}{12}\frac{\ln(1-k)-3\ln2}{\arctan\sqrt{{\displaystyle \frac{1+J_{z}}{1-J_{z}}}}}-{\displaystyle \frac{\ln2}{2}}+O(1/\ln(a))}\label{eq:SG14}\end{equation}
 The leading correction to this expression comes from the $O(\epsilon)$
term of equation (\ref{eq:XYZ11}). The constant $J_{z}$ is connected
to $\beta$ by\begin{equation}
J_{z}=-\cos\pi\left(1-\frac{\beta^{2}}{8\pi}\right)\label{eq:SG15}\end{equation}
 thus using this property and the scaling expression (\ref{eq:SG11})
we can write down the entanglement entropy as\begin{equation}
{\cal S}_{sG}={\displaystyle \frac{1}{6}\ln\left(\frac{1}{Ma}\right)+\frac{1}{6}\ln\left(\frac{\sin\left[\pi\left(1-\frac{\beta^{2}}{8\pi}\right)\right]}{\left(1-\frac{\beta^{2}}{8\pi}\right)}\right)+O(1/\ln(a))}\label{eq:SG16}\end{equation}
 This result confirms the general theory due to (\cite{Calabrese2}, \cite{Doyon}),
in the limit where the system is bipartite in two infinite intervals,
with the central charge equal to 1, as it should, because the sine-Gordon
model can be considered a perturbation of a $c=1$ conformal field
theory (described by a free massless boson compactified on a circle
of radius $\sqrt{\pi}/\beta$) by a relevant operator of (left) conformal
dimension $\beta^{2}/8\pi$. We can write \begin{equation}
{\cal S}_{sG}\approx{\displaystyle \frac{1}{6}\ln\left(\frac{1}{Ma}\right)+U(\beta)\qquad a\to0}\label{eq:SG17}\end{equation}
 where the constant term $U(\beta)$ takes the value \begin{equation}
U(\beta)={\displaystyle \frac{1}{6}\ln\left(\frac{\sin\left[\pi\left(1-\frac{\beta^{2}}{8\pi}\right)\right]}{\left(1-\frac{\beta^{2}}{8\pi}\right)}\right)}\label{eq:SG18}\end{equation}

At $\beta^{2}=4\pi$, when the sine-Gordon model becomes the free
Dirac fermion theory, it assumes the value $U(\sqrt{4\pi})=\frac{1}{6}\log2=0.11552453...$,
while at $\beta^{2}=8\pi$, where the theory becomes a relevant perturbation
of the WZW conformal model of level 1 by its operator of left dimension
$\frac{1}{4}$, it becomes $U(\sqrt{8\pi})=\frac{1}{6}\log\pi=0.19078814...$.\\
 We notice that formula (\ref{eq:SG16}) yields the exact value of
the {\it overall} constant $U(\beta)$, since it has been derived from equation
(\ref{eq:XYZ11}) which is exact up to terms $O(\epsilon)$. As
mentioned in the introduction and as observed by many authors
(\cite{Bombelli}, \cite{Callan}, \cite{Schumacher}, \cite{Nishino1}), $U$ should contain a
contribution from the Affleck Ludwig boundary term as well as a
model-dependent constant that will be discussed in the next chapter.

\section{Renyi entropy in terms of theta functions}

Expression (\ref{eq:renyi4bis}) for the entropy is exact in the thermodynamic limit and it is quite convenient for numerical evaluations. However, to better understand the behavior of the entropy we can use a more explicitly analytical expression. This is done by recognizing that (\ref{eq:renyi4bis}) is essentially a sum of two q-series in the parameter $q= \eu^{-\epsilon}$. These series can be conveniently expressed in terms of Jacobi theta functions, which can be defined by the products
\bea
   \theta_1 (z,q) & = & 2 q^{1/4} \sin z \prod_{j=1}^\infty
   \left( 1 - q^{2j} \right) \left[ 1 - 2 q^{2j} \cos (2z) + q^{4j} \right] \; ,
   \label{theta1def} \\
   \theta_2 (z,q) & = & 2 q^{1/2} \cos z \prod_{j=1}^\infty
   \left( 1 - q^{2j} \right) \left[ 1 + 2 q^{2j} \cos (2z) + q^{4j} \right] \; ,
   \label{theta2def} \\
   \theta_3 (z,q) & = & \prod_{j=1}^\infty
   \left( 1 - q^{2j} \right) \left[ 1 + 2 q^{2j-1} \cos (2z) + q^{4j-2} \right] \; ,
   \label{theta3def} \\
   \theta_4 (z,q) & = & \prod_{j=1}^\infty
   \left( 1 - q^{2j} \right) \left[ 1 - 2 q^{2j-1} \cos (2z) + q^{4j-2} \right] \; .
   \label{theta4def}
\eea

It is then easy to see that we have the identity
\be
   \prod_{j=1}^\infty \left( 1 + q^{2j} \right) =
   \left( { \theta_2^2 (0,q) \over 4 q^{1/2} \theta_3 (0,q) \theta_4 (0,q) } \right)^{1/6} \; .
\ee
With this help, we can write (\ref{eq:renyi4bis}) as
\bea
   {\cal S}_{\alpha} & = & {\displaystyle\frac{\alpha}{\alpha-1}}
   \ln \prod_{j=1}^{\infty} \left ( 1 + \eu^{-2j\epsilon}  \right)
   + {\displaystyle\frac{1}{1-\alpha}}
   \ln \prod_{j=1}^{\infty} \left ( 1 + \eu^{-2j\alpha\epsilon}  \right)
   \nonumber  \\
   & = & {\alpha \over 6(\alpha -1)}
   \ln{\theta_2^2 (0,q) \over \theta_3 (0,q) \theta_4 (0,q) }
   \nonumber \\
   & + & {1 \over 6(1 - \alpha)}
   \ln{\theta_2^2 (0,q^{\alpha}) \over \theta_3 (0,q^{\alpha}) \theta_4 (0,q^{\alpha}) } - {1 \over 3} \ln 2\; .
   \label{eq:modular2}
\eea
and the Von Neumann entropy as
\bea
   {\cal S} & = & {\displaystyle\frac{1}{6}}
   \ln {\theta_2^2 (0,q) \over \theta_3 (0,q) \theta_4 (0,q) }
   - {\displaystyle\frac{1}{6}} {\de \over \de \alpha} \left. \ln
   {\theta_2^2 (0,q^{\alpha}) \over \theta_3 (0,q^{\alpha}) \theta_4 (0,q^{\alpha}) } \right|_{\alpha =1}
   -\frac{\ln(2)}{3}
   \nonumber \\
   & = & \frac{1}{6} \ln {\theta_2^2 (0,q) \over
   4 \theta_3 (0,q) \theta_4 (0,q) }-\frac{\ln(q)}{2}\left(2\theta_{3}^{4}(0,q)-\theta_{2}^{4}(0,q)\right)
   \label{vonNeumannTheta}
\eea

These expressions for the entropies are completely equivalent to the series we started from and might even appear more complicated.
However, the advantage of expressing them in terms of elliptic function will become apparent in the section (\ref{sec:essential}), where this formulation will allow us to study their behavior.

\section{Analytical extension of Baxter's solution}
\label{sect:newparameters}
In this section we will show that, using some simple transformations for the elliptic functions, we can provide some more compact parametrizations for the XYZ model. In particular we will show how to cover the entire plane $(J_{y},J_{z})$ by means of a single parametrization formula, without using the Baxter rearrangement procedure.
To streamline the computation, it is convenient to perform a Landen transformation on (\ref{eq:XYZ3bis}) and to switch to new parameters $(u,l)$
\be
   l \equiv {2 \sqrt{k} \over 1 + k} \; , \qquad \qquad
   u \equiv (1+k) \lambda \; ,
   \label{landen}
\ee
so that\footnote{See, for instance, table 8.152 of (\cite{gradshteyn})}
\bea
   \Gamma & = & {1 + k \; \sn^2 (\ii \lambda ; k) \over 1 - k \; \sn^2 (\ii \lambda ; k) }
   = {1 \over \dn ( \ii u; l ) } \; , 
   \label{Gnd} \\
   \Delta & = & - { \cn (\ii \lambda ; k) \; \dn (\ii \lambda ; k) \over
   1 - k \; \sn^2 (\ii \lambda ; k) }
   = - { \cn (\ii u ; l) \over \dn (\ii u ; l) } \; .
   \label{Dcd} 
\eea
The natural domain of $(\lambda,k)$ corresponds to 
\be
   0 \le u \le 2 K(l')  \qquad \qquad
   0 \le l \le 1 \; .
   \label{zlnatdomain}
\ee
We also have the following identities
\be
\label{eps1}
   l' = \sqrt{1 - l^2} = { 1 - k \over 1 +k}, \quad
   K (l) = (1 +k) K (k), \quad
   K (l') = {1 + k \over 2} K (k'),
\ee
from which it follows that the Landen transformation doubles the elliptic parameter $\tau(k) = \ii {K(k') \over K(k)} = 2 \ii {K(l') \over K(l)} = 2 \tau(l)$.
Yet, we can re-write $\epsilon$ of (\ref{eq:eps}) as:
\be
\label{eps2}
   \epsilon = \pi {\lambda \over 2K(k)} = \pi {u \over 2K(l)}.
\ee
Now using (\ref{ellid3})) we have  
\begin{equation}
    \label{ste8bis}
    l = \displaystyle{\sqrt{ {1-\Gamma^{2} \over \Delta^{2}-\Gamma^{2} } } } \; .
\end{equation}
Furthermore, elliptic functions can be inverted in elliptic integrals and from (\ref{Dcd}) we have 
\begin{equation}
    \label{ste8tris}
    \ii u = \int_{-1}^{-\Gamma} { \de t \over \sqrt{ (1 - t^2)( 1 - l^{\prime 2} t^2)} }
    = \ii \bigg[ K(l') - F (\arcsin \Gamma; l') \bigg]  \; .
\end{equation}
Together, (\ref{ste8bis}, \ref{ste8tris}) invert (\ref{Gnd}, \ref{Dcd}) and, together with (\ref{st11}, \ref{st12}), give the value of Baxter's parameters for equivalent points of the phase diagram. It is important to notice that these identities ensure that the domain (\ref{zlnatdomain}) maps into $|\Gamma| \le 1$ and $\Delta \le -1$ and vice-versa.

The idea of the analytical continuation is to use (\ref{ste8bis}) to extend the domain of $l$ beyond its natural regime. To this end, let us start from a given region, let say ${\rm I}_a$ in fig. \ref{fig:phasediagram}, with $|J_z| \le 1$ and $J_y \ge 1$. By (\ref{st12}) we have:
\be
   \hskip -1.5cm 
   J_y (u,l) = - \Delta = {\cn ( \ii u ; \,l) \over \dn ( \ii u ; \,l ) } \; , 
   \qquad \qquad
   J_z (u,l) = -\Gamma = - {1 \over \dn ( \ii u ; \,l ) } \; ,
   \label{Iaparameter}
\ee
and thus
\begin{equation}
    \label{lzI}
   l = \sqrt{ {1 - J_z^2 \over J_y^2 - J_z^2 } }  \; ,
   \qquad \qquad
    \ii u= \int_{-1}^{J_z} { \de t \over \sqrt{ (1 - t^2)( 1 - l^{\prime 2} t^2)} } \; .
\end{equation}
Notice that within region ${\rm I}_a$ we can recast the second identity as
\be
   u = K (l')  + F (\arcsin J_z; l') \; ,
\ee
which makes $u$ explicitly real and $0 \le u \le 2 K(l')$. 

\begin{figure}
\centering
\includegraphics[width=8cm]{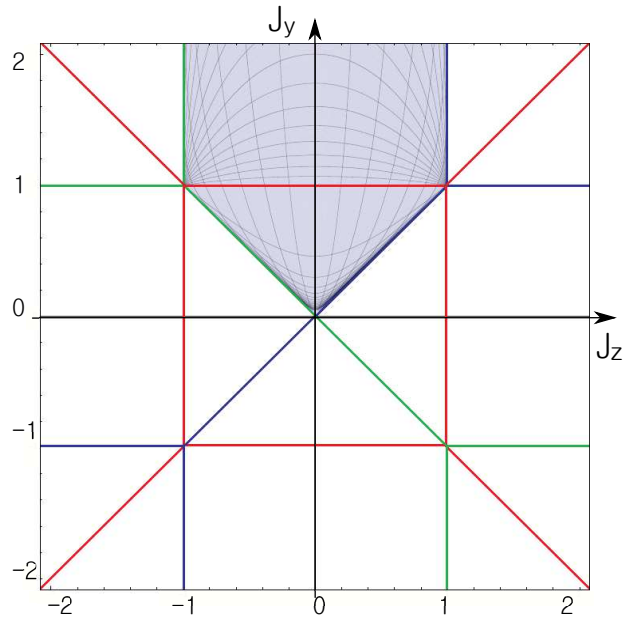}
\caption{Contour lines of the parametrization of regions ${\rm I}_a$ and ${\rm II}_a$.}
\label{fig:param1}
\end{figure}

We now take (\ref{lzI}) as the definition of $l$ and use it to extend it over the whole phase diagram.
For instance, in region ${\rm II}_a$ ($0 \le J_y \le 1$, $J_z^2 \le J_y^2$), using (\ref{lzI}) we find $l >1$. To show that this analytical continuation gives the correct results, we write $l =1 / \tilde{l}$, so that $0 \le \tilde{l} \le 1$ and use (\ref{jacobiSTS}) in (\ref{Iaparameter}), which gives
\be
    J_y (u,l) = {\dn (\ii \tilde{u} ; \,\tilde{l} ) \over \cn (\ii \tilde{u} ; \,\tilde{l}  ) }\;, 
   \qquad \qquad 
   J_z (u, l) = -{1 \over \cn ( \ii \tilde{u} ;\tilde{l}  ) } \; , 
\ee
which reproduces Baxter's definitions in (\ref{st11}):  $\Gamma = -{J_z \over J_y}$, $\Delta = - {1 \over J_y}$. Note that we also rescale the argument $\tilde{u} = l u$, so that $0 \le \tilde{u} \le 2 K(\tilde{l}')$. This shows that we can use (\ref{Iaparameter}) to cover both regions ${\rm I}_a$ and ${\rm II}_a$, by letting $0 \le l < \infty$ (note that $l=1$ corresponds to the boundary between the two regions). Figure \ref{fig:param1} shows the contour lines obtained with this parametrization.

We can proceed similarly for the rest of the phase diagram. However, the second part of (\ref{lzI}) needs some adjustment.
Since elliptic integrals are multivalued functions of their parameters, to determine the proper branch of the integrand one starts in the principal region and follows the analytical continuation. The result of this procedure can be summarized as
\bea
   \ii u & = & \ii \bigg[ K(l') + F (\arcsin J_z; l') \bigg] 
   + \bigg[ 1 - \sgn(J_y) \bigg]  K (l) 
   \nonumber \\
   & = & F (\arcsin J_y; l) - K(l) + \ii \bigg[ 1 + \sgn(J_z) \bigg]  K (l')  \; ,
  \label{zparam}
\eea
where all the elliptic integrals here are taken at their principal value. We did not find a single parametrization without discontinuities in the whole phase diagram. Close to $J_z \simeq 0$, the first line of (\ref{zparam}) is continuous, while the second has a jump. The opposite happens for $J_y \simeq 0$, but, due to the periodicity properties of (\ref{Gnd}, \ref{Dcd}), both expressions are proper inversions of (\ref{Iaparameter}) valid everywhere. 

\begin{figure}
   \centering
	\includegraphics[width=10cm]{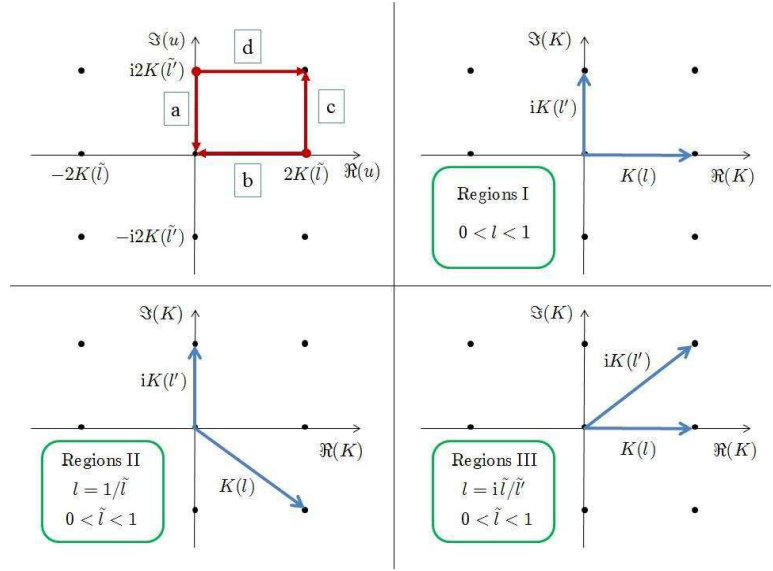}
   \caption{On the top left, in regions of type $a$, $b$, $c$, and $d$, the argument of the elliptic functions runs along one of the sides of the rectangle. In the remaining quadrants, the directions of the quater-periods $K(l)$ and $\ii K(l')$ in ${\rm I}$, ${\rm II}$, and ${\rm III}$ regions.}
   \label{fig:paths}
\end{figure}

Thus, we accomplished to invert (\ref{Iaparameter}) and to assign a pair of $(u,l)$ to each point of the phase diagram $(J_y, J_z)$, modulo the periodicity in $u$ space. The analysis of these mappings shows an interesting structure. From (\ref{lzI}) it follows that that regions ${\rm I}$'s have $0 \le l \le 1$, while ${\rm II}$'s have $l \ge 1$, which can be written as $l = 1 / \tilde{l}$, with $0 \le \tilde{l} \le 1$. Finally, regions ${\rm III}$'s have purely imaginary $l = \ii \, \tilde{l}/\tilde{l'}$, with $0\le \tilde{l} \le 1$. In each region, the argument runs along one of the sides of a rectangle in the complex plane of sides $2 K(\tilde{l})$ and $\ii 2 K(\tilde{l}')$. Thus, we can write
\be
   J_y (z,l) =  {\cn ( \zeta \: u(z) ; \,l) \over \dn ( \zeta \: u(z) ; \,l ) }, \quad 
   J_z (u,l) =  {1 \over \dn ( \zeta \: u (z) ; \,l ) },
   \quad  0 \le z \le \pi,
   \label{zlparameter}
\ee
where $\zeta =1, \tilde{l}, \tilde{l}'$ in regions ${\rm I}$'s,  ${\rm II}$'s, and ${\rm III}$'s respectively. Moreover, $u (z) = \ii (\pi - z) {2 \over \pi} K(\tilde{l}')$ for regions of type $a$;  $u (z) = (\pi - z) {2 \over \pi} K(\tilde{l})$ for type $b$'s; $u (z) =2 K(\tilde{l}) +  \ii z {2 \over \pi} K(\tilde{l}')$ for regions of type $c$; and $u (z) = \ii 2 K(\tilde{l}') + z {2 \over \pi} K(\tilde{l})$ for regions of type $d$. In figure \ref{fig:paths} we draw these four paths for the argument and the directions of the primary periods of the elliptic functions for regions of type ${\rm I}$, ${\rm II}$, and ${\rm III}$. Figure \ref{fig:contours} shows the contour lines of $(u,l)$ in $(J_y, J_z)$ plane, as given by (\ref{zlparameter}).

It is known that only purely real or purely imaginary arguments (modulo a half periodicity) ensure the reality of an elliptic function (and thus that the coupling constant are real and the Hamiltonian hermitian). We will show that this regular structure for the the path of the argument is actually connected with the symmetries of the model. The validity of this analytical continuation is checked, like we just did for region ${\rm II}_a$, using the relations connecting elliptic functions of different elliptic parameters (see \ref{app:elliptic} or (\cite{lawden})). These relations are rooted in the modular invariance of the torus on which elliptic functions are defined. In the next section, we are going to show explicitly the action of the modular group, in covering the phase diagram of the XYZ model. 

Before we proceed, we remark that, while it is possible that the validity of Baxter parametrization outside of its natural domain has been observed before by other author's analysis, we did not find any reference to it in the literature and we believe that we are the first to show in details how to construct the complete extension to the whole phase diagram, which is condensed in Table \ref{transformationtable} in the next section.

\section{The action of the modular group}
In the previous section, we showed that Baxter's parametrization of the parameters of the XYZ chain can be analytically extended outside of the principal regime to cover the whole phase-diagram and thus that any path in the $(J_y,J_z)$ ${\mathbb{R}}^2$ space corresponds to a continuous path of $(u,l)$ in ${\mathbb{C}}^2$. This extension is due to the fact that the analytical continuation of elliptic functions in the complex plane can be related to the action of the modular group. We refer to (\cite{lawden}) for a detail explanation of the relation between elliptic functions and the modular group and to Appendix A for a collection of useful identities. 

Points related by a modular transformations correspond to the same $(\Gamma,\Delta)$ point in the mapping to the eight-vertex model (\ref{st11},\ref{st12}). We now want to show that the modular group can be connected to the symmetries of the model that we discussed in section (\ref{sec:symmetries}). Let us start again in region ${\rm I_a}$, but to make the modular structure more apparent, this time let us write the parametrization in terms of theta functions using (\ref{jacobidef})
\be
  J_{y}= \frac{\theta_{3}(0|\tau)}{\theta_{2}(0|\tau)}
  \frac{\theta_{2} \left[ (z - \pi) \tau|\tau \right]}{\theta_{3} \left[ (z - \pi) \tau|\tau \right]} \; , 
  \qquad \quad 
  J_{z}= \frac{\theta_{3}(0|\tau)}{\theta_{4}(0|\tau)}
  \frac{\theta_{4} \left[ (z - \pi) \tau|\tau \right]}{\theta_{3} \left[ (z - \pi) \tau|\tau \right]} \; ,
\label{jthetas}
\ee
where $0 \le z \le \pi$, and we used (\ref{thetaperiod}).

As the Jacobi elliptic functions are doubly-periodic, the torus on which they are defined can be viewed as the quotient $\mathbb{C}/\mathbb{Z}^2$. In region ${\rm I_a}$, for $0 \le l \le 1$, the fundamental domain is the rectangle with sides on the real and imaginary axis, of length respectively $\omega_1 = 4 K(l)$ and $\omega_3 = \ii 4K(l')$, and the path followed by $z$ is a half-period along the imaginary axis. A modular transformation turns the rectangle into a parallelogram constructed in the same two-dimensional lattice, but should leave the path of the argument untouched. To this end, we extend the representation of the modular group to keep track of the original periods along the axis. In ${\rm I_a}$, let us introduce the two vectors $\sigma_R = {\omega_1 \over \omega_1} = 1$ and $\sigma_I = {\omega_3 \over \omega_1} = \tau$ (we remind that, when using theta functions, every length is normalized by the periodicity on the real axis, i.e. $\omega_1$). Later, we will also need to extend the modular group and thus we introduce the additional complex number $\phi = -\pi \sigma_I$, which is a sort of an initial phase in the path of the argument. 

With these definitions, we write (\ref{jthetas}) as
\be
  J_{y}= \frac{\theta_{3}(0|\tau)}{\theta_{2}(0|\tau)}
  \frac{\theta_{2} \left[ z \, \sigma_I + \phi |\tau \right]}{\theta_{3} \left[ z \, \sigma_I + \phi  |\tau \right]} \; , 
  \qquad \quad 
  J_{z}= \frac{\theta_{3}(0|\tau)}{\theta_{4}(0|\tau)}
  \frac{\theta_{4} \left[ z \, \sigma_I + \phi  |\tau \right]}{\theta_{3} \left[ z \, \sigma_I + \phi |\tau \right]} \; .
\label{thetaJs}
\ee
\begin{figure}
   \dimen0=\textwidth
   \advance\dimen0 by -\columnsep
   \divide\dimen0 by 3
   \noindent\begin{minipage}[t]{\dimen0}
   {\centering \qquad \quad Regions ${\rm I} \quad \qquad$} 
   \includegraphics[width=\columnwidth]{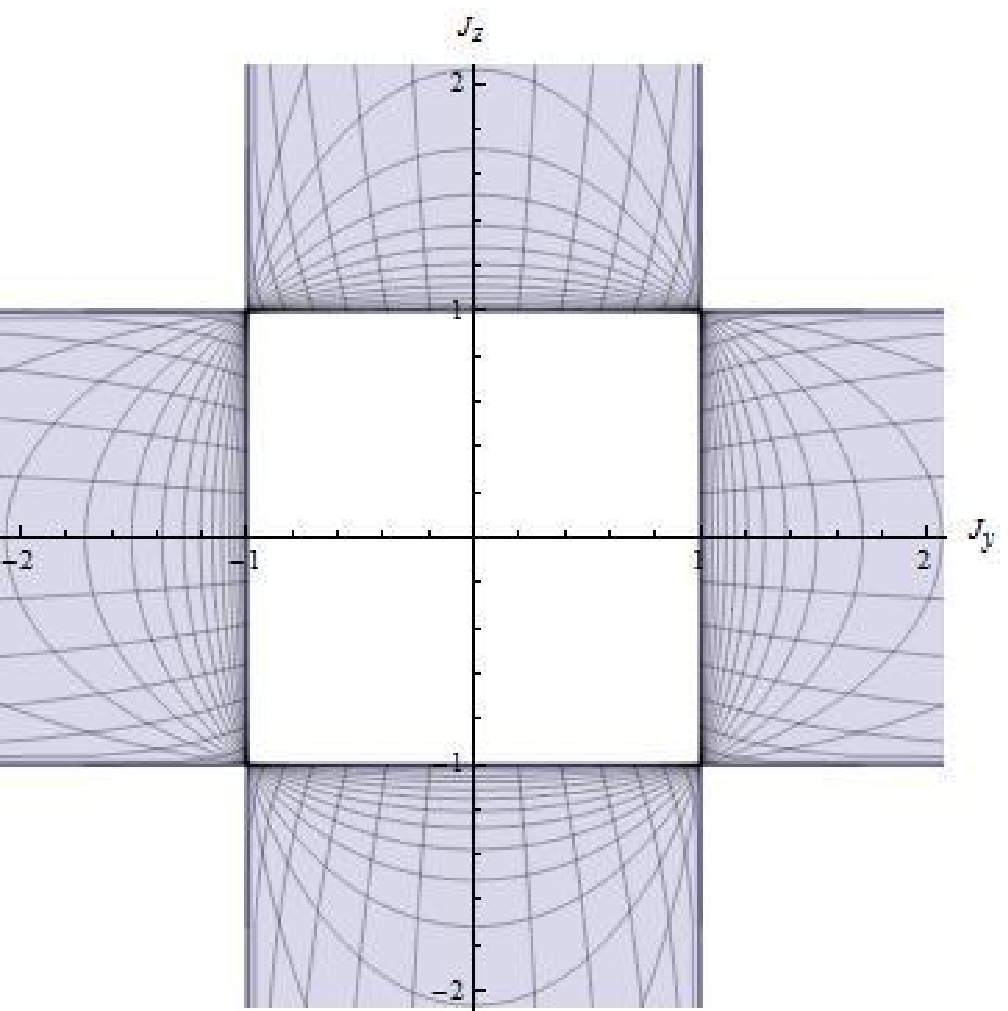}
   \end{minipage}
   \hfill
   \begin{minipage}[t]{\dimen0}
   {\centering \qquad \quad Regions ${\rm II} \quad \qquad$} 
   \includegraphics[width=\columnwidth]{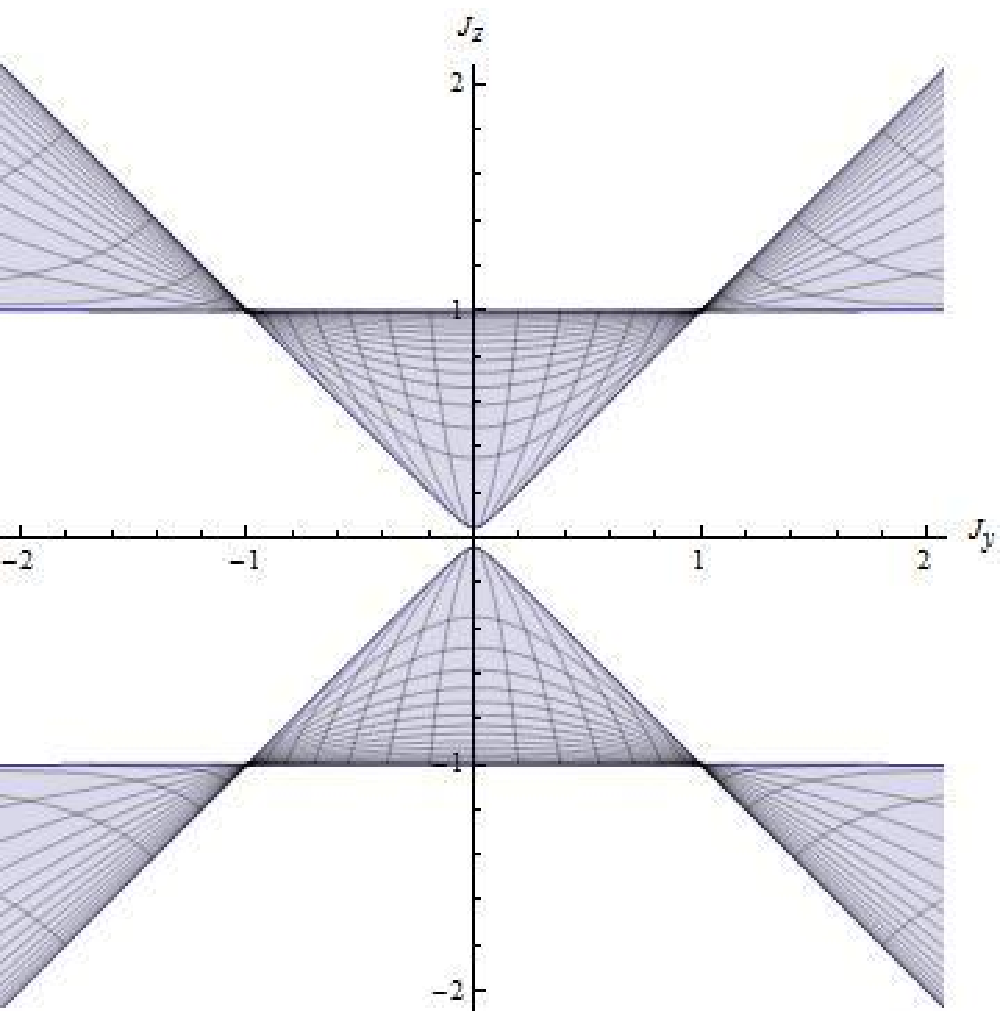}
   \end{minipage}
   \hfill
   \begin{minipage}[t]{\dimen0}
   {\centering \qquad \quad Regions ${\rm III} \quad \qquad$} 
   \includegraphics[width=\columnwidth]{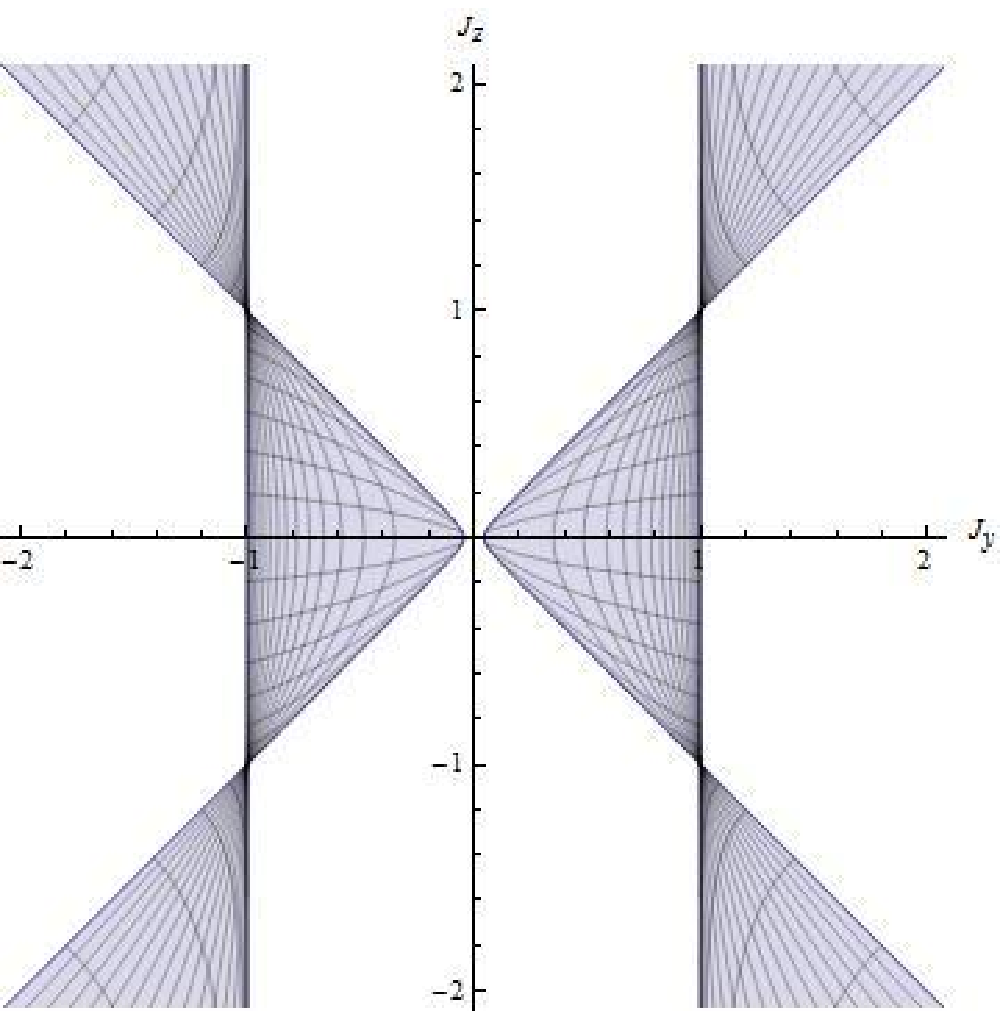}
   \end{minipage}
   \begin{minipage}[t]{\textwidth}
\caption{Contour plots of the $(u,l)$ lines in the different regions of the phase diagram, as given by (\ref{zlparameter}).}
   \label{fig:contours}
   \end{minipage}
\end{figure}

The two generators of the modular group act as
\be
    {\bf T} \left( \begin{array}{c} \tau \cr \sigma_R \cr \sigma_I \cr \phi \end{array} \right) =
    \left( \begin{array}{c} \tau + 1\cr \sigma_R \cr \sigma_I \cr \phi \end{array} \right) \; , \qquad \qquad
    {\bf S} \left( \begin{array}{c} \tau \cr \sigma_R \cr \sigma_I \cr \phi \end{array} \right) =
    \left( \begin{array}{c} - {1 \over \tau} \cr  {\sigma_R \over \tau} \cr  {\sigma_I \over \tau} \cr  {\phi \over \tau} \end{array} \right) \; .
    \label{STgendef}
\ee
A general modular transformation which gives $\tau' = {c + d \, \tau \over a + b \, \tau}$ also does 
\be
  \sigma_R \to {1 \over a + b \, \tau} = d - b \, \tau', \quad 
  \sigma_I \to {\tau \over a + b\, \tau} = -c + a \, \tau', \quad 
  \phi \to {\phi \over a + b \, \tau},
\ee
i.e. it expresses the original real and imaginary periods in terms of the two new periods.

For instance, applying the ${\bf S}$ transformation ${\bf S} \tau = \tau_S = -{1 \over \tau}$  to region ${\rm I_a}$ we have
\bea
\label{SAction}
   {\bf S} J_y & = &
   \frac{\theta_{3}(0|\tau_S)}{\theta_{2}(0|\tau_S)}
  \frac{\theta_{2} \left[ z - \pi |\tau_S \right]}{\theta_{3} \left[ z - \pi |\tau_S \right]} 
  = \frac{\theta_{3}(0|\tau)}{\theta_{4}(0|\tau)}
  \frac{\theta_{4} \left[ (z - \pi) \tau |\tau \right]}{\theta_{3} \left[ (z - \pi) \tau |\tau \right]} = J_z \; ,
  \\ 
  {\bf S} J_z & = &
  \frac{\theta_{3}(0|\tau_S)}{\theta_{4}(0|\tau_S)}
  \frac{\theta_{4} \left[ z - \pi |\tau_S \right]}{\theta_{3} \left[ z - \pi |\tau_S \right]} 
  = \frac{\theta_{3}(0|\tau)}{\theta_{2}(0|\tau)}
  \frac{\theta_{2} \left[ (z - \pi) \tau |\tau \right]}{\theta_{3} \left[ (z - \pi) \tau |\tau \right]} = J_y \; ,
\eea

which is equivalent to the Baxter's elliptic parametrization of region ${\rm I}_b$ and shows that
\be
   {\bf S} = \tilde{\bf R}_{y,z} = {\bf R}_x \; .
\ee

The other generator of the modular group ${\bf T} \tau = \tau_T = \tau +1$ gives
\bea
\label{TAction}
   {\bf T} J_y & = & 
   \frac{\theta_{3}(0|\tau_T)}{\theta_{2}(0|\tau_T)}
  \frac{\theta_{2} \left[ (z - \pi) \tau |\tau_T \right]}{\theta_{3} \left[ (z - \pi) \tau |\tau_T \right]} 
   =  \frac{\theta_{4}(0|\tau)}{\theta_{2}(0|\tau)}
  \frac{\theta_{2} \left[ (z - \pi ) \tau | \tau \right]}{\theta_{4} \left[  (z - \pi ) \tau | \tau \right]}  
  = {J_y \over J_z} \, ,\\ 
  {\bf T} J_z & = & 
  \frac{\theta_{3}(0|\tau_T)}{\theta_{4}(0|\tau_T)}
  \frac{\theta_{4} \left[ ( z - \pi ) \tau |\tau_T \right]}{\theta_{3} \left[ ( z - \pi ) \tau |\tau_T \right]} 
  = \frac{\theta_{4}(0|\tau)}{\theta_{3}(0|\tau)}
  \frac{\theta_{3} \left[  (z - \pi ) \tau | \tau \right]}{\theta_{4} \left[  (z - \pi ) \tau |\tau \right]}   
  = {1 \over J_z} \; ,
\eea

which maps ${\rm I}_a$ into region ${\rm III}_a$, in agreement with (\ref{st12}) and gives 
\be
   {\bf T} = \tilde{\bf R}_{x,z} = {\bf R}_y \; ,
\ee
as can be seen by rescaling the couplings $J_x$, $J_y$, and $J_z$ by $1 /J_z$.

Finally, we introduce an additional operator ${\bf P}$:
\be
    {\bf P} \left( \begin{array}{c} \tau \cr \sigma_R \cr \sigma_I \cr \phi \end{array} \right) =
    \left( \begin{array}{c} \tau \cr \sigma_R \cr \sigma_I \cr \phi + \pi \left[ \sigma_R + \sigma_I \right] \end{array} \right) \; ,
    \label{Pdef}
\ee
which shifts by half a period in both directions the origin of the path of the argument. Applying it to ${\rm I}_a$ gives
\bea
\label{PAction}
   {\bf P} J_y & = & 
   \frac{\theta_{3}(0|\tau)}{\theta_{2}(0|\tau)}
  \frac{\theta_{2} \left[ z \tau + \pi |\tau \right]}{\theta_{3} \left[ z \tau + \pi |\tau \right]} 
  = - J_y  \, ,\\ 
  {\bf P} J_z & = & 
  \frac{\theta_{3}(0|\tau)}{\theta_{4}(0|\tau)}
  \frac{\theta_{4} \left[ z \tau + \pi |\tau \right]}{\theta_{3} \left[ z \tau + \pi |\tau \right]} 
  = - J_z \; ,
\eea

which covers ${\rm I}_c$ and leads to the identification 
\be
   {\bf P} = \tilde{\bf P}_{y,z} = {\bf P}_x \; .
\ee

The three operation ${\bf T}$, ${\bf S}$ and ${\bf P}$ generate a group, with is the direct product $\mathfrak{G} = {\rm PSL} (2,\mathbb{Z}) \otimes \mathbb{Z}_2$. Thus, the group laws are
\be
  ({\bf S} \cdot {\bf T})^{3}=\mathbb{I}, \quad
  {\bf S}^2 = \mathbb{I}, \quad
  {\bf P}^2 = \mathbb{I}, \quad
  {\bf P} \cdot {\bf S} = {\bf S} \cdot {\bf P}, \quad
  {\bf P} \cdot {\bf T} = {\bf T} \cdot {\bf P}.
\ee
These are the same group laws satisfied by the generators of the symmetries of the XYZ chain (\cite{Evangelisti6}), as discussed in the introduction, with the addition of ${\bf T}^2 = 1$. In fact, for our purposes, ${\bf T} = {\bf T}^{-1}$ as can be easily check, due to the periodicity properties of the elliptic functions and the fact that the relation between $\tau$ and $l$ is not single-valued.

\begin{table}
\hskip -1.0cm\begin{tabular}{| c | c | c | c | c | c | c || c | c | c | c  | c | c | c |}
\hline
\multirow{2}{*}{\scriptsize{Region}} & {\scriptsize{Modular}} & \multirow{2}{*}{$\sigma_R$} & \multirow{2}{*}{$\sigma_I$} & \multirow{2}{*}{$\phi$} & \multirow{2}{*}{$\tau$} & \multirow{2}{*}{$l$} & 
\multirow{2}{*}{\scriptsize{Region}} & {\scriptsize{Modular}} & \multirow{2}{*}{$\sigma_R$} & \multirow{2}{*}{$\sigma_I$} & \multirow{2}{*}{$\phi$} & \multirow{2}{*}{$\tau$} & \multirow{2}{*}{$l$} \cr
& {\scriptsize{Generator}} &  &  & &  &  & 
& {\scriptsize{Generator}} &  &  & &  &  \cr
\hline
${\rm I}_a$ & {\bf I} & $1$ & $\tau$ & $-\pi \tau$ & $\tau$ & $\tilde{l}$ & 
${\rm I}_c$ & {\bf P} & $1$ & $\tau$ & $\pi$ & $\tau$ & $\tilde{l}$ \cr
\hline
${\rm I}_b$ & {\bf S} & ${1 \over \tau}$ & $1$ & $-\pi $ & $-{1 \over \tau}$ & $\tilde{l}'$ & 
${\rm I}_d$ & {\bf SP} & ${1 \over \tau}$ & $1$ & ${\pi \over \tau} $ & $-{1 \over \tau}$ & $\tilde{l}'$ \cr
\hline
${\rm III}_a$ & {\bf T} & $1$ & $\tau$ & $-\pi \tau $ & $ \tau + 1$ & $\ii {\tilde{l} \over \tilde{l}'}$ & 
${\rm III}_c$ & {\bf TP} & $1$ & $\tau$ & $\pi $ & $ \tau + 1$ & $\ii {\tilde{l} \over \tilde{l}'}$ \cr
\hline
${\rm II}_b$ & {\bf TS} & ${1 \over \tau}$ & $1$ & $-\pi$ & ${\tau -1 \over \tau}$ & ${1 \over \tilde{l}'}$ & 
${\rm II}_d$ & {\bf TSP} & ${1 \over \tau}$ & $1$ & ${\pi \over \tau} $ & ${\tau -1 \over \tau}$ & ${1 \over \tilde{l}'}$\cr
\hline
${\rm III}_b$ & {\bf ST} & ${1 \over \tau + 1}$ & ${\tau \over \tau +1}$ & ${- \pi \tau \over \tau +1} $ & ${-1 \over \tau+1}$ & $\ii {\tilde{l}' \over \tilde{l}}$ &
${\rm III}_d$ & {\bf STP} & ${1 \over \tau + 1}$ & ${\tau \over \tau +1}$ & $ {\pi \over \tau +1} $ & ${-1 \over \tau+1}$ & $\ii {\tilde{l}' \over \tilde{l}}$ \cr
\hline
${\rm II}_a$ & {\bf TST} & ${1 \over \tau + 1}$ & ${\tau \over \tau +1}$ & ${- \pi \tau \over \tau +1} $ & ${\tau \over \tau+1}$ & ${1 \over \tilde{l}}$ &
${\rm II}_c$ & {\bf TSTP} & ${1 \over \tau + 1}$ & ${\tau \over \tau +1}$ & ${\pi \over \tau +1}$ & ${\tau \over \tau+1}$ & ${1 \over \tilde{l}}$ \cr
 & {\bf STS} & ${1 \over \tau - 1}$ & ${\tau \over \tau -1}$ & ${- \pi \tau \over \tau -1} $ & ${\tau \over \tau-1}$ & ${1 \over \tilde{l}}$ &
 & {\bf STSP} & ${1 \over \tau - 1}$ & ${\tau \over \tau -1}$ & ${\pi \over \tau -1}$ & ${\tau \over \tau-1}$ & ${1 \over \tilde{l}}$ \cr
\hline
\end{tabular}
\caption{List of the action of modular transformations, according to (\ref{STgendef}, \ref{Pdef}), in mapping the phase diagram, starting from ${\rm I}_a$ with the parametrization given by (\ref{thetaJs})}
\label{transformationtable}
\end{table}

We collect in table \ref{transformationtable} the action of the different transformations generated by ${\bf T}$, ${\bf S}$ and ${\bf P}$ in mapping region ${\rm I}_a$ in the rest of the phase diagram. It is clear that $\mathfrak{G}$ is isomorphic to the group generated by $\tilde{\bf R}_{y,z}$, $\tilde{\bf R}_{x,z}$, and $\tilde{\bf P}_{y,z}$, as we anticipated, and thus, for instance, ${\bf S} \cdot {\bf T} =  \tilde{\bf R}_{y,z} \cdot \tilde{\bf R}_{x,z} =  {\bf R}_{x,z} \cdot {\bf R}_{y,z}$. 
A few additional comments are in order. In the previous section we gave a prescription for the path of the argument around a rectangle, see (\ref{zlparameter}) and figure \ref{fig:contours}. This structure is explained by table \ref{transformationtable}: the ${\bf S}$ generator interchanges the role of the axes and thus the path that runs along the imaginary axis in the new basis runs along the real one, consistently with the prescription given for regions of type $b$ and $d$. The action of ${\bf P}$ is to shift the path and thus interchanges regions of type $a$ with $c$ and $b$ with $d$ (and viceversa). The role of ${\bf P}$ can be thought of as that of promoting the modular group to its discrete $\mathbb{Z}_2 $-affinization.

In conclusions, we see that the whole line $l^2 \in (- \infty, \infty)$ can be used in (\ref{zlparameter}), which divides in three parts, one for each of the regions of type $I$, $II$, or $III$. It is known, that, for each point on this line, only if the argument of the elliptic functions is purely imaginary or purely real (modulo a half periodicity) we are guaranteed that the parameters of the model are real. Hence the two (four) paths, related by ${\bf S}$ (and ${\bf P}$) duality. The paths of the argument are compact, due to periodicity. Extending this mapping by considering a generic complex argument $u$, would render the Hamiltonian non-Hermitian, but would preserve the structure that makes it integrable.

\section{Analysis of the entropy}

For a quantitative analysis of the Renyi entropy it is now convenient to use the definition of the parameter $\epsilon$ given in (\ref{eps2}), as a function of the parameters $(z,l)$. 
As it is clear from (\ref{eq:renyi4}), the Renyi entropy can be seen as a function of this parameter only. Then, it is a simple, monotonically decreasing function, such that
\be
   \lim_{\epsilon \to 0} {\cal S}_\alpha = \infty, \qquad \qquad
   \lim_{\epsilon \to \infty} {\cal S}_\alpha = 0.
\ee

However, the behavior of the entropy as a function of the physical parameters of the XYZ model $J_y$ and $J_z$ is more complicated, as they define $z$ and $l$ and thus $\epsilon$. It is worth remarking that the expressions for the entropy given so far apply also for all the other integrable models generated by the R-matrix of the eight-vertex model. For instance, for the XY model one has $\epsilon = \pi  I(l')/I(l)$ and for the XXZ model $\epsilon = \cosh^{-1} J_z$. Compared to this example, the entropy of the XYZ model is more interesting, since it is truly two-dimensional with its independent $l$ and $z$ dependence.

To study the entropy in the $(J_y,J_z)$ phase diagram, we thus need to use (\ref{st11}, \ref{st12}) and to invert (\ref{Iaparameter}) to express   $(z,l)$ as a function of the two parameters $(\Gamma, \Delta)$.
The result was already written in formula (\ref{ste8bis}), that we recall here: 
\begin{equation}
\label{ste8}
    \begin{cases}
     l = \displaystyle{\sqrt{ {1-\Gamma^{2} \over \Delta^{2}-\Gamma^{2} } } }\\
     {\rm dn}(\ii u;l) = \displaystyle{1\over \Gamma}
    \end{cases}
\end{equation}

\begin{figure}[t]
   \begin{center}
      \includegraphics[scale=0.35]{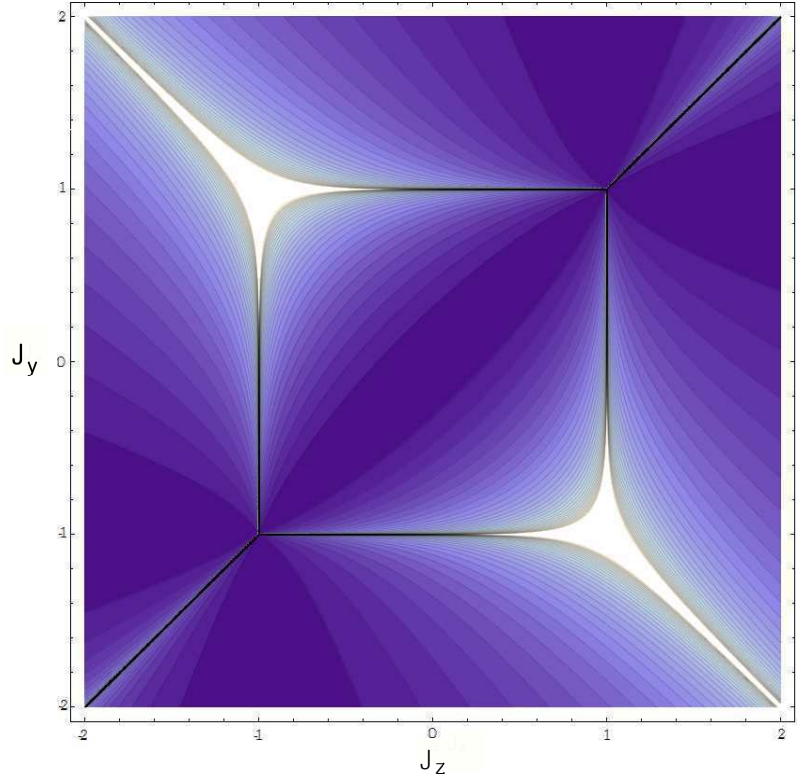}
      \par
   \end{center}
   \caption{Curves of constant entropy of the XYZ model in the ($J_{y},J_{z}$) plane. Regions of similar colors have similar entropy values and the line where colors change are the lines of constant entropy. The brighter is the color, the bigger is the entropy.}
   \label{fig:isoentropy}
\end{figure}

In figure (\ref{fig:isoentropy}) we plot the Von Neumann entropy in the $(J_{y},J_{z})$ plane, for $-2 < J_{y} < 2, -2 < J_{z} < 2$.
This is a contour plot, so regions of similar colors have similar values of the entropy and the line dividing regions of different colors are examples of line of constant entropy.
We immediately see that the entropy draws the same phase diagram we presented in figure \ref{fig:phasediagram}, being singular (diverging) on the critical lines and vanishing on the ferromagnetic Ising lines $(J_y=1,J_z>1)$, $(J_y>1, J_z=1)$, $(|J_y|<1,J_z=J_y)$ and the $J_y \leftrightarrow - J_y$ symmetric ones.

The four tri-critical points $(J_y = \pm 1, J_z = \pm 1)$ are immediately recognized as more interesting and we shall study the entropy in their neighbourhoods in detail.
Let us first look at the conformal point $(J_y = 1, J_z = - 1)$ and use (\ref{ste8}) to write $l$ and $z$.
If we take
\be
   \Gamma = 1 - \rho \cos \phi, \qquad
   \Delta = -1 - \rho \sin \phi, \qquad \qquad
   0 \le \phi \le {\pi \over 2},
\ee
we have
\be
\begin{cases}
   l  =  \left(\tan(\phi)+1\right)^{-1/2}+ {\rm O} \left( \rho \right), \\
   {\rm snh} \left( u ; l\right)  = 
   \sqrt{2\text{cos}(\phi)+2 \text{sin}(\phi)}\,\sqrt{\rho}
   + {\rm O} \left( \rho^{3/2} \right) 
\end{cases}
   \label{vanishinglambda}.
\ee
Thus, $l$ is not defined at the tri-critical point, since its value as $\rho \to 0$ depends on the direction $\phi$ of approach (for $\phi=0$, $l=1$ and for $\phi=\pi/2$, $l=0$).
From (\ref{vanishinglambda}) we see that $z \to 0$ as $\rho \to 0$ and, using $\,\sn (u,l) \simeq u + {\rm O} \left( u^2 \right)$, we find
\be
   u = \sqrt{2\text{cos}(\phi)+2 \text{sin}(\phi)}\,\sqrt{\rho} 
   + {\rm O} \left( \rho^{3/2} \right)
   \label{vanishinglambdabis}.
\ee
Thus, $\epsilon \sim \sqrt{\rho}$ and in any neighbourhood of the conformal point the entropy is diverging.
In particular, on the direction $\phi=0$, since ${\rm K}(1) \to \infty$, the entropy is divergent for any $\rho$ (as expected, since this is a critical line).
\begin{figure}[t]
   \begin{center}
      \includegraphics[scale=0.35]{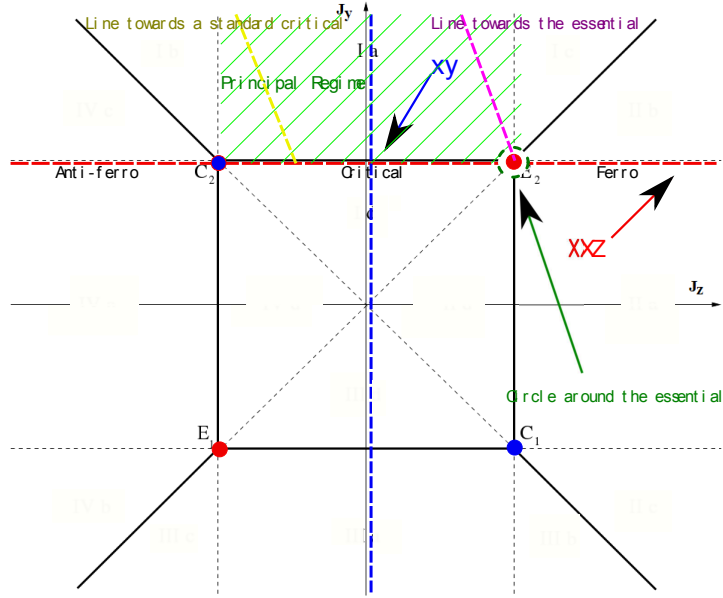}
      \par
   \end{center}
   \caption{Sketch of the numerical paths along which we check out the validity of our results. Comparisons between analytical and numerical results along the different lines drawn in this picture are shown in figures (\ref{fig:xxz}), (\ref{fig:xyz}), (\ref{fig:essential}) and (\ref{fig:essential2}).}
   \label{fig:num_paths}
\end{figure}

Expanding around the other tri-critical point as
\be
   \Gamma = - 1 + \rho \cos \phi, \qquad
   \Delta = -1 - \rho \sin \phi, \qquad \qquad
   0 \le \phi \le {\pi \over 2},
\ee
we find
\be
\begin{cases}
   l  =  \left(\tan(\phi)+1\right)^{-1/2}+ {\rm O} \left( \rho \right), \\
   {\rm snh} \left( u ; l\right)  = 
   -\sqrt{2\text{cos}(\phi)+2 \text{sin}(\phi)}\,\sqrt{\rho}
   + {\rm O} \left( \rho^{3/2} \right)
   
\end{cases}
   \label{singularlambda}\;.
\ee
Thus, $l$ has exactly the same behavior as in the previous case, while ${\rm snh}(u)$ remains unchanged apart from the sign. However in the neighbour of this point we must use the following expansion to extract the limit of $u$, $\,\sn (u,l) \simeq (u-2\ii{\rm K'}) + {\rm O} \left( (u-2\ii{\rm K'})^2 \right)$. This implies $u \sim 2{\rm K'}$ close to $E_{1}$ and thus $\epsilon \sim K(l')/K(l)$, which can take any non-negative value from 0 to $\infty$ for $\phi$ that goes from 0 to $\pi/2$. As we mentioned, the behaviour of the entropy is completely determined by $\epsilon$, thus the Renyi entropy has an essential singularity at $E_{1,2}$, and can take any positive real value in any neighbourhood of it. 
\begin{figure}[t]
   \begin{center}
      \includegraphics[scale=0.35]{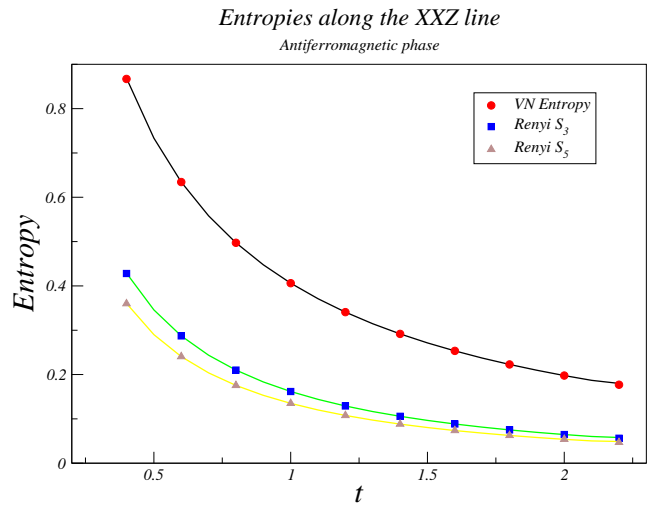}
      \par
   \end{center}
   \caption{Comparison between analytical (solid lines) and numerical results (DMRG data points) along the XXZ line of figure (\ref{fig:num_paths}), for the Von Neumann (VN) and Renyi ($S_{3},S_{5}$) entropies. $t$ is a coordinate which parametrizes the coupling constants: $J_z=-1-t$ and $J_y=1$. The logarithm in the definition of the entropies has been chosen to base 2, while the numerical data are extrapolated in the $L \to\infty$ limit. }
   \label{fig:xxz}
\end{figure}
In conclusion, we can say that the points $C_{1,2}$ are  {\em ordinary} critical points, in the sense that in vicinity of this point the Von Neumann block entropy has the universal behavior found in (\cite{Calabrese2}). A detailed study of the entropy near these conformal critical points (and also along the conformal critical lines will be presented in the next chapter).
\begin{figure}[t]
   \begin{center}
      \includegraphics[scale=0.35]{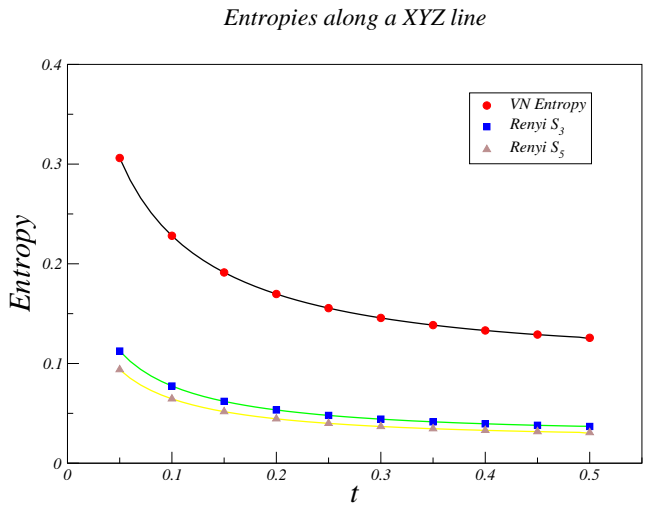}
      \par
   \end{center}
   \caption{Comparison between analytical (solid lines) and numerical results (DMRG data points) along a XYZ line of figure (\ref{fig:num_paths}), for the Von Neumann (VN) and Renyi ($S_{3},S_{5}$) entropies. Here $t$ is defined through the equations: $J_z=2/3-t/2$ and $J_y=1+\sqrt{3}t/2$. The logarithm in the definition of the entropies has been chosen to base 2, while the numerical data are extrapolated in the $L \to\infty$ limit.}
   \label{fig:xyz}
\end{figure}
The behavior of the entropy around the tri-critical points $E_{1,2}$ is clearly quite different. It can be seen directly from the plot that it is direction dependent, since it is an accumulation point for iso-entropy curves. Moreover, it shows a strong discontinuity crossing it along one of the critical lines, since it goes suddenly from diverging to vanishing. In the next sections we will describe the nature of the transition around these points and show that there the entropies have an essential singularity. Another point with the same singular behavior was found at the bi-critical point of the anisotropic XY model in a transverse magnetic field in (\cite{Franchini}) and was named {\em essential critical point} for the singular behavior of the entropy there.

\begin{figure}[t]
\begin{center}
\includegraphics[scale=0.35]{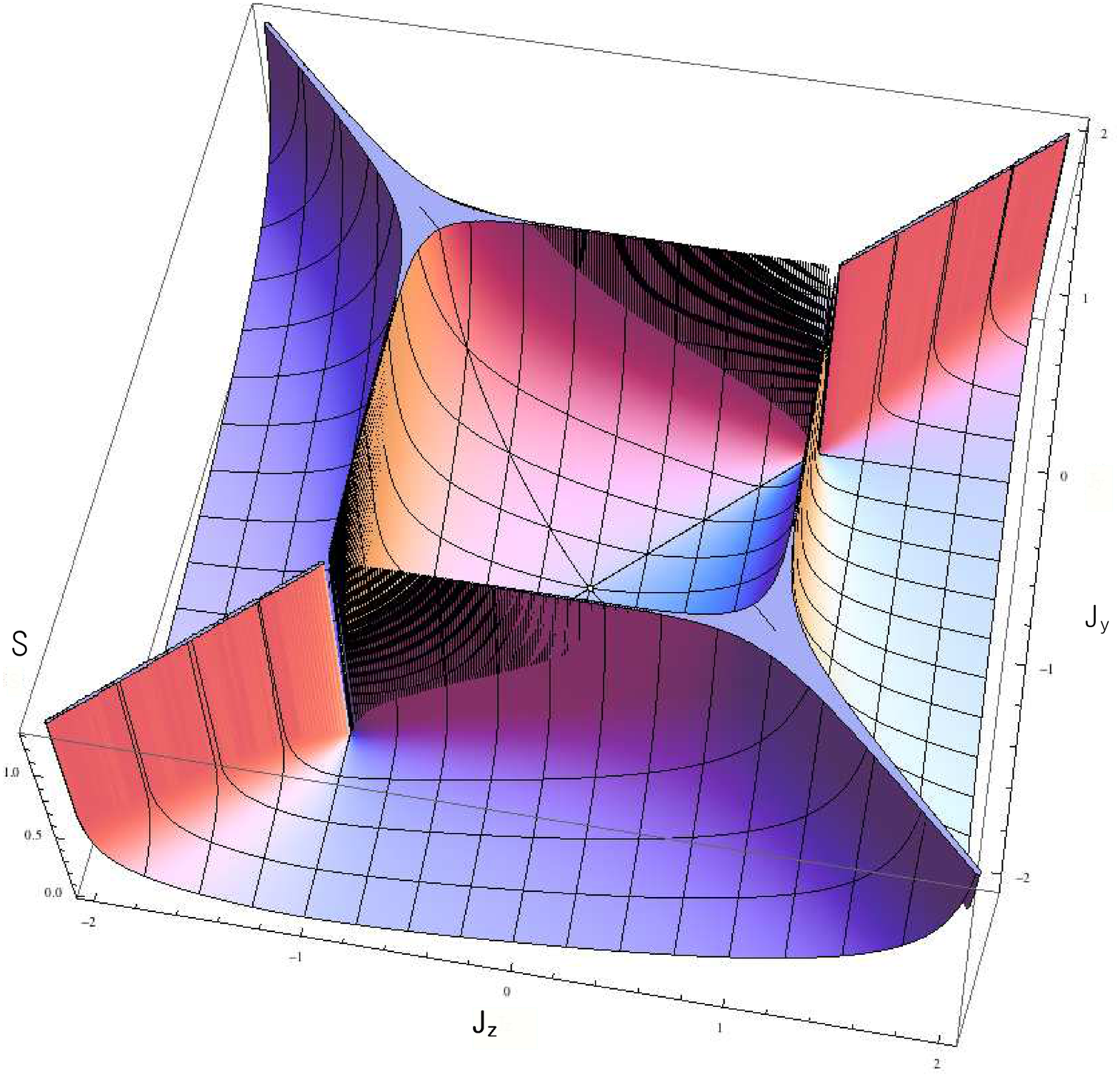}
\par\end{center}
\caption{Three-dimensional plot of the Von Neumann block entropy as a function of the coupling constants of the XYZ model. The entanglement entropy diverges when the system is critical.}
\label{fig:entropy}
\end{figure}

\section{The essential critical points}
\label{sec:essential}
\begin{figure}[t]
\begin{center}
\includegraphics[scale=0.25]{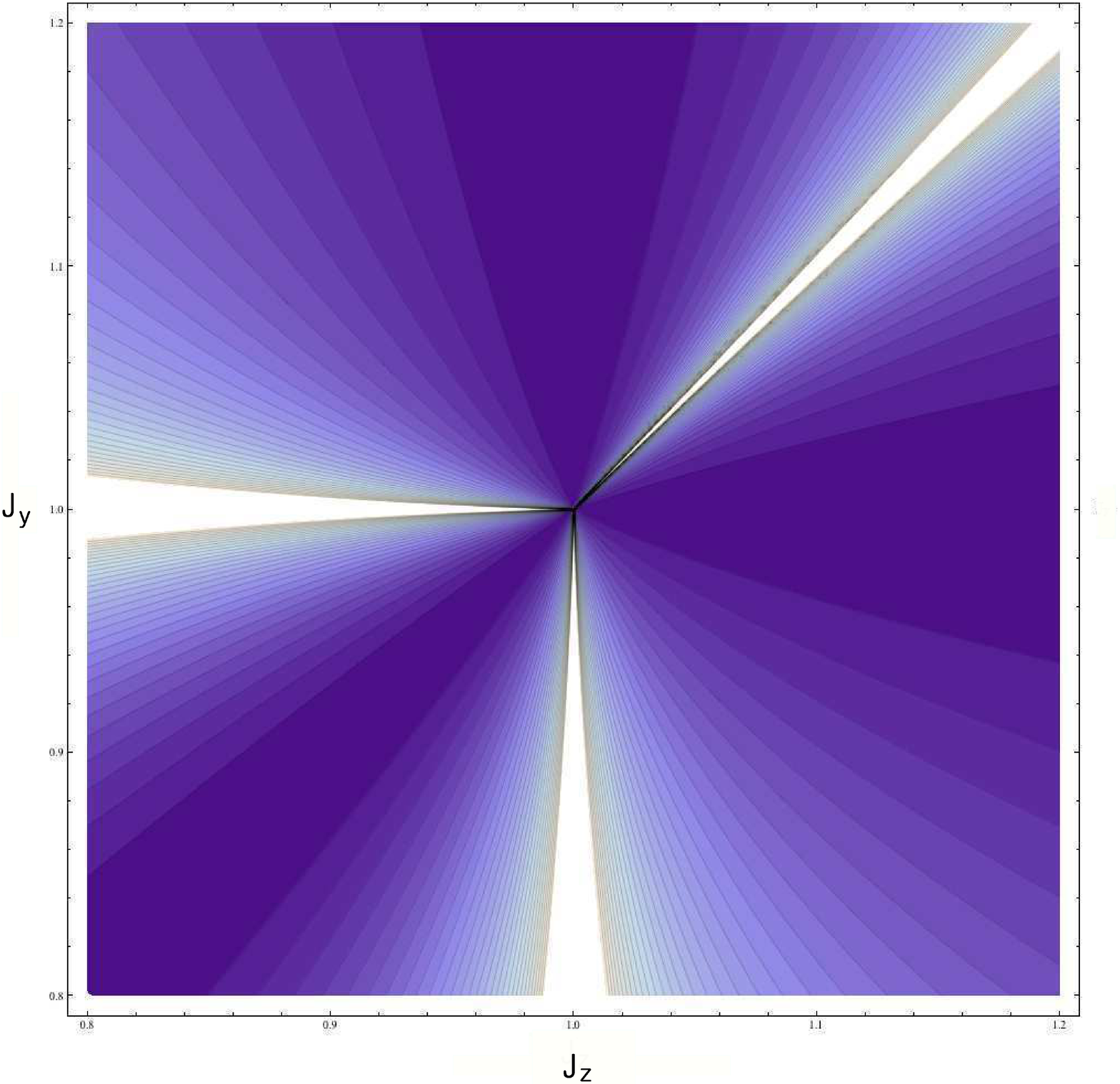}
\par\end{center}
\caption{Curves of constant entropy of the XYZ model in the vicinity of $(J_{y}=1,J_{z}=1)$. We can see that near the essential critical point the lines of constant entropy grow denser. }
\label{fig:isoentropy2}
\end{figure}

Let us now analyze in more details the properties of the ferromagnetic tri-critical points E$_{1,2}$.
Just looking at a blowout of the entropy plot close to this point in figure \ref{fig:isoentropy2} it is evident that depending on the direction of approach the entropy can take any positive value arbitrarily close to it.

This point is very different from all other points along the critical line, because it is not conformal, since low energy excitations are magnons with a quadratic dispersion relation $\varepsilon (q) = 1 - \cos q$. This implies a breakdown of the usual conformal predictions by Calabrese and Cardy (\cite{Calabrese2}).
We showed in the previous sections the singular behavior of the entropy close to this point, since it can take any positive, real value arbitrarily close to it. Moreover, from figure \ref{fig:isoentropy2} one sees that from any point in the phase diagram one can reach one of the non-conformal points following a curve of constant entropy. A point with the same singular behavior for the entropy was already described in (\cite{Franchini}) as the bi-critical point of the XY model and it was also a non-conformal point with a quadratic dispersion relation.
\begin{figure}[t]
   \begin{center}
      \includegraphics[scale=0.35]{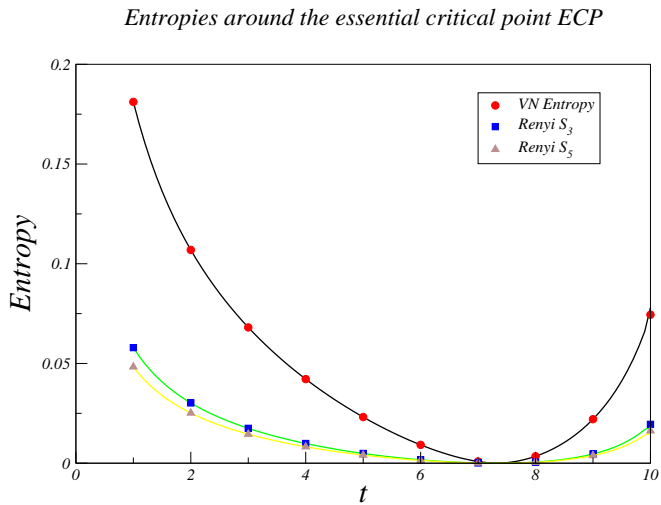}
      \par
   \end{center}
   \caption{Analytical (solid lines) and numerical results (DMRG data points) for the Von Neumann and Renyi entropies along a path surrounding the essential critical point in regime $I_{a}$ (see figure (\ref{fig:num_paths})). The parameter $t$ parametrizes the couplings as: $J_z=1-1/5 \cos(3/44\,\pi t)$ and $J_y=1+1/5 \sin(3/44\,\pi t)$. Note that here the logarithm in the definition of the entropies has been chosen to base 2, while the numerical data are extrapolated in the $L \to\infty$ limit.}
   \label{fig:essential} 
\end{figure}
So far, we studied this point by means of the expressions that we have from the underlying integrable structure. However, it is possible to understand the singular behavior of the entropy in more qualitative, physical terms by looking at the structure of the ground state wavefunction.
At the isotropic point, the Hamiltonian is given by
\begin{equation}
   \hat{H}_{XXX} =
   -{\displaystyle \sum_{n=1}^N} \vec{\sigma}_n \cdot \vec{\sigma}_{n+1} \; .
   \label{eq:XXXH}
\end{equation}
Since $SU(2)$ is unbroken at this point, the Hamiltonian commutes with the generators
\be
   S^2 \equiv {1 \over 4} \sum_{n=1}^N \vec{\sigma}_n \cdot \vec{\sigma}_n \; , \qquad
   S^z \equiv {1 \over 2} \sum_{n=1}^N \vec{\sigma}_n^z \; , \qquad
   S^\pm \equiv {1 \over 2} \sum_{n=1}^N \vec{\sigma}_n^\pm \; .
\ee
Thus, the ground state is $N+1$-fold degenerate, since it is in the representation $S^2 = {N \over 2} \left( {N \over 2} +1 \right)$. This corresponds to a basis of ferromagnetic states, aligned along the different axis. Which state is realized depends on the spontaneous symmetry breaking: a (classical) state aligned in a particular direction will have no entanglement, but a superposition of such states will have an entanglement growing with the number of (classical) states involved.

Hence, we understand that on one of the Ising ferromagnetic lines, the easy-axis anisotropy singles out a classical state as the ground state. Thus, along one of these lines, close to the critical point, the entropy will exactly vanish. In the paramagnetic phases, we have an easy plane anisotropy and thus the ground state is constructed by a large superposition of classical states, yielding a large entropy, diverging in the thermodynamic limit, as prescribed by CFT. In the rest of the phase diagram, we will have an intermediate situation. This qualitative picture helps in interpreting the strong discontinuity in crossing the essential critical point.

Note that the picture is different at the anti-ferromagnetic tri-critical point, since there the ground state always belongs to a subspace without finite magnetization and hence it does not go under such a strong discontinuity crossing the tri-critical point. In fact, while a classical ferromagnetic state is an eigenvector of the Hamiltonian anywhere on a ferromagnetic Ising line, a classical Neel state approximates an eigenstate of the AFM Ising line only far from the isotropic point.
From the examples of the XYZ chain and of the XY model (\cite{Franchini2007}), we can infer that points of discontinuous phase transitions show a characteristic singular behavior of the entanglement entropy, which can be easily detected numerically and might be used as an efficient tool to distinguish, for example, between first and higher order phase transitions in more complicated models (\cite{degliesposti2003}, \cite{franca2008}). It is important to remark that ${\cal S}_\alpha$ and the correlation length show very different behaviors close to $E_{1,2}$, since, as it is expected for any phase transition, $\xi$ diverges as one approaches the critical lines/points, without showing the essential singularity characteristic of the entropy (\cite{Ercolessi2011}). Thus, close to the discontinuous points $E_{1,2}$ the entropy formula derived in (\cite{Calabrese2}) is no longer valid: indeed it applies in the vicinity of all conformal points of the critical lines (including the BKT points $C_{1,2}$), while it fails if we are close to non-conformal points, such as $E_{1,2}$ where the spectrum of excitations has a quadratic dispersion relation for small momenta.\\
Let us now try to understand more quantitatively the behaviour of the Renyi entropy in proximity of the essential critical point. Suppose we approach the point E$_{1}$ following a straght line in the $(J_{y},J_{z})$-plane:
\be
\label{eq:essential1}
\begin{cases}
J_{z}=1-t \\
J_{y}=1+m\cdot t,
\end{cases}
\ee
where $t$ is a parameter measuring the distance from E$_{1}$ along a straight line, and $m$ is an angular coefficient such that the path belongs to the principal regime of figure (\ref{fig:phasediagram}). When $t\to 0$ we approach the critical point. Expanding the expressions (\ref{ste8}) for $z$ and $l$ in this limit, we obtain the following expansion for $\epsilon$ (when expanding $z$ and $l$ in the intermediate steps we used the {\em Lagrange inversion theorem}):
\be  
\label{eq:essential2}
\epsilon=\frac{\pi I\left([(1+m)^{-\frac{1}{2}}]'\right)}{I\left((1+m)^{-\frac{1}{2}}\right)}-\frac{\pi(1+m)^{\frac{1}{2}}}{\sqrt{2}\,I\left((1+m)^{-\frac{1}{2}}\right)}\,t^{\frac{1}{2}}+O(t).
\ee
Now we want to plug this expression into formula (\ref{eq:modular2}) to obtain the asymptotic behaviour of the Renyi entropy:
\be
\label{eq:essential3}
{\cal S}_{\alpha}={\displaystyle\frac{\alpha\ln(A)-\ln(A_{\alpha})}{6(\alpha-1)}}+\left(\frac{B}{A}+\frac{B_{\alpha}}{A_{\alpha}}\right)\,t^{\frac{1}{2}}+O(t),
\ee
where we clearly see that the entropy saturates to a constant value as we approach the essential critical point. The constants $A$ and $B$ are given by:
\be
\label{eq:essential3}
A={\displaystyle\frac{\theta_{2}(0,{\rm e}^{a})^{2}}{\theta_{3}(0,{\rm e}^{a})\theta_{4}(0,{\rm e}^{a})}},\qquad a=\frac{\pi I\left([(1+m)^{-1/2}]'\right)}{I\left((1+m)^{-1/2}\right)},
\ee
and
\be
\label{eq:essential4}
\begin{array}{lcl}
B & = & -{\displaystyle\left[\theta_{2}(0,{\rm e}^a) \theta_{4}(0,{\rm e}^a) \theta_{3}'(0,{\rm e}^a)+\theta_{2}(0,{\rm e}^a) \theta_{3}(0,{\rm e}^a) \theta_{4}'(0,{\rm e}^a)\right.} \\
& - & {\displaystyle\left. 2\theta_{3}(0,{\rm e}^a) \theta_{4}(0,{\rm e}^a) \theta_{2}'(0,{\rm e}^a)\right]}{\displaystyle\frac{b\,{\rm e}^a \theta_{2}(0,{\rm e}^a)}{\theta_{3}(0,{\rm e}^a)^2 \theta_{4}(0,{\rm e}^a)^2}},
\end{array}
\ee
where $\theta'$ is the derivative of the Jacobi theta function with respect to its second argument and
\be
\label{eq:essential5}
b=-\frac{\pi(1+m)^{1/2}}{\sqrt{2}\,I((1+m)^{-1/2}}.
\ee
$A_{\alpha}$ and $B_{\alpha}$ are obtained from $A$ and $B$ by replacing $a\to\alpha a$ and $b\to\alpha b$. Approaching the essential critical point along a straight line the correlation length diverges as:
\be
\label{eq:essential6}
\xi^{-1}\propto t^{1/2},
\ee
therefore the Renyi entropy goes like:
\be
\label{eq:essential7}
{\cal S}_{\alpha}\approx c_{\alpha}^{1}(m) + c_{\alpha}^{2}(m)\,\xi^{-1/2},
\ee
where $c_{\alpha}^{1}$ and $c_{\alpha}^{2}$ are constants depending on the slope $m$ of the line approaching the critical point. This asymptotic behaviour is very different from the one obtained approaching a conformal critical point and holds true even if we get close to the essential critical point following a parametrization curve with $\tau$ constant.  
\begin{figure}[t]
  \begin{center}
     \includegraphics[scale=0.35]{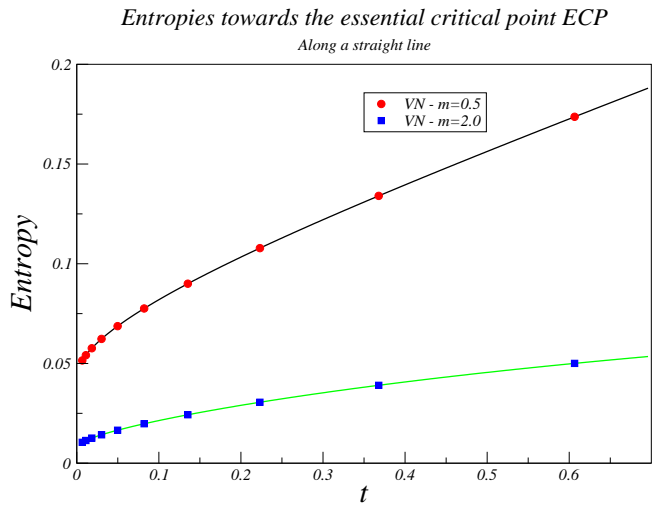}
     \par
   \end{center}
   \caption{Analytical (solid lines) and numerical results (DMRG data points) for the Von Neumann entropy along a straight path towards the essential critical point in regime $I_{a}$ (see figure (\ref{fig:num_paths})). The parameter $t$ parametrizes the couplings as defined in equation (\ref{eq:essential1}). Note that here the logarithm in the definition of the entropy has been chosen to base 2, while the numerical data are extrapolated in the $L \to\infty$ limit.}
   \label{fig:essential2} 
\end{figure}

\section{Conclusions}
The solution of the XYZ chain is based on its relation with the eight-vertex model. However, the relation between the parameters $(\Gamma, \Delta)$ of the later is not one to one with the coupling $(J_x, J_y, J_z)$ (\ref{eq:XYZ1}) of the former (and with the Boltzmann weights of the classical model as well) and the mapping changes in the different regions of the phase diagram (\ref{st11}, \ref{st12}). We have shown that this re-arrangement procedure is reproduced by the {\it natural} analytical extension of the solution valid in a given region. We also gave a prescription that relates the parameter $(u,l)$ of the elliptic functions and the physical parameter of the XYZ model in the whole of the phase diagram and an inversion formula valid everywhere.

The main interest of this analytical continuation lies in its connection with the action of the modular group. This connection allowed us to show that a certain abelian (``affine'') extension of the modular group realizes in parameter space the physical symmetries of the model. This symmetries are in the form of {\it dualities}, since they connect points with different coupling, related by some spin rotation. 

Modular invariance plays a central role in the structure and integrability of Conformal Field Theories in $1 + 1$ dimension. On its critical lines, the XYZ model is described by a $c=1$ CFT and thus, in the scaling limit, its partition function is modular invariant in real space. Our results show that, even in the gapped phase, it is also a modular invariant in parameter space, due to the symmetries of the model. Elliptic structures are common in the solution of integrable models and it is tempting to speculate that their modular properties do encode in general their symmetries and thus the class of integrability-preserving relevant perturbations that drive the system away from criticality. \\

We have also studied the bipartite Renyi entropy of the 1-D $XYZ$ spin chain in its phase diagram, using exact analytical expressions derived from the integrability of the model. The entropy diverges on the critical lines: close to conformal points the divergence is logarithmical in the correlation length with power-law corrections (this point will be discussed extensively in chapter \ref{chap3}). At the non-conformal points the entropy has an essential singularity.  We argued that this may be a characteristic feature of discontinuous  phase transition points that could allow to easily numerically discriminate between first and higher order phase transitions.



\chapter{Unusual corrections to the entanglement entropy}
\label{chap3}
\ifpdf
    \graphicspath{{Chapter2bis/Chapter2bisFigs/PNG/}{Chapter2bis/Chapter2bisFigs/PDF/}{Chapter2bis/Chapter2bisFigs/}}
\else
    \graphicspath{{Chapter2bis/Chapter2bisFigs/EPS/}{Chapter2bis/Chapter2bisFigs/}}
\fi
In this chapter we study analytically the corrections to the leading terms in the Renyi entropy of a massive lattice theory, showing significant deviations from na\"ive expectations. In particular, we show that finite size and finite mass effects give rise to different contributions (with different exponents) and thus violate a simple scaling argument. In the specific, we look at the entanglement entropy of a bipartite XYZ spin-$1/2$  chain in its ground state. When the system is divided into two semi-infinite half-chains, we have an analytical expression of the R\'enyi entropy as a function of a single mass parameter. In the scaling limit, we show that the entropy as a function of the correlation length formally coincides with that of a bulk Ising model. This should be compared with the fact that, at criticality, the model is described by a $c=1$ conformal field theory and the corrections to the entropy due to  finite size effects show exponents depending on the compactification radius of the theory. If the lattice spacing is retained finite, the relation between the mass parameter and the correlation length generates new subleading terms in the entropy, whose form is path-dependent in phase-space and whose interpretation within a field theory is not available yet. These contributions arise as a consequence of the existence of stable bound states and are thus a distinctive feature of truly interacting theories, such as the XYZ chain.\\
This chapter covers the content of the paper ``{\em Correlation Length and Unusual Corrections to the Entanglement Entropy}'', E. Ercolessi, S. Evangelisti, F. Franchini and F. Ravanini, arXiv:1201.6367, Phys. Rev. B, {\bf 85} (2012).

\section{The problem}
\markboth{\MakeUppercase{\thechapter. Unusual corrections to the entanglement entropy}}{\thechapter. Unusual corrections to the entanglement entropy}

Let us consider the formal expression of the Renyi entropy of a bipartite quantum system:
\be
   {\cal S}_{\alpha} \equiv {1 \over 1-\alpha} \ln \Tr \rho_A^\alpha \; ,
   \label{eq:renyi1}
\ee
Varying the parameter $\alpha$ in (\ref{eq:renyi1}) gives us access to a lot of information on $\rho_A$, including its full spectrum (\cite{Calabrese1}, \cite{Franchini1}).

For gapped systems, the entanglement entropy satisfies the so-called {\it area law}, which means that its leading contribution for sufficiently large subsystems is proportional to the area of the boundary separating system $A$ from $B$. In $1+1$ dimensional systems, the area law implies that the entropy asymptotically saturates to a constant (the boundary between regions being made just by isolated points).

Critical systems can present deviations from the simple area law. In one dimension, in particular, the entanglement entropy of systems in the universality class of a conformal field theory (CFT) is known to diverge logarithmically with the subsystem size (\cite{Holzey}, \cite{Calabrese2}). From CFT, a lot is known also about the subleading corrections, which, in general, take the {\em unusual} form (\cite{calabrese2010}, \cite{camplostrini2010}, \cite{cardy2010} and \cite{essler2010})
\be
   {\cal S}_\alpha (\ell) = {c + \bar{c} \over 12} \left( 1 + {1 \over \alpha} \right) \ln {\ell \over a_0} + c'_\alpha
   + b_\alpha (\ell) \,\ell^{-2 h/\alpha} + \ldots \; ,
   \label{Sncriticalexp}
\ee
where $c$ is the central charge of the CFT, $\ell$ is the length of subsystem $A$, $a_0$ is a short distance cutoff, $c'_\alpha$ is a non-universal constants, $b_{\alpha}(\ell)$ is a periodic function of $\ell$  and $h$ is the scaling dimension of the operator responsible for the correction (relevant or irrelevant, but not marginal, since these operators generate a different kind of correction, which will be discussed later). This result is achieved using replicas, and thus, strictly speaking, requires $\alpha$ to be an integer. Moreover, it should be noted that in Ref. (\cite{cardy2010}) the corrections are obtained from dimensionality arguments, by regularizing divergent correlations by an ultra-violet cut-off $a_0$. Thus, technically, the subleading contributions in (\ref{Sncriticalexp}) are extracted from scaling properties and are all of the form $\ell/a_0$.

Determining the exponents of the corrections is important both in fitting numerics (where often really large $\ell$ are unobtainable) and also for a better understanding of the model. For instance, the scaling exponent $h$ also determines the large $n$ limit of the entropy ({\it single copy entanglement}) (\cite{calabrese2010}). Moreover, especially for $c=1$ theories, $h$ provides a measure of the compactification radius of the theory (\cite{essler2010}) and thus of the decaying of the correlation functions. Up to now, this conjecture has been checked in a variety of critical quantum spin chains models (\cite{fagotti}, \cite{alcaraz2011}, \cite{dalmonte2011}).

Moving away from a conformal point, in the gapped phase universality still holds for sufficiently small relevant perturbations. Simple scaling arguments guarantee that the leading terms survive, but with the correlation length $\xi$ replacing the infra-red length-scale $\ell$. Recent results, based on exactly solvable models, indicate the appearance of the same kind of unusual corrections to the Renyi
	 entropies, which are now functions of the correlation length $\xi$ with the same exponent $h$ (\cite{calabrese2010}):
\be
   {\cal S}_\alpha = {c \over 12} \left( {1 + \alpha \over \alpha} \right) \ln {\xi \over a_0} + A_\alpha
   + B_\alpha \xi^{-h/\alpha} + \ldots \; .
   \label{Sexp}
\ee
There is a factor of $2$ difference in each term between (\ref{Sncriticalexp}) and (\ref{Sexp}), due to the fact that the first is a bulk theory (with both chiralities in the CFT), while the latter is expected to be akin to a boundary theory, were only one chirality in the light-cone modes effectively survives.

In this chapter, we are going to investigate the subleading terms in the Renyi entropy of the one-dimensional XYZ model. In the scaling limit we find that the entropy is (modulo a multiplicative redefinition of the correlation length)
\bea
    \hskip-2.0cm {\cal S}_\alpha & = & {1 + \alpha \over 12 \alpha} \ln {\xi \over a_0} - {1 \over 2} \ln 2
    \label{Sxiexp}\\
    && -{1 \over 1 - \alpha} \sum_{n=1}^\infty \sigma_{-1} (n) \left[ \left( {\xi \over a_0} \right)^{-{2n \over \alpha}}
    - \left( {\xi \over a_0} \right)^{-{4n \over \alpha}} \right]
    \nonumber \\
    &&  + {\alpha \over 1 - \alpha} \sum_{n=1}^\infty \sigma_{-1} (n) \left[ \left( {\xi \over a_0} \right)^{-2n}
    - \left( {\xi \over a_0} \right)^{-4n} \right] \: ,
    \nonumber
\eea
where $\sigma_{-1} (n)$ is a {\it divisor function}, defined by (\ref{wndef}).

While the leading term correctly reproduces a $c=1$ central charge, the interpretation of the scaling exponents in the subleading addenda is less straightforward. Comparing the exponent of the first correction with (\ref{Sexp}) would indicate $h=2$, i.e. a marginal operator. This term cannot be due to the same operator acting at the critical point since, as shown in Ref. (\cite{cardy2010}), marginal operators give rise to logarithmic corrections. Moreover, the critical XXZ chain is known to have relevant fields with $h=K$ (the Luttinger parameter) and the opening of a gap in the XYZ model implies the presence of a $h <2$ operator. 
We will show that the operator content that can be extracted from (\ref{Sxiexp}) matches  that of a {\em bulk Ising} model  and the first correction can be interpreted as arising from the energy field. Notice that the leading, logarithmic term can thus be equally interpreted as $ {c + \bar{c} \over 12}$ with $c = \bar{c} = {1 \over 2}$.

	Furthermore, if we include also lattice effects, which vanish in the strict scaling limit, additional corrections appear in (\ref{Sxiexp}) and, while they are less important than the dominant one for sufficiently large $\alpha$, they can be relevant for numerical simulations in certain ranges of $\alpha$. These corrections turn out to be path dependent (probably due to the action of different operators) and many kind of terms can arise, such as $\xi^{-2h}$, $\xi^{-2h/\alpha}$, $\xi^{-(2-h)}$, $\xi^{-2(1+1/\alpha)}$ or even $1/\ln(\xi)$. In light of Ref.(\cite{cardy2010}), some of these terms were to be expected, but the others still lack a field theoretical interpretation, which might be possible by applying a reasoning similar to that of Ref.(\cite{cardy2010}) for a sine-Gordon model.

\section{Correlation length of the XYZ spin chain}

Consider the quantum spin-$\frac{1}{2}$ ferromagnetic XYZ chain, which is described by the Hamiltonian defined in (\ref{eq:XYZ1}). 
In Fig.~\ref{fig:phasediagram} we draw a cartoon of the phase diagram of the XYZ chain. As we have already seen, the model is symmetric under reflections along the diagonals in the $(J_z, J_y)$ plane. The system is gapped in the whole plane, except for six critical half-lines/segments: $J_z = \pm 1$, $|J_y| \le 1$; $J_y= \pm 1$, $|J_z| \le 1$ and $J_z = \pm J_y$, $|J_z| \ge 1$. All of these lines correspond to the paramagnetic phase of an XXZ chain, but with the anisotropy along different directions. Thus, in the scaling limit they are described by a $c=1$ CFT, with compactification radius varying along the line. We will use $0 \le \beta \le \sqrt{8 \pi}$ (the sine-Gordon parameter of the corresponding massive theory) to parametrize the radius. The critical segments meet three by three at four ``tricritical'' points, two of which are conformal, whilst the remaining are not (see the previous chapter for further details about this important distinction). 

\begin{figure}[t]
   \begin{center}
   \includegraphics[width=8cm]{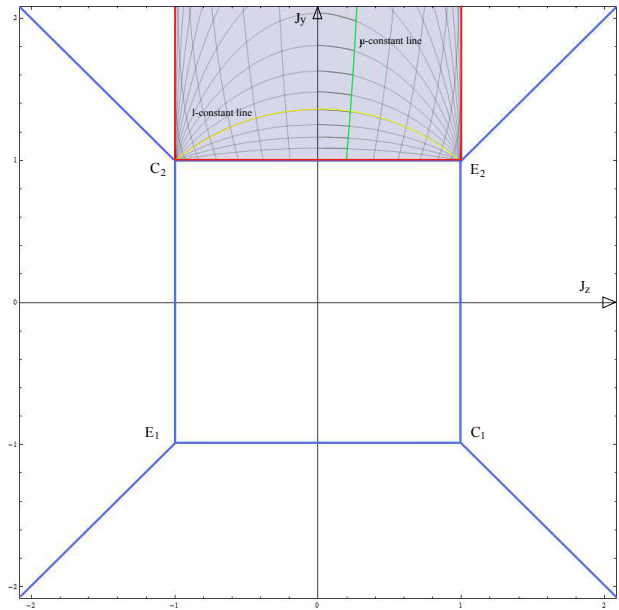}
   \end{center}
   \caption{(Colored online) Phase Diagram of the XYZ model in the $(J_z, J_y)$ plane. The blue solid lines --$J_z = \pm 1$, $|J_y| \le 1$; $J_y= \pm 1$, $|J_z| \le 1$ and $J_z = \pm J_y$, $|J_z| \ge 1$-- correspond to the critical phase of a rotated XXZ chain. Out of the four ``tricritical'' points, $C_{1,2}$ are conformal and $E_{1,2}$ are not. The area within the red rectangle is the portion of the phase diagram we study in this article, and it is completely equivalent to the principal regime of the eight-vertex model. The yellow and green lines are the curves of constant $l$ and $\mu$ respectively in the $(J_z,J_y)$ plane, according to the parametrization (\ref{ellipticparam}).}
   \label{fig:phasediagram}
\end{figure}
In studying the XYZ chain, we observed in (\ref{eq:XYZ4a}) that one can take advantage of the fact that (\ref{eq:XYZ1}) commutes with the transfer matrices of the the zero-field eight-vertex model (see for example Refs. (\cite{Sutherland} and  \cite{Baxter})) and thus the two systems can be solved simultaneously. The solution of the latter is achieved through the parametrization of $J_y, J_z$ in terms of elliptic functions
\begin{equation}\begin{array}{lcl}
   J_y & = & -\Delta \equiv \frac{\mbox{cn}(\ii \lambda)\;\mbox{dn}(\ii \lambda)}{1-k\;\mbox{sn}^{2}(\ii \lambda)} \; \\
   J_z & = & -\Gamma \equiv -\frac{1+k\;\mbox{sn}^{2}(\ii \lambda)}{1-k\;\mbox{sn}^{2}(\ii \lambda)} \; ,
   \label{eq:XYZ3bis}
	\end{array}
\end{equation}
where $(\Gamma,\Delta)$ are again the well-known Baxter parameters (\cite{Baxter}), and $\mbox{sn}(x)$, $\mbox{cn}(x)$ and $\mbox{dn}(x)$ are Jacobian elliptic functions of parameter $k$.
$\lambda$ and $k$ are parameters, whose natural domains are
\begin{equation}
   0<k<1 \; , \qquad 0 \le \lambda \le I(k') \; ,
   \label{eq:XYZ4}
\end{equation}
$I(k')$ being the complete elliptic integral of the first kind of argument $k' \equiv \sqrt{1-k^{2}}$.

The definition of $(\Delta,\Gamma)$ itself is particularly suitable to describe the anti-ferroelectric phase of the eight-vertex model (also referred to as the {\it principal regime}), corresponding to $\Delta\le-1$ and $|\Gamma|\le1$. However, using the symmetries of the model and the freedom under the rearrangement of parameters, it can be applied to the whole of the phase diagram of the spin Hamiltonian (for more details see Ref. (\cite{Baxter})). For the sake of simplicity, in this chapter we will focus only on the {\em rotated} principal regime: $J_y \ge 1$, $|J_z|\le 1$, see Fig.~\ref{fig:phasediagram}.

Before we proceed, it is more convenient to switch to an elliptic para\-me\-trization equivalent to (\ref{eq:XYZ3bis}), where we introduce a slightly different notation with respect to the one used in (\ref{landen}):
\be
   l \equiv {2 \sqrt{k} \over 1 + k}, \qquad
   \mu \equiv \pi {\lambda \over I(k')}=2u.
\ee
The elliptic parameter $l$ corresponds to a gnome $\tau \equiv \ii {I(l') \over I(l)} = \ii {I(k') \over 2 I(k)}$, which is half of the original, while $u$ was introduced in (\ref{landen}).  The relation between $k$ and $l$ is known as {\it Landen transformation}. Note that $0 \le \mu \le \pi$.

In terms of these new parameters, we have
\be
   \Gamma = {1 \over \dn [ 2 \ii I(l') \mu / \pi  ; l ] } \; , \qquad
   \Delta = - { \cn [2 \ii I(l') \mu / \pi ; l] \over \dn [2 \ii I(l') \mu / \pi ; l] } \; .
   \label{ellipticparam}
\ee

Curves of constant $l$ always run from the AFM Heisenberg point at  $\mu=0$ to the isotropic ferromagnetic point at $\mu= \pi$. For $l=1$ the curve coincides with one of the critical lines discussed above, while for $l=0$ the curve run away from the critical one to infinity and then back. In Fig.~\ref{fig:phasediagram} we draw these curves for some values of the parameters.

For later convenience, we also recall that
\be
  x \equiv \exp \left[ - \pi {\lambda \over 2 I (k)} \right] = \eu^{\ii \mu \tau} .
  \label{qxdef}
\ee
Using Jacobi's theta functions, we introduce the elliptic parameter $k_1$ connected to $x$, i.e.
\be
  k_1 \equiv {\theta_2^2 (0, x) \over \theta_3^2 (0,x)}
  =  {x^{1\over 2} \over 4} {(-1;x^2)^4_\infty \over (-x;x^2)^4_\infty}
  \equiv k (x)  \; ,
  \label{k1def}
\ee
or, equivalently, $\pi {I(k'_1) \over I(k_1)} = - \ii \mu \tau$ (i.e., $l$ is to $\tau$ what $k_1$ is to $\mu \tau / \pi$). In (\ref{k1def}) we also used the {\em q}-Pochhammer symbol
\be
   (a;q)_n \equiv \prod_{k=0}^{n-1} (1 - a q^k) \; .
\ee

The correlation length and the low-energy excitations of the XYZ chain were calculated in Ref. (\cite{McCoy}).
There are two types of excitations. The first can be characterized as free quasi-particles (spinons). The lowest band is a 2-parameter continuum with
\bea
  \Delta E_{\rm free} (q_1,q_2) & = & - J \: {\sn [2 I(l') \mu /\pi;l'] \over I(l) } \; I(k_1) \times \\
  \label{scaling7}
  && \hskip -1cm \left( \sqrt{1 - k_1^2 \cos^2 q_1} + \sqrt{1 - k_1^2 \cos^2 q_2} \right) \; .
  \nonumber
\eea
The energy minimum of these state is achieved for $q_{1,2}=0, \pm \pi$ and gives a mass gap
\bea
  \hskip -0.6cm \Delta E_{\rm free} & = & 2 J  {1 \over I(l)} \sn \left[ 2 I(l') {\mu  \over \pi} ; l' \right] \; I(k_1) k'_1 \; .
  \label{freegap}
\eea

For $\mu > \pi/2$, in addition to the free states just discussed, some bound states become progressively stable.
They are characterized by the following dispersion relation
\bea
  \label{scaling11}
  \Delta E_s (q) & = & - 2 J
  {\sn [2 I(l') \mu /\pi;l'] \over I(l) } \:
  { I(k_1) \over \sn(s y;k'_1)}
  \nonumber \\
  & \times & \sqrt{1 - \dn^2 (s y;k_{1}') \, \cos^2 {q \over 2}}
  \nonumber \\
  & \times & \sqrt{1 - \cn^2 (s y;k_{1}') \, \cos^2 {q \over 2}} \; ,
\eea
where $y \equiv \ii I(k_1) \tau \left( {\mu \over \pi} - 1 \right)$ and $s$ counts the number of quasi-momenta in the string state. In the scaling limit, these bound states become breathers.
The mass-gap for the bound states is (setting $q=0$ above)
\bea
  \Delta E_s & = & \Delta E_{\rm free} \: \sn (s y;k'_1) \; ,
  \label{boundgap}
\eea
from which one sees that for $\mu > \pi/2$ the $s=1$ bound state becomes the lightest excitation.

\begin{figure}[t]
\begin{center}
   \includegraphics[width=8cm]{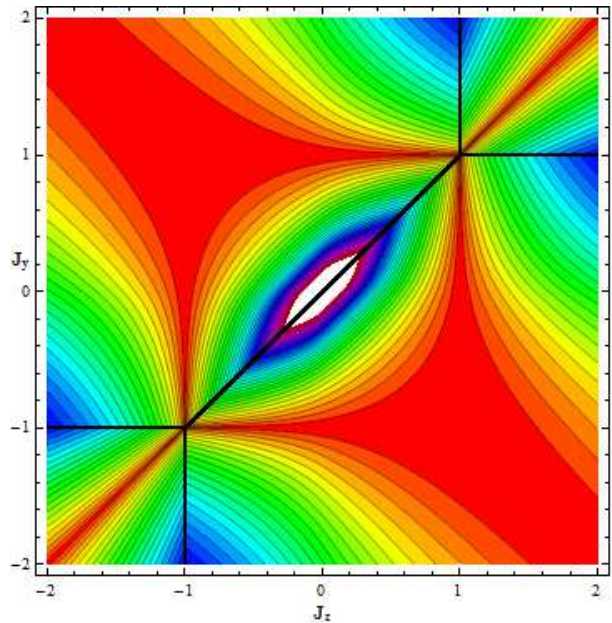}
   \caption{Curves of constant correlation length of the XYZ model in the ($J_z,J_y$) plane. Regions of similar colors correspond to the same  correlation length values and red colors are associated to higher values of $\xi$. It is worth stressing that the correlation length does not show any essential critical behavior, whilst the block entropy shows it in proximity of points E$_{1}$ and E$_{2}$.}
	\end{center}
   \label{fig:correlation_length}
\end{figure}

The correlation length for the XYZ chain (Fig.\ref{fig:correlation_length}) was also calculated in Ref. (\cite{McCoy}) and it is given by:
\be
   \xi^{-1} = {1 \over a_0} \left\{ \begin{array}{ll}
        -\frac{1}{2}\ln k_{2}
        & 0 \leq \mu \leq \frac{\pi}{2} \; ,\cr
        -\frac{1}{2} \ln \frac{k_{2}} {\dn^{2} \left[ \ii 2 I(k_{2}) {\tau \over \pi} \left( \mu  - { \pi \over 2} \right) ; k'_{2} \right]}
        & \frac{\pi}{2} < \mu \leq \pi \; , \cr
   \end{array} \right.
   \label{xibound}
\ee
where $a_0$ is a short distance cut-off, such as the lattice spacing, that sets the length unit and the new parameter $k_2$ is the Landen transformed of $k_1$:
\be
  k_2 \equiv k (x^2) = {1 - k'_1 \over 1 + k'_1} \; .
  \label{k2def}
\ee

The first behavior in (\ref{xibound}) is due to the free particles states, while the $s=1$ bound state is responsible for the second \footnote{In Ref. (\cite{McCoy}) the correlation length close to the critical line is calculated for periodic boundary conditions as that of the $s = 1$ bound state in the disordered phase of the 8-vertex model. In the ordered phase this state is forbidden by superselection rules and one has to use the $s = 2$ bound state (both of which become breathers in the scaling limit). The CTM construction implies fixed boundary conditions and thus we can use the $s = 1$ breather, as the lightest excitation in the attractive regime of the sine-Gordon line of the {\em XYZ} model.}.

\section{The reduce density matrix in terms of characters}

The bipartite entanglement entropy for the ground state of the XYZ chain was calculated in (\ref{eq:modular2}), in the limit where the infinite chain is partitioned in two (semi-infinite) half lines. For this configuration, the reduced density matrix can be computed as the product of the four corner transfer matrices (CTM) of the corresponding eight-vertex model (\cite{Peschel1}, \cite{Nishino1} and  \cite{Nishino2}). In (\ref{eq:renyi2}) it was shown that it can be written as (here for simplicity we drop the subscript $R$):
\be
   \rho = {1 \over {\cal Z}} \bigotimes_{j=1}^\infty
                \left( \begin{array}{cc}
                    1 & 0 \\
                    0 & x^{2j}
                \end{array} \right) \; ,
   \label{XYZrho}
\ee
where $\mathcal{Z} \equiv (-x^2,x^2)_\infty$, that is, the partition function of the eight-vertex model, is the normalization factor that ensures that $\Tr \rho =1$.
Thus we have
\be
   \label{eq:renyi3}
   \mbox{Tr} \rho^{\alpha} = { \left( -x^{2 \alpha}; x^{2 \alpha} \right)_\infty \over
   \left( -x^2;x^2 \right)^\alpha_\infty}
\ee
and for the R\'enyi
	 entropy
\be
   {\cal S}_{\alpha} =
   \frac{\alpha}{\alpha-1} \sum_{j=1}^{\infty} \ln \left ( 1 + x^{2 j} \right)
   + \frac{1}{1-\alpha} \sum_{j=1}^{\infty} \ln \left ( 1 + x^{2 j \alpha} \right) \; .
   \label{eq:renyi4}
\ee

The structure (Eqs. \ref{XYZrho}, \ref{eq:renyi4}) for the reduced density matrix of the half-line is common to all integrable, local spin-$1/2$ chains (\cite{Baxter}) and thus the entanglement spectrum of these models is the same and only depends on $x$, which in this context is usually parametrized as $x = \eu^{-\epsilon}$. For the XYZ chain, $\epsilon = - \ii \mu \tau$.

From (\ref{eq:renyi4}), one can see that the Renyi
	 entropy is a monotonically decreasing function of $\epsilon$:
\be
   \lim_{\epsilon \to 0} {\cal S}_\alpha = \infty \; , \qquad \qquad
   \lim_{\epsilon \to \infty} {\cal S}_\alpha = 0 \; .
\ee
Using (\ref{qxdef}) and (\cite{Baxter})
\be
   \label{ste8}
   l = \sqrt{ {1-\Gamma^{2} \over \Delta^{2}-\Gamma^{2} } } \;, \qquad
   \dn \left[ 2 \ii I(l') {\mu \over \pi} ;l \right] = {1\over \Gamma}
\ee
we can plot the entanglement entropy in the phase diagram of the XYZ model. In Fig \ref{fig:isoentropy} we show a contour plot of the Von Neumann entropy in the $(J_z,J_y)$ plane, from which one can clearly see the different behavior of the conformal and non-conformal points.

{\em q}-products of the form (\ref{eq:renyi3}) give easy access to the spectral distribution of the reduced density matrix (\cite{Franchini1}), since
\be
   (-q,q)_\infty = \prod_{k=1}^\infty (1 + q^k) = 1 + \sum_{n=1}^\infty p^{(1)} (n) q^n \; ,
   \label{partition1}
\ee
where $p^{(1)} (n)$ is the number of partitions of $n$ in distinct positive integers. Also note that, since
\be
  \prod_{k=1}^\infty (1 + q^k) = \prod_{k=1}^\infty (1 - q^{2k -1})^{-1} \; ,
  \label{prodrel}
\ee
$p^{(1)} (n) = p_{\cal O} (n)$, that is,  the number of partitions of $n$ into positive {\it odd} integers.

Moreover, one can recognize ${\cal Z}$ to be formally equal to the character of the spin field of the Ising CFT.
To show this, we write
\be
   {\cal Z} \left( q=x^2 \right)= \prod_{j=1}^\infty (1 + q^j)
   = \prod_{j=1}^\infty {1 - q^{2j} \over 1 - q^j}
\ee
and we use the Euler's formula for pentagonal numbers (which is a consequence of the Jacobi triple-product Identity) (\cite{Wittaker})
\bea
  \prod_{j=1}^\infty \left( 1 - q^{2j} \right)
  & = & \sum_{n=-\infty}^\infty (-1)^n q^{n(3n-1)}
  \label{pentagonal}
  \\
  & = & \sum_{n=-\infty}^\infty \left[ q^{2n(6n-1)} - q^{(2n+1)(6n+2)} \right]
   \nonumber
\eea
to recognize that
\be
   {\cal Z} = x^{-{1 \over 12}} \chi_{1,2}^{\rm Ising} \left( \ii \epsilon / \pi \right) \; ,
   \label{ZchiIsing}
\ee
where $ \chi_{1,2}^{\rm Ising} (\tau)$ is the character of the spin $h_{1,2} = 1/16$ operator of a $c=1/2$ CFT:
\bea
   \hskip-0.5cm \chi_{r,s}^{(p,p')} (\tau) & = &
   {q^{-1/24} \over \prod_{j=1}^\infty (1 - q^j)} \sum_ {n=-\infty}^\infty \left[
   q^{[2 p p' n + r p - s p']^2 \over 4 p p'} \right.
   \nonumber \\
   && \qquad \qquad \qquad \left.
   -  q^{[2 p p' n + r p + s p']^2 \over 4 p p'} \right] \; ,
   \label{characterdef}
\eea
with $q \equiv \eu^{2 \ii \pi \tau}$ and $(p,p')=(4,3)$ for the Ising minimal model.
We have
\be
   \mbox{Tr} \hat{\rho}^{\alpha} = { \chi_{1,2}^{\rm Ising} (\ii \alpha \epsilon / \pi) \over
   \left[ \chi_{1,2}^{\rm Ising} (\ii \epsilon / \pi) \right]^\alpha  }  \: .
   \label{isingrho}
\ee

As we discussed above, the critical line of the XXZ chain is approached for $l \to 1$, that is, for $\tau \to 0$ and $x \to 1$.
On this line, excitations are gapless and in the scaling limit the theory can be described by a conformal field theory with central charge $c=1$. To each $0<\mu<\pi$ it correspond a different point on the critical line, with sine-Gordon parameter $\beta^2 = 8 \pi \left( 1 - {\mu \over \pi} \right)$ (\cite{Luther}). The two endpoints are exceptions, since $\mu=0,\pi$ identify the same points for every $l$. However, while in the conformal one we still have $x \to 1$, close to the ferromagnetic point, around $\mu = \pi$, both $x$ and $q$ can take any value between $0$ and $1$. Hence a very different behavior of the entanglement entropy follows. (\cite{Ercolessi2011},  \cite{Franchini})

In order to study the asymptotic behavior of the entanglement entropy close to the conformal points, it is convenient to use the dual variable
\be
   \tx \equiv \eu^{- \ii {\pi^2 \over \mu \tau}} = \eu^{- {\pi^2 \over \epsilon}} \; .
   \label{qxprimedef}
\ee
which is such that  $\tx \to 0$ as $l \to 1$.

Expressions like (\ref{eq:renyi3}) involving q-products can be written in terms of elliptic theta functions as was done in (\ref{eq:modular2}). To study the conformal limit, one performs a modular transformation that switches $x \to \tx$.
Since
\be
   k (\tx) = k' (x) = k'_1 = { (x;x^2)^4_\infty \over (-x;x^2)^4_\infty } \; ,
\ee
and using (\ref{k1def}), we have
\bea
   \left( -x^{2 \alpha}; x^{2 \alpha} \right)_\infty & = &
   \left[ {k^2 \left( x^\alpha \right) \over 16 x^\alpha k' \left( x^\alpha \right) } \right]^{1/12}
   \nonumber \\
   = \left[ { k^{\prime 2} \left( \tx^{1/\alpha } \right)
   \over 16 x^\alpha k \left( \tx^{1/\alpha} \right) } \right]^{1/12}
   & = & { \left( \tx^{1/\alpha} ;  \tx^{2/\alpha} \right)_\infty \over
   2^{1/2} x^{\alpha / 12} \tx^{1/24\alpha} } \; .
\eea
Thus
\be
   \Tr \rho^\alpha = 2^{{\alpha -1 \over 2}} \tx^{\alpha^2 -1 \over 24\alpha}
   { \left( \tx^{1/\alpha};  \tx^{2/\alpha} \right)_\infty \over
   \left( \tx; \tx^2 \right)^\alpha_\infty } \; .
   \label{dualrho}
\ee

The modular transformation that allowed us to switch from $x$ to $\tx$ is the same one that connects characters in minimal model of inverse temperature. For the spin operator of the Ising model we have
\be
   \chi_{1,2}^{\rm Ising} (\tau) = {1 \over \sqrt{2}} \left[  \chi_{1,1}^{\rm Ising} (-1/ \tau)
   - \chi_{2,1}^{\rm Ising} (-1 / \tau) \right] \; .
\ee
Using (\ref{characterdef}) and the identities (\ref{prodrel} and  \ref{pentagonal}) one can prove that
\be
   \chi_{1,2}^{\rm Ising} \left( \ii \epsilon / \pi \right) = {1 \over \sqrt{2}} \tx^{-{1 \over 24}} \left( \tx; \tx^2 \right)_\infty \; ,
   \label{dualqprodZ}
\ee
which agrees with (\ref{dualrho}) and implies
\be
    \Tr \hat{\rho}^\alpha = 2^{\alpha -1 \over 2} { \chi_{1,1}^{\rm Ising} (\ii \pi / \alpha \epsilon)
    -  \chi_{2,1}^{\rm Ising} (\ii \pi / \alpha \epsilon)
    \over \left[  \chi_{1,1}^{\rm Ising} (\ii \pi / \epsilon)
    -  \chi_{2,1}^{\rm Ising} (\ii \pi / \epsilon) \right]^\alpha } \; .
    \label{dualrhochar}
\ee
This agrees with what conjectured in Ref.(\cite{calabrese2010} and \cite{Saleur}), but with the important difference that the characters in (\ref{dualrhochar}) are $c=1/2$ and do not belong to the infrared $c=1$ bulk description of the XYZ chain. This Ising character structure for the CTM of the eight-vertex model was already noticed, 1see Ref. (\cite{Cardy1}). We also notice that, being only a formal equivalence, eq. (\ref{ZchiIsing}) does not imply any underlying Virasoro algebra at work for CTM (as far as we know) and it is thus important to recognize that these manipulations stand on more general mathematical concepts.

\section{Expansion of the entanglement entropy}
\label{sec:expansion}

Close to the conformal points, Eq. (\ref{eq:renyi4}) is just a formal series, since $x \simeq 1$. However, using (\ref{dualrho}), it is straightforward to write a series expansion for the R\'enyi
	 entropy (\ref{eq:renyi1}) in powers of $\tx \ll 1$ (\cite{Evangelisti4}):
\bea
    \hskip-2.0cm {\cal S}_\alpha & = & - {1 + \alpha \over 24 \alpha} \ln \tx - {1 \over 2} \ln 2
    \label{Stxexp}\\
    && -{1 \over 1 - \alpha} \sum_{n=1}^\infty \sigma_{-1} (n) \left[ \tx^{n \over \alpha}
    - \alpha \tx^n - \tx^{2n \over \alpha} + \alpha \tx^{2n} \right] \: ,
    \nonumber
\eea
where the coefficients
\be
   \sigma_{-1} (n) \equiv {1 \over n} \sum_{\substack{ j < k=1 \\ j \cdot k =n}}^\infty (j + k)
   + \sum_{\substack{ j=1 \\ j^2 =n}}^\infty {1 \over j} = {\sigma_1 (n) \over n}
   \label{wndef}
\ee
is a divisor function (\cite{apostolbook}) and takes into account the expansion of the logarithm over a q-product and play a role similar to the partitions of integers in (\ref{partition1}).
It is worth noticing that the constant term $\ln(2^{-1/2})\equiv \ln(S_{1/16}^{0})$ - where $S_{1/16}^{0}$ is an element of the modular S-matrix of the Ising model - is the contribution to the entropy due to the boundary (\cite{Affleck}).

The $\alpha \to 1$ yields the Von Neumann entropy:
\bea
    \hskip-2.0cm {\cal S} & = & - {1 \over 12} \ln \tx - {1 \over 2} \ln 2
    \label{SVNtxexp}\\
    && - \sum_{n=1}^\infty \sigma_{-1} (n) \left[ n \left(  \tx^n - 2 \tx^{2n} \right) \ln \tx
    + \tx^{n} - \tx^{2n} \right] \: .
    \nonumber
\eea
We see that, contrary to what happens for $\alpha>1$, all subleading  terms -which are powers of $\tx$- acquire a logarithmic correction, which strictly vanishes only at the critical points.

\subsection{Scaling limit}

Comparing with (\ref{Sexp}) and coherently with (\ref{isingrho}), (\ref{Stxexp}) and (\ref{SVNtxexp}) can be identified with the expansion of a $c=1/2$ theory. However, the parameter of this expansion, $\tx$, has meaning only within Baxter's parametrization of the model (\ref{eq:XYZ3bis}). To gain generality, the entropy is normally measured as a function of a {\it universal} parameter, such as the {\it correlation length} or the {\it mass gap}.

In the scaling limit, up to a multiplicative constant, one has:
\be
   \tx \approx \left( {\xi \over a_0} \right)^{-2} \approx \left( {\Delta E \over J} \right)^2 \; ,
   \label{xileading}
\ee
so that, substituting this into (\ref{Stxexp}), we get (\ref{Sxiexp}). Relation (\ref{xileading}) is crucial in turning the leading coefficient in the entropy of  a $c=1/2$ entropy such as (\ref{Stxexp}) into that of a $c=1$ theory, but it also doubles all the exponents of the subdominant corrections.
It is reasonable to assume that this $c=1$ model is some sort of double Ising, but its operator content does not seem to match any reasonable $c=1$ model, since only even exponent states exist.
Moreover, comparing (\ref{Sxiexp}) with (\ref{Sexp}), one would conclude that a $h=2$ operator is responsible for the first correction. It was argued in Ref. (\cite{cardy2010}) that a marginal field gives rise to logarithmic corrections in the entropy, thus, either this corrections is due to descendant of the {\it identity} (namely, the stress-energy tensor), or we should think of it as a $2h/\alpha$, with $h=1$.

In fact, we can write the partition function of the eight-vertex model as a bulk Ising model (i.e., quadratic in characters). Starting from (\ref{dualqprodZ}), we have
\bea
  \hskip -.5cm  {\cal Z} & = & {1 \over \sqrt{2}} x^{-{1 \over 12}} \xi^{{1 \over 12}}
  \prod_{k=1}^\infty \left( 1 - \xi^{1-2k} \right) \left( 1 + \xi^{1-2k} \right)
  \nonumber \\
  & = & {x^{-{1 \over 12}} \over \sqrt{2}} \left[ \chi^{\rm Ising}_{1,1} \left( {\ii \over \pi} \ln \xi \right)
  + \chi^{\rm Ising}_{2,1} \left( {\ii \over \pi} \ln \xi \right) \right]
  \nonumber \\
  && \qquad \times
  \left[ \bar{\chi}^{\rm Ising}_{1,1} \left( {\ii \over \pi} \ln \xi \right)
  - \bar{\chi}^{\rm Ising}_{2,1} \left( {\ii \over \pi} \ln \xi \right) \right]
  \label{bulkIsing} \\
  & = & {x^{-{1 \over 12}} \over \sqrt{2}} \left[ \left| \chi_{0}  \right|^2 - \left| \chi_{1/2} \right|^2
  - \chi_0 \bar{\chi}_{1/2} + \chi_{1/2} \bar{\chi}_0 \right] \; .
  \nonumber
\eea
This formulation provides a simple explanation of the Renyi entropy expansion (\ref{Sxiexp}) and its operator content. In fact,  it interprets the first correction as the Ising energy operator, and not as a descendant of the identity.\\
It also means that the prefactor in front of the logarithm in the entropy can be interpreted as ${c + \bar{c} \over 12}$ with $c = \bar{c} = {1 \over 2}$.

No fundamental reason is known for which CTM spectra (and partition functions) of integrable models can be written as characters in terms of the mass parameter $x$ or $\tilde{x}$. This is the case  also for Eq. (\ref{bulkIsing}).  Thus, so far, we can only bring forth this observation while any connection with some underlying Virasoro algebra remains to be discovered. To the contrary, sufficiently close to a critical point, the CTM construction can be seen as a boundary CFT and thus its character structure as function of the size of the system is dictated by the neighboring fix point. (\cite{Cardy1})

We are led to conclude that, in the scaling limit, the entropy can be written as a function of two variables: ${\cal S}_\alpha \left( {\ell \over a_0}, {\xi \over a_0} \right)$. When the inverse mass is larger that the subsystem size, we have the usual expansion of the form (\ref{Sncriticalexp}). But when the correlation length becomes the infrared cut-off scale, apparently a different expansion is possible, which, unlike (\ref{Sexp}), can contain terms with different exponents, like in (\ref{Sxiexp}). Hence, while the leading universal behavior has always the same numerical value and scales like the logarithm of the relevant infra-red scale, the exponents of the corrections might be in principle different for terms in $\ell$ and in $\xi$.

In the scaling limit, $a_0 \to 0$, $J \to \infty$, and $\tx \to 0$ in such a way to keep physical quantities finite. In this limit, only the scaling relation (\ref{xileading}) survives. However, at any finite lattice spacing $a_0$, there will be corrections which feed back into the entropy and that can be relevant for numerical simulations. To discuss these subleading terms we have to consider two regimes separately.

\subsection{Free Excitations: $0\leq \mu \leq \frac{\pi}{2}$}

For $0\leq \mu \leq \frac{\pi}{2}$ the lowest energy states are free, with dispersion relation (\ref{scaling7}).
To express the entropy as a function of the correlation length we need to invert (\ref{xibound}). This cannot be done in closed form. So we have to first expand (\ref{xibound}), finding
\bea
   {a_0 \over \xi } & = & 4 \tx^{1/2} \sum_{n=0}^\infty \sigma_{-1} (2n+1) \: \tx^{n}
   \nonumber \\
   & = & 4 \tx^{1/2} + {16 \over 3} \tx^{3/2} +  {24 \over 5} \tx^{5/2} + \ldots
   \label{xifree1}
\eea
and  then to invert this (by hand) to the desired order:
\be
   \tx = {1 \over 16} {a_0^2 \over \xi^2} \left[ 1 - {1 \over 6} \: {a_0^2 \over \xi^2}
   + {7 \over 144} \: {a_0^4 \over \xi^4} + \Ord \left( \xi^{-6} \right) \right] \: .
	\label{eq:inv}
\ee
We get:
\bea
   {\cal S}_\alpha & = & {1 + \alpha \over 12 \alpha} \ln {\xi \over a_0}
   + {1 - 2 \alpha \over 6 \alpha} \ln 2
   \nonumber \\
   && + B_\alpha \xi^{-{2 \over \alpha}} + C_\alpha \xi^{-2{ 1 + \alpha \over \alpha}}
   + B'_\alpha \xi^{-{4 \over \alpha}} + \ldots
   \nonumber \\
   && - \alpha B_\alpha \xi^{-2} - \alpha B'_\alpha \xi^{-4} + \ldots
   \label{Sfreeexp}
\eea
where the coefficients only depend on $\alpha$ and contain the proper power of $a_0$ to keep each term dimensionless [for instance $B_\alpha= {1 \over \alpha -1} \left( {a_0 \over 4} \right)^{2/\alpha}$].
We note that a new term has appeared (and more will be seen at higher orders) and that it is not of the forms discussed in Ref.(\cite{camplostrini2010}).

It is worth noticing that if we express the mass-gap (\ref{freegap}) as function of $\tx$, we would get a different series expansion and thus different corrections to  the entropy. These subleading terms would have a different form, compared to (\ref{Sfreeexp}), and even be path dependent on how one approaches the critical point. We prefer not to dwell into these details now, postponing the description of this kind of path-depended behavior with the bound state's correlation length to the next section.

\subsection{Bound states: $\frac{\pi}{2} < \mu < \pi $}

For $\mu > \pi/2$, bound states become stable, and the lightest excitation becomes the $s=1$ state with dispersion relation (\ref{scaling11}). Accordingly, the expression for the correlation length is different in this region from before. Using the dual variable $\tx$ and the formulation of elliptic functions as infinite products, we can write (\ref{xibound}) as
\bea
   \hskip-2cm{a_0 \over \xi } & = & \ln { \left( - \tx^{1+ \tau \over 2}; \tx \right)_\infty \left( - \tx^{1- \tau \over 2}; \tx \right)_\infty \over
   \left( \tx^{1+ \tau \over 2}; \tx \right)_\infty \left( \tx^{1- \tau \over 2}; \tx \right)_\infty }
   \label{xibound1} \\
   & = & 4 \sum_{n=1}^\infty \sum_{k=1}^\infty {1 \over 2k -1} \cos \left[ {\pi^2 \over 2 \mu} (2k-1) \right] \tx^{(2n-1)(2k-1) \over 2} \:.
   \nonumber
\eea

\begin{figure}[t]
   \begin{center}
   \includegraphics[width=8cm]{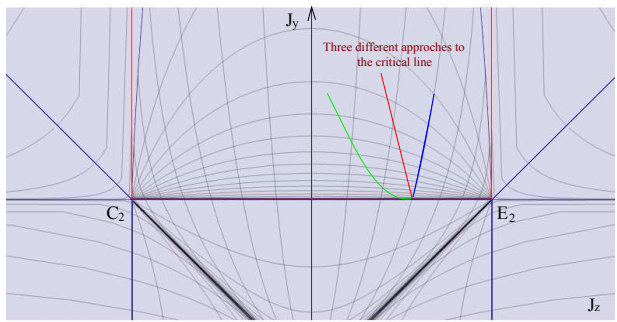}
   \end{center}
   \caption{Three different ways to approach the critical line: along the $\mu$-constant lines (blue line), which is  referred to as renormalization group flow in the text; along straight lines (red line); along lines that approach the criticality with zero derivative (green line).}
   \label{fig:approching}
\end{figure}

The major difference in (\ref{xibound1}) compared to (\ref{xifree1}) is that the bound state correlation length does not depend on $\tx$ alone, but separately on $\mu$ and $\tau$. This means that in inverting (\ref{xibound1}) to find $\tx$ as a function of $\xi$, we have to first specify a relation between $\mu$ and $\tau$, i.e. to choose a path of approach to the critical line. We will follow three different paths, that are shown in  Fig.~\ref{fig:approching}.

\begin{enumerate}[(a)]

\item Renormalization group flow: The first natural path is (represented in the blue line in Fig.~\ref{fig:approching}):
\be
   \tau=\ii s \; ,  \qquad  \qquad
   \mu = \mu_0 \; ,
  \label{scaling12}
\ee
where $s \to 0$ guides our approach to the gapless point. This path, keeping $\mu$ fixed, corresponds to the RG flow. In the scaling limit, the XYZ chain is described by a sine-Gordon model, where $\mu$ is proportional to the compactification radius (\cite{Luther}). Thus, assuming (\ref{scaling12}) means changing the bare mass scale, without touching $\beta$. In the $(J_z,J_y)$ plane, this path asymptotically crosses the critical line with slope $m = -2 / \cos \mu_0$.
Substituting this in (\ref{xibound1}) we get:
\be
  \label{scaling13}
   {a_0 \over \xi} =  4 g (\mu_0) \: \tx^{1/2} + {16 \over 3} g^3 (\mu_0) \: \tx^{3/2} +\Ord \left( \tx^{5/2} \right)  \; ,
\ee
where $ g(\mu) \equiv \cos {\pi^2 \over 2 \mu }$. Comparing (\ref{scaling13}) with the free case (\ref{xifree1}) we immediately conclude that the entropy retains an expansion similar to (\ref{Sfreeexp}), with the difference that all the coefficients now depend on $\mu_0$ and thus change along the critical line:
\bea
  \hskip-1cm{\cal S}_{\alpha} & \simeq & \frac{1 + \alpha}{12 \alpha} \ln \xi  + A_{\alpha} (\mu_0) + B_\alpha (\mu_0) \xi^{-{2 \over \alpha}}
  \nonumber \\
  && - \alpha B_\alpha (\mu_0) \xi^{-2} + C_\alpha (\mu_0) \xi^{-2 - { 2 \over \alpha}} + \ldots
  \label{scaling14}
\eea

\item  Straight lines in $(J_z,J_y)$ space: Let us now approach a conformal critical point exactly linearly in the $(J_z,J_y)$ plane:
\be
   \label{scaling24}
   J_y = 1+m \cdot s \; , \qquad \qquad
   J_z = s - \cos \mu_0  \; .
\ee
This path corresponds to the following parametrization of $\tau$ and $\mu$ (an example of which is the red line in Fig.~\ref{fig:approching}):
\be
   \label{scaling31}
   \begin{cases}
   \tau = -\ii \frac{\pi}{\ln(s)} + \Ord \left( \frac{1}{\ln^{2} s }\right) \; , \\
   \mu = \mu_0 + r(m,\mu_0) \cdot s + \Ord (s^{2}) \; ,
   \end{cases}
\ee
where $ r (m,\mu) \equiv {2 + m \cos \mu \over 2 \sin \mu }$.
Thus, in the limit $s \to 0$, the entropy parameter $\tx$ vanishes like $\tx \propto s^{\pi/\mu_0}$.
Using (\ref{scaling31}) in (\ref{xibound1})
\be
  \label{scaling34}
  {a_0 \over \xi } \simeq  4 g ( \mu_0 )  \tx^{1 \over 2}
  +  4 r (m,\mu_0)  g' (\mu_0) \tx^{\frac{1}{2}+\frac{\mu_0}{\pi}} + \ldots
\ee
Inverting this relation, we arrive at the following expansion of the R\'enyi
	 entropy along (\ref{scaling24})
\bea
   \label{scaling36}
   \hskip-2cm{\cal S}_{\alpha} & \simeq & \frac{1 + \alpha}{12 \alpha} \ln \xi + A_{\alpha} (\mu_0) \\
   & & + B_{\alpha} (\mu_0) \xi^{-2/\alpha}  + D_{\alpha} (m,\mu_0) \xi^{-2\mu_0/\pi} \ldots
   \nonumber
\eea
We notice that the last term yields a new type of correction, with a non-constant exponent, which varies with $\mu$. This term is of the form $\xi^{-(2-h)}$, where $h$ here is the scaling dimension of the vertex operator ${\rm e}^{\ii\beta\phi}$ of the underlying sine-Gordon theory. To the best of our knowledge, this is the first time that such a correction in the Renyi entropy of gapped systems is discussed and it also differs from those discussed in Ref. (\cite{camplostrini2010}). It is surprising to see the appearance of the operator content of the underlying sine-Gordon field theory in the exponents of the expansion of $\mathcal{S}_{\alpha}$ on the lattice, whilst these operators do not enter in the scaling limit (\ref{Sxiexp}). We do not have a satisfactory understanding of this result, but we believe that an approach similar to that of Ref.(\cite{cardy2010}) might clarify the point. 

Since $\frac{\pi}{2}< \mu_0 < \pi$, the exponent of this new correction ranges between $1$ and $2$ and always dominates over the $\xi^{-2}$ term in (\ref{scaling14}) and competes with the correction $\xi^{-2/\alpha}$ for $\alpha < 2$.
Notice, however, that for $m_0 = -2 / \cos \mu_0 $, $r (m_0,\mu_0) =0$ and thus $D_\alpha (m_0,\mu_0) =0$: this new correction disappears. This is precisely the slope that corresponds to the RG flow we considered before. Also, $g'(\pi/2) = 0$ and thus the coefficient in front of this new correction vanishes at the crossing point between the free and the bound state. Thus, we can conclude that this new term is turned on by the existence of a bound state and it is a clear signature of a truly interacting theory. Moreover, the path one chooses to approach criticality selects a scaling limit in which irrelevant operators can be generated and these can modify the perturbative series that defines the correlation length, leading to something like we observed.

\item  Straight lines in $(l,\mu)$ space: Finally, as a generalization of the first case, let us consider the straight line
\be
   \label{scaling18}
   \tau=\ii s \: , \qquad \qquad
   \mu = \mu_0 + r \cdot s \: ,
\ee
where $r$ is the slope in the $(\tau,\mu)$ plane. Interestingly, this trajectory maps into a curve in the $(J_z,J_y)$-plane approaching the critical point with zero derivative, i.e. with a purely quadratic relation in a small enough neighborhood of the conformal point (see for instance the green line in Fig.~\ref{fig:approching}).
Putting (\ref{scaling18}) in (\ref{xibound1}) we have
\bea
   \hskip -1.5cm{a_0 \over \xi } & \hskip-0.2cm= & \hskip-0.2cm 4 g (\mu_0) \tx^{1/2}
   +  4 r {\textstyle {\pi^2 \over \mu_0 }}  g' ( \mu_0)  {\tx^{1/2} \over \ln \tx}
   \nonumber \\
   && + {16 \over 3} g^3 ( \mu_0) \tx^{3/2} {\textstyle + \Ord \left( {\tx^{1/2} \over \ln^2 \tx},  {\tx^{3/2} \over \ln \tx} \right)  } \; .
   \label{scaling21}
\eea
We notice the appearance of a strange logarithmic correction in the expansion. Inverting (\ref{scaling21}) and plugging it into the entropy we get
\be
  \label{scaling23}
  {\cal S}_{\alpha} = \frac{1 + \alpha}{12 \alpha} \ln \xi  + A_\alpha (\mu) +\frac{E_{\alpha}(r,u)}{\ln \xi } + \ldots \: ,
\ee
This kind of logarithmic corrections were found also in Ref. (\cite{calabrese2010}), but only in studying the limit $\alpha \to\infty$, i.e. the so-called single-copy entropy. They were also predicted in Ref. (\cite{camplostrini2010}) as signatures of marginal operators in a CFT, but with a different power. From our results, it seems that this kind of very unusual corrections may appear at finite $\alpha$'s, by selecting a proper path  approaching the critical point. This can be due to the presence of a marginally-relevant operator in the theory, generated when reaching the critical line through a zero-slope curve and corresponding to a changing of the compactification radius.
\end{enumerate}

\section{Conclusions}

In this chapetr by using the example of the integrable XYZ chain, we proved that, for a massive model, the study of the corrections to the entanglement entropy as a function of the correlation length requires a separate analysis from the one that yields the entropy as function of the subsystem size.

For the bipartite Renyi entropy of the XYZ model of a semi-infinite half-line we found, in the scaling limit and as a function of the correlation length, the universal form (\ref{Sxiexp}), where all subleading contributions are explicitly written, thanks to a novel formulation of the reduced density matrix in terms of {\em q}-products. We argued that these corrections could be interpreted in light of a previously unnoticed bulk Ising structure of the CTM formulation of the model. This means that corrections as a function of the correlation length have different exponents compared to those depending on the length of the subsystem, unlike what was expected from previous studies. This also implies that the coefficient ${c + \bar{c} \over 12}$ of the logarithmic leading term has the same value both for $c = \bar{c} = 1/2$ of the bulk Ising formulation in the mass parameter and for the $c=1$, $\bar{c}=0$ of the critical chiral free boson model in the subsystem size.

In this respect, it is also interesting to note that the reduced density matrix $\hat{\rho}$ of (\ref{XYZrho}) can be written as (\cite{Peschel2}, \cite{Peschel4},  \cite{Peschel5} and  \cite{Peschel1}):
\be
\label{thermal}
\hat{\rho}\propto {\rm e}^{-H_{{\rm CTM}}}, \qquad H_{{\rm CTM}}=\sum_{j=1}^{\infty}2\epsilon_{j}\eta_{j}^{\dag}\eta_{j}
\ee
where $(\eta_{j}^{\dag},\eta_{j})$ are (Majorana) fermionic creation and annihilation operators for single particle states with eigenvalue $2\epsilon_{j}=2j\epsilon$ (note that $H_{{\rm CTM}}$ is not the Hamiltonian of the subsystem $A$).
This representation strongly supports the interpretation that the $c=1$ theory is constructed in terms of $c=1/2$ (Majorana) characters.

In Refs.(\cite{camplostrini2010}  and \cite{alcaraz2011}) it was shown that the first correction in the Renyi entropy of a critical XXZ chain as a function of the subsystem size $\ell$ goes like $\ell^{-2 K / \alpha}$, where $K$ is the Luttinger parameter of the model. This fact has also been checked in many other critical $c=1$ quantum spin chain models via DMRG simulations (\cite{fagotti}, 	\cite{alcaraz2011} \cite{dalmonte2011}).
	When going to  the corresponding massive model, assuming that the operators responsible for the corrections remain the same, the simple scaling prescription (\cite{calabrese2010}) would give a term of the type $\xi^{-K/\alpha}$. The results presented here would then indicate an improbable fixed value for the Luttinger parameter $K=2$. This exponent for the massive XXZ chain was observed before (take, for instance, Ref. \cite{calabrese2010}), but its nature has not been discussed.
	Instead, consistently with Ref. (\cite{alcaraz2011}), we found in Sec. \ref{sec:expansion} that the operator responsible for this correction is the energy of the underlying bulk Ising model.
	As pointed out in Ref.(\cite{cardy2010}), the leading correction in $\ell$ is of the form $\ell^{-2K/\alpha}$ in the one-interval case and $\ell^{-K/\alpha}$ for the half-line, whereas for two intervals, in the Ising case, the exponent acquires an additional factor of $2$, which counts the number of twist fields at the edge of the interval (\cite{fagotticala}). It would be interesting to perform a calculation for one interval with two boundary points in our case too, to check whether a doubling of the exponent in the correlation length would happen in this case as well.
	We also observe that our results stem from the study of the formally conformal  structure emerging in the Renyi entropy of the ground state. It would be of great interest to perform a similar analysis for some excited states.

We also showed that, if one takes into account lattice effects, there is a proliferation of new standard and {\em unusual} corrections, which in general are path-dependent, and can assume the forms: $\xi^{-2h}$, $\xi^{-2h/\alpha}$, $\xi^{-(2-h)}$, $\xi^{-2(1+1/\alpha)}$, or even $1/\ln(\xi)$, where $h$ is the scaling dimension of a relevant operator of the critical bulk theory.
We conjecture that the last logarithmic correction is a consequence of some marginally-relevant operator in the theory.
In addition each of the previous terms appears multiplied by a point-dependent and $\alpha$-dependent coefficient that can vanish in even large parametric regions, preventing the corresponding unusual correction from showing up in the scaling limit of the Renyi entropy.
We should remark that the same analysis, carried out in terms of the mass-gap instead of the correlation length, would generate even different corrections. These differences will be analyzed elsewhere for the XYZ chain, but are a general feature of lattice models that have to be taken into account in numerical analysis.

In Ref. (\cite{camplostrini2010}) it was given a list of possible subleading contributions to the Renyi entropy for a CFT. From a na\"ive scaling argument, one could expect the same kind of terms to appear in a massive theory sufficiently close to criticality. However, some of the corrections we observed do not fit this expectation. This could be due to strictly ultraviolet effects that cannot be captured by a QFT or to a need to improve the scaling argument. We believe that an analysis similar to that carried out for critical systems in Ref. (\cite{cardy2010}) could be successfully applied to a massive sine-Gordon theory, by introducing an ultra-violet cut-off (the lattice spacing) to regularize divergent integrals and extract the corrections from the counterterms\footnote{An interesting alternative field theoretical approach to massive models has been presented in: \\
	J.L. Cardy, O.A. Castro-Alvaredo, and B. Doyon, J. Stat. Phys. \textbf{130} (2007) 129,  O.A. Castro-Alvaredo, and B. Doyon, J. Phys. A \textbf{41}  (2008) 275203, B. Doyon, Phys. Rev. Lett. \textbf{102} (2009) 031602.} .
This type of check would be dual to  what we have presented here: while in our approach we started from a lattice model to infer its universal behavior, in the other, one would start from a field theory to understand the origin of the correction. In conclusion, we believe that further analysis are needed to provide a consistent field theoretical interpretation of the corrections arising in massive models, especially for the relevancy of such problem in numerical studies.



\chapter{The Ising chain in a transverse field}
\label{chap_ising}
\ifpdf
    \graphicspath{{Chapter3/Chapter3Figs/PNG/}{Chapter3/Chapter3Figs/PDF/}{Chapter3/Chapter3Figs/}}
\else
    \graphicspath{{Chapter3/Chapter3Figs/EPS/}{Chapter3/Chapter3Figs/}}
\fi
In this chapter we will review the quantum properties of the transverse field Ising chain (TFIC). The idea is that of providing the reader with a general description of this quantum model and its main features, being the TFIC a crucial paradigm for quantum critical behaviour. Here we will focus on its properties at equilibrium, both at zero and non-zero temperature. We will see how different techniques can be used to compute correlators in different regimes, with emphasis on the semi-classical method, which will be the main subject of the second part of this thesis. In fact, in the next chapter we will generalize this simple but at the same time powerful technique to the out-of-equilibrium case. \\
The introduction to the TFIC presented here has been inspired by chapter $4$ of the book  ``{\em Quantum phase transitions}'' (first edition) written by Subir Sachdev (see also ``{\em Dynamics and transport near quantum-critical points}'', Subir Sachdev, Proceedings of the Nato Advanced Study Institute on Dynamical Properties of Unconventional Magnetic Systems, Geilo, Norway, April 1997). The semi-classical approach presented here was first introduced by S. Sachdev and A.P. Young in the article ``{\em Low temperature relaxation dynamics of the Ising chain in transverse field}'', Physical Review Letters, {\bf 78}, 2220 (1997). 

\section{General properties}
\markboth{\MakeUppercase{\thechapter. The Ising chain in a transverse field}}{\thechapter. The Ising chain in a transverse field}

Let us consider the TFIC Hamiltonian in one dimension:
\be
\label{chap4_eq:1}
H_{\rm I}=-J{\displaystyle\sum_{j}^{N}[\sigma_{j}^{z}\sigma_{j+1}^{z}+g\sigma_{j}^{x}]}.
\ee
As we have discussed in the chapter \ref{cap:2} and will establish more rigorously in this chapter, $H_{\rm I}$ exhibits a phase transition at zero temperature between an ordered phase with the $Z_{2}$ symmetry broken and a quantum paramagnetic phase with a unique ground state where the symmetry remains unbroken. The universality class of this quantum phase transition is that of the 2D classical Ising model, thanks to the quantum-classical mapping between these two models. \\
An interesting quantity to study is the dynamic two-point correlations of the order parameter $\sigma^{z}$, which is defined as:
\be
\label{chap4_eq:2}
\begin{array}{lcl}
C(x_{i},t) & \equiv &  \langle\sigma^{z}(x_{i},t)\sigma(0,0)\rangle\\
& = & {\rm Tr}\left({\it e}^{-H_{\rm I}/T}\,{\rm e}^{i H_{\rm I}t}\,\sigma_{x_{i}}^{z}\,{\it e}^{-i H_{\rm I}t}\,\sigma_{0}^{z}\right)/{\cal Z},
\end{array}
\ee
where ${\cal Z}={\rm Tr}({\it e}^{-H_{\rm I}/T})$ is the partition function, $x_{i}=ia$ is the spatial coordinate of the $i$th spin with $a$ the lattice spacing. Here $t$ is the real physical time. It is sometimes convenient to consider correlations at imaginary time $\tau$, which is defined by the analytical continuation $it \to \tau$:
\be
\label{chap4_eq:3}
C(x_{i}, \tau)= {\rm Tr}\left({\it e}^{-H_{\rm I}/T}\,{\it e}^{H_{\rm I}\tau}\,\sigma_{x_{i}}^{z}\,{\it e}^{- H_{\rm I}\tau}\,\sigma_{0}^{z}/{\cal Z}\right),
\ee
in order to think of it as the correlator of the classical 2D Ising model on an infinite strip of width $1/T$ and periodic boundary conditions along the {\em imaginary} time direction. Another important quantity we can consider is the {\em dynamic structure factor} $S(k,w)$, which is simply defined as the Fourier trasform of $C(x,t)$, namely:
\be
\label{chap4_eq:4}
S(k,w)=\int dx\int dt\,C(x,t)\,{\it e}^{-i(kx-wt)}.
\ee
This quantity happens to be particularly useful because it is directly proportional to the cross section in scattering experiments in which the probe couples to $\sigma^{z}$. If we integrate over the energy we have that the cross section is proportional to the {\em equal-time} structure factor $S(k)$, namely:
\be
\label{chap4_eq:5}
S(k)=\int \frac{dw}{2\pi} S(k,w),
\ee
which is also the Fourier transform of $C(x,0)$\footnote{The identity $(\sigma_{i}^{z})^{2}=1$ implies that $C(0,0)=1$ and consequently leads to the following sum rule for the dynamic structure factor: $\int\frac{dkdw}{(2\pi)^2}S(k,w)=1$}. Finally another quantity that it is worth recalling is the {\em dynamic susceptibily} $\chi(k,w_{n})$, which can be computed as:
\be
\label{chap4_eq:5}
\chi(k,w_{n})=\int_{0}^{1/T}\,d\tau\,\int dx\,C(x,\tau)\,{\it e}^{-i(kx-w_{n}\tau)},
\ee 
where $w_{n}=2\pi nT$, with $n$ integer, is the Matsubara imaginary frequency. The analytical continuation to real frequencies may be done by $iw_{n}\to w + i\delta$, where $\delta$ is a positive and infinitesimal. The dynamic susceptibility measures the response of the magnetization $\sigma^{z}$ to an external field that couples linearly to $\sigma^{z}$. From (\ref{chap4_eq:5}) and (\ref{chap4_eq:4}) it is pretty clear that there is a relationship between $S(k,w)$ and $\chi(k,w)$, called {\em fluctuation-dissipation} theorem, and it will be discussed in detail in chapter (\ref{chap_final}).
We will begin this chapter by describing a simple physical picture of the ground state of the TFIC by examining the large- and small-$g$ limits, where two very different physical scenarios emerge. The exact solution, which will be discussed later, shows that there is a critical point at $g=1$ but that the qualitative features of the ground state in the two limiting cases describe very well the properties of both phases for any $g\neq 1$. Only at the critical point $g=1$ the system  has real different properties at $T=0$. 

\subsection{Strong coupling}
\label{sec:strong}
\begin{figure}[t]
   \begin{center}
   \includegraphics[width=8cm]{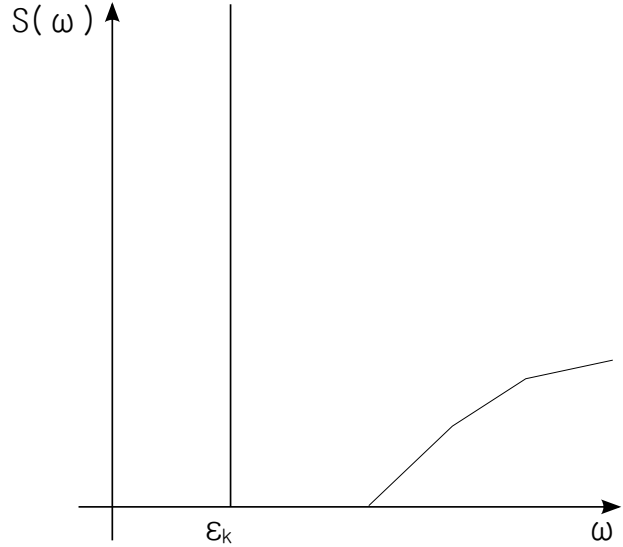}
   \end{center}
   \caption{Sketch of the dynamic structure factor $S(k,w)$ of $H_{\rm I}$ as a function of $w$ at zero temperature and small momentum. A quasiparticle delta function at $w=\epsilon_{k}$ and a three-particle continuum at higher frequencies are shown in the picture. There are additional $n$-particle continua for $n\geq 5$ and odd, which are not displayed here.}
   \label{chap4_fig:1}
\end{figure}

In the $g\to\infty$ limit the ground state of the TFIC is a quantum paramagnet, invariant under the $Z_2$ symmetry $\sigma_{i}^{z} \to -\sigma_{i}^{z}$, with exponentially decaying $\sigma_{z}$ correlators (\cite{Sachdev1}). In this limit we can also list the exact eigenstates. The lowest excited states are:
\be
\label{chap4_eq:6}
|i\rangle = |\leftarrow\rangle_{i}{\displaystyle\prod_{j\neq i}}|\rightarrow\rangle_{j},
\ee 
which is obtained by flipping the state on site $i$ to the other direction on $\sigma^{x}$. All such states are degenerate, and we will refer to them as the {\em single particle} states. In the same way the next degenerate manifold of states are the two-particle states $|i,j\rangle$, where we have flipped the spins at sites $i$ and $j$, and so forth to the general $n$-particle states. If we limit ourselves to  first order perturbations in $1/g$ we can neglet the mixing between states with a different number of particles and just analyse how the degeneracy within each subspace is modified. In the case of the one particle states, the exchange term $\sigma_{i}^{z}\sigma_{i+1}^{z}$
in the Hamiltonian is not diagonal in the basis of the $|\leftarrow\rangle$, $|\rightarrow\rangle$ states and leads only to the off-diagonal matrix element:
\be
\label{chap4_eq:7}
\langle i|H_{\rm I}|i+1\rangle = -J
\ee
which has the effect of shifting the particle between nearest neighbor sites. As what happens in the tight-binding models of solid state physics, the Hamiltonian is 
therefore diagonalized by Fourier-transformed basis:
\be
\label{chap4_eq:8}
|k\rangle = {\displaystyle\frac{1}{\sqrt{N}}}{\displaystyle\sum_{j}{\it e}^{ikx_{j}}|j\rangle},
\ee
where $N$ is the total number of sites. This eigenstate has energy:
\be
\label{chap4_eq:9}
\epsilon_{k}=Jg[2-(2/g)\cos(ka)+O(1/g^{2})],
\ee
where $a$ is the lattice spacing and where we have added an overall constant to the original Hamiltonian to make the energy of the ground state zero. \\
Let now focus on the two-particle states. At $g=\infty$ this subspace is spanned by the following states:
\be
\label{chap4_eq:10}
|i,j\rangle=|\leftarrow\rangle_{i}|\leftarrow\rangle_{j}{\displaystyle\prod_{p\neq i,j}}|\rightarrow\rangle_{p},
\ee
where $i\neq j$. Noticing that $|i,j\rangle = |j,i\rangle$ we can restrict our attention to the case $i>j$, which also means that this state is symmetric under the exchange of the particle positions, and so we can treat particles as bosons. At first order in $1/g$ these states will be mixed by the matrix element (\ref{chap4_eq:7}):
this will couple $|i,j\rangle$ to $|i\pm 1,j\rangle$ and $|i,j\pm 1\rangle$ for all $i > j+1$, while $|i,i-1\rangle$ will couple only to $|i+1,i-1\rangle$ and $|i,i-2\rangle$. When $i$ and $j$ are well separeted we can ignore this last case and consider the two particle independent of each other. Let say that the particles acquire momenta $k_1$ and $k_2$ and the total energy of this two-particle state will be $E_{k}=\epsilon_{k_{1}}+\epsilon_{k_{2}}$ with total momentum $k=k_1+k_2$. However when the two particle approach each other we have to consider the mixing between these momentum states arising from the restrictions in the matrix elements noted above. In practice we have to study the scattering problem associated with the dynamics of these excitations along the quantum Ising chain. The scattering of two incoming particles with momenta $k_1$ and $k_2$ will conserve total energy and total momentum. In the limit of small momenta, these conservation laws allow only one solution in $d=1$: The momenta of the particle in the final state are also $k_1$ and $k_2$. The uniqueness of the solution is a special feature of the one-dimensional case, which is lost in higher dimensions. Following this reasoning we can conclude that the wavefunction of the two-particle state will have the form for $i \gg j$:
\be
\label{chap4_eq:11}
\left({\it e}^{i(k_{1}x_{i}+k_{2}x_{j})}+S_{k_{1}k_{2}} {\it e}^{i(k_{2}x_{i}+k_{1}x_{j})}\right)|i,j\rangle,
\ee
where $S_{k_{1}k_{2}}$ is the $S$-matrix for the two-particle scattering, which for the TFIC is $S_{k_{1}k_{2}}=-1$ for all momenta $k_1,k_2$ (\cite{Sachdev1}). Notice that, for fixed total momentum $k$, there is still an arbitraruness in the single-particle momenta $k_{1,2}$, and therefore the total energy $E_{k}$ can take a range of values. There is thus no definite energy momentum relation, instead we have a  ``two-particle continuum'', with a threshold at $2\epsilon_{0}$. The same considerations can be used when dealing with  the $n$-particle continua, which have thresholds at $n\epsilon_{0}$. At higher oreder in $1/g$ we have to take into consideration the mixing between states with different particle number, such as the following matrix element:
\be
\label{chap4_eq:12}
\langle 0|H_{\rm I}|i,i+1\rangle = -J,
\ee
which introduces a coupling between $n$ and $n+2$ particle states. However, also at higher orders qualitative features of the spectrum does not change, and we will have renormalized one-particle states with a definite energy-momentum relationship and  renormalized $n\geq 2$ particle continua with thresholds at $n\epsilon_{0}$. Furthermore it has to be noticed that the integrability and stability of the one-particle states is not modified at any order in $1/g$: The one particle state with energy $\epsilon_{k}$ is the lowest energy state with momentum $k$, and this protects it from decay. \\
As already noted below equation (\ref{chap4_eq:11}), the $S$-matrix of the TFIC assumes a very simple form for all values of the momenta and at all orders in $1/g$ (the number of outgoing particles is always equal to the number of ingoing ones, being the model integrable).\\
The features of the spectrum described above have important consequences for the dynamic structure factor $S(k,w)$. If one inserts a complete sets of states between the operators  in (\ref{chap4_eq:4}) we see that at $T=0$:
\be
\label{chap4_eq:13}
 S(k,w)=2\pi{\displaystyle\sum_{s}}|\langle\sigma^{z}(k)\rangle|^{2}\delta(w-E_{s}),
\ee
where the sum over $s$ extends over all eigenstates of $H_{\rm I}$ with energy $E_{s}>0$, therefore at zero temperature we have that $S(k,w)$ is non-zero only for $w>0$
(the energy of the ground state was chosen to be zero). The dynamical susceptibility can be obtained from the structure factor thanks to the fluctuation-dissipation theorem, ${\rm Im}\chi(k,w)=({\rm sgn}(w)/2)S(k,w)$. The eigenstates and energies described above allow us to simply deduce the main features of $S(k,w)$, which are sketched in figure (\ref{chap4_fig:1}). The operator $\sigma^{z}$ flips a single site, hence the matrix element in (\ref{chap4_eq:13}) is nonzero for the single particle states: only the state with momentum $k$ contributes, and there is a delta function contribution to $S(k,w)\sim \delta(k,w)$ (which represents the quasi-particle peak). In the limit $g=\infty$ this peak comprises the entire spectral density and saturates the sum rule of equation $\ref{chap4_eq:13}$. For smaller $g$ the quasiparticle amplitude decreases and the multiparticle states also contribute the the spectral density. The mixing between the one- and three-particle states which occurs in general for $q\neq\infty$ means that the next contribution to $S(k,w)$ appears above the three-particle threshold $w>3\epsilon_{0}$; in this case the contribution is no longer a delta-function because there is a continuum of such states. Therefore the shape of $S(k,w)$ is rather a smooth function of $w$ (see figure (\ref{chap4_fig:1})). Similarly there are continua above higher odd number particle thresholds, and only odd number of particle contribute due to the fact that the matrix element in (\ref{chap4_eq:13}) vanishes for even numbers of particles. 

\subsection{Weak coupling}
The $g=0$ ground states are those states where all the spins are aligned along the positive (or negative) $z$-direction. They are twofold degenerate and posses long-range correlations in the magnetic order parameter $\sigma^{z}$:
\be
\label{chap4_eq:14}
{\displaystyle\lim_{|x|\to \infty}}C(x,0)=N_{0}^{2}\neq 0,
\ee
where $C(x,t)$ is the correlator defined in (\ref{chap4_eq:2}). The spontaneous magnetization $N_{0}$ is equal to $\pm\langle\sigma_{z}\rangle$ in the two ground states, corresponding to spontaneous breaking of the $Z_{2}$ symmetry.  The excited states can be described in terms of elementary domain wall (or kinks) excitations.
For instance the state:
\be
\label{chap4_eq:15}
\dots|\uparrow\rangle_{i}|\uparrow\rangle_{i+1}|\uparrow\rangle_{i+2}|\downarrow\rangle_{i+3}|\downarrow\rangle_{i+4}|\downarrow\rangle_{i+5}\dots
\ee
has one domain wall, that is a pair of spins with opposite momenta. At $g=0$ the energy of this state is clearly $2J$ (and for a state with $n$ domain walls it would be $2J\times n$). What happens when $g$ is small but non zero is very similar to what happens in the paramagnetic phase when $g$ is large but finite: The domain walls becomes particles, which can hop and form momentum eigenstates with excitation energy:
\be
\label{chap4_eq:16}
\epsilon_{k}=J(2-2g\cos(ka)+O(g)^{2}).
\ee
In the ferromagnetic phase the spectrum can be interpreted in terms of $n$-particle states, although it must be said that the interpretation of the particle in this case is very different from that in the large-$g$ limit. Again, the perturbation theory in $g$ only mixes states that differ by an even number of particles, although the matrix element in (\ref{chap4_eq:13}) is non zero only for states with an even number of particles. 
The structure factor $S(k,w)$ will have a delta function at $k=0$ and $w=0$, from the term in  (\ref{chap4_eq:13}) where $s$ is one of the ground states, which indicates the presence of long-range order. In addition, there is no single particle contribution, and the first finite $w$ spectral density is represented by the continuum above the two-particle threshold. To summarize we can say that:
\be
\label{chap4_eq:17}
S(k,w) = (2\pi)^{2}N_{0}^{2}\delta(k)\delta(w)+{\rm continua\,of\,even\,particles}.
\ee 
The $S$-matrix for the collision of two domain walls can be computed in a perturbation theory in $g$, and the results are very similar to those found in the paramagnetic phase where we expanded in $1/g$: there is no particle production, and $S_{kk'}=-1$ at all momenta to all orders in $g$.  

\section{Exact spectrum}
All the qualitative considerations of the previous sections allow us to develop an sound physical picture. We now want to set up a formalism that will lead us to an exact determination of many physical properties of the TFIC, and that will eventually confirm the approximate methods we have used to study the limiting cases for $g>1$, $g<1$. In this section we basically follow the approach and notation of (\cite{Essler}). \\
The essential point in this exact solution is the Jordan-Wigner transformation, which is a powerful mapping between models with spin-1/2 degrees of freedom and spinless fermion. The central observation here is that there is a simple mapping between the Hilbert space of a system with a spin-1/2 degree of freedom per site and that of a spinless fermion hopping between sites with single orbitals. We could associate the spin-up state with an empty orbital on the site and a spin-down state with an occupied one. If the canonical fermion operator $c_{i}$ annihilates a spinless fermion on site $i$, then this intuitive mapping immediately implies the operator relation:
\be
\label{chap4_eq18}
\sigma_{1}^{z}=1-2c_{i}^{\dag}c_{i}.
\ee
Although this equivalence works for a single site, we cannot extend it to the case of many sites, because while two fermionic operators on different sites anticommute, two spin operators commute. The solution to this dilemma was found by Jordan and Wigner, and it will be shown in the next to sections for the case of periodic and open boundary conditions on the TFIC.

\subsection{Periodic boundary conditions}

Let us consider the Ising ferromagnet with Hamiltonian given by (\ref{chap4_eq:1}), where $g$ is dimensionless and we impose periodic boundary consitions:
\be
\label{chap4_eq:20}
\sigma_{N+1}^{\alpha}\equiv\sigma_{1}^{\alpha}.
\ee
As we said the quantum Ising chain can be mapped to a model of spinless fermions by the Jordan-Wigner transformation:
\be
\label{chap4_eq:21}
\begin{array}{lcl}
\sigma_{j}^{x} & = & 1-2c_{j}^{\dag}c_{j}\\
\sigma_{j}^{z} & = & -{\displaystyle\prod_{l=1}^{j-1}}(1-2c_{l}^{\dag}c_{l})(c_{j}+c_{j}^{\dag}).
\end{array}
\ee
In terms of these spinless fermions the Hamiltonian (\ref{chap4_eq:1}) reads:
\be
\label{chap4_eq:22}
\begin{array}{lcl}
H_{\rm I} & = & -J{\displaystyle\sum_{j=1}^{N-1}}c_{j}^{\dag}[c_{j+1}+c_{j+1}^{\dag}]+{\rm h.c.}\\
& - & Jg {\displaystyle\sum_{j=1}^{N}}(1-2c_{j}^{\dag}c_{j})\\
& - & J{\it e}^{i\pi\hat{N}}(c_{N}-c_{N}^{\dag})(c_{1}+c_{1}^{\dag}),
\end{array}
\ee
where $\hat{N}$ is the {\em number} operator:
\be
\label{chap4_eq:23}
\hat{N}={\displaystyle\sum_{j=1}^{N}}c_{j}^{\dag}c_{j}.
\ee
Before diagonalizing the full Hamiltonian it is crucial to observe that:
\be
\label{chap4_eq:24}
[H,{\it e}^{i\pi\hat{N}}]=0,
\ee
which means that the Hamiltonian is in fact block-diagonal in the eigenbasis of ${\it e}^{i\pi\hat{N}}$. Therefore we can split the Hilbert space as:
\be
\label{chap4_eq:25}
{\cal H}={\cal H_{+}}\oplus{\cal H_{-}}.
\ee
A basis of ${\cal H_{+}}$ is given by states of the form:
\be
\label{chap4_eq:26}
|m_{1},\dots,m_{2n}\rangle={\displaystyle\prod_{j=1}^{2n}}c_{m_{j}}^{\dag}|0\rangle,\qquad\qquad n=0,\dots,N/2,
\ee
while ${\cal H_{-}}$ is spanned by states of the form:
\be
\label{chap4_eq:27}
|m_{1},\dots,m_{2n-1}\rangle={\displaystyle\prod_{j=1}^{2n-1}}c_{m_{j}}^{\dag}|0\rangle,\qquad\qquad n=1,\dots,N/2.
\ee 
The corresponding blocks of the Hamiltonian are ($\sigma=\pm$):
\be
\label{chap4_eq:28}
\begin{array}{lcl}
H_{\sigma} & = & -J{\displaystyle\sum_{j=1}^{N-1}}c_{j,\sigma}^{\dag}[c_{j+1,\sigma}+c_{j+1,\sigma}^{\dag}]+{\rm h.c.}\\
 & + & 2gJ{\displaystyle\sum_{j=1}^{N}}c_{j,\sigma}^{\dag}c_{j,\sigma}-gJN,
\end{array}
\ee
where the fermionic creation/annihilation operators are subject to the boundary conditions:
\be
\label{chap4_eq:29}
\begin{array}{lcl}
c_{N+1,-} & = & c_{1,-}\\
c_{N+1,+} & = & -c_{1,+}.
\end{array}
\ee
Let us denote the constant by $C=-gJN$. The Hamiltonians $H_{\pm}$ may now be diagonalized by a Fourier transform followed by a Bogoliubov rotation. The momentum spaces Fermi operators with the appropriate boundary conditions are:
\be
\label{chap4_eq:30} 
c_{\sigma}(k_{\sigma,n})=\frac{1}{\sqrt{N}}{\displaystyle\sum_{j=1}^{N}}c_{j}{\it e}^{-ik_{\sigma,n}j},
\ee
where
\be
\label{chap4_eq:31}
\begin{array}{lcl}
k_{-,n} & = & {\displaystyle\frac{2\pi n}{N}}, \qquad\qquad\qquad n=-\frac{N}{2}+1,\dots,\frac{N}{2},\\
k_{+,n} & = & {\displaystyle\frac{2\pi (n+1/2)}{N}}, \qquad\quad n=-\frac{N}{2},\dots,\frac{N}{2}-1.
\end{array}
\ee
In either sector we then carry out a Bogoliubov transformation:
\be
\label{chap4_eq:32}
\left(\begin{array}{c}
        c_{\sigma}(k_{\sigma,n})\\
        c_{\sigma}^{\dag}(-k_{\sigma,n})\end{array}\right)=R_{g}(k_{\sigma,n})\left(\begin{array}{c}
        \alpha_{k_{\sigma,n}}\\
        \alpha^{\dag}_{-k_{\sigma,n}}\end{array}\right),
\ee
where the matrix $R_{g}$ is given by:
\be
\label{chap4_eq:33}
R_{g}(k)=\left(\begin{array}{cc}
        \cos(\theta_{k}/2) & i\sin(\theta_{k}/2)\\
        i\sin(\theta_{k}/2) & \cos(\theta_{k}/2)\end{array}\right).
\ee
In the $-$ sector we have to treat the modes with momenta $k_{0}$ and $k_{-N/2}=-\pi$ separately as they do not get paired up under the Bogoliubov transformation. The Bogoliubov angle is found to be:
\be
\label{chap4_eq:34}
\begin{array}{lcl}
\theta_{k} & = & {\rm arctan}\left[\frac{\sin(ka)}{\cos(ka)-g}\right]\\
& + & \pi\theta_{H}(1-g){\rm sgn}(k)\theta_{H}(|k|-{\rm arccos(g)}),
\end{array}
\ee
where $\theta_{H}$ is the Heaviside step function and where we always choose the principal branch of the arctan. 
With this choice of the Bogoliubov angle we find:
\be
\label{chap4_eq:35}
H_{+}={\displaystyle\sum_{n=-N/2}^{N/2-1}}\epsilon_{g}(k_{+,n})\alpha_{k_{+,n}}^{\dag}\alpha_{k_{+,n}}+C_{+},
\ee
where the dispersion relation is:
\be
\label{chap4_eq:36}
\epsilon_{g}(k)=2J\sqrt{1+g^2-2g\cos(k)}.
\ee
The constant contribution is found to be:
\be
\label{chap4_eq:37}
C_{+}=C+{\displaystyle\sum_{n=0}^{N/2-1}}[-\epsilon_{g}(k_{+,n})-2J\cos(k_{+,n})+2gJ].
\ee
The analogous calculation for $H_{-}$ gives:
\be
\label{chap4_eq:38}
\begin{array}{lcl}
H_{-} & = & {\displaystyle\sum_{n=-N/2-1,\, n\neq 0}^{N/2-1}}\epsilon_{g}(k_{-,n})\alpha_{k_{-,n}}^{\dag}\alpha_{k_{-,n}}\\
& - & 2J(1-g)\alpha_{0}^{\dag}\alpha_{0}+2J(1+g)\alpha_{-\pi}^{\dag}\alpha_{-\pi}+C_{-},
\end{array}
\ee
where the constant contribution is equal to:
\be
\label{chap4_eq:39}
C_{-}=C+{\displaystyle\sum_{n=1}^{N/2-1}}[-\epsilon_{g}(k_{-,n})-2J\cos(k_{-,n})+2gJ].
\ee
Before proceeding it is useful to introduce the {\em two} vacua:
\be
\label{chap4_eq:40}
\alpha_{k_{\sigma,n}}|0,\sigma\rangle=0,\qquad \forall\,k_{\sigma,n}.
\ee
The ground state in the $+$ sector is simply the $+$ vacuum with energy:
\be
\label{chap4_eq:41}
E_{GS,+}=C_{+}.
\ee
In the $-$ sector the lowest energy state in the ordered phase $g<1$ is:
\be
\label{chap4_eq:42}
\alpha_{0}^{\dag}|0,-\rangle
\ee
with energy $E_{GS,-}=C_{-}-2J(1-g)$. Let us now consider the case of large $N$:
\be
\label{chap4_eq:43}
\begin{array}{lcl}
C_{+}-C_{-} & = & -{\displaystyle\sum_{n=0}^{N/2-1}}2j\left[\cos\left(\frac{2\pi}{N}(n+\frac{1}{2})\right)-\cos\left(\frac{2\pi}{N}n\right)+\epsilon(k_{+,n})\right.\\
&  & -\left.\epsilon(k_{-,n})\right]-[\epsilon(0)+2J(1-g)]\\
& = & J{\displaystyle\int_{0}^{\pi}}dx \sin(x)-\frac{1}{2}{\displaystyle\int_{0}^{\pi}}dx\epsilon'(x)-[\epsilon(0)+2J(1-g)]+O(N^{-1}).
\end{array}
\ee
In the ordered phase we have $\epsilon(0)=2J(1-g)$ and hence:
\be
\label{chap4_eq:44}
C_{+}-C_{-}=-2J(1-g)+O(N^{-1}).
\ee
As we expected this result shows that in the infinite volume of the ordered phase limit we have two degenerate ground states. On the other hand, in the disordered phase we have $\epsilon(0)=2J(g-1)$ and therefore we obtain:
\be
\label{chap4eq:45}
C_{+}-C_{-}=O(N^{-1}),
\ee
which means that there is a unique ground state, which lies in the $+$ sector. A basis of the full Hilbert space is provided by:
\be
\label{chap4_eq:46}
\begin{array}{lcl}
|n_{1},\dots,n_{2m},+\rangle & = & {\displaystyle\prod_{j=1}^{2m}}\alpha_{k_{+,n_{j}}}^{\dag}|0,+\rangle,\\
|n_{1},\dots,n_{2m+1},-\rangle & = & {\displaystyle\prod_{j=1}^{2m+1}}\alpha_{k_{-,n_{j}}}^{\dag}|0,-\rangle.
\end{array}
\ee
As $\sigma_{j}^{z}$ changes the fermion number by one, we conclude that it can have a non-zero expectation value only between states that are linear combinations of states involving the $+$ and the $-$ sectors. In the ordered phase the two ground states are therefore:
\be
\label{chap4_eq:47}
|0\rangle_{\rm NS}\pm |0\rangle_{\rm R},
\ee
where we define
\be
\label{chap4_eq:48}
\begin{array}{lcl}
|0\rangle_{\rm NS}  & = &  |0,+\rangle, \\
|0\rangle_{\rm R} & = & \alpha_{0}^{\dag}|0,-\rangle.
\end{array}
\ee
In order to simplify the notation we can carry out a particle-hole transformation on the $\alpha_0$ fermion $\alpha_0 \to \alpha_{0}^{\dag}$. After this we have for both the $R$ and $NS$ sectors:
\be
\label{chap4_eq:49}
\begin{array}{lcl}
|k_{1},\dots,k_{2m}\rangle_{R} & = & {\displaystyle\prod_{k_{j}\in R}^{2m}}\alpha_{k_{j}}^{\dag}|0\rangle_{R},\\
|p_{1},\dots,p_{2m}\rangle_{NS} & = & {\displaystyle\prod_{p_{j}\in NS}^{2m}}\alpha_{p_{j}}^{\dag}|0\rangle_{NS}.
\end{array}
\ee
We can now understand why it is not trivial to determine the expectation value of $\sigma_{j}^{z}$. Firstly we have:
\be
\label{chap4_eq:50}
[\sigma_{j}^{z},{\it e}^{i\pi\hat{N}}]\neq 0,
\ee
so $\sigma_{j}^{z}$ connects the two sectors. Still, we can express $\sigma_{j}^{z}$ in terms of our Bogoliubov fermions of either sector. However, the problem is that the action of say the $+$ Boliubov fermions on the $-$ vacuum is non-trivial:
\be
\label{chap4_eq:51}
\alpha_{k_{+,n}}|0,-\rangle \neq 0.
\ee
Therefore even the matrix element $\langle 0,+|\sigma_{1}^{z}|0,-\rangle$ cannot be simply computed by using Wick's theorem. \\
In the disordered phase the ground state is $|0\rangle_{NS}$, while a complete set of states is then given by:
\be
\label{chap4_eq:52}
\begin{array}{lcl}
|k_{1},\dots,k_{2m+1}\rangle_{R} & = & {\displaystyle\prod_{k_{j}\in R}^{2m+1}}\alpha_{k_{j}}^{\dag}|0\rangle_{R},\\
|p_{1},\dots,p_{2m}\rangle_{NS} & = & {\displaystyle\prod_{p_{j}\in NS}^{2m}}\alpha_{p_{j}}^{\dag}|0\rangle_{NS}.
\end{array}
\ee 
The above structure of the spectrum confirms the approximate considerations of the previous section: We have found that the particles are in fact free fermions, and two fermions will not scatter even when the get close to each other. Alternatively they can be considered as hard-core bosons, which have an $S$-matrix that does not allow particle production and that equals $-1$ at all momenta. We will see that the latter point of view is much more useful, as the bosonic particles have a simple, local interpretation in terms of the underlying spin excitations: For $g\gg 1$ the bosons are simply spins oriented in the $|\leftarrow\rangle$ direction, whereas for $g \ll 1$ they are domain walls between two ground states. The fermionic representation is useful for certain technical manipulations, but the bosonic point of view is much more useful for making physical arguments, as we shall see. 

\subsection{Open boundary conditions}

Let us now consider the Ising ferromagnet on an open chain with $N$ sites:
\be
\label{chap4_eq:53}
H_{\rm I}=-J{\displaystyle\sum_{j}^{N-1}[\sigma_{j}^{z}\sigma_{j+1}^{z}+g\sigma_{j}^{x}]}.
\ee
In terms of the spinless Jordan-Wigner fermions the Hamiltonian reads:
\be
\label{chap4_eq:54}
\begin{array}{lcl}
H_{\sigma} & = & -J{\displaystyle\sum_{j=1}^{N-1}}c_{j}^{\dag}[c_{j+1}+c_{j+1}^{\dag}]+{\rm h.c.}\\
 & - & gJ{\displaystyle\sum_{j=1}^{N}}(1-2c_{j}^{\dag}c_{j}).
\end{array}
\ee
The appropriate Bogoliubov transformation can be worked out following Lieb, Shulz and Mattis:
\be
\label{chap4_eq:55}
\alpha_{k}={\displaystyle\sum_{j=1}^{N}}\frac{\phi_{k,j}+\psi_{k,j}}{2}c_{j}+\frac{\phi_{k,j}-\psi_{k,j}}{2}c_{j}^{\dag},
\ee
where we have introduced
\be
\label{chap4_eq:56}
\begin{array}{lcl}
\phi_{k,n} & = & N_{k}[g\sin(nk)-\sin((n-1)k)],\\
\psi_{k,n} & = & {\displaystyle\frac{2JN_{k}}{\epsilon(k)}}\sin(nk).
\end{array}
\ee
The normalization factor is given by 
\be
\label{chap4_eq:57}
N_{k}^{2}=\frac{8J^{2}}{\epsilon(k)^{2}}\frac{1}{N+1}\left(1-\frac{\sin(2Nk)}{2g(N+1)\sin(k)}\right)^{-1},
\ee
while as before $\epsilon_{g}(k)=2J\sqrt{1+g^2-2g\cos(k)}$. The $k$'s are solutions to the following set of quantization conditions:
\be
\label{chap4_eq:58}
\sin(Nk)-g\sin((N+1)k)=0.
\ee
The solutions of this equation are all real with one exception in the case of $0<g<1$. Finally the Hamiltonian becomes:
\be
\label{chap4_eq:59}
H_{\rm I}={\displaystyle\sum_{k}}\epsilon(k)\alpha_{k}^{\dag}\alpha_{k}+C.
\ee
The two lowest energy states in the ordered phase are the fermion vacuum $|0\rangle$ and the boundary bound state
\be
\label{chap4_eq:60}
|k_{0}\rangle = \alpha_{k_{0}}^{\dag}|0\rangle,
\ee
where $k_{0}$ fulfils the equation:
\be
\label{chap4_eq:61}
\sinh(Nk_{0})-g\sinh((N+1)k_{0})=0.
\ee
For large $N$ and $k_{0}>0$ we obtain:
\be
\label{chap4_eq:62}
k_{0}=-i\ln(g)+O({\it e}^{-\gamma N}),
\ee
therefore we have:
\be
\label{chap4_eq:63}
\epsilon(k_{0})\approx 2J\sqrt{1+g^2-g(g+1/g)}=0,
\ee
which shows that the two states are degenerate up to terms exponentially small in the system size. Let us return to the quantization condition (\ref{chap4_eq:58}). They can be written in logarithmic form as:
\be
\label{chap4_eq:64}
k_{n}=\frac{\pi n}{N}+\frac{1}{N}{\rm arctan}\left(\frac{g\sin(k_{n})}{1-g\cos(k_{n})}\right),
\ee
where $n=1,2,\dots,N-1$. Remember that for large $N$ we can turn sums into integrals using the Euler-Maclaurin sum formula:
\be
\label{chap4_eq:65}
\frac{1}{N}{\displaystyle\sum_{j=1}^{N-1}}f(k_{n})={\displaystyle\int_{0}^{\pi}}\frac{dk}{\pi}f(k)+O(N^{-1}).
\ee 
The boundary phase shift does not affect the $O(1)$ term of such expressions. \\
It is now convenient to introduce the operators:
\be
\label{chap4_eq:66}
A_{j}=c_{j}+c_{j}^{\dag}\qquad B_{j}=c_{j}-c_{j}^{\dag},
\ee
which satisfy the following commutation relations:
\be
\label{chap4_eq:67}
\begin{array}{lcl}
 \{A_{j},A_{l}\}& = &  2\delta_{j,l}\\
\{B_{j},B_{l}\} & = &  -2\delta_{j,l}\\
\{A_{j},B_{l}\} & = & 0. 
\end{array}
\ee
Using these operators we can re-write the magnetization operator as:
\be
\label{chap4_eq:68}
\sigma_{n}^{z}=-{\displaystyle\sum_{j=1}^{n-1}}A_{j}B_{j}A_{n},
\ee
while the fact that the $\phi_{k}$'s and $\psi_{k}$'s form orthonormal set of vectors we have:
\be
\label{chap4_eq:69}
\begin{array}{lcl}
A_{j} & = & \sum_{k}\phi_{k}[\alpha_{k}+\alpha_{k}^{\dag}] \\
B_{j} & = & \sum_{k}\psi_{k}[\alpha_{k}-\alpha_{k}^{\dag}].  
\end{array}
\ee
We are now ready to compute two-point functions of $A$'s and $B$'s. From the anticommutation relations of the $A_{j}$'s and $B_{j}$'s we can easy compute the following correlators:
\be
\label{chap4_eq:70}
\begin{array}{lcl}
 \langle 0|A_{j}A_{l}|0\rangle & = & \delta_{j,l}, \\
 \langle 0|B_{j}B_{l}|0\rangle & = & -\delta_{j,l}, 
\end{array}
\ee
Now we want to calculate the correlator between $A_j$ and $B_j$:
\be
\label{chap4_eq:71}
\begin{array}{lcl}
 \langle 0|A_{j}B_{l}|0\rangle & = & -{\displaystyle\sum_{n=0}^{N-1}}\phi_{k,j}\psi_{k,l}\\
  & \simeq & {\displaystyle\int_{-\pi}^{\pi}\frac{dk}{2\pi}\frac{(g-{\it e}^{-ik})[{\it e}^{ik(j-l)}-{\it e}^{ik(j+l)}]}{\sqrt{1+g^2-2g\cos(k)}}}.
\end{array}
\ee
Here we have assumed that $j,l \ll N$. In particular this assumption allows us to drop the contribution from the $k_0$ mode, and the remaining integral can in principle be carried out to yield a regularized generalized hypergeometric function using the integral:
\be
\label{chap4_eq:72}
\begin{array}{l}
 {\displaystyle\int_{-\pi}^{\pi}\frac{dk}{2\pi}}{\rm e}^{ikm}{\sqrt{1+g^2-2g\cos(k)}}\\
={\displaystyle\frac{_{3}F_{2}(\frac{1}{2},\frac{1}{2},1;1-m,1+m;\frac{4g}{(1+g)^{2}})}{(1+g)\Gamma(1-m)\Gamma(1+m)}}.
\end{array}
\ee
The magnetization at site $n$ is given by:
\be
\label{chap4_eq:73}
\begin{array}{lcl}
\langle 0|\sigma_{n}^{z}|k_{0}\rangle  = & - &\frac{1}{2}{\displaystyle\sum_{l=1}^{N}}\phi_{k_{0},l}\langle 0|{\displaystyle\sum_{j=1}^{n-1}}A_{j}B_{j}A_{n}A_{l}|\rangle\\
& + & \frac{1}{2}{\displaystyle\sum_{l=1}^{N}}\psi_{k_{0},l}\langle 0|{\displaystyle\sum_{j=1}^{n-1}}A_{j}B_{j}A_{n}B_{l}|\rangle.
\end{array}
\ee
Using Wick's theorem we find that:
\be
\label{chap4_eq:74}
\begin{array}{l}
\langle 0|{\displaystyle\sum_{j=1}^{n-1}}A_{j}B_{j}A_{n}A_{l}|0\rangle={\rm det}(G_{1}),\\
\langle 0|{\displaystyle\sum_{j=1}^{n-1}}A_{j}B_{j}A_{n}B_{l}|0\rangle=\theta_{H}(l+1-n){\rm det}(G_{2}),
\end{array}
\ee
where we have introduced the matrices:
\be
\label{chap4_eq:75}
\begin{array}{l}
(G_1)_{ij}=(G_2)_{ij}=\langle A_{j}B_{i}\rangle,\qquad i=1,\dots,n-1,\\
(G_1)_{n,j}=\delta_{j,l},\qquad (G_{2})_{n,j}=-\delta_{ij}.
\end{array}
\ee
So apart from the last row both $G_1$ and $G_2$ are sums of Toeplitz and Hankel matrices. 

\section{Equal-time Correlations of the order \newline parameter at finite $T$}
\label{sec:corrT}
In this section we shall outline one of the possible techniques to compute equal-time correlations of the order parameter at finite temperature.
We will use the symbol $\langle\hat{O}\rangle$ to indicate the statistical average at temperature $T$ of the operator $\hat{O}$, and we will also assume all the operators involved in the matrix element to be equal-time. Using the fermionic representation of $\sigma_{i}^{z}$ we can write:
\be
\label{chap4_eq:76}
\begin{array}{lcl}
\langle\sigma_{i}^{z}\sigma_{i+n}^{z}\rangle & = & \left\langle(c_{i}^{\dag}+c_{i})\left[{\displaystyle\prod_{j=i}^{i+n-1}}(c_{j}^{\dag}+c_{j})(c_{j}^{\dag}-c_{j})   \right](c_{i+n}^{\dag}+c_{i+n})\right\rangle \\
& = & \left\langle(c_{i}^{\dag}-c_{i})\left[{\displaystyle\prod_{j=i+1}^{i+n-1}}(c_{j}^{\dag}+c_{j})(c_{j}^{\dag}-c_{j})   \right](c_{i+n}^{\dag}+c_{i+n})\right\rangle.
\end{array}
\ee
Let us notice that the string in the previous equation only extends between the sites $i$ and $i+n$, with the operators on site to the left of $i$ heving cancelled between the two strings. Now, using the notation introduced in (\ref{chap4_eq:66}) we have:
\be
\label{chap4_eq:77}
\langle\sigma_{i}^{z}\sigma_{i+n}^{z}\rangle=\langle B_{i}A_{i+1}B_{i+1}\dots A_{i+n-1}B_{i+n-1}A_{i+n}\rangle.
\ee
Since these expectation values are computed with respect to a free Fermi theory, the term on the right-hand side can be computed by the finite-temperature Wick's theorem, which connects it to a sum over products of expectation values of pairs of operators, which can be easily calculated:
\be
\label{chap4_eq:78}
\begin{array}{lcl}
\langle A_{i}A_{j}\rangle & = &  \delta_{ij}\\
\langle B_{i}B_{j}\rangle & = &  -\delta_{ij}\\
\langle B_{i}A_{j}\rangle & = & -\langle A_{j}B_{i}\rangle = D_{i-j+1},
\end{array}
\ee
where we have introduced  $D_{n}$:
\be
\label{chap4_eq:79}
D_{n}\equiv{\displaystyle\int_{0}^{2\pi}\frac{d\phi}{2\pi}}{\it e}^{-in\phi}\tilde{D}({\it e}^{i\phi}),
\ee
and
\be
\label{chap4_eq:80}
\tilde{D}(z={\it e}^{i\phi})\equiv\left(\frac{1-gz}{1-g/z}\right)^{1/2}{\rm tanh}\left[\frac{J}{T}((1-gz)(1-g/z))^{1/2}\right].
\ee
In order to determine $\langle B_{i}A_{j}\rangle$ we have used the representation (\ref{chap4_eq:32}) and evaluated expectation values of the $\alpha_{k}$ under the free fermion Hamiltonian defined in (\ref{chap4_eq:25}), in the thermodynamic limit. Collecting the terms in the Wick expansion we find:
\be
\label{chap4_eq:81}
\langle\sigma_{i}^{z}\sigma_{i+n}^{z}\rangle=T_{n}\equiv \left|\begin{array}{cccc}
        D_{0}   & D_{-1} & \dots   & D_{-n+1}   \\
        D_{1}   &        &         &  \\
				\cdot   &        &         &  \\
				\cdot   &        &  D_{0}  & D_{-1}  \\
				D_{n-1} &        &  D_{1}  & D_{0} \end{array}\right|.
\ee
Now the problem is to evaluate the determinant $T_{n}$: To obtain the universal scaling limit answer we need to take the limit $n\to \infty$. The expression for $T_{n}$ belongs to a special class of determinants known as Toeplitz determinants, and the limit $n\to\infty$ can indeed be computed in closed form using a quite sophisticated mathematical theory. In the next section we will see how to use the same techniques to write an algorithm to compute the non-equal time correlations at zero temperature, which will be the main target of the next chapters. 
However the details of these techniques will not be discussed here, but refer the reader to the literature (\cite{Sachdev1}). 

\section{Time-dependent correlation function after a quench at $T=0$}

In this section we will introduce the protocol for a quench in the TFIC, and see how the to compute time-dependent correlators in this framework.
Let us consider the situation where we quench the transverse field from $hJ$ to $h'J$. The system then starts out in the ground state $|0\rangle$ of
the original Hamiltonian, defined by:
\be
\label{chap4_eq:82}
\alpha_{k}|0\rangle = 0.
\ee
The time evolution of this state is:
\be
\label{chap4_eq:83}
{\it e}^{-iH't}|0\rangle,
\ee
where $H'$ is the new Hamiltonian corresponding to the transverse field $h'J$. The latter can be diagonalized in terms of new Bogoliubov fermions:
\be
\label{chap4_eq:84}
H(h')={\displaystyle\sum_{k}}\epsilon_{h'}(k)\left[\beta_{k}^{\dag}\beta_{k}-\frac{1}{2}\right].
\ee 
The new and old Bogoliubov fermions are related by:
\be
\label{chap4_eq:85}
\left(\begin{array}{c}
\beta_{k}\\
\beta_{-k}^{\dag}
\end{array}\right) = U(k)\left(\begin{array}{c}
\alpha_{k}\\
\alpha_{-k}^{\dag}
\end{array}\right)
\ee
where the matrix $U$ is given by:
\be
\label{chap4_eq:86}
U(k)=R_{h'}^{T}(k)R_{h}(k)=\left(\begin{array}{cc}
\cos\left(\frac{\theta_{k}-\theta'_{k}}{2}\right) & i\sin\left(\frac{\theta_{k}-\theta'_{k}}{2}\right)\\
 i\sin\left(\frac{\theta_{k}-\theta'_{k}}{2}\right) & 
\cos\left(\frac{\theta_{k}-\theta'_{k}}{2}\right) 
\end{array}\right).
\ee
This relationship allows us to link the two vacua. Starting with:
\be
\label{chap4_eq:87}
|0'\rangle = {\displaystyle\sum_{n=0}^{\infty}\sum_{k_{1},\dots,k_{n}}}f_{k_{1},\dots,k_{n}}\alpha_{k_{1}}^{\dag}\dots\alpha_{k_{n}}^{\dag}|0\rangle,
\ee
and then imposing the condition:
\be
\label{chap4_eq:88}
0=\beta_{k}|0'\rangle = \left[U_{11}(k)\alpha_{k}+U_{12}(k)\alpha_{-k}^{\dag}\right],
\ee
we conclude that up to a normalization factor the new vacuum cna be expressed in terms of the old one as:
\be
\label{chap4_eq:89}
|0'\rangle=\exp\left(-\frac{i}{2}{\displaystyle\sum_{k}}{\rm tan}\left[\frac{\theta_{k}-\theta'_{k}}{2}\right]\alpha_{k}^{\dag}\alpha_{-k}^{\dag}\right)|0\rangle.
\ee

This kind of state is sometimes called boundary state or squeezed-state, expecially in the context of integrable quantum field theory.\\
Solving the Heisenberg equation of motion we have:
\be
\label{cahp4_eq:90}
\beta_{k}(t)={\it e}^{-i\epsilon_{h'}t}\beta_{k}.
\ee
This allows us to derive an expression for the transverse magnetization $\sigma_{j}^{z}$ in terms of the ``old'' Bogoliubov fermions, which represents a first pedagogical example of time-dependent expectation value:
\be
\label{chap4_eq:91}
\sigma_{j}^{x}(t)=-\frac{1}{L}{\displaystyle\sum_{k,p}}{\it e}^{-i(k-p)j}(\alpha_{k}^{\dag},\alpha_{-k})S(k,p,t)\left(\begin{array}{c}
\alpha_{p}\\
\alpha_{-p}^{\dag}
\end{array}\right)
\ee
where $S(k,p,t)$ is a $2\times 2$ matrix:
\be
\label{chap4_eq:92}
S(k,p,t)=U^{T}(k)D_{h'}^{\dag}(k)R_{h'}^{\dag}(k)\sigma^{z}R_{h'}(p)D_{h'}(p)U(p).
\ee
with:
\be
\label{chap4_eq:93}
D_{h'}(k)=\left(\begin{array}{cc}
{\it e}^{i\epsilon_{h}(k)t} & 0 \\
  0 &  {\it e}^{-i\epsilon_{h}(k)t}
\end{array}\right).
\ee
Let us now determine the time evolution of the expectation value of $\sigma_{j}^{x}$ after the quench, namely:
\be
\label{chap4_eq:94}
\langle\sigma_{j}^{x}(t)\rangle = \langle 0|{\it e}^{iH(h')t}\sigma_{j}^{x}{\it e}^{-iH(h')t}|0\rangle,
\ee
which, using translational invariance, can immediately re-written as:
\be
\label{chap4_eq:95} 
\begin{array}{lcl}
\langle\sigma_{j}^{x}(t)\rangle & = &  1-2{\displaystyle\sum_{j}}\langle 0|{\it e}^{iH(h')t}c_{j}^{\dag}c_{j}{\it e}^{-iH(h')t}|0\rangle \\
 &  = & 1-{\displaystyle\frac{2}{N}\sum_{k}}\langle 0|{\it e}^{iH(h')t}c^{\dag}(k)c(k){\it e}^{-iH(h')t}|0\rangle.
\end{array}
\ee
This can be evaluated as follows:
\be
\label{chap4_eq:96}
\begin{array}{lcl}
\langle\sigma_{j}^{x}(t)\rangle & = & -{\displaystyle\frac{1}{N}\sum_{k}}\langle 0|{\it e}^{iH(h')t}(c^{\dag}(k),c(-k))\sigma^{z}\left(\begin{array}{c}
c_{k}\\
c_{-k}^{\dag}
\end{array}\right){\it e}^{-iH(h')t}|0\rangle\\
& = & -{\displaystyle\frac{1}{N}\sum_{k}}\langle 0|{\it e}^{iH(h')t}(\beta_{k}^{\dag},\beta_{k})R_{h}^{\dag}\sigma^{z}R_{h}\left(\begin{array}{c}
\beta_{k}\\
\beta_{-k}^{\dag}
\end{array}\right){\it e}^{-iH(h')t}|0\rangle\\
& = & -{\displaystyle\frac{1}{N}\sum_{k}}\langle 0|(\alpha_{k}^{\dag},\alpha_{-k})S(k,t)\left(\begin{array}{c}
\alpha_{k}\\
\alpha_{-k}^{\dag}
\end{array}\right)|0\rangle.
\end{array}
\ee
After a little algebra we find the following result:
\be
\label{chap4_eq:97}
\begin{array}{lcl}
\langle\sigma_{j}^{x}(t)\rangle &  = &  {\displaystyle\frac{1}{N}\sum_{k}}S_{22}(k,t)\\
& = & {\displaystyle\int_{-\pi}^{\pi}\frac{dk}{2\pi}}\left[\cos(\theta_{k})\cos^{2}(\epsilon_{h'}(k)t)+\cos(\theta_{k}-2\theta'_{k})\sin^{2}(\epsilon_{h'}(k)t)\right].
\end{array}
\ee
Let us now focus on the computation of the two-point  correlation function of the order parameter of the TFIC, following the approach of (\cite{Fagotti2}).
Let us consider again the Ising ferromagnet on an open chain with $L$ sites\footnote{Here we prefer to use $L$ instead of $N$ to indicate the total number of spins, in order to avoid confusing it with $N_{k}$, another parameter we shall define later.}:
\be
H=-\sum_{j=1}^{L-1}J\Bigl[\sigma_j^z\sigma_{j+1}^z+h\sigma_j^x\Bigr].
\ee
The Hamiltonian can be written in terms of the Majorana fermions $A_j$ and $B_j$, where the spin operator $\sigma^{z}$ reads as:
\be
\begin{aligned}
\sigma_l^z=\prod_{j<l}(A_j B_j)  A_l \\ 
\end{aligned}
\ee
Because of this, the spin correlation functions can be obtained by means of the Wick's theorem both at finite temperature and at temperature zero, when the system is in the ground state. 
In addition, Wick's theorem can be also used when we consider the evolution of the system after a quantum quench in which both the initial and the final Hamiltonians are Ising ferromagnets, \emph{i.e.} a quench in which the magnetic field is abruptly changed from $Jh_0$ to $Jh$.

We are interested in the time dependent two-point  correlation function of the order parameter
\be\label{eq:corr}
\braket{\sigma_n^z(t_2)\sigma^z_{n+\ell}(t_1)}\qquad \qquad \braket{\sigma^z_{n+\ell}(t_2)\sigma^z_{n}(t_1)},
\ee
where $t_2\geq t_1$ and as usual we indicated with $\sigma^x(t)$ the time dependent operator in the Heisenberg picture
\be
\sigma^z(t)=e^{i H t}\sigma^z e^{-i H t}.
\ee
Now we consider the first expectation value of \eqref{eq:corr}. In terms of the Majorana fermions we get
\be
\braket{\sigma_n^z(t_2)\sigma^z_{n+\ell}(t_1)}=\braket{\prod_{j<n}(A_j(t_2)B_j(t_2) )  A_n(t_2)\prod_{j<n+\ell}(A_j(t_1) B_j(t_1))  A_{n+\ell}(t_1)}.
\ee
This expectation value can be written as the Pfaffian of a matrix (up to a sign):
\be
\braket{\sigma_n^z(t_2)\sigma^z_{n+\ell}(t_1)}=\mathrm{Pf}[\Gamma],
\ee
where $\Gamma$ is the $(4n+2\ell-2)\times (4n+2\ell-2)$ skew-symmetric matrix, with the upper triangular part given by
\be\label{eq:Gamma}
\Gamma_{i j}=\begin{cases}
\begin{cases}
\braket{A_{\frac{i+1}{2}}(t_2)A_{\frac{j+1}{2}}(t_2)}&i,j\text{ odd}\\
\braket{A_{\frac{i+1}{2}}(t_2)B_{\frac{j}{2}}(t_2)}&i \text{ odd},j\text{ even}\\
\braket{B_{\frac{i}{2}}(t_2)A_{\frac{j+1}{2}}(t_2)}&i \text{ even},j\text{ odd}\\
\braket{B_{\frac{i}{2}}(t_2)B_{\frac{j}{2}}(t_2)}&i,j\text{ even}\\
\end{cases}
&i< j<2n\\
\begin{cases}
\braket{A_{\frac{i}{2}-n+1}(t_1)A_{\frac{j}{2}-n+1}(t_1)}&i,j\text{ even}\\
\braket{A_{\frac{i}{2}-n+1}(t_1)B_{\frac{j+1}{2}-n}(t_1)}&i \text{ even},j\text{ odd}\\
\braket{B_{\frac{i+1}{2}-n}(t_1)A_{\frac{j}{2}-n+1}(t_1)}&i \text{ odd},j\text{ even}\\
\braket{B_{\frac{i+1}{2}-n}(t_1)B_{\frac{j+1}{2}-n}(t_1)}&i,j\text{ odd}\\
\end{cases}
&2n \leq i< j\\
\begin{cases}
\braket{A_{\frac{i+1}{2}}(t_2)A_{\frac{j}{2}-n+1}(t_1)}&i\text{ odd},j\text{ even}\\
\braket{A_{\frac{i+1}{2}}(t_2)B_{\frac{j+1}{2}-n}(t_1)}&i,j\text{ odd}\\
\braket{B_{\frac{i}{2}}(t_2)A_{\frac{j}{2}-n+1}(t_1)}&i,j\text{ even}\\
\braket{B_{\frac{i}{2}}(t_2)B_{\frac{j+1}{2}-n}(t_1)}&i\text{ even},j\text{ odd}\\
\end{cases}
&i< 2n, 2n \leq j
\end{cases}
\ee
The other correlation function of \eqref{eq:corr} can be obtained by substituting $n$ with $n+\ell$ in the matrix above.
Thus, the basic ingredients to compute the two-point correlation function are the fermionic correlations
\be
\braket{A_l(t_2) A_n(t_1)}, \quad \braket{B_l(t_2) B_n(t_1)},\quad\text{and}\quad\braket{A_l(t_2) B_n(t_1)}
\ee
and the corresponding equal-time ones.
Before the quench operators $A$'s are only correlated with operators $B$'s, indeed
\be
\braket{A_l(0) A_n(0)}=\delta_{l n}\qquad \braket{B_l(0) B_n(0)}=-\delta_{l n}\, .
\ee 
In order to write the other fermionic correlations before and after the quench it is convenient to define the functions
\be\label{eq:phipsi}
\begin{aligned}
\begin{aligned}
\phi_{k_j,n}^{(h)}&=(-1)^n N_{k_j} \Bigl[\sin((n-1)k_j)-h \sin(n k_j)\Bigr]\\
\psi_{k_j,n}^{(h)}&=(-1)^{L-j}\phi_{k_j,L+1-n}
\end{aligned}&&&&
\begin{aligned}
&j=0,\dots, L-1\\ 
&n=1,\dots, L\, .
\end{aligned}
\end{aligned}
\ee
where
\be
N_k^2=\frac{2}{L + h^2 (1 + L) - h (1 + 2 L) \cos{k}}
\ee
and $k_j$ are the $L$ momenta, solutions of the equation
\be\label{eq:momenta}
\sin(L k_j)-h\sin((L+1)k_j)=0\, .
\ee
We observe that in the disordered phase ($h>1$) all momenta are real. We also write eq. \eqref{eq:momenta} in logarithmic form
\be
k_j=\frac{\pi j}{L}+\frac{1}{L}\arctan\Bigl(\frac{h \sin(k_j)}{1-h \cos(k_j)}\Bigr)\, .
\ee
The following orthogonality conditions hold:
\be
\sum_{n}\phi_{k_j,n}\phi_{k_{j^\prime},n}=\delta_{j j^\prime}\qquad \sum_{j}\phi_{k_j,n}\phi_{k_{j},l}=\delta_{l n}
\ee
and analogously for $\psi$.

Let us now define
\be
G\equiv \braket{A(0)\otimes B(0)}
\ee
the fermionic correlations at the initial time; $Jh_0$ is  the magnetic field before the quench $Jh_0\rightarrow Jh$ and  $\braket{v\otimes w}$ denotes the matrix with elements
\be
\braket{v\otimes w}_{l n}=\braket{v_l w_n}\, .
\ee
In particular we have
\be
G_{l n}=-\sum_{j=0}^{L-1}\phi^{(h_0)}_{k_j,l}\psi^{(h_0)}_{k_j,n}\, ,
\ee
where $(h_0)$ indicates that $\phi$ and $\psi$ are the functions defined in eq.~\eqref{eq:phipsi} for the magnetic field $Jh_0$ (the magnetic field modifies the momenta). 
We now consider the fermionic correlations after the quench.
For simplifying the notation we define the auxiliary matrices
\be\label{eq:defQ}
\begin{aligned}
Q^{\phi\phi}_{l n}(t)&=\sum_{j=0}^{L-1}\phi_{k_j,l}^{(h)}\phi_{k_j,n}^{(h)}\cos(\varepsilon_{k_j}t)\\
Q^{\psi\phi}_{l n}(t)={Q^{\phi \psi}}^T_{l n}(t)&=\sum_{j=0}^{L-1}\psi_{k_j,l}^{(h)}\phi_{k_j,n}^{(h)}\sin(\varepsilon_{k_j}t)\\
Q^{\psi\psi}_{l n}(t)&=\sum_{j=0}^{L-1}\psi_{k_j,l}^{(h)}\psi_{k_j,n}^{(h)}\cos(\varepsilon_{k_j}t)
\end{aligned}
\ee
and the vectors
\be
[\vec\Phi_j]_l=\phi_{k_j,l}^{(h_0)}\qquad[\vec\Psi_j]_l=\psi_{k_j,l}^{(h_0)}\, ,
\ee
which satisfy the relations
\be\label{eq:relations}
\begin{aligned}
G\vec\Psi_j&=-\vec\Phi_j\\
G^T\vec\Phi_j&=-\vec\Psi_j\, .
\end{aligned}
\ee
The evolution after the quench of the operators $A$ and $B$ in  the Heisenberg picture can be written as follows
\be
\begin{aligned}
A(t)&=Q^{\phi\phi}A+i Q^{\phi\psi}B\\
B(t)&=i Q^{\psi\phi}A+Q^{\psi\psi}B\, ,
\end{aligned}
\ee
and hence the two-point functions become:
\begin{multline}
\braket{A(t_2)\otimes A(t_1)}=\braket{(Q^{\phi\phi}(t_2)A+i Q^{\phi\psi}(t_2)B)\otimes(A Q^{\phi\phi}(t_1) +i B  Q^{\psi\phi}(t_1))}=\\
Q^{\phi\phi}(t_2)Q^{\phi\phi}(t_1)+Q^{\phi\psi}(t_2)Q^{\psi\phi}(t_1)+ i Q^{\phi\phi}(t_2) G Q^{\psi\phi}(t_1)-iQ^{\phi\psi}(t_2) G^TQ^{\phi\phi}(t_1) 
\end{multline}
\begin{multline}
\braket{B(t_2)\otimes B(t_1)}=\braket{(i Q^{\psi\phi}A+Q^{\psi\psi}B)\otimes(i AQ^{\phi\psi}+BQ^{\psi\psi})}=\\
-Q^{\psi\phi}(t_2)Q^{\phi\psi}(t_1)-Q^{\psi\psi}(t_2)Q^{\psi\psi}(t_1)+ i Q^{\psi\phi}(t_2) G Q^{\psi\psi}(t_1)-iQ^{\psi\psi}(t_2) G^TQ^{\phi\psi}(t_1) 
\end{multline}
\begin{multline}
\braket{A(t_2)\otimes B(t_1)}=\braket{(Q^{\phi\phi}(t_2)A+i Q^{\phi\psi}(t_2)B)\otimes(i AQ^{\phi\psi}+BQ^{\psi\psi})}=\\
i Q^{\phi\phi}(t_2)Q^{\phi\psi}(t_1)-iQ^{\phi\psi}(t_2)Q^{\psi\psi}(t_1)+ Q^{\phi\phi}(t_2) G Q^{\psi\psi}(t_1)+Q^{\phi\psi}(t_2) G^TQ^{\phi\psi}(t_1) 
\end{multline}
\begin{multline}
\braket{B(t_2)\otimes A(t_1)}=\braket{(i Q^{\psi\phi}A+Q^{\psi\psi}B)\otimes(A Q^{\phi\phi}(t_1) +i B  Q^{\psi\phi}(t_1))}=\\
i Q^{\psi\phi}(t_2)Q^{\phi\phi}(t_1)-iQ^{\psi\psi}(t_2)Q^{\psi\phi}(t_1)- Q^{\psi\phi}(t_2) G Q^{\psi\phi}(t_1)-Q^{\psi\psi}(t_2) G^TQ^{\phi\phi}(t_1) 
\end{multline}
By substituting these correlations into eq. \eqref{eq:Gamma} we obtain the two-point function of the order parameter.
The orthogonality relations could be useful to simplify the fermionic correlations when $t_2=t_1$ or $t_1=0$.
These results will be used to compute the dynamic correlators at $T=0$ numerically, and the resulting Mathematica code based upon them is listed in appendix (\ref{appB}). Using this numerical code we will check our analytical results on the dynamics of the TFIC, that will be based on a semi-classical approach, as we shall see in the next chapter.
\begin{figure}[t]
   \begin{center}
   \includegraphics[width=7cm]{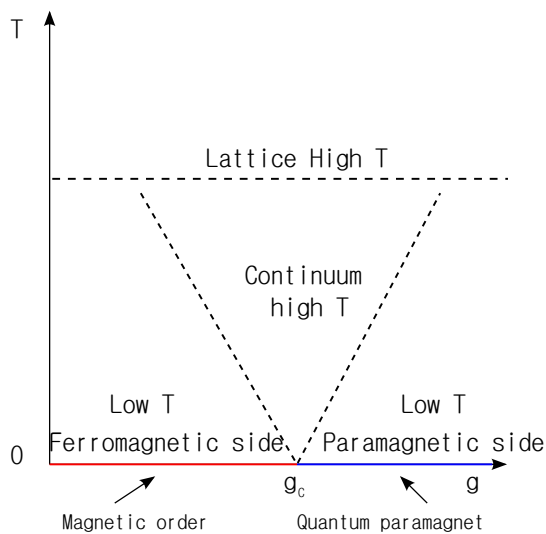}
   \end{center}
   \caption{Finite-$T$ phase diagram of the TFIC, as a function of the coupling $g$ and temperature $T$. There is a quantum phase transition at $g=g_{c}=1$. Magnetic long-range order ($N_{0}\neq 0$) is present only at zero temperature and $g<g_{c}$, while the ground state for $g>g_{c}$ is a quantum paramagnet. The dashed lines are crossovers at $|\Delta|\sim T$. The low-$T$ regions on both the ferromagnetic and paramagnetic side are studied in section (\ref{sec:semic}), whilst the continuum high-$T$ region is not discussed here because its description would require a different toolbox (a detailed analysis of this conformal critical point can be found in (\cite{Sachdev1})). Its properties are universal and determined by the continuum theory which describes this quantum critical point. Finally there is also a ``lattice high-$T$'' region with $T\gg J$ where properties are not universal and determined by the lattice scale Hamiltonian.}
   \label{chap4_fig:2}
\end{figure}
\section{The semi-classical method at finite $T$}
\label{sec:semic}

In this section we will introduce the semi-classical method to study real time, $T\neq 0$ correlations of the one-dimensional TFIC. 
This method was first introduced by S. Sachdev and P. Young (\cite{Young}), and the following review is based upon chapter $4$ of (\cite{Sachdev1}) and (\cite{Sachdevreview}).
The semi-classical method will be used here to obtain the exact asymptotics of order parameter correlations in the two low $T$ regions on either side of its quantum critical point (see figure (\ref{chap4_fig:2})). We will see that in these low $T$ regimes the spin correlation function can be expressed as a product of two factors. The first one, arising from quantum effects, gives the $T=0$ value of the correlation function, and the second one, which comes from a {\em classical} theory, describes the effects of temperature.   

\subsection{Ferromagnetic side}
\label{sec:ferr}
\begin{figure}[t]
   \begin{center}
   \includegraphics[width=7cm]{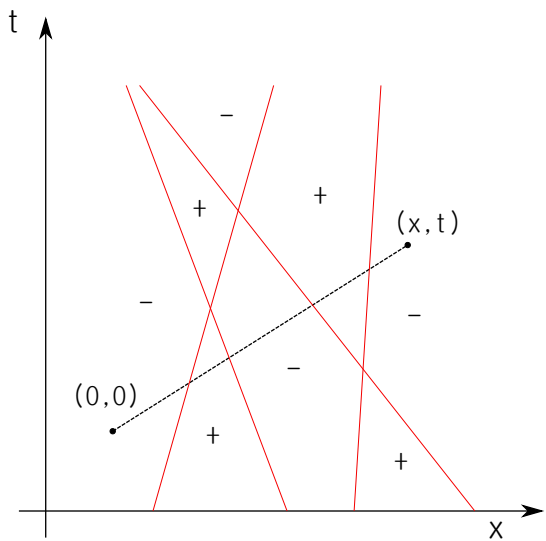}
   \end{center}
   \caption{A typical semi-classsical configuration contributing to the double path integral for $\langle\sigma^{z}(x,t)\sigma^{z}(0,0)\rangle$. Red lines are thermally excited particles that propagate forward and backwards in time. The $\pm$ are significant only in the ferromagnetically ordered phase, and denote the orientation of the order parameter. In the paramagnetic phase the black line is a particle propagating only forward in time from $(0,0)$ to $(x,t)$.}
   \label{chap4_fig:3}
\end{figure}

\begin{figure}[t]
   \begin{center}
   \includegraphics[width=9cm]{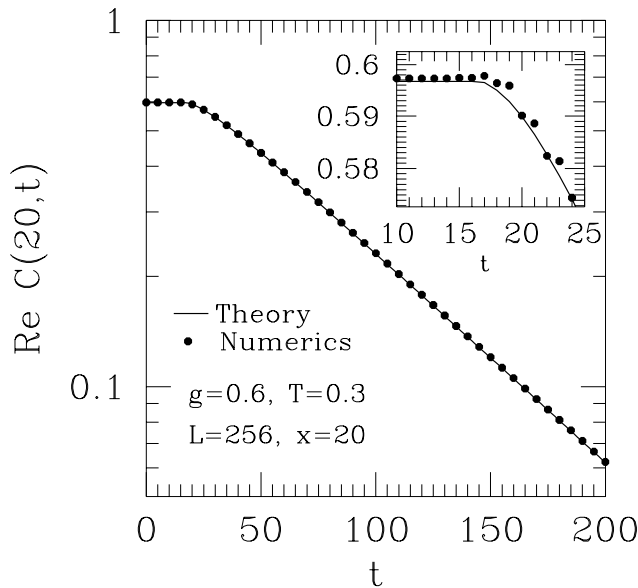}
   \end{center}
   \caption{The points show numerical data for the TFIC in the ferromagnetic region, obtained for a lattice size of $L=256$ with free boundary conditions (from (\cite{Young})). Printed with permission of the authors.}
   \label{chap4_fig:4}
\end{figure}

Let us start by analysing the low-$T$ magnetically ordered side. The excitations consist of particles (the so called kinks) whose main separation ($\xi\sim {\it e}^{\Delta/T}$, where $\xi$ is the correlation length and $\Delta$ the mass-gap) is much larger than their de Broglie wavelength ($\sim (c^{2}/\Delta T)^{1/2}$, where $c$ is the maximal propagation velocity in the system) as $T \to 0$, which corresponds exactly to the canonical condition for the applicability of classical physics.  The energy of a domain wall when the momentum is small is $\Delta+c^{2}k^{2}/2\Delta$,  and hence the density $\rho$ can be written by using classical Boltzman statistics as:
\be
\label{chap4_eq:98}
\rho={\displaystyle\int}\frac{dk}{2\pi}{\it e}^{-(\Delta+c^{2}k^{2}/2\Delta)/T}=\left(\frac{T\Delta}{2\pi c^{2}}\right)^{1/2}{\it e}^{-\Delta/T}.
\ee
Now if we computed the correlator $C(x,0)$ using the determinant technique which was introduced in section (\ref{sec:corrT}), we would get:
\be
\label{chap4_eq:99}
C(x,0)=N_{0}^{2}{\it e}^{-|x|/\xi},
\ee
where  $N_{0}^{2}\propto\Delta^{1/4}$ and $\xi^{-1}= \left(\frac{2T|\Delta|}{\pi c^{2}}\right)^{1/2}{\it e}^{-\Delta/T}$, therefore $\xi=1/2\rho$. This result can also be understood if we assume that the domain walls are classical point particles, distributed independently with a density $\rho$. Consider now a system of size $L\gg |x|$, and say that we heve $M$ thermally excited particles, so that $\rho=M/L$. Let $q$ be the probability for a particle to be located inbetween the points $0$ and $x$, that is $q=|x|/L$.  
Now the probability that a given set of $j$ particles are the only ones between the points $0$ and $x$ is the $q^{j}(1-q)^{M-j}$. As in this picture each particle reverses the orientation of the ground state, we have: 
\be
\label{chap4_eq:100}
\sigma^{z}(x,0)\sigma^{z}(0,0)=N_{0}^{2}(-1)^{j},
\ee
and summing over all possibilities we obtain:
\be
\label{chap4_eq:101}
\begin{array}{lcl}
C(x,0) & = & N_{0}^{2}{\displaystyle\sum_{j=0}^{M}}(-1)^{j}q^{j}(1-q)^{M-j}{\displaystyle\frac{M!}{j!(M-j)!}}\\
& = & N_{0}^{2}(1-2q)^{M}\approx N_{0}^{2}{\it e}^{-2qM}=N_{0}^{2}{\it e}^{-2\rho |x|},
\end{array}
\ee
which is exactly the result we would have obtained if we had done the whole calculation using the determinant technique of section (\ref{sec:corrT}).
The semi-classical picture can also be used to compute unequal-time correlations, where one has to take into account collisions between the particles. Even though the particles are very dilute, in one dimension they cannot really avoid each other, and neighbor particles will always eventually collide (this is not the case in higher dimensions where sufficiently dilute particles can be treated as noninteracting). During any collision particles are certainly closer than their di Broglie wavelengths, and therefore we must treat the scattering process quantum mechanically. To do that we will consider the two-particle $S$-matrix, and due to the diluteness of the particles in the system we will consider collisions of only pairs of particles. \\
In order to study dynamic correlations, let us reexamine the explicit expression for $C(x,t)$ in (\ref{chap4_eq:2}). This quantity can be evaluated exactly using some simple physical arguments. The key idea is that classical mechanics emerges from quantum mechanics as a stationary phase evaluation of a first-quantized Feyman path integral. Let us notice that the integral is over a set of trajectories moving forward in time, representing the operator ${\it e}^{-iHt}$, and a second set moving backwards in time, corresponding to the action of  ${\it e}^{iHt}$. In the semi-classical limit, stationary phase is achieved when the backwards paths are simply the reverse of the forward ones, and both sets represent the classical trajectories. An example of such paths is shown in figure (\ref{chap4_fig:3}).
Let us now notice that the classical trajectories remain straight lines across collisions because the momenta before and after the scatetring process are the same. This observation is a direct consequence of conservation of total momentum ($k_{1}+k_{2}=k'_{1}+k'_{2}$) and energy ($\epsilon_{k_{1}}+\epsilon_{k_{2}}=\epsilon_{k'_{1}}+\epsilon_{k'_{2}}$) in each two-particle collision, which has the unique solution $k_{1}=k'_{1}$ and $k_{2}=k'_{2}$ (or a permutation of the indices, but this last solution do not need to be considered separately because the particles are identical) in one dimension. Moreover for each collision the amplitude of the path acquire a phase $S_{k_{1}k_{2}}$ along the forward path and its complex conjugate along the backwards path. Therefore the net factor after the collision is $|S_{k_{1},k_{2}}|^{2}=1$. These two facts imply that the trajectories of the particles are simply independently distributed straight lines, with a uniform density $\rho$ along the $x$-axis, with an inverse slope given by:
\be
\label{chap4_eq:102}
v_{k}\equiv\frac{d\epsilon_{k}}{dk},
\ee
where the value of the momenta are chosen with the Boltzmann probability density ${\it e}^{-\epsilon/T}$. 
In practice now computing dynamic correlatos is an exercise in classical probabilities. As each particle line is the boundary between domains with opposite orientations of the spins, the value of $\sigma^{z}(x,t)\sigma^{z}(0,0)$ is the square of the magnetization renormalized by quantum fluctuations ($N_{0}^{2}$) times $(-1)^{j}$, where $j$ is the number of different trajectories inserting the black line in figure (\ref{chap4_fig:3}). What is left now is the average of $N_{0}^{2}(-1)^{j}$ over the classical ensemble of paths defined above. Let us choose again a system size $L\gg |x|$ with $M$ particles, the probability $q$ that a given particle with velocity $v_{k}$ is between the points $(0,0)$ and $(x,t)$ is:
\be
\label{chap4_eq:103}
q=\frac{|x-v_{k}t|}{L}.
\ee 
Averaging over velocities and then evaluating the summation in (\ref{chap4_eq:101}) gives us the following result:
\be
\label{chap4_eq:104}
C(x,t)=N_{0}^{2}\exp\left(-{\displaystyle\int\frac{dk}{\pi}}{\it e}^{-\epsilon_{k}/T}|x-v_{k}t|\right).
\ee
The equal-time or equal-space form of the relaxation function $R(x,t)$ is simply given by:
\be
\label{chap4_eq:104}
\begin{array}{lcl}
R(x,0)={\it e}^{-|x|/\xi},\quad
R(0,t)={\it e}^{-|t|/\tau},
\end{array}
\ee
whilst for general $x,t$ the function $R$ also is a monotonically decreasing function with increasing $|x|$ or $|t|$, but the decay is no longer exponential. 
The spatial correlation length $\xi$ was given in (\ref{chap4_eq:99}), while we can determine $\tau$ from (\ref{chap4_eq:104}):
\be
\label{chap4_eq:105}
\begin{array}{lcl}
\tau^{-1} & = & {\displaystyle\frac{2}{\pi}\int_{0}^{\infty}}dk\frac{d\epsilon_{k}}{dk}{\it e}^{-\epsilon_{k}/T}\\
& = & {\displaystyle\frac{2}{\pi}\int_{|\Delta|}^{\infty}}d\epsilon_{k}{\it e}^{-\epsilon_{k}/T}\\
& = & {\displaystyle\frac{2T}{\pi}}{\it e}^{-|\Delta|/T}.
\end{array}
\ee
The correlator $C(x,t)$ can also be written as:
\be
\label{chap4_eq:106}
C(x,t)=N_{0}^{2}\phi_{R}\left(\frac{x}{\xi},\frac{t}{\tau}\right),
\ee
where the relaxation function $R(x,t)$ satisfies the scaling form:
\be
\label{chap4_eq:107}
R(x,t)=\phi_{R}\left(\frac{x}{\xi},\frac{t}{\tau}\right).
\ee
This scaling function is valid only for $T\ll \Delta$, and can be explicitly written as:
\be
\label{chap4_eq:108}
\ln\phi_{R}(\bar{x},\bar{t})=-\bar{x}{\rm erf}\left(\frac{\bar{x}}{\bar{t}\sqrt{\pi}}\right)-\bar{t}{\it e}^{-\bar{x}^{2}/(\pi\bar{t}^{2})}.
\ee
All these results have been compared with exact numerical computations and the agreement was found to be perfect (see figure (\ref{chap4_fig:4}) and also (\cite{Young}, \cite{Sachdev1})).  
\subsection{Paramagnetic side}
\label{sec:par}
We have already argued in the strong-coupling analysis of section (\ref{sec:strong}) that an important feature of the spectral density of the quantum paramagnet was the quasi-particle delta function which was shown in figure (\ref{chap4_fig:1}). It is natural to expect that the leading term in the large-$x$ dependence of the correlator at zero temperature is determined simply by the contribution of this pole, plus the relativistic invariance of the continuum theory which can be defined in the proper scaling limit (for more details see (\cite{Sachdev1})). The dynamical suceptibilty must therefore have the form:
\be
\label{chap4_eq:109}
\chi(k,w)={\displaystyle\frac{A}{c^{2}k^{2}+\Delta^{2}-(w-i\delta)^{2}}}+\dots,\qquad T=0,
\ee
where $A$ is the quasi-particle residue and $\delta$ is a positive infinitesimal. The continuum of excitations above the three-particle threshold is represented by the ellipses in (\ref{chap4_eq:109}). Let us use (\ref{chap4_eq:109}) to deduce the $T=0$ equal-time correlations. This can be easily done by analytically continuing (\ref{chap4_eq:109}) to imaginary frequencies $w_{n}$, and then using the inverse of definition (\ref{chap4_eq:5}), which gives:
\be
\label{chap4_eq:110}
\begin{array}{lcl}
C(x,0) & = & A{\displaystyle\int\frac{dw}{2\pi}\int\frac{dk}{2\pi}\frac{{\it e}^{-ikx}}{w^{2}+c^{2}k^{2}+\Delta^{2}}}\\
& = & {\displaystyle\frac{A}{\sqrt{8\pi c|\Delta||x|}}{\it e}^{-\Delta|x|/c}}, \qquad |x|\to\infty\quad {\rm at}\quad T=0,
\end{array}
\ee
To determine the value of $A$ one has to do a microscopic lattice calculation, which would give (see for instance (\cite{McCoy})):
\be
\label{chap4-eq:111}
A=2cJ^{-1/4}|\Delta|^{1/4}.
\ee
We immediately see that the residue vanishes at the critical point $\Delta=0$, where the quasiparticle picture breaks down, and we will have to consider a new and  completely different structure of excitations. Nonetheless the above is a complete description of the correlations and excitations of the quantum paramagnetic ground state.  Let us turn to the dynamic properties at $T>0$, where there will be a small density of quasiparticle excitations that will behave classically for the same reason as in section (\ref{sec:ferr}): their main distance is much larger that their de Broglie wavelength. The scattering processes of these thermally excited quasiparticle will lead to a broadening of the delta function pole in (\ref{chap4_eq:109}). The specific form of this broadening can be computed exactly in the limit of low temperature $T\ll |\Delta|$ using a semi-classical approach similar to that employed fro the ordered side. Here the key observation is that we may consider the operator $\sigma^{z}$ to be given by:
\be
\label{chap4_eq:112}
\sigma^{z}(x,t)=\sqrt{A}(\psi(x,t)+\psi^{\dag}(x,t))+\dots,	
\ee
where $\psi^{\dag}$ is the operator that creates a single-particle excitation, and the ellipses represent multiparticle creation or annihilation terms, which turns out to be subdominant in the long-time limit. This representation may also be linked to $g\gg 1$ picture discussed before, in which the single particle excitations were $|\to\rangle$ spins: The $\sigma^{z}$ operator flips spins between the $\pm x$ directions, and therefore creates and annihilates quasiparticles. 
Because the computation  of the nonzero $T$ relaxation is best done in real space and time, let us first write down the $T=0$ correlations in this framework. We define $K(x,t)$ as the $T=0$ correlator of the order parameter (in the continuum and relativistic limit):
\be
\label{chap4_eq:113}
\begin{array}{lcl}
K(x,t) & \equiv &  \langle\sigma^{z}(x,t)\sigma^{z}(0,0)\rangle_{T=0}\\
& = & {\displaystyle\int\frac{dk}{2\pi}\frac{c|\Delta|^{1/4}}{J^{1/4}\epsilon_{k}}{\it e}^{i(kx-\epsilon_{k}t)}}\\
& = & {\displaystyle\frac{|\Delta|^{1/4}}{J^{1/4}\pi}}K_{0}\left(|\Delta|(x^{2}-c^{2}t^{2})^{1/2}/c\right),
\end{array}
\ee
where $K_{0}$ is the modified Bessel function. This result has been obtained by the Fourier transform of (\ref{chap4_eq:109}). The first step of the previous calculation is valid in general (also for the lattice theory), whilst the second one makes use of the relativistic invariance of the continuum theory (see (\cite{Sachdev1})).  
Our main interest now is the $T>0$ properties of the correlations within the light-cone ($x<ct$), where the correlations are large and oscillatory (corresponding of the propagation of real particle) and display semi-classical dynamics.
\begin{figure}[t]
   \begin{center}
   \includegraphics[width=9cm]{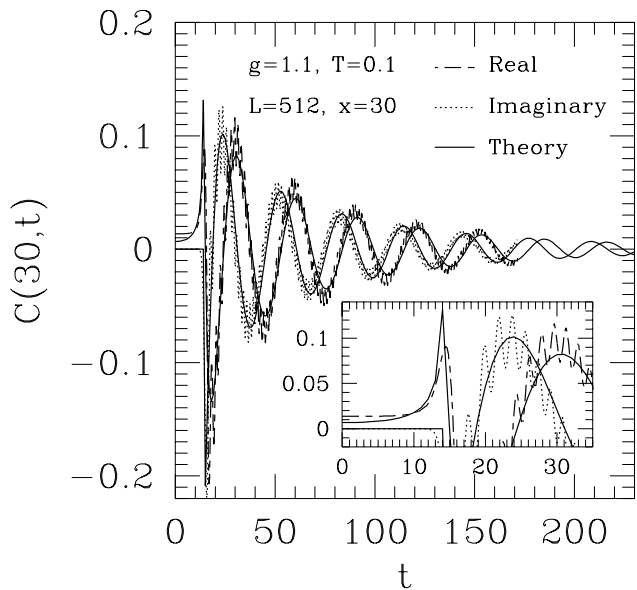}
   \end{center}
   \caption{The numerical data for the TFIC with $J=1$, in the paramagnetic region for a lattice of $L=512$ with free boundary conditions (from (\cite{Young})). The numerical has a ringing at high frequency, which is due to the upper energy cut-off in the dispersion relation, that is present in the lattice model but not in the continuum theory. The inset shows a part of the same data on a larger scale for sake of clarity. Printed with permission of the authors.}
   \label{chap4_fig:5}
\end{figure} 
Let us now focus on the $T\neq 0$ evolution, in the same semi-classical path integral approach that was employed earlier in section (\ref{sec:ferr}). As before we are dealing with semi-classical particles, even though the physical characterization of these particles is now quite different: They are quasiparticles excitations above a quantum paramagnet, no longer domain walls between magnetically ordered regions. Nevertheless the path-integral representation of (\ref{chap4_eq:2}) leads to two sets of paths: 
one forward and the other backward in time. However there is a peculiar trajectory that propagates only forward in time, that is the trajectory representing the particle created by the first $\sigma_{0}^{z}$ and annihilated by the second one. The inverse process will not be considerated here, because the probability for the first $\sigma^{z}$ to destroy an already present one is negligible, being the density of thermally excited particle exponentially small. Yet, as in the semi-classical limit, the forward and backward trajectories of the excitations are expected to be the same; the particle on the trajectory created by the first $\sigma_{0}^{z}$ must be annihilated at the $\sigma_{i}^{z}$, otherwise the initial and final states in the trace of (\ref{chap4_eq:2}) will not be the same. \\
This reasoning can be represented in a spacetime snapshot of the paths that is the same as in figure (\ref{chap4_fig:3}), but its physical meaning is now very different. The $\pm$ signs in the domains should be ignored. In the absence of any thermally excited particle the black line represents the propagation of a free particle moving from $(0,0)$ to $(x,t)$, and will contribute the $T=0$ Feyman propagator above, $K(x,t)$, to $\langle\sigma^{z}(x,t)\sigma^{z}(0,0)\rangle$. The scattering of the background thermally excited particles (the red lines in figure (\ref{chap4_fig:3})) introduces factors of the $S$ matrix at each collision; as the black line only propagates forward in time, the $S$ matrix elements for collisions between this line and the red ones are not neutralized by their complex conjugate trajectory. All other collisions occur both forward and backward in time, therefore they contribute $|S_{k_1,k_2}|^{2}=1$. \\
Using the low momentuum value $S_{k_1,k_2}=-1$, we see that the contribution to  $\langle\sigma^{z}(x,t)\sigma^{z}(0,0)\rangle$ from the set of trajectories equals $(-1)^{j}K(x,t)$, where $j$ is the number of red lines intersecting the black one. Remarkably the $(-1)^{j}$ factor is exactly the same we found in the magnetically ordered phase case, although for very different physical reasons. Computing the average over all trajectories we obtain the low-$T$ dynamic correlator in the paramagnetic side:
\be
\label{chap4_eq:114}
C(x,t)= K(x,t)R(x,t),
\ee
where the relaxation function is the same as in (\ref{chap4_eq:107}). Notice that this result clearly displays the separation in scales at which quantum and thermal effects act. Quantum fluctuations determine the oscillatory,complex function $K(x,t)$, which gives the $T=0$ value of the correlator. 
Exponetial relaxation of spin correlations occurs at longer scales $\sim\xi,\tau$ and is controlled by the classical motion of particles. This classical realization is expected to broaden the quasiparticle pole with widths of order $\xi^{-1}$ and $\tau^{-1}$ in momentum and energy space. \\
These analytical prediction have been tested numerically (see figure (\ref{chap4_fig:5}) and also the article (\cite{Young})) and again the agreement is very good. Same differences appear outside the light-cone, but this is outisde the domain of validity of the semi-classical approach. We will find a similar behaviour in the next chapter, when we will deal with dynamic correlations  after a quantum quench.



\chapter{Semi-classical theory for quantum quenches in the O(3) \newline non-linear sigma-model}
\label{chap4}
\ifpdf
    \graphicspath{{Chapter4/Chapter4Figs/PNG/}{Chapter4/Chapter4Figs/PDF/}{Chapter4/Chapter4Figs/}}
\else
    \graphicspath{{Chapter4/Chapter4Figs/EPS/}{Chapter4/Chapter4Figs/}}
\fi

In this chapter we use the semi-classical approach to study the non-equilibrium dynamics of the O(3) non-linear sigma model. For a class of quenches defined below in the text, we obtain the order-parameter dynamical correlator in the thermodynamic limit. In particular we predict quench-dependent  relaxation times and correlation lengths. The approach developed for the O(3) non-linear sigma model can also be applied to the transverse field Ising chain, where the semi-classical results can be directly compared to both the exact and the numerical ones, revealing the limits of the method. \\
This chapter basically covers the content of the analogous article ``{\em Semi-classical theory for quantum quenches in the O(3) non-linear sigma-model}'' written by S. Evangelisti, arXiv:1210.4028, submitted to Journal of Statistical Mechanics.

\section{Introduction to the problem}
\markboth{\MakeUppercase{\thechapter.  Semi-classical theory for the O(3) non-linear sigma-model}}{\thechapter. Semi-classical theory for the O(3) non-linear sigma-model}

The out-of-equilibrium physics of many-body systems has attracted a lot of interest in recent years, not least because of the experimental realizations of
ultracold atomic gases in optical lattices (\cite{Greiner}, \cite{Kinoshita}, \cite{Hofferberth}, \cite{Trotzky}, \cite{Cheneau}, \cite{Gring}).  
These experiments have observed the dynamics of many-body systems on a long time scale after a quantum quench, finding essentially unitary time-evolution.  In three-dimensional systems fast relaxation towards a thermal steady state has been observed. On the contrary, in quasi-one-dimensional systems the relaxation process is normally much slower and leads to a peculiar non-thermal stationary state (\cite{Kinoshita}). These results have led to a huge theoretical push (\cite{Hofferberth}, \cite{Trotzky}, \cite{Cheneau}, \cite{Gring}, \cite{Polkovnikov}, \cite{Rigol}, \cite{Rigol2}, \cite{Calabrese3}, \cite{Cazalilla1}, \cite{Cazalilla2}, \cite{Cazalilla3}, \cite{Cazalilla4}, \cite{Barthel}, \cite{Rossini1}, \cite{Rossini2}, \cite{Fioretto1}, \cite{Biroli}, \cite{Banulus}, \cite{Gogolin}, \cite{Rigol3}, \cite{Mossel}, \cite{Caux}, \cite{Manmana}) to address fundamental questions such as whether there is an asymptotic stationary state, and, if it exists, which {\em ensemble} characterizes it. The belief is that observables of non-integrable systems effectively thermalize, which implies that their stationary state is characterized by a thermal Gibbs ensemble. Numerical works on non-integrable systems confirm this expectation, even if some contradictory results point out that some issues have not been completely understood yet (\cite{Biroli},\cite{Banulus}, \cite{Gogolin}, \cite{Kollath}).  On the other hand, in integrable systems, because of the existence of {\em local} integrals of motion, the stationary state is expected to be described by a generalized Gibbs ensemble (GGE), where each mode associated with a conseved quantity is characterized by its own temperature. So far results on integrable systems have been focussed on free fermion models, such as the transverse field Ising chain and the quantum XY chain (\cite{McCoy1}, \cite{McCoy2}, \cite{McCoy3}, \cite{Igloi4}, \cite{Sengupta}, \cite{Fagotti1}, \cite{Silva1}, \cite{Silva2}, \cite{Venuti1}, \cite{Venuti2}, \cite{Igloi1}, \cite{Foini}, \cite{Igloi2}, \cite{Calabrese4}, \cite{Fabian}, \cite{Calabrese5}, \cite{Calabrese6}, \cite{Igloi3}). \\
For the transverse field Ising chain in thermal equilibrium Sachdev and Young first introduced a semi-classical description of the physical properties of the model in terms of ballistically moving quasi-particles (\cite{Young}). This approach turned out to be incredibly accurate in predicting the temperature dependence of correlation length, relaxation time and in general to compute the order-parameter two point function in the ferromagnetic and paramagnetic phases. For global quenches this technique has been used to describe the dynamics of the transverse field Ising chain and the quantum XY model, with great accuracy (\cite{Igloi4}, \cite{Igloi1}, \cite{Igloi2}, \cite{Igloi3}). In these works quantitative features of the relaxation process  have been explained with a quasi-particle picture, which had been introduced before in particular to study the evolution of the entanglement entropy (\cite{Fagotti1}, \cite{Calabrese8}, \cite{Calabrese7}, \cite{Eisler}). In practice the quench injects an extensive amount of energy into the system, which creates quasi-particles homogeneously in space, that then move ballistically with constant velocity. Because of momentum conservation, these quasi-particles are created in pairs with opposite momenta and are quantum entangled (within each pair). Their dynamics can be treated classically as long as they do not collide. But since collisions are unavoidable in one dimension,  every scattering process has to be treated quantum mechanically. We shall use this semi-classical approach to calculate the dynamical correlation functions analytically for the O(3) non-linear sigma model. The dynamics of this model have already been studied at finite temperature in equilibrium (\cite{Sachdev3}, \cite{Sachdev1} and \cite{Rapp2}), where the order-parameter correlation function shows a {\em universal} form. Here we shall study its behaviour after a quantum quench, preparing the system in a state which is not an eigenstate of the Hamiltonian. In addition a section will be dedicated to the case of the transverse field Ising chain, whose order-parameter two-point correlation function can be derived straightforwardly from the general one obtained for the O(3) non-linear sigma model.\\
The chapter is organized as follows: first we introduce the model and the general form of the initial states we shall consider throughout the whole article, 
before introducing and commenting the main result of the paper, concerning the O(3) non-linear sigma model. Then we describe the basic ideas behind the semi-classical technique and see in detail how to apply them in the present case.

\section{The Model}
\label{s:model}%
Let us start by considering the one-dimensional $O(3)$ quantum rotor chain, which is described by the following Hamiltonian:
\begin{equation}
\label{eq:1}
\hat{H}^{\rm rotor}=\frac{Jg}{2}\sum_{i}\hat{L}_{i}^{2}-J\sum_{i}\hat{n}_{i}\hat{n}_{i+1},
\end{equation}
where $\hat{n}_{i}$ is the position operator of the rotor on site $i=1,\dots,L$ with the constraint $\hat{n}_{i}^{2}=1$\,$\forall_{i}$, and $\hat{L}_{i}=\hat{n}_{i}\times \hat{p}_{i}$ is the angular momentum operator. $J$ is an overall energy scale and $g$ is a positive coupling constant. The operators which appear in the Hamiltonian satisfy the usual commutation relations $[\hat{L}_{i}^{\alpha},\hat{L}_{i}^{\beta}]=i\epsilon_{\alpha\beta\gamma}\hat{L}_{i}^{\gamma}$ and $[\hat{L}_{i}^{\alpha},\hat{n}_{i}^{\beta}]=i\epsilon_{\alpha\beta\gamma}\hat{n}_{i}^{\gamma}$, where $\alpha,\beta,\gamma$ represent the three spatial directions. 
The continuum limit of this lattice model is the  O(3) non-linear sigma model (${\rm nl\sigma m}$), whose Lagrangian density reads:
\begin{equation}
\label{eq:1quatris}
\mathcal{L}={\displaystyle\frac{1}{2\tilde{g}}(\partial_{\mu}\tilde{n}_{i})^{2}},\quad \tilde{n}_{i}^{2}=1,
\end{equation}
where $\tilde{n}_{i}=\tilde{n}_{i}(x,t)$ are three scalar fields and $\tilde{g}$ is a (bare) coupling constant. Here we have already set the maximal propagation velocity $c=1$.  This model is O(3)-symmetric, renormalizable and asymptotically free, and it has three massive particles in the O(3)-multiplet.  The exact $\mathcal{S}$-matrix of this model is known (\cite{Zamolodchikov}), with any scattering event involving no particle production and the general $n$-particle $\mathcal{S}$-matrix factorizes into a product of two-particle amplitudes. 
It is worth noticing that the O(3) non-linear sigma model also provides a description of the low-energy excitations of the one-dimensional antiferromagnetic $S=1$ Heisenberg chain (\cite{Haldane}), whose dynamical correlation functions' lineshape can be measured experimentally. 
In the $g\gg 1$ limit, in the ground state of Hamiltonian (\ref{eq:1}) all rotors must be in the $L_{i}^{2}=0$ state to minimize the kinetic energy.  The low-energy excitations above the ground state form a triplet with quantum numbers $L_{i}^{z}=(-1,0,1)$.  It is worth remarking that the structure of the low-energy spectrum is the same for any $g > 0$ and the system has a gap $\Delta(g)$. A finite gap is a necessary condition to apply a semiclassical approximation, as we shall see in the next section (for a deeper introduction to the semi-classical method and its range of applicability see (\cite{Sachdev1}) and references therein).\\
In this article we study the dynamical correlator of the order parameter $\tilde{n}^{z}$ after having prepared the system in a squeezed coherent state, namely\footnote{Here for a matter of simplicity we ignore the existence of zero-rapidity terms in the definition of $|\psi\rangle$. In addition it is known that an integrable boundary state as written in equation (\ref{eq:2}) is typically not normalizable, because the amplitude $K$ does not go to zero for large momenta. 
Therefore one has to introduce an extrapolation time in order to make the norm of the state finite, as was done for instance in (\cite{Fioretto1}).}:
\begin{equation}
\label{eq:2}
|\psi\rangle = \exp{\left (\sum_{a, b}\int_0^{\infty}\frac{dk}{2\pi}\, K^{ab}(k)Z^{\dag}_{a}(-k)Z^{\dag}_{b}(k) \right)}|0\rangle,
\end{equation}
where $|0\rangle$ is the ground state of the model, $K^{ab}(k)$ is the amplitude relative to the creation of a pair of particles with equal and opposite momenta, $Z^{\dag}_{a}(k)$ are creation operators of an excitation with quantum number $a=(-1,0,1)$ (the $z$-component of the angular momentum) and momentum $k$.
The main reason for choosing such an initial state comes from its relation with boundary integrable states: as shown by (\cite{Calabrese3},\cite{Calabrese7}) some dynamical problems can be mapped into an equilibrium boundary problem defined in a strip geometry, where the initial state $|\psi\rangle$ acts as a boundary condition. In integrable field theories the most natural boundary states preserve the integrability of the bulk model, or in other words, do not spoil the integrals of motion. These boundary states were originally studied by Ghoshal and Zamolodchikov (\cite{Goshal1}), and are supposed to capture the universal behaviour of all quantum quenches in integrable models. Furthermore in (\cite{Fioretto1}) it was shown that any quantum quench of an integrable field theory with this kind of initial states leads to a steady state which is described by a GGE ensemble. Precisely they show that long time limit of the one point function of a local operator can be described by a generalized Gibbs ensemble\footnote{A rigorous proof that the LeClair-Mussardo series work for the one-point functions in the case of a quench has been obtained in (\cite{Pozsgay}).}. In short, their result is strong evidence that Rigol {\em et} al.'s conjecture does hold for integrable field theories. \\
The amplitude $K^{ab}(k)$ in expression (\ref{eq:2}) is a regular function which must satisfy a set of constraints that depend on the $\mathcal{S}$-matrix, such as crossing equations, boundary unitarity and boundary Yang Baxter equation. Different solutions of these equations form the set of integrable boundary conditions of the theory.\footnote{For instance for the case of the O(3)  non-linear sigma model with free and fixed boundary conditions these states have been studied in (\cite{Goshal2}).} \\
The integrable boundary states belong to the class (\ref{eq:2}), but the semi-classical approach applies to the larger class of quenches whose initial states are expressed by a {\em coherent superposition of particle pairs}.
The order parameter $\tilde{n}^{z}(x)$ may be written as (\cite{Sachdev1}):
\begin{equation}
\label{eq:2tris}
\tilde{n}^{z}(x,t)\propto (Z^{\dag}_{0}(x,t)+Z_{0}(x,t))+\dots
\end{equation}
where $Z_{0}(x,t)$ is the (time-dependent) Fourier transform of $Z_{0}(k)$ and the ellipses represent multiparticle creation or annihilation terms, which will be considered negligible because they are subdominat in the long-time limit. 
Relation (\ref{eq:2tris}) comes from the observation that the operator $\tilde{n}^{z}(x)$ either creates a quasiparticle ($L_{i}=1$) with $L_{i}^{z}=0$ at position $x$ with some velocity $v$ or destroys one already present in the system, as will be discussed in the section (\ref{s:semi-classics}).\\
The quench protocol is the following: At time $t=0$ we prepare the system in the initial state (\ref{eq:2}) and  for $t>0 $ this state evolves according to the Hamiltonian (\ref{eq:1}):
\begin{equation}
\label{eq:3}
|\psi(t)\rangle= \exp{(-i\hat{H}t)}|\psi\rangle
\end{equation}
On the other side the time-evolution of an operator is written as:
\begin{equation}
\label{eq:4}
\hat{O}(x, t)=\exp{(i\hat{H}t)}\hat{O}(x)\exp{(-i\hat{H}t)}.
\end{equation}
We are going to analyze the two-point correlator:
\begin{equation}
\label{eq:5}
C^{\rm nl\sigma m}(x_{1}, t_{1}; x_{2}, t_{2})=\langle \psi |\tilde{n}^{z}(x_{2}, t_{2})\tilde{n}^{z}(x_{1}, t_{1})| \psi\rangle.
\end{equation}
The autocorrelation function is obtained for $x_1=x_2=x'$, whereas the equal-time correlation function is defined by imposing $t_1=t_2=t'$, even though in this chapter the emphasis will be given to the non-equal time case. 

\section{Summary and discussion of the results}
\label{s:results}%
In this section we summarize the main result of the chapter. The semi-classical method is based upon the existence of a small parameter, namely the average density of  excitation pairs with quantum numbers $(a,b)$ $n^{ab}(k)$, in the initial state $|\psi\rangle$:
\begin{equation}
\label{eq:res1}
\frac{\langle\psi|n^{ab}(k)|\psi\rangle}{\langle\psi|\psi\rangle}\equiv f^{ab}_k\approx|K^{ab}(k)|^{2},
\end{equation}
which is valid in the limit ${\rm max}_{k}\,|K^{ab}(k)|^{2}\ll 1$, $\forall a,b$. This limit also defines what we call {\em small quench}. The explicit computation of (\ref{eq:res1}) will be carried out in section (\ref{app:occ_numb}).
In the limit of small quenches we obtain the following expressions for the (non-equal time) two-point function of the order-parameter of the O(3) non-linear sigma model (\ref{eq:5}) in the limit $T\equiv t_2+t_1 \to \infty$:
\begin{equation}
\label{eq:res1bis}
C^{\rm nl\sigma m}(x;t)=C_{\rm propag}(x,t)\,R(x;t),
\end{equation}
where $C_{\rm propag}(x,t)$ corresponds to the coherent propagation of a quasipaticle, while $R(x;t)$ is the relaxation function, which describes the scattering with the excited quasiparticles. By definition $t\equiv t_2-t_1$ and $x=x_2-x_1$ (without any loss of generality we assume both $(x,t)$ to be positive). The pre-factor in (\ref{eq:res1bis}) reads:
\begin{equation}
\label{eq:res1tris}
C_{\rm propag}(x,t)\propto K_{0}(\Delta\sqrt{x^{2}-t^{2}}),
\end{equation}
where $K_{0}$ is the modified Bessel function of the second kind and $\Delta$ is the mass-gap.
The relaxation function $R(x;t)$ is given by (\cite{Evangelisti5}):								
\begin{equation}
\label{eq:res2}
\begin{array}{l}
R(x; t)={\displaystyle\int_{-\pi}^{\pi}\frac{d\phi}{2\pi}}\exp{\left(-t(1-\cos\phi){\displaystyle\int_{0}^{\infty}\frac{dk}{\pi}f_k\, v_{k}\Theta[v_{k}t-x]}\right)}\\
\\
\times\,\exp{\left(-x(1-\cos\phi){\displaystyle\int_{0}^{\infty}\frac{dk}{\pi}f_k\,\Theta[x-v_{k}t]}\right)}\cos\left(x\sin\phi{\displaystyle\int_{0}^{\infty}\frac{dk}{\pi}f_k}\right)\\
\\
\times\,{\displaystyle\frac{1+2\cos(\phi)(P^{00}-1)Q+\cos(2\phi)(Q^{2}-P^{00})-P^{00}Q^{2}}{1-2P^{00}\cos(2\phi)+(P^{00})^{2}}}.
\end{array}
\end{equation}
$Q$ is defined by  $Q= \sum_{\lambda=\pm 1,0}P^{0\lambda}$, $f_{k}= \sum_{a,b}f_{k}^{ab}$, $v_k$ is the velocity of the particle with momentum $k$ and $P^{ab}=\int_{0}^{\infty}\frac{dk}{2\pi}P^{ab}(k)$, where
\begin{equation}
\label{eq:res2bis}
P^{ab}(k)=\frac{{\displaystyle f_{k}^{ab}}}{{\displaystyle\sum_{a,b}\int_{0}^{\infty}\frac{dk}{2\pi}f_{k}^{ab}}}.
\end{equation}
Formula (\ref{eq:res2}) is the main result of this chapter, and it describes the long-time behaviour of {\em any} initial state of the form (\ref{eq:2}). Nonetheless if we want the state $|\psi\rangle$ to respect  the O(3)-symmetry 
we must impose $P^{00}=P^{1,-1}=P^{-1,1}=1/3$ ($Q=1/3$), and all other probabilities equal to zero.\\
The most general expression for the relaxation function $R(x; t, T)$ for generic values of the times $(t_1, t_2)$ will be given in the section (\ref{app:general}). 
Formula (\ref{eq:res2}) defines the quench-specific time and length scales:
\be
\label{eq:res3}
\tau^{-1}  =  {\displaystyle\int_{0}^{\infty}\frac{dk}{\pi}v_{k}\,f_k}\qquad\qquad\xi^{-1}  =  {\displaystyle\int_{0}^{\infty}\frac{dk}{\pi}f_k}.
\ee
We also note that the correlator of equation (\ref{eq:res2}) is {\em never} thermal unless one prepares the system at $t=0$ in a thermal inital state. This result confirms the belief that one-dimensional integrable systems relax towards a peculiar non-thermal distribution, namely the generalized Gibbs ensemble (GGE). 

\section{The Semi-classical Theory}\label{s:semi-classics}%
The main idea behind the semi-classical approach to the computation of out-of-the-equilibrium correlators is basically encoded in the following representation of the correlator (\ref{eq:5}): 
\begin{equation}
\label{eq:5bis}	
\begin{array}{l}								
C^{\rm nl\sigma m}(x_{1}, t_{1}; x_{2}, t_{2})\approx {\displaystyle\sum_{\{ \lambda_{\nu}\}}\int\prod_{\nu}^{N}dx_{\nu}\prod_{\nu}^{N}dk_{\nu}}\\
\times \left[P(\{x_{\nu},k_{\nu},\lambda_{\nu}\})\langle \{ x_{\nu},k_{\nu},\lambda_{\nu}  \} |\tilde{n}^{z}(x_{2}, t_{2})\tilde{n}^{z}(x_{1}, t_{1}) |\{x_{\nu},k_{\nu},\lambda_{\nu}\}\rangle\right],
\end{array}
\end{equation}
where the function $P(\{x_{\nu},k_{\nu},\lambda_{\nu}\})$ is the probability density of having quasiparticles at postion $x_{\nu}$ at time $t=0$ with momentum $k_{\nu}$ and quantum number $\lambda_{\nu}$.  The matrix element in the equation represents the value of the correlator once we specify the particular configuration $\{x_{\nu},k_{\nu},\lambda_{\nu}\}$.  $N$ in (\ref{eq:5}) is the total number of particles, and of course in the above equation an average over $N$ should be taken. The summation over $\{ \lambda_{\nu}\}$ averages over all possible quantum numbers of the $N$ particles, which are the three possible values of $L_{i}^{z}=(-1,0,1)$. We will start by assuming the system to be finite with free boundary conditions, and only later will we take the thermodynamic limit.
In spite of this we will assume the system to be translationally invariant, corrispondingly corrections due to the presence of boundaries will be considered negligible (for large enough systems). Therefore the correlator of equation (\ref{eq:5}) becomes a function of $|x_{2}-x_{1}|$,  whilst the same cannot be said for the time dependence, as was true in the equilibrium case.\\
In equation (\ref{eq:5bis}) we have substituted a complicated time-dependent matrix element with a sum over all possibile initial states, which can be represented by pairs of ballistically-moving quasiparticles. This technique has already been used extensively to compute finite-temperature correlators (see for example (\cite{Rapp2}, \cite{Sachdev1}, \cite{Rapp1})), and dynamical correlation functions (\cite{Igloi1}, \cite{Igloi2}). The general idea is the following: each quasiparticle carries a momentum $k$ and a quantum number $\lambda_{\nu}$, and it is created at time $t=0$ together with a partner with equal and opposite momentum. 
If the squared modulus of Fourier transform $|\tilde{K}^{ab}(x)|^{2}$ of the amplitude $K^{ab}(k)$ is a fast enough decreasing function as $|x|\to \infty$, the probability of creating these quasiparticles far apart from each other becomes negligible. For this reason in the follow we will always assume quasiparticles to be created in pairs in the same position $x_{\nu}$, where the index $\nu$ labels the quasiparticle pairs. \\
We start by considering this {\em gas} of quasiparticles to be very dilute by tuning the amplitudes $K^{ab}(k)$, therefore the quasiparticle  are entangled only within each single Cooper pair.
This means that the probability of a particular initial state $\{x_{\nu}, k_{\nu}, \lambda_{\nu} \}$ can be factorized in a product of single-pair probabilities:
\begin{equation}
\label{eq:42bis}
\begin{array}{lcl}
P(\{ x_{\nu}, k_{\nu}, \lambda_{\nu} \}) & = &  {\displaystyle \prod_{\nu=1}^{M}}\, P(x_{\nu}, k_{\nu}, \lambda_{\nu}) \\
& = &  {\displaystyle\frac{1}{L^{M}}} {\displaystyle \prod_{\nu=1}^{M} P(k_{\nu}, \lambda_{\nu}) },
\end{array}
\end{equation}
where in the last equality we made use that the system is homogeneous in space (corrections will be present close to the boundaries, but they are negligible in the thermodynmic limit). The quantity $\frac{1}{L} P(k_{\nu}, \lambda_{\nu})$ is the probability for a single pair to be created at a certain position $x_{\nu}$ with momenta $(k_{\nu}, -k_{\nu})$ and quantum numbers $(\lambda_{\nu}, \lambda'_{\nu})$, which is graphically represented in figure (\ref{fig:1}). 
Without any loss of generality we can suppose $x_1=0$, thanks to the (quasi)-translational invariance of the system.\\
\begin{figure}[t]
   \begin{center}
   \includegraphics[width=4cm]{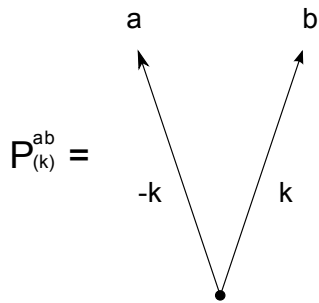}
	 \end{center}
   \caption{Graphical representation of the probability function $P^{ab}(k)$. }
\label{fig:1}
\end{figure}
Calculating the matrix elements for the operator $\tilde{n}^{z}$ ($\propto \cos\theta$) with the first few sherical harmonics, one finds that either creates a quasiparticle with $L_{i}^{z}=\lambda=0$ ($L_{i}=1$) at $x=x_1$ and $t=t_1$ with some velocity $v$ or destroys one already present in the initial state $\{ x_{\nu}, k_{\nu}, \lambda_{\nu} \}$. We assume the latter to be negligible,  because the excitations are very dilute in space and therefore their density is very small (in other words we will use $K^{ab}(k)$ as an expansion parameter). 
This adjoint particle can be created by $\tilde{n}^{z}(x_{1}, t_{1})$ {\em either} inbetween two different quasiparticle pairs {\em or} within two quasiparticle that belong to the same Cooper pair (i.e., two quasiparticles that originated in the same point $x_{\nu}$ at time $t=0$). A possible dynamical scenario is pictured in figure (\ref{fig:2}). 
Due to the collisions with the other excited particles there are only certain configurations where the quantum mechanical overlap in $\langle \psi |\tilde{n}^{z}(x_2,t_2)\tilde{n}^{z}(x_1,t_1)| \psi\rangle$ will be non-zero. Similar to other models the $O(3)$ non-linear sigma model in the long-wavelength limit has  a purely reflective scattering matrix:
\begin{equation}
\label{eq:6}
\mathcal{S}_{\lambda_{1} \lambda_{2}}^{\lambda'_{1} \lambda'_{2}} \,\longrightarrow \,(-1)\,\delta_{\lambda_{1}}^{\lambda'_{2}}\delta_{\lambda_{2}}^{\lambda'_{1}}.
\end{equation}
This means that in this limit the particles are impenetrable and the sequence of $\{\lambda_{\nu} \}$ {\em does not} change in time (see figure (\ref{fig:1})). Here we are implicitly assuming that the momentum distribution $f_k$ is a peaked function around zero. 
The operator $\tilde{n}^{z}(x_1,t_1)$ creates a quasiparticle with a probability amplitude $e_{0}(v)$, where the subscript $_{0}$ refers to the quantum number of the created particle and $v$ is its velocity. This particle, together with the other quasiparticles, propagates under the action of $\exp{(-i\hat{H}t)}$ and collides with them (in one-dimensional systems collisions amongst particles can never be ignored). The $\mathcal{S}$-matrix takes on an exchange form (\ref{eq:6}), and therefore particles only exchange their velocities while conserving their internal quantum numbers (see figure (\ref{fig:2})).\\
As a consequence, at any time $t\geq t_1$ precisely one of the particles will have the velocity $v$ of the particle which was created by  $\tilde{n}^{z}(x_1,t_1)$, and will be at position $x_1+v(t-t_1)$. This very particle must be annihilated at time $t_2$ by $\tilde{n}^{z}(x_2,t_2)$, otherwise the final state after the action of $\exp{(i\hat{H}t)}$ will be orthogonal to the initial one.  The probability amplitude that this particle is annihilated is proportional to $(e_{\lambda'}(v))^{*}{\rm e}^{ik(x_2-x_1)}$, where $\lambda'$ is the quantum number of the particle that is removed by $\tilde{n}^{z}(x_2,t_2)$. As we shall see below, $\lambda'$ has to be equal to zero.  This request is automatically guaranteed by another requirement, namely that the internal quantum numbers in the {\em final} state be exactly the same as those of the {\em initial} one. If this does not happen the final state is orthogonal to the initial one, and that particular configuration of quasiparticles does not contribute to the average (\ref{eq:5bis}).\\
\begin{figure}[t]
   \begin{center}
   \includegraphics[width=7cm]{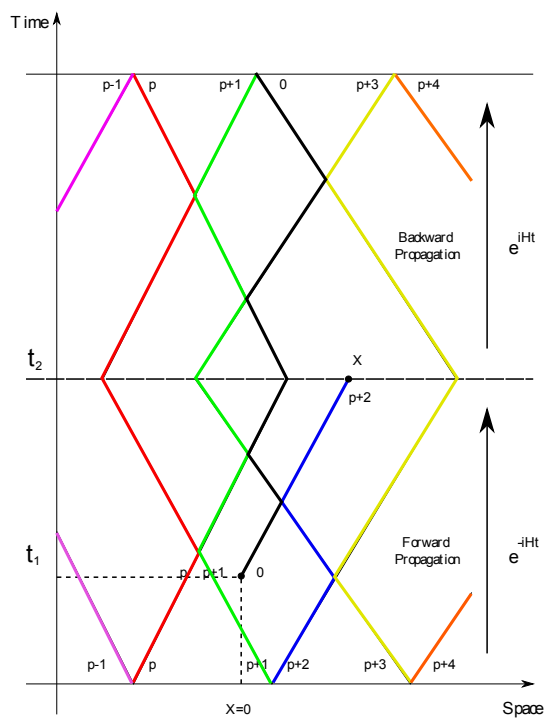}
	 \end{center}
   \caption{Propagation and scattering of quasiparticles in the semi-classical picture of the $O(3)$ non-linear sigma model. Lines drawn with different colours represent trajectories of different particles. The two states before and after the forward and backward time evolution must be identical, and this imposes constraints on the quantum numbers of the particles. Notice that in this example the particle with label $p+2$ is removed at time $t_2$. }
\label{fig:2}
\end{figure}
Let us now consider the case in which at time $t=0$ the point $x_1=0$ is located inbetween particle $\lambda_p$ and $\lambda_{p+1}$ (see figure (\ref{fig:2})). Consider now a line connecting the points $(0, 0)$ and $(0, t_1)$ (this line does not correspond to any real particle trajectory).  Define $n' = N'_{+}-N'_{-}$ as the number of intersections between this line and the other lines which correspond to particle trajectories, where $N'_{+}$ are intersections from the right and $N'_{-}$ those from the left.   By definition $n'$ is given by:
\begin{equation}
\label{eq:43bis}
\begin{array}{lcl}
n' & =  & {\displaystyle\sum_{\nu}^{N}}\left\{\Theta[0-x_{\nu}(t_1)]-\Theta[0-x_{\nu}]\right\}\\
 & = & {\displaystyle\sum_{\nu}^{M}}\left\{\Theta[0-x_{\nu}^{1}(t_1)] +\Theta[0-x_{\nu}^{2}(t_1)]
 -    2\Theta[0-x_{\nu}] \right\},
\end{array}
\end{equation}
where the labels $^{(1,2)}$ correspond to the trajectories of two quasiparticle originated at the same point, and $N$ is the number of pairs $N=M/2$. $\Theta$'s are Heaviside step functions. Then at time $t=t_1$ the operator $\tilde{n}_{z}(0, t_1)$ creates a quasiparticle ($L_{\nu}=1$) with $L_{\nu}^{z}=\lambda_{\nu}=0$ at $x_1=0$ with some momentum $k$. The adjoint particle will be created inbetween particles $\lambda'_{p+n'}$ and $\lambda'_{p+n'+1}$. Suppose that $n'\geq 0$. The generalization of the following results to the case of negative $n'$ is straightforward. The scattering $\mathcal{S}$-matrix takes on the exchange form (\ref{eq:6}), therefore during the motion particles only exchange their velocities while conserving their internal quantum numbers. This also means that the order of quantum number in each configuration from the left to the right never changes in time. The action of the operator $\tilde{n}_{z}(0, t_1)$ on the initial sequence of quantum numbers may be written as:
\begin{equation}
\label{eq:44bis}
\begin{array}{lcl}
\tilde{n}^{z}(0, t_1) & : &  \{\dots,\lambda_{p+n'}, \lambda_{p+n'+1},\dots \}\\
& \mapsto &  \{\dots,\lambda_{p+n'}, \lambda_{0},  \lambda_{p+n'+1},\dots \}.
\end{array}
\end{equation} 
At time $t=t_2$ the operator $\tilde{n}_{z}(x, t_2)$ will remove the particle with label $p+n'+n$ from the set of excitations, where $n=N_{+}-N_{-}$ is the difference between the number of intersections from the right and those from the left of the line connecting points $(0, t_1)$ and $(x, t_2)$. This quantity is given by:
\bea
\label{eq:45bis}
 n  &=&   {\displaystyle\sum_{\nu}^{N}}\left\{\Theta[x-x_{\nu}(t_2)]-\Theta[0-x_{\nu}(t_1)]\right\},\nonumber\\
    &=&   {\displaystyle\sum_{\nu}^{M}}\left\{\Theta[x-x_{\nu}^{1}(t_2)] +\Theta[x-x_{\nu}^{2}(t_2)]
     - \Theta[0-x_{\nu}^{1}(t_1)]- \Theta[0-x_{\nu}^{2}(t_1)]\right\}.
\eea
This number tells us which particle is moving along this line at time $t=t_2$, particle that will be removed by the action of   $\tilde{n}_{z}(x, t_2)$:
\begin{equation}
\label{eq:46bis}
\begin{array}{lcl}
\tilde{n}^{z}(x, t_2) & : &  \{\dots,\lambda_{p+n'}, \lambda_0, \lambda_{p+n'+1},\dots, \lambda_{p+n'+n-1}, \lambda_{p+n'+n}, \lambda_{p+n'+n+1},\dots \} \\
& \mapsto &  \{\dots,\lambda_{p+n'}, \lambda_0, \lambda_{p+n'+1},\dots, \lambda_{p+n'+n-1}, \lambda_{p+n'+n+1},\dots \} .
\end{array}
\end{equation} 
Comparing the very last sequence of quantum numbers with the initial one (that at time $t=0$), we end up with the following constraint on the set of quantum numbers:
\begin{equation}
\label{eq:46tris}
\boxed{
\lambda_{0}\equiv\lambda_{p+n'+1}\equiv\lambda_{p+n'+2}\dots\equiv\lambda_{p+n'+n}}
\end{equation}
In practice we have a sequence of $n$ quantum numbers that must all be equal to $\lambda_{0}=0$,  which starts at position $p+n'$. 
To identify the contribution of a particular configuration of quasiparticles we must consider the phase factors.  
 Quasiparticles in the initial state generate a phase factor $\exp{(-it_{2}\sum_{\nu}\epsilon(v_{\nu}))}$ under the action of $\exp{(-it_{2}\hat{H})}$ ($\epsilon(v)$ being the particle energy). This phase factor, however, completely cancels under the action of  $\exp{(it_{2}\hat{H})}$, except for the quasiparticle added by the operator $\tilde{n}^{z}(0, t_1)$. This gives a factor $\exp{(-i(t_2 -t_1)\epsilon(v))}$. Moreover every collision results in a sign change of the many-body wave function, but all these signs cancel under the forward and backward propagations, except those that are associated with collisions with the {\em extra} particle. These give an extra sign $(-1)^{N_{+}+N_{-}}$, which can be conveniently re-expressed as:
\begin{equation}
\label{eq:47bis}
(-1)^{N_{+}+N_{-}}=(-1)^{N_{+}-N_{-}}=(-1)^{n}.
\end{equation}
Collecting all these pieces together we obtain the following expression for the general correlation function (\ref{eq:5}):
\begin{equation}
\label{eq:48bis}
C^{\rm nl\sigma m}(x_1, t_1; x_2, t_2)  = \left( {\displaystyle\int(e_{0}(k)^{*}e_{0}(k)){\rm e}^{-i(k(x_2-x_1)-\epsilon(k)(t_2-t_1))}dk} \right)  \times R(x_1, t_1; x_2, t_2),
\end{equation}
where $e_{0}(k)$ is the probability creation amplitude and $R(x_1, t_1; x_2, t_2)$ is the relaxation function, which is given by:
\bea
\label{eq:49bis}
 && R(x_1, t_1; x_2, t_2) = \langle{\displaystyle\sum_{n=-\infty}^{+\infty}\sum_{n'=-\infty}^{+\infty}}(-1)^{n}\delta_{n',\sum_{\nu_1}[\dots]}\delta_{n,\sum_{\nu_2}[\dots]}\\
& \, & \times\left[\delta_{\lambda_{0},\lambda_{p+n'+1}}  \delta_{\lambda_{p+n'+1},\lambda_{p+n'+2}}\dots\delta_{\lambda_{p+n'+n-1},\lambda_{p+n'+n}} \right]\rangle_{\{ x_{\nu}, k_{\nu}, \lambda_{\nu}\}},
\eea
where the sums with indices $\nu_1$ and $\nu_2$ are given by equations (\ref{eq:43bis}) and (\ref{eq:45bis}) respectively. The $(-1)^{n}$ appearing in the previous equation is exactly that introduced in (\ref{eq:47bis}). 
From the previous average we can immediately see that only those configurations which have a  sequence of $n$ quantum numbers all equal to $\lambda_0\equiv 0$ contribute to the correlator (\ref{eq:5}). 
We now face the problem of computing an average of the form $\langle \hat{O}\rangle_{\{x_{\nu}, k_{\nu}, \lambda_{\nu}\}}$, where $\hat{O}$ is a generic local operator, and the distribution probability takes into account all possible {\em initial} conditions. Solving this problem will enable us to compute the correlator (\ref{eq:5bis}).  Due to the assumption of equation (\ref{eq:42bis}) we can write the average as:
\begin{equation}
\label{eq:50bis}
{\displaystyle\int_{-L/2}^{L/2}\frac{dx_1}{L}\frac{dx_2}{L}\dots\frac{dx_M}{L}
\sum_{\stackrel{a^{(1)}, b^{(1)}}{k_1>0}}\sum_{\stackrel{a^{(2)}, b^{(2)}}{k_2>0}}\dots\sum_{\stackrel{a^{(M)}, b^{(M)}} {k_M>0}}
\prod_{\nu=1}^{M}P(k_{\nu}, \lambda_{\nu})\,\hat{O}(\{x_{\nu}, k_{\nu}, \lambda_{\nu}\})}.
\end{equation}
We will consider separately the cases in which $(n', n)$ are even and/or  odd numbers.
\begin{itemize}
\item ($n'=$ even and $n=$ even): when $n'$ is an even integer the adjoint particle is created inbetween two different quasiparticle pairs, thus we have a sequence of $n/2$ quasiparticle pairs which are all forced to have their quantum numbers equal to $\lambda_{0}\equiv 0$. 
The average over the set of quantum numbers can be written as:
\be
\begin{array}{lcl}
\label{eq:51bis}
 P_{1}(\{k_{\nu}\}) &=& {\displaystyle\sum_{a^{(1)}, b^{(1)}}\sum_{a^{(2)}, b^{(2)}}\dots\sum_{a^{(M)}, b^{(M)}}\prod_{\nu=1}^{M}}\,P^{a^{\nu}, b^{\nu}}(k_{\nu})\\
&& \times\left[\delta_{0,\lambda_{p+n'+1}}  \delta_{\lambda_{p+n'+1},\lambda_{p+n'+2}}\dots\delta_{\lambda_{p+n'+n-1},\lambda_{p+n'+n}}\right],
\end{array}
\ee
 where we have introduced the notation $P^{a^{\nu}, b^{\nu}}(k_{\nu})\equiv P(k_{\nu}, \lambda_{\nu})$ (see figure (\ref{fig:5})). 

It is worth noticing that the labels in the Kronecker deltas in the previous equation refer to the quasiparticles, while the labels in the summations and in the integrals refer to quasiparticle pairs. In order not to mix up these two different notations we recognize that in this case $p+n'$ is an even number, therefore particle with label $p+n'+1$ belongs to the $\frac{p+n'}{2}+1$ pair, and so on. From this pair to the right we have a sequence of $n/2$ quasiparticle pairs all with quantum numbers $\lambda=0$. In addition the probability distribution factorizes into single-pair probabilities.
Let us remember that the average is taken over all possible initial conditions, therefore $k_{\nu}$ specifies the value of the momentum of the particle at time $t=0+\epsilon$, where $\epsilon$ is a small positive quantity.  Roughly speaking $k_{\nu}$ is the value of the momentum of a particle right after time $t=0$ and before the first scattering process.  Equation (\ref{eq:51bis}) can be easily evaluated to give:
\be
\label{eq:52bis}
\begin{array}{lcl}
 P_{1}(\{k_{\nu}\}) &=& P(k_{1})\dots P\left( k_{\frac{p+n'}{2}}\right)\underbrace{P^{00}\left( k_{\frac{p+n'}{2}+1}\right)\dots P^{00}\left (k_{\frac{p+n'}{2}+\frac{n}{2}}\right)}_{\frac{\left|n\right|}{2}{\rm times}}\\
& & \times P\left(k_{\frac{p+n'}{2}+\frac{n}{2}+1}\right)\dots P(k_{M}),
\end{array}
\ee
where $P(k)=\sum_{a, b}P^{ab}(k)$.
\begin{figure}[t]
   \begin{center}
   \includegraphics[width=4cm]{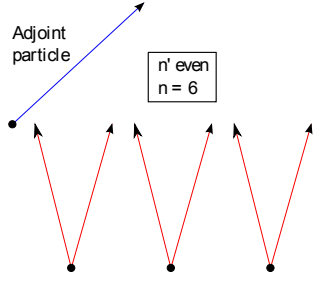}
	 \end{center}
   \caption{Computation of $P_{1}(\{k_{\nu} \})$: an example with $n=6$. Red trajectories represent those particles whose quantum number must be equal to $\lambda_{0}=0$. }
\label{fig:5}
\end{figure}

\item($n'=$ even and $n=$ odd): also in this case the adjoint particle is created inbetween two different pairs, but when $n$ is odd the result of equation (\ref{eq:51bis}) becomes:
\begin{equation}
\label{eq:53bis}
\begin{array}{lcl}
P_{2}(\{ k_{\nu}\}) &  = & P(k_{1})\dots P\left ( k_{\frac{p+n'}{2}}\right)\underbrace{P^{00}\left( k_{\frac{p+n'}{2}+1}\right)\dots P^{00}\left (k_{\frac{p+n'}{2}+\frac{n-1}{2}}\right)}_{\frac{|n|-1}{2} {\rm times}} \\
& \times & {\displaystyle\sum_{\lambda=0,\pm 1}} P^{0\lambda}\left( k_{\frac{p+n'}{2}+\frac{n-1}{2}+1}\right) P\left ( k_{\frac{p+n'}{2}+\frac{n-1}{2}+2}\right )\dots P(k_{M}).
\end{array}
\end{equation}
In practice the last particle of the sequence of $\lambda_{\nu}=0$ belongs to a pair whose right-side partner can carry all possible quantum numbers.

\item($n'=$ odd and $n=$ even): let us consider the case in which $n'$ is an odd integer. In this case the adjoint particle is created within a quasiparticle pair, and the average over quantum numbers gives ($n\ne 0$):
\begin{equation}
\label{eq:54bis}
\begin{array}{lcl}
 P_{3}(\{ k_{\nu}\}) & = & P(k_{1})\dots P\left ( k_{\frac{p+n'-1}{2}}\right)
{\displaystyle\sum_{\lambda=0,\pm 1}} P^{\lambda 0}\left( k_{\frac{p+n'-1}{2}+1}\right)\\
& \times & \underbrace{P^{00}\left( k_{\frac{p+n'-1}{2}+2}\right)\dots P^{00}\left (k_{\frac{p+n'-1}{2}+\frac{n}{2}}\right)}_{\frac{|n|}{2}-1 {\rm times}}\\
& \times & {\displaystyle\sum_{\lambda=0,\pm 1}} P^{0\lambda}\left( k_{\frac{p+n'-1}{2}+\frac{n}{2}+1}\right) P\left ( k_{\frac{p+n'-1}{2}+\frac{n}{2}+2}\right )\dots P(k_{M}).
\end{array}
\end{equation}
The case $n =0$ simply gives a sequence of $M$ different $P(k_{\nu})$'s.
When $n'$ is odd and $n$ even in the sequence of quantum numbers with $\lambda_{\nu}=0$ the first and the last one belong to pairs whose partner has no constraint on its quantum number $\lambda$. This is the reason for the factors $\sum_{\lambda} P^{0\lambda}(k)$ and $\sum_{\lambda} P^{\lambda 0}(k)$ in the previous expression. 

\item($n'=$ odd and $n=$ odd): in this case the average over the set of quantum numbers gives:
\begin{equation}
\label{eq:55bis}
\begin{array}{lcl}
P_{4}(\{ k_{\nu}\}) &  = & P(k_{1})\dots P\left ( k_{\frac{p+n'-1}{2}}\right)
{\displaystyle\sum_{\lambda=0,\pm 1}} P^{\lambda 0}\left( k_{\frac{p+n'-1}{2}+1}\right)\\
& \times & \underbrace{P^{00}\left( k_{\frac{p+n'-1}{2}+2}\right)\dots P^{00}\left (k_{\frac{p+n'-1}{2}+\frac{n-1}{2}+1}\right)}_{\frac{|n|-1}{2} {\rm times}}\\
& \times & P\left ( k_{\frac{p+n'-1}{2}+\frac{n-1}{2}+2}\right )\dots P(k_{M}).
\end{array}
\end{equation}
\end{itemize}
We are now ready to write down the global expression for the quantum number average:
\begin{equation}
\label{eq:56bis}
\begin{array}{l}
{\displaystyle\sum_{a^{(1)}, b^{(1)}}\sum_{a^{(2)}, b^{(2)}}\dots\sum_{a^{(M)}, b^{(M)}}\prod_{\nu=1}^{M}}\,P^{a^{\nu}, b^{\nu}}(k_{\nu})\, \delta_{0,\lambda_{p+n'+1}} \delta_{\lambda_{p+n'+1},\lambda_{p+n'+2}}\dots\delta_{\lambda_{p+n'+n-1},\lambda_{p+n'+n}}\\
={\displaystyle\frac{1+(-1)^{n'}}{2}}\left[{\displaystyle\frac{1+(-1)^{n}}{2}} P_{1}(\{k_{\nu}\})+ {\displaystyle\frac{1-(-1)^{n}}{2}} P_{2}(\{k_{\nu}\})\right]\\
+{\displaystyle\frac{1-(-1)^{n'}}{2}}\left[{\displaystyle\frac{1+(-1)^{n}}{2}} P_{3}(\{k_{\nu}\})+ {\displaystyle\frac{1-(-1)^{n}}{2}} P_{4}(\{k_{\nu}\})\right].
\end{array}
\end{equation}
This expression is a function of $(n, n')$ and the set of quantum momenta of a particular quasiparticle configuration.  It is worth noticing that $n$ and $n'$ enter the expression for $P_{i}$, $i=1,\dots,4$ in different ways. While $n$ tells us how long the sequence is, $n'$ gives us information about the starting point of the sequence and can easily be absorbed into the definition of $p$. As we shall see, for the actual computation of the relaxation function $R$ the starting position of the sequence is not important, what matters is only the length of the sequence itself. 
\begin{figure}[t]
   \begin{center}
   \includegraphics[width=4cm]{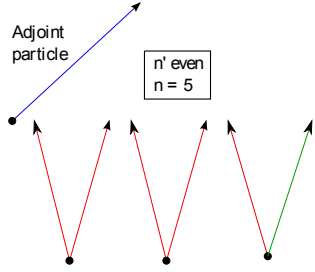}
	 \end{center}
   \caption{Computation of $P_{2}(\{k_{\nu} \})$: an example with $n=5$.}
\label{fig:5bis}
\end{figure}
We must now compute the average of the previous quantity over all possible initial positions and momenta. Namely we must take the following average:
\begin{equation}
\label{eq:57bis}
{\displaystyle\int_{-L/2}^{L/2}\frac{dx_1}{L}\dots\frac{dx_M}{L}\sum_{ k_{1}>0}\dots\sum_{ k_{M}>0}}\,\,
{\displaystyle\sum_{n=-\infty}^{+\infty}}(-1)^{n}\delta_{n, \sum_{\nu_{2}}}\left [\dots\right]{\displaystyle\sum_{n'=-\infty}^{+\infty}}\delta_{n', \sum_{\nu_{1}}}\left [\dots\right],
\end{equation}
where the quantity in the square brackets is given by equation (\ref{eq:56bis}). We can split this average into four pieces, let us call them $A, B, C$ and $D$, one for each $P_{i}(\{ k_{\nu}\})$ respectively, where $i=1,\dots,4$. We shall do the computation of the term which contains $P_{1}(\{k_{\nu} \})$ in some detail; the other cases are straightforward modifications.\\
Let us compute explicity $A$, the first addend of the right side of equation (\ref{eq:56bis}), where the first summation over $n'$ gets immediately canceled out  by the corresponding Kronecker delta because $n'$ itself enters the function $P_{1}(\{k_{\nu} \})$ trivially, therefore we have:
\begin{equation}
\label{eq:59bis}
\begin{array}{lcl}
A&=&{\displaystyle\int_{-L/2}^{L/2}\frac{dx_1}{L}\dots\frac{dx_M}{L}\sum_{ k_{1}>0}\dots\sum_{ k_{M}>0}}\,\,
{\displaystyle\sum_{n=-\infty}^{+\infty}}(-1)^{n}\delta_{n, \sum_{\nu_{2}}}\left [\dots\right]\\
&& \times \left( {\displaystyle\frac{1+(-1)^{\sum_{\nu_{1}}[\dots]}}{2}} {\displaystyle\frac{1+(-1)^{n}}{2}} P_{1}(\{k_{\nu}\}) \right),
\end{array}
\end{equation}
where $\sum_{\nu_{1}}[\dots]$ is specified by (\ref{eq:43bis}). $A$ can itself be splitted into two pieces, say $A_{1}$ and  $A_{2}$, which correspond to the two contributions to the integration of the terms into the bracket, namely:
\begin{equation}
\label{eq:60bis}
{\displaystyle\frac{1}{2}+\frac{(-1)^{\sum_{\nu_{1}}}}{2}} \longrightarrow A_1 + A_2.
\end{equation}
Let us start by computing the contribution of $A_1$. By making use of the following integral representation of the Kronecker delta
\be
\label{int_rep}
\delta_{n,\sum_{\nu}[\dots]} = \frac{1}{2\pi}{\displaystyle\int_{-\pi}^{\pi}}d\phi\,{\rm e}^{i\phi(n-\sum_{\nu}[\dots])},
\ee
\begin{figure}[t]
   \begin{center}
   \includegraphics[width=4cm]{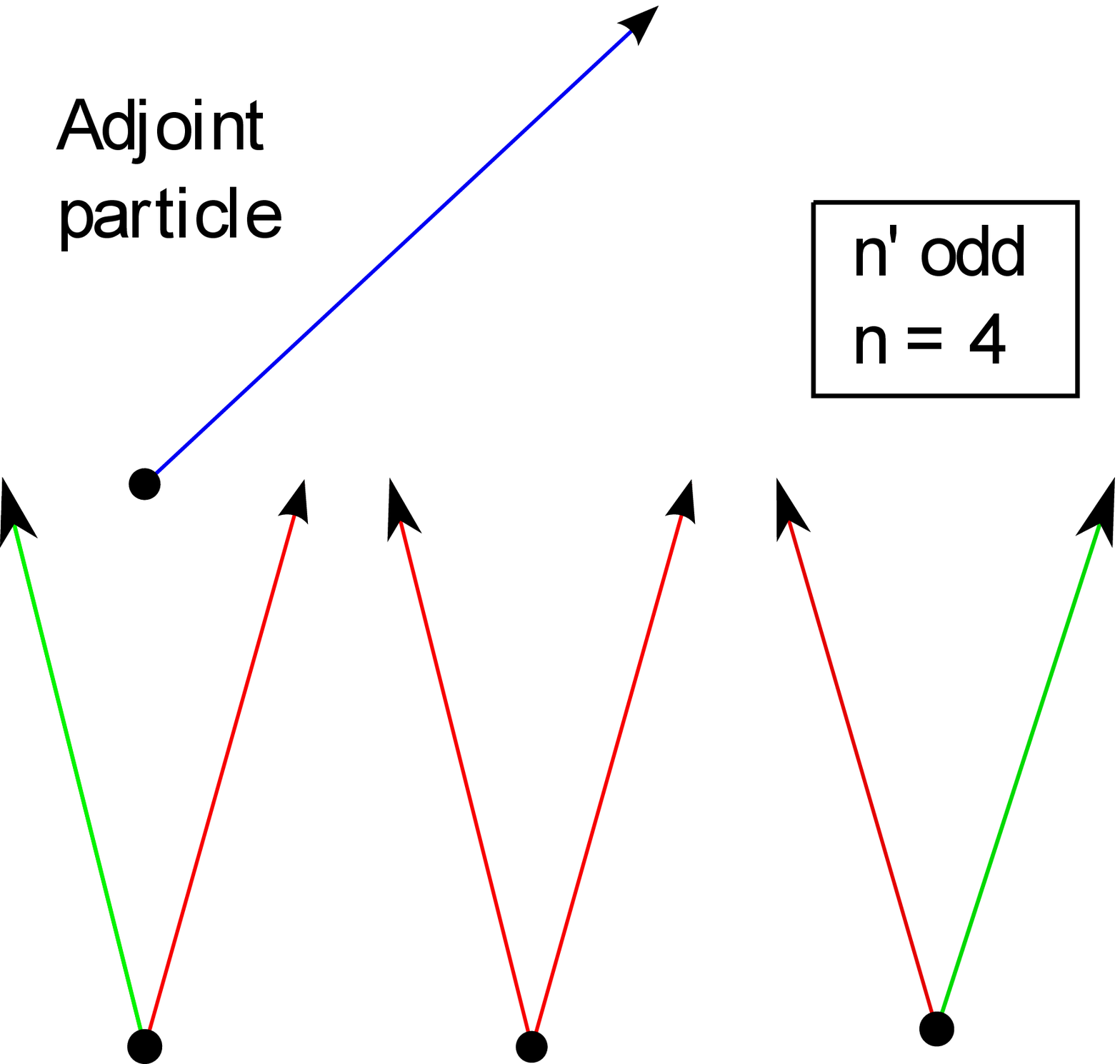}
	 \end{center}
   \caption{Computation of $P_{3}(\{k_{\nu} \})$: an example with $n'$ odd and $n=4$.}
\label{fig:5tris}
\end{figure}
 we can write $A$ as:
\begin{equation}
\label{eq:61bis}
\int_{-\pi}^{\pi}\frac{d\phi}{2\pi}{\displaystyle\sum_{n=-\infty}^{+\infty}}(-1)^{n}{\rm e}^{i n\phi}{\displaystyle\frac{1+(-1)^{n}}{2}}{\displaystyle\sum_{ k_{1}>0}\dots\sum_{ k_{M}>0}}\,\,{\displaystyle\int_{-L/2}^{L/2}\frac{dx_1}{L}\dots\frac{dx_M}{L}}\,{\rm e}^{-i\phi\sum_{\nu_{2}}[\dots]}\,\frac{P_{1}(\{k_{\nu}\})}{2}.
\end{equation}
Now we want to compute the spatial integral, remembering that the sum $\sum_{\nu_{2}}[\dots]$ is specified in equation (\ref{eq:45bis}). We begin by considering $x>0$, $t
_{2}>t_{1}$ and $x<v_{max}(t_2-t_1)$, where $v_{max}$ is the maximal velocity of the excitations of the model (in the section (\ref{app:general}) we will show the general result for arbitrary times and distances). We will refer to this situation as the within-the-light-cone case.
The spatial integral factorizes in a straightforward way, thus we can write:
\begin{equation}
\label{eq:63bis}
\begin{array}{l}
{\displaystyle\int_{-L/2}^{L/2}\frac{dx_1}{L}\dots\frac{dx_M}{L}}\,{\rm e}^{-i\phi\sum_{\nu_{2}}[\dots]}={\displaystyle\prod_{\nu=1}^{M}}\left\{ 1-\frac{2v_{k_{\nu}}t}{L}(1-\cos\phi)\Theta[v_{k_{\nu}}t-x]\right.\\
\\
\left.-\frac{2ix\sin\phi}{L}\Theta[v_{k_{\nu}}t-x]-\frac{2x}{L} \left(i\sin\phi+2\sin^{2}(\phi/2)\right)\Theta[x-v_{k_{\nu}}t] \Theta[v_{k_{\nu}}T-x]\right.\\
\\
-\left.\left(\frac{v_{k_{\nu}}T}{L}(1+\cos(2\phi)-i\sin(2\phi)-\cos\phi+i\sin\phi)\right.\right.\\
\\
+ \left.\left.\frac{x}{L}(i\sin(2\phi)+2\sin^{2}\phi) \right) \Theta[x-v_{k_{\nu}}T]\right\},
\end{array}
\end{equation}
where $v_{k_{\nu}}$ is the particle velocity defined by $v_k\equiv\frac{\partial\epsilon(k)}{\partial k}$, $t\equiv t_2-t_1$ and $T\equiv t_2 + t_1$. 
Let us notice the following definition of the probability $P(k)$:
\be
\label{eq.def}
P(k)\equiv{\displaystyle\frac{f_k}{M}},\qquad M=\sum_{k>0}f_k, 
\ee
where $f_k$ is the occupation number of the $k$-mode. Furthermore, the following identities can be useful in taking the thermodynamic limit:
\begin{equation}
\label{eq:67bis}
{\displaystyle\frac{{\displaystyle\sum_{k>0}P(k) v_k}}{L}}\equiv{\displaystyle\frac{{\displaystyle\sum_{k>0}f_k v_k}}{L{\displaystyle\sum_{k>0}f_k}}}\longrightarrow{\displaystyle\frac{{\displaystyle\int_{0}^{\infty}\frac{dk}{2\pi}f_k v_k}}{M}},\quad{\displaystyle\frac{{\displaystyle\sum_{k>0}P(k)}}{L}}\equiv{\displaystyle\frac{{\displaystyle\sum_{k>0}f_k}}{L{\displaystyle\sum_{k>0}f_k}}}\longrightarrow{\displaystyle\frac{{\displaystyle\int_{0}^{\infty}\frac{dk}{2\pi}f_k}}{M}}.
\end{equation}
\begin{figure}[t]
   \begin{center}
   \includegraphics[width=4cm]{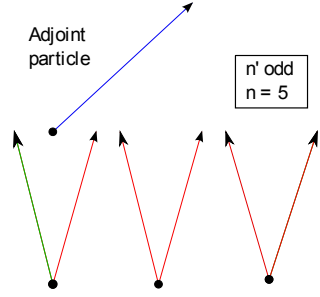}
	 \end{center}
   \caption{Computation of $P_{4}(\{k_{\nu} \})$: an example with $n'$ odd and $n=5$.}
\label{fig:5quatris}
\end{figure}
Let us now compute the average over the momenta of the previous result, taking the limit $L,M \to\infty$ ($M/L\equiv\rho$ fixed):
\begin{equation}
\label{eq:64bis}
{\displaystyle\lim_{M,L\to\infty}}{\displaystyle\sum_{ k_{1}>0}\dots\sum_{ k_{M}>0}}P_{1}(\{k_{\nu}\}){\displaystyle\prod_{\nu=1}^{M}}\left\{\dots\right\},
\end{equation}
where $P_{1}(\{k_{\nu}\})$ is given by (\ref{eq:52bis}) and the $\{\dots\}$ are the right-hand terms of equation (\ref{eq:63bis}). This expression can be evaluated to yield:
\begin{equation}
\label{eq:69bis}
\begin{array}{l}
{\displaystyle\lim_{M,L\to\infty}}\,{\displaystyle\sum_{ \{k_{\nu}>0\}}P_{1}(\{k_{\nu}\})}{\displaystyle\prod_{\nu=1}^{M}}\left( \dots\right) \\
=\exp{\left\{-t(1-\cos\phi){\displaystyle\int_{0}^{\infty}}\frac{dk}{\pi}f_k\, v_{k}\Theta[v_{k}t-x]-2x\sin^{2}(\phi/2){\displaystyle\int_{0}^{\infty}}\frac{dk}{\pi}f_k\,\Theta[x-v_{k}t]\Theta[v_{k}T-x]\right\}}\\
\times\exp{\left\{-T(1+\cos(2\phi)-i\sin(2\phi)-2\cos\phi+2i\sin\phi){\displaystyle\int_{0}^{\infty}}\frac{dk}{2\pi}f_k\,\Theta[x-v_{k}T]\right\}}\\
\times\exp{\left\{-i x\sin\phi{\displaystyle\int_{0}^{\infty}}\frac{dk}{\pi}f_k\,(\Theta[v_{k}t-x]+\Theta[x-v_{k}t]\Theta[v_{k}T-x])\right\}}\\
\times\exp{\left\{-x(2\sin^{2}\phi+i\sin(2\phi)){\displaystyle\int_{0}^{\infty}}\frac{dk}{2\pi}f_k\,\Theta[x-v_{k}T]\right\}}(P^{00})^{|n|/2},\\ \\
\end{array}
\end{equation}
where $P^{00}=\int_{0}^{\infty}\frac{dk}{2\pi}P^{00}(k)$ and $P^{00}(k)$ is now a probability {\em density}. 
We can now take the limit when $T$ is very large ($x/T \ll v_{max}$), assuming that $T\int_{0}^{\infty}\frac{dk}{2\pi}f_k v_k\,\Theta[x-v_{k}T]\to 0$, so the previous expression simplifies, giving:
\begin{equation}
\label{eq:69tris}
\begin{array}{lcl}
\dots & = &  \exp{\left\{-t(1-\cos\phi){\displaystyle\int_{0}^{\infty}}\frac{dk}{\pi}f_k\, v_{k}\Theta[v_{k}t-x]-2x\sin^{2}(\phi/2){\displaystyle\int_{0}^{\infty}}\frac{dk}{\pi}f_k\,\Theta[x-v_{k}t]\right\}}\\
& \times &  \exp{\left\{-i x\sin\phi{\displaystyle\int_{0}^{\infty}}\frac{dk}{\pi}f_k\right\}}\,(P^{00})^{|n|/2}.
\end{array}
\end{equation}
 Before computing the sum over $n$ it is convenient to obtain the analogy of equation (\ref{eq:69tris}) for the case $A_{2}$, and see which contributions survive in the limit of $T$ very large . For $A_2$ we start with:
\begin{equation}
\label{eq:70bis}
{\displaystyle\sum_{ k_{1}>0}\dots\sum_{ k_{M}>0}}\,{\displaystyle\int_{-L/2}^{L/2}\frac{dx_1}{L}\dots\frac{dx_M}{L}}\,{\rm e}^{-i\phi\sum_{\nu_{2}}[\dots]}\,P_{1}(\{k_{\nu}\})\,(-1)^{\sum_{\nu_{1}}[\dots]}, 
\end{equation}
then by repeating the same steps we end up with:
\begin{equation}
\label{eq:72bis}
\begin{array}{l}
{\displaystyle\lim_{M\to\infty}}\,{\displaystyle\sum_{ k_{1}>0}\dots\sum_{ k_{M}>0}}P_{1}(\{k_{\nu}\}){\displaystyle\prod_{\nu=1}^{M}}\left( \dots\right) \\
=\exp{\left(-(T-t\cos\phi){\displaystyle\int_{0}^{\infty}\frac{dk}{2\pi}f_k v_k}-i x\sin\phi{\displaystyle\int_{0}^{\infty}\frac{dk}{2\pi}f_k\,\Theta[v_k t-x]}\right)}\\
\times\exp{\left(-i t\sin\phi{\displaystyle\int_{0}^{\infty}\frac{dk}{2\pi}f_k v_k\,\Theta[x-v_k t]}\right)}\,(P^{00})^{|n|/2}.
\end{array}
\end{equation}
We immediately see that this contribution is exponentially decreasing as $T$ becomes large, therefore we neglect it in this limit.
We have already computed all the ingredients we need to do the sum over $n$, recalling that only even values of $n$ give non-zero contribution, as can be seen from (\ref{eq:61bis}). Performing the summation, the global contribution from term $A$ to the relaxation function (\ref{eq:49bis}) is given by:
\begin{equation}
\label{eq:73bis}
\begin{array}{l}
 A =  {\displaystyle\lim_{\substack{L, M\to\infty}}}\left\langle
{\displaystyle\sum_{n=-\infty}^{+\infty}}(-1)^{n}\delta_{n, \sum_{\nu_{2}}[\dots]}\left( \frac{1+(-1)^{\sum_{\nu_{1}}}}{2} \frac{1+(-1)^{n}}{2} P_{1}(\{k_{\nu}\}) \right)\right\rangle_{\{x_{\nu},k_{\nu}\}}\\
 = {\displaystyle\int_{-\pi}^{+\pi}\frac{d\phi}{4\pi}}\,\,{\rm e}^{-t(1-\cos\phi)\int_{0}^{\infty}\frac{dk}{\pi}f_k\, v_{k}\Theta[v_{k}t-x]}\,{\rm e}^{-2x\sin^{2}(\phi/2)\int_{0}^{\infty}\frac{dk}{\pi}f_k\,\Theta[x-v_{k}t]}\\
\times \cos\left(x\sin\phi\int_{0}^{\infty}\frac{dk}{\pi}f_k\right){\displaystyle\frac{1-(P^{00})^{2}}{1-2P^{00}\cos(2\phi)+(P^{00})^{2}}},
\end{array}
\end{equation}
which in the large $T$ limit depends only on $t$. Repeating all the steps we did for the term containing $P_{1}(\{k_{\nu}\})$, we can then compute the contributions from the terms containing $P_{2}(\{k_{\nu}\})$,  $P_{3}(\{k_{\nu}\})$ and $P_{4}(\{k_{\nu}\})$ (namely $B$, $C$, and $D$). Here we list the results of these computations (always in the limit $x/T \ll v_{max}$):
\begin{equation}
\label{eq:74bis}
\begin{array}{l}
 B = {\displaystyle\lim_{\substack{L, M\to\infty}}}\left\langle
{\displaystyle\sum_{n=-\infty}^{+\infty}}(-1)^{n}\delta_{n, \sum_{\nu_{2}}[\dots]}\left( \frac{1+(-1)^{\sum_{\nu_{1}}}}{2} \frac{1-(-1)^{n}}{2} P_{2}(\{k_{\nu}\}) \right)\right\rangle_{\{x_{\nu},k_{\nu}\}}\\
={\displaystyle\int_{-\pi}^{+\pi}\frac{d\phi}{4\pi}}\, {\rm e}^{-t(1-\cos\phi)\int_{0}^{\infty}\frac{dk}{\pi}f_k\, v_{k}\Theta[v_{k}t-x]}\,\, {\rm e}^{-2x\sin^{2}(\phi/2)\int_{0}^{\infty}\frac{dk}{\pi}f_k\,\Theta[x-v_{k}t]}\\
\times\cos\left(x\sin\phi\int_{0}^{\infty}\frac{dk}{\pi}f_k\right){\displaystyle\frac{2\cos\phi(P^{00}-1)}{1-2P^{00}\cos(2\phi)+(P^{00})^{2}}}{\displaystyle\sum_{\lambda}P^{0\lambda}},
\end{array}
\end{equation}
and
\begin{equation}
\label{eq:74bistris}
\begin{array}{l}
C={\displaystyle\lim_{\substack{L, M\to\infty}}}\left\langle
{\displaystyle\sum_{n=-\infty}^{+\infty}}(-1)^{n}\delta_{n, \sum_{\nu_{2}}[\dots]}\left( \frac{1-(-1)^{\sum_{\nu_{1}}}}{2} \frac{1+(-1)^{n}}{2} P_{3}(\{k_{\nu}\}) \right)\right\rangle_{\{x_{\nu},k_{\nu}\}}\\
={\displaystyle\int_{-\pi}^{+\pi}\frac{d\phi}{4\pi}}\,{\rm e}^{-t(1-\cos\phi)\int_{0}^{\infty}\frac{dk}{\pi}f_k\, v_{k}\Theta[v_{k}t-x]}\,\, {\rm e}^{-2x\sin^{2}(\phi/2)\int_{0}^{\infty}\frac{dk}{\pi}f_k\,\Theta[x-v_{k}t]}\\
\times\cos\left(x\sin\phi\int_{0}^{\infty}\frac{dk}{\pi}f_k\right){\displaystyle\frac{1+2\cos(2\phi)[(\sum_{\lambda}P^{0\lambda})^{2}-P^{00}]-2P^{00}( \sum_{\lambda}P^{0\lambda})^{2}+(P^{00})^{2}}{1-2P^{00}\cos(2\phi)+(P^{00})^{2}}},
\end{array}
\end{equation}
where we have assumed $P^{0\lambda}\equiv P^{\lambda 0}$, and finally
\begin{equation}
\label{eq:74bisquatris}
\begin{array}{l}
 D={\displaystyle\lim_{\substack{L, M\to\infty}}}\left\langle
{\displaystyle\sum_{n=-\infty}^{+\infty}}(-1)^{n}\delta_{n, \sum_{\nu_{2}}}\left( \frac{1-(-1)^{\sum_{\nu_{1}}}}{2} \frac{1-(-1)^{n}}{2} P_{4}(\{k_{\nu}\}) \right)\right\rangle_{\{x_{\nu},k_{\nu}\}}\\
={\displaystyle\int_{-\pi}^{+\pi}\frac{d\phi}{4\pi}}\,{\rm e}^{-t(1-\cos\phi)\int_{0}^{\infty}\frac{dk}{\pi}f_k\, v_{k}\Theta[v_{k}t-x]}\,\, {\rm e}^{-2x\sin^{2}(\phi/2)\int_{0}^{\infty}\frac{dk}{\pi}f_k\,\Theta[x-v_{k}t]}\\
\times\cos\left(x\sin\phi\int_{0}^{\infty}\frac{dk}{\pi}f_k\right){\displaystyle \frac{2\cos\phi(P^{00}-1)}{1-2P^{00}\cos(2\phi)+(P^{00})^{2}}}{\displaystyle\sum_{\lambda}P^{0\lambda}}.
\end{array}
\end{equation}
Plugging all the pieces together,  $A$+$B$+$C$+$D$, we end up with the expression (\ref{eq:res2}) for the relaxation function :\\
\begin{equation}
\label{eq:74tris}
\begin{array}{l}
R(x; t)={\displaystyle\int_{-\pi}^{\pi}\frac{d\phi}{2\pi}}\exp{\left(-t(1-\cos\phi){\displaystyle\int_{0}^{\infty}\frac{dk}{\pi}f_k\, v_{k}\Theta[v_{k}t-x]}\right)}\\
\\
\times\,\exp{\left(-x(1-\cos\phi){\displaystyle\int_{0}^{\infty}\frac{dk}{\pi}f_k\,\Theta[x-v_{k}t]}\right)}\cos\left(x\sin\phi{\displaystyle\int_{0}^{\infty}\frac{dk}{\pi}f_k}\right)\\
\\
\times\,{\displaystyle\frac{1+2\cos(\phi)(P^{00}-1)Q+\cos(2\phi)(Q^{2}-P^{00})-P^{00}Q^{2}}{1-2P^{00}\cos(2\phi)+(P^{00})^{2}}},
\end{array}
\end{equation}
where $Q= \sum_{\lambda=\pm 1,0}P^{0\lambda}$.
It would be useful to have an obvious comparison between this semi-classical result and a direct GGE computation of the same correlator for the case of the O(3) non-linear sigma model, as was done for the transverse field Ising model and the XY chain in transverse field. The new approach to these kinds of problems introduced in (\cite{Caux1},\cite{Caux2} and \cite{Tvelik}) may be a missing tool in this respect.\\
Formula (\ref{eq:74tris}) can be tested in several ways, for instance by choosing a thermally-populated initial state, with $f_{k}\propto {\rm e}^{-\beta k^{2}}$, and the same probability for each quantum number  to appear. This check can be done both analytically and numerically. In the former case, starting from expression (\ref{eq:74tris}) and plugging in a thermal distribution for $f_{k}$ and the quantum numbers, we find the same universal analytical result of Reference (\cite{Rapp1}), whereas in the latter case  we have performed a numerical average in the same spirit as that in Reference (\cite{Damle}). The results are shown in figure (\ref{fig:numerical}), which indicates a perfect agreement between the analitycal and numerical prediction. 
\begin{figure}[t]
   \begin{center}
   \includegraphics[width=12cm]{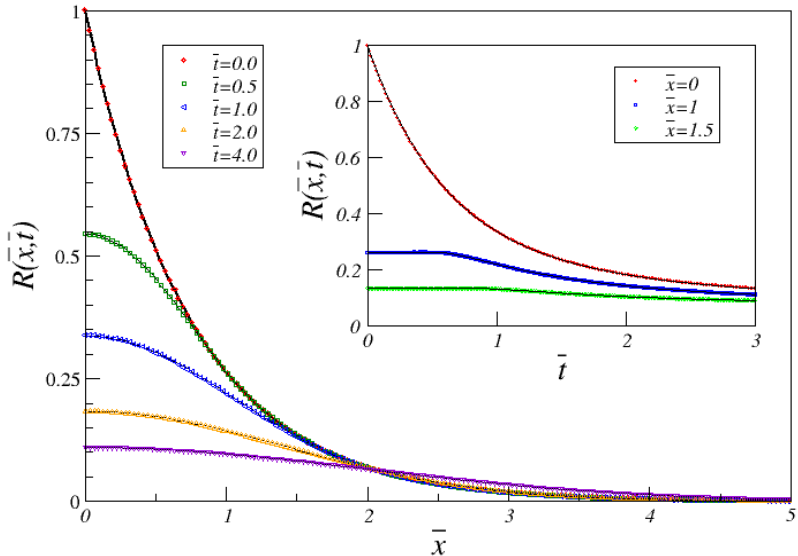}
	 \end{center}
   \caption{The relaxation function $R(\bar{x}, \bar{t})$ for $f_{k}= A\,{\rm e}^{-\beta k^{2}}$ (where $A=0.1$ and $\beta=0.8$ have been used in this case) and $P^{a,b}= 1/9$ for any choice of quantum numbers. Space and time are measured in unit of $\xi$ and $\tau$ respectively. Data points are numerical results and the solid line is the analytical prediction for the two-point function of a thermal initial state, see Eq. (\ref{eq:74tris}).}
\label{fig:numerical}
\end{figure}

\section{General expression of the relaxation function $R(x, t, T)$}
\label{app:general}

In this esction we show the form of the relaxation function for the O(3) non-linear sigma model when the times $T$ and $t$ are arbitrary, with the only constraint $T>t>0$. This expression can be written as: 
\be
\label{eq:74quatris}
 R(x; t, T)  =  R_{1}(x; t, T) + R_{2}(x; t, T),
\ee
where the first addend reads as:
\begin{equation}
\label{eq:74penta}
\begin{array}{l}
R_{1}(x; t, T)={\displaystyle\int_{-\pi}^{\pi}\frac{d\phi}{2\pi}}\exp{\left(-t(1-\cos\phi){\displaystyle\int_{0}^{\infty}\frac{dk}{\pi}f_k\, v_{k}\Theta[v_{k}t-x]}\right)} \\
\times \exp{\left(-T(1+\cos2\phi-2\cos\phi){\displaystyle\int_{0}^{\infty}\frac{dk}{2\pi}f_k v_k\,\Theta[x-v_{k}T]}\right)}\\
\times\exp{\left(-x(1-\cos\phi){\displaystyle\int_{0}^{\infty}\frac{dk}{\pi}f_k\,\Theta[x-v_{k}t]\Theta[v_k T-x]}\right)} \\
\times\exp{\left(-x(1-\cos2\phi){\displaystyle\int_{0}^{\infty}\frac{dk}{2\pi}f_k\,\Theta[x-v_{k}T]}\right)}\\
\times\cos\left\{x\sin2\phi{\displaystyle\int_{0}^{\infty}\frac{dk}{2\pi}f_k\Theta[x-v_k T] + x\sin\phi\int_{0}^{\infty}\frac{dk}{\pi}f_k\Theta[v_k T-x]}\right. \\
+\left.2T(\sin\phi-\sin2\phi) {\displaystyle\int_{0}^{\infty}\frac{dk}{2\pi}f_k v_k\,\Theta[x-v_k T]}\right\}\\
\times{\displaystyle\frac{1+2\cos(\phi)(P^{00}-1)Q+\cos(2\phi)(Q^{2}-P^{00})-P^{00}Q^{2}}{1-2P^{00}\cos(2\phi)+(P^{00})^{2}}}\\,
\end{array}
\end{equation}
and the second term is given by
\begin{equation}
\label{eq:74pentabis}
\begin{array}{l}
R_{2}(x; t, T)={\displaystyle\int_{-\pi}^{\pi}\frac{d\phi}{2\pi}}\exp{\left(-2(T-t\cos\phi){\displaystyle\int_{0}^{\infty}\frac{dk}{2\pi}f_k\, v_{k}\Theta[v_{k}T-x]}\right)}  \\
\times\exp{\left(-x(1-\cos2\phi){\displaystyle\int_{0}^{\infty}\frac{dk}{2\pi}f_k\,\Theta[x-v_{k}T]}\right)}\\
\times\exp{\left([-T(1+\cos2\phi) +2t\cos\phi]{\displaystyle\int_{0}^{\infty}\frac{dk}{2\pi}f_k v_k\,\Theta[x-v_{k}T]}\right)}\\
\times\cos\left\{2x\sin\phi{\displaystyle\int_{0}^{\infty}\frac{dk}{2\pi}f_k\Theta[v_k t -x]+2t\sin\phi\int_{0}^{\infty}\frac{dk}{2\pi}f_k v_k\Theta[x-v_k t]\Theta[v_k T -x]} \right.\\
+ \left. (-T\sin2\phi+t\sin\phi){\displaystyle\int_{0}^{\infty}\frac{dk}{2\pi}f_k v_k\,\Theta[x-v_k T]+x\sin2\phi\int_{0}^{\infty}\frac{dk}{2\pi}f_k\,\Theta[x-v_k T]}\right\}\\
\times {\displaystyle\frac{P^{00}Q^{2}-(P^{00})^{2}-\cos(2\phi)(Q^{2}-P^{00})}{1-2P^{00}\cos(2\phi)+(P^{00})^{2}}}.
\end{array}
\end{equation}
This formula represents the leading order of the semi-classical approach to the $n_z-n_z$ correlator in the $O(3)$ non-linear sigma model, and it is worth mentioning that it has been tested numerically, finding perfect agreement.  From this expression all subcases can be derived straightforwardly. 

\section{Occupation numbers and mode \newline probabilities}  %
\label{app:occ_numb}

In this section we compute the occupation numbers and the probability distributions for each mode, starting from the definition of the squeezed coherent state (\ref{eq:2}) and the operator algebra of the operators $(Z, Z^{\dag})$. Let us recall the definition of the initial state $|\psi\rangle$:
\begin{equation}
\label{eq:76}
|\psi\rangle = \exp{\left (\sum_{a, b}\int_{0}^{\infty} \frac{dk}{2\pi}\, K^{ab}(k)Z^{\dag}_{a}(-k)Z^{\dag}_{b}(k) \right)}|0\rangle,
\end{equation}
The $Z$ operators obey the Zamolodchikov-Faddeev (ZF) algebra, namely:
\begin{equation}
\label{eq:77}
\begin{array}{l}
Z_{a}(k_1)Z_{b}(k_2)-\mathcal{S}_{ab}^{cd}(k_1, k_2)Z_{d}(k_2)Z_{c}(k_1)=0\\
Z_{a}^{\dag}(k_1)Z_{b}^{\dag}(k_2)-\mathcal{S}_{ab}^{cd}(k_1, k_2)Z_{d}^{\dag}(k_2)Z_{c}^{\dag}(k_1)=0\\
Z_{a}(k_1)Z_{b}^{\dag}(k_2)-\mathcal{S}_{cb}^{ad}(k_1, k_2)Z_{d}^{\dag}(k_2)Z_{c}(k_1)=\delta_{ab}\,\delta(k_1-k_2).
\end{array}
\end{equation}
Intuitively this means that the exchange of two quasiparticle is realized by the two-particle scattering matrix $\mathcal{S}(k_1, k_2)$. In the present case we assume the form (\ref{eq:6}) for the $\mathcal{S}$ matrix, and thus the ZF algebra becomes: 
\begin{equation}
\label{eq:78}
\begin{array}{l}
Z_{a}(k_1)Z_{b}(k_2)+Z_{a}(k_2)Z_{b}(k_1)=0\\
Z_{a}^{\dag}(k_1)Z_{b}^{\dag}(k_2)+Z_{a}^{\dag}(k_2)Z_{b}^{\dag}(k_1)=0\\
Z_{a}(k_1)Z_{b}^{\dag}(k_2)=[-\sum_{c}Z_{c}^{\dag}(k_2)Z_{c}(k_1)+\delta(k_1-k_2)]\delta_{ab}.
\end{array}
\end{equation}
The ground state $|0\rangle$ satisfies $Z_{a}(k)|0\rangle =0$, $\forall{a}$ $\forall{k}$. From equations (\ref{eq:77}) it is clear that this algebra encodes the fact that the order of quantum number is preserved in time. The main quantities we want to compute are the occupation numbers $f_{k}^{ab}$, which are defined by:
\begin{equation}
\label{eq:79}
\frac{\langle\psi|Z_{a}^{\dag}(-k)Z_{b}^{\dag}(k)Z_{b}(k)Z_{a}(-k)|\psi\rangle}{\langle\psi|\psi\rangle}\equiv f_{k}^{ab}.
\end{equation}
This quantity is exactly the number of quasiparticle pairs of kind $(a, b)$ with momenta $(-k, k)$ (see figure (\ref{fig:5})), and it is directly connected to the probabilities $P^{\lambda\lambda'}$ we defined in the body of the chapter. 
The idea is to treat the $K$-matrix as an expansion parameter, but working from the beginning in the thermodynamic limit the squeezed state does not have a good expansion. Divergences appear in the expansion terms, in the form of squared Dirac delta-functions. In general there are two different methods of regularizing these divergences: one directly regulates the integral expressions in the infinite volume whereas the other operates through subtracting divergences in a large, finite volume. These two techniques were studied and compared in (\cite{Fabian1}), after having been proposed in (\cite{Takaz1} and \cite{Takaz2}). In the follow we shall use the second method.\\
Let us start by analysing the expansion of the denominator of (\ref{eq:79}). First Taylor expand the squezeed state as:
\begin{equation}
\label{eq:84}
|\psi\rangle= (1+ \sum_{\stackrel {k>0}{a, b}}K^{ab}(k)Z^{\dag}_{a}(-k)Z^{\dag}_{b}(k)  + \frac{1}{2!}(\sum_{\stackrel {k>0}{a, b}}K^{ab}(k)Z^{\dag}_{a}(-k)Z^{\dag}_{b}(k))^{2}+\dots)|0\rangle,
\end{equation}
therefore we have:
\begin{equation}
\label{eq:85}
\langle\psi|\psi\rangle=1+Z_1+Z_2+\dots,
\end{equation}
where 
\begin{equation}
\label{eq:86}
Z_1= \sum_{\stackrel {k>0}{a, b}}\sum_{\stackrel {\xi>0}{c, d}} (K^{ab}(k))^{*}K^{cd}(\xi)\langle 0| Z_{b}(k)Z_{a}(-k)Z_{c}^{\dag}(-\xi)Z_{d}^{\dag}(\xi)|0\rangle = \sum_{\stackrel {k>0}{a, b}}|K^{ab}(k)|^{2}.
\end{equation} 
As expected this term is proportional to the volume $L$. the higher orders will be proportional to $L^{2}$, $L^{3}$ {\em et cetera}. Let us consider the expansion of the numerator of (\ref{eq:79}):
\begin{equation}
\label{eq:87}
\langle\psi| Z_{\alpha}^{\dag}(-\xi)Z_{\beta}^{\dag}(\xi)Z_{\beta}(\xi)Z_{\alpha}(-\xi)|\psi\rangle = W_1+\frac{1}{4} W_2+\dots,
\end{equation}
where the first term $W_1$is given by:
\begin{equation}
\label{eq:88}
\begin{array}{lcl}
W_1 & = & {\displaystyle\sum_{\stackrel {k_1>0}{a, b}}\sum_{\stackrel {k_2>0}{c, d}}}\langle 0| Z_{c}(k_2)Z_{d}(-k_2)Z_{\alpha}^{\dag}(-\xi)Z_{\beta}^{\dag}(\xi)Z_{\beta}(\xi)Z_{\alpha}(-\xi)Z_{a}^{\dag}(-k_1)Z_{b}^{\dag}(k_1)|0\rangle \\
& = & |K^{\alpha\beta}(\xi)|^{2}.
\end{array}
\end{equation}
This quantity is finite, also in the infinite volume limit $L\to\infty$, whilst $W_2$ is proportional to $L$. We now need the expression of $W_2$ to see how these divergences cancel against the normalization of the boundary state. By definition we have:
\begin{equation}
\label{eq:89}
\begin{array}{l}
 W_2 = {\displaystyle\sum_{\stackrel {k_1>0}{a, b}}\sum_{\stackrel {k_2>0}{c, d}}\sum_{\stackrel {k_3>0}{e, f}}\sum_{\stackrel {k_4>0}{g, h}}\langle 0| Z_{a}(k_1)Z_{b}(-k_1)Z_{c}(k_2)Z_{d}(-k_2)Z_{\alpha}^{\dag}(-\xi)Z_{\beta}^{\dag}(\xi)}\\
 {\displaystyle \times Z_{\beta}(\xi)Z_{\alpha}(-\xi)Z_{e}^{\dag}(-k_3)Z_{f}^{\dag}(k_3)Z_{g}^{\dag}(-k_4)Z_{h}^{\dag}(k_4)|0\rangle}
\end{array}
\end{equation}
After a bit of algebra we end up with the following result for $W_2$:
\begin{equation}
\label{eq:89bis}
 W_2 =  {\displaystyle 4|K^{\alpha\beta}(\xi)|^{2}\sum_{\stackrel {k>0}{a, b}}|K^{ab}(k)|^{2}+2|K^{\alpha\beta}(\xi)|^{2}\sum_{ a, b}|K^{ab}(\xi)|^{2}},
\end{equation}
where we immediately see that the first term is of order $L$. In particular the first term of (\ref{eq:89}) can be written as $4 W_1 Z_1$, thus the first correction in equation (\ref{eq:79}) is given by:
\begin{equation}
\label{eq:90}
\frac{\langle\psi|Z_{a}^{\dag}(-k)Z_{b}^{\dag}(k)Z_{b}(k)Z_{a}(-k)|\psi\rangle}{\langle\psi|\psi\rangle}={\displaystyle\frac{W_1(1+Z_1)}{1+Z_1}}+\dots=W_1 + o(W_1).
\end{equation}
We see that the divergent term indeed cancels against the normalization factor, and this happens order by order. We conclude that as long as $K^{ab}(k)$ is small we can calculate the matrix elements like (\ref{eq:79}) by expanding the squeezed state. The final result can be written as:
\begin{equation}
\label{eq:80}
\boxed{
f_{k}^{ab}=|K^{ab}(k)|^{2}+o(|K^{ab}(k)|^{2})}
\end{equation}
which explains the name {\em amplitudes} for the matrix elements $K^{ab}(k)$. The momentum occupation numbers $f_k$ are given by:
\begin{equation}
\label{eq:81}
f_{k}= \sum_{a,b}f_{k}^{ab},
\end{equation}
while for the probability densities $P^{ab}(k)$ -which are defined in the thermodynamic limit- we have:
\begin{equation}
\label{eq:82}
P^{ab}(k)=\frac{{\displaystyle f_{k}^{ab}}}{{\displaystyle\sum_{a,b}\int_{0}^{\infty}\frac{dk}{2\pi}f_{k}^{ab}}},
\end{equation}
which makes the normalization condition $\sum_{a,b}\int_{0}^{\infty}\frac{dk}{2\pi}P^{ab}(k)=1$ explicit. The probability $P^{ab}$ is defined by $P^{ab}=\int_{0}^{\infty}\frac{dk}{2\pi}P^{ab}(k)$.

\section{The dynamics of the transverse field Ising chain}  %
\label{sec:Ising}
\subsection{The paramagnetic phase}  
\label{sec:subIsing1}
\begin{figure}[t]
\begin{center}
\includegraphics[width=10cm]{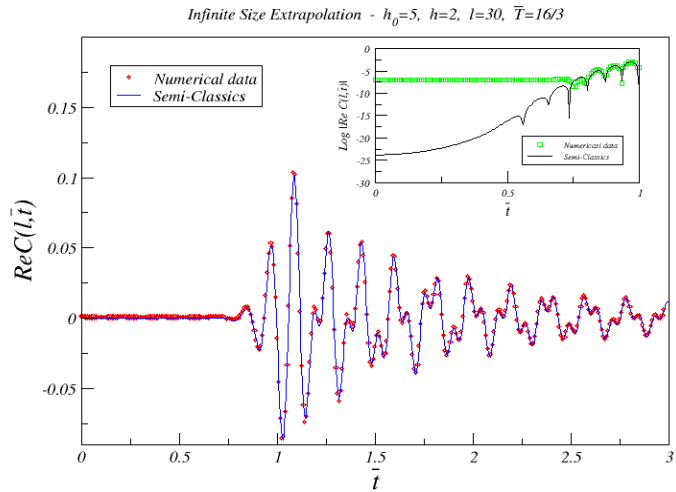}
\end{center}
\caption{The real part of the non-equal-time two point function after a quantum quench in the disordered phase, from $h_0 = 5$ to $h=2$. By definition $\bar{t}=t/t_f$ and $\bar{T}=T/t_f$. Data points represent the numerical data extrapolated in the thermodynamic limit ($L \to \infty$), while the solid line is given by equation (\ref{eq:75}). If we look at the inset in the plot, we notice that the semi-classical formula does not capture well the behaviour of the correlator outside of the light-cone (the same plot for the imaginary part shows a better agreement, even if at very small values of $\bar{t}$ numerical errors do not allow us to compare analytical result with the numerical ones).}
\label{Fig:numerics1bis}
\end{figure}

\begin{figure}[t]
\begin{center}
\includegraphics[width=10cm]{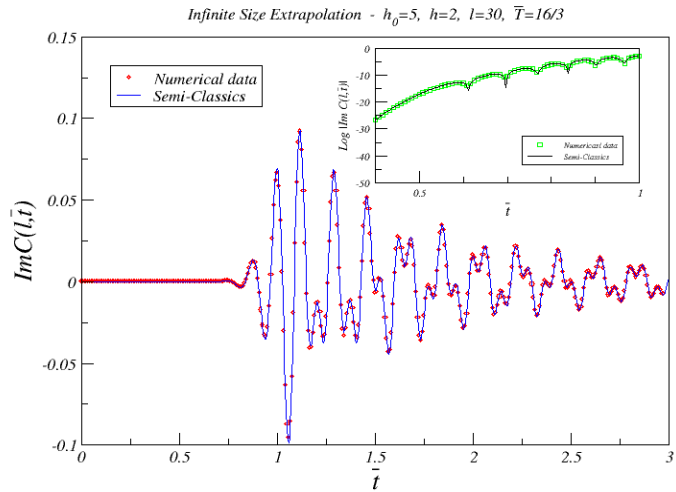}
\end{center}
\caption{The imaginary part of the non-equal-time two point function after a quantum quench in the disordered phase, from $h_0 = 5$ to $h=2$. By definition $\bar{t}=t/t_f$ and $\bar{T}=T/t_f$. Data points represent the numerical data extrapolated in the thermodynamic limit ($L \to \infty$), while the solid line is given by equation (\ref{eq:75}).}
\label{Fig:numerics1}
\end{figure}

\begin{figure}[t]
\begin{center}
\includegraphics[width=10cm]{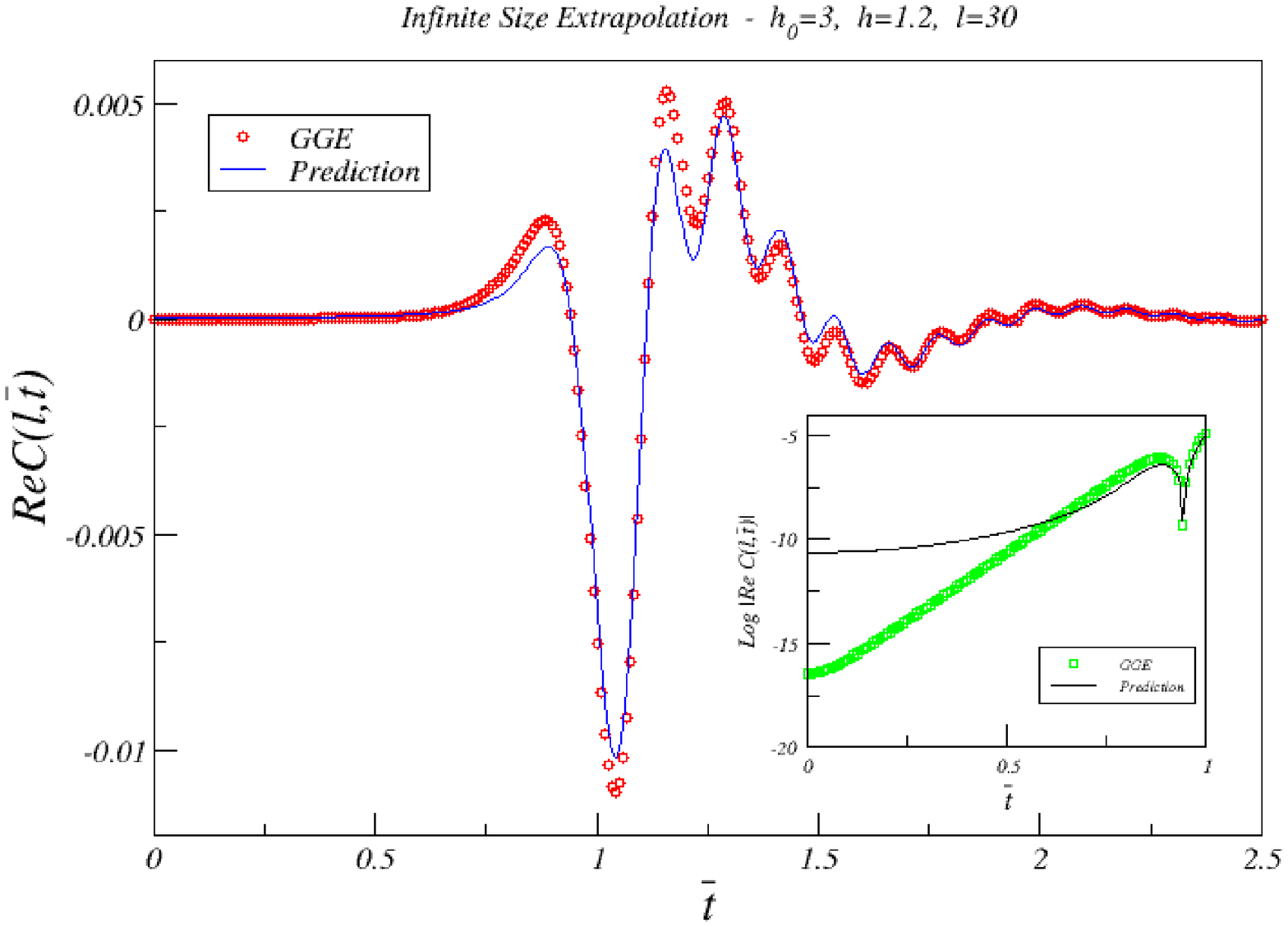}
\end{center}
\caption{The real part of the non-equal-time two point function after a quantum quench in the disordered phase, from $h_0 = 3$ to $h=1.2$. By definition $\bar{t}=t/t_f$ and $\bar{T}/t_f\gg 1$. Data points represent the numerical data extrapolated in the thermodynamic limit ($L \to \infty$), while the solid line is given by equation (\ref{eq:75}). The same considerations made for (\ref{Fig:numerics1}) hold true here. In addition we notice that as this quench is closer to the critical point, the agreement is only qualitative at small $\bar{t}$, despite the substitution $f_k \to -1/2 \log|\cos(\Delta_k)|$. To improve the agreement we should correct the expression for the pre-factor of equation (\ref{eq:75}), following the exact results of reference (\cite{Evangelisti2}).}
\label{Fig:numerics2}
\end{figure}

\begin{figure}[t]
\begin{center}
\includegraphics[width=10cm]{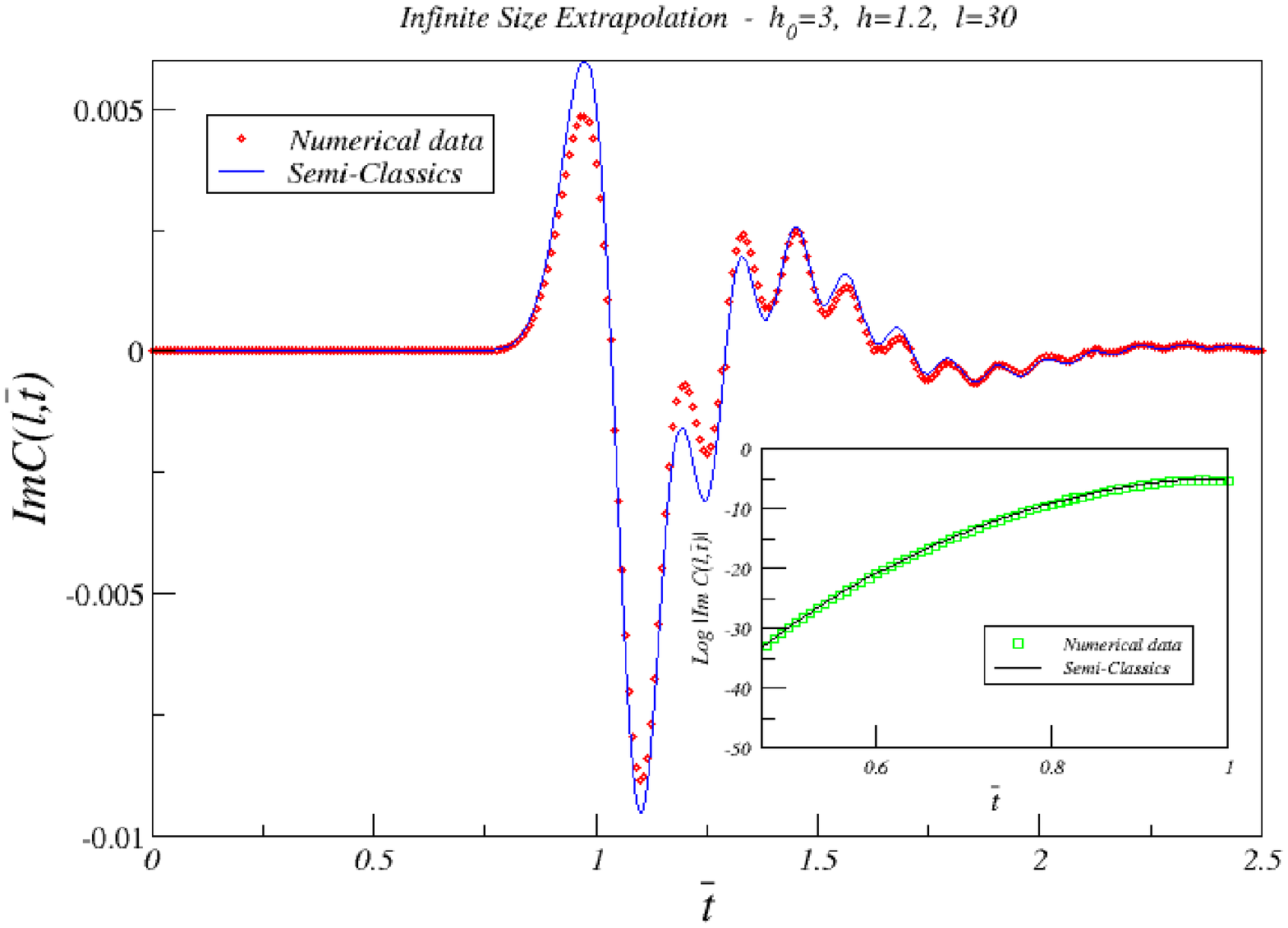}
\end{center}
\caption{Theimaginary part of the non-equal-time two point function after a quantum quench in the disordered phase, from $h_0 = 3$ to $h=1.2$. By definition $\bar{t}=t/t_f$ and $\bar{T}/t_f\gg 1$. Data points represent the numerical data extrapolated in the thermodynamic limit ($L \to \infty$), while the solid line is given by equation (\ref{eq:75}). The same considerations made for (\ref{Fig:numerics1}) hold true here.}
\label{Fig:numerics2bis}
\end{figure}

Another important check of formula (\ref{eq:74tris}) is represented by the transverse field Ising chain limit (TFIC). Briefly, we now focus on the dynamics after a quantum quench in the TFIC, whose Hamiltonian reads:
\begin{equation}
\label{eq:Ising_ham}
\hat{H}=-J\left({\displaystyle\sum_{i=1}^{L-1}\sigma_{i}^{x}\sigma_{i+1}^{x}}-h{\displaystyle\sum_{i=1}^{L}\sigma_{i}^{z}}\right),
\end{equation}
where $\sigma_{i}^{\alpha}$ are the Pauli matrices at site $i$, $J>0$ is the energy scale, $h$ the transverse magnetic field, and we impose free boundary conditions (for the moment we assume the system to be large but {\em finite}).
The goal is to determine the dynamical order-parameter two-point function:
\be
\label{eq:ising_cor}
C^{\rm Ising}(x, T, t)=\langle \psi |\sigma_{x_{2}}^{x}(t_{2})\sigma_{x_{1}}^{x}(t_{1})| \psi\rangle,
\ee
where $x=x_2-x_1$, $T=t_1+t_2$ and $t=t_2-t_1$. The action of the operator $\sigma_{i}^{x}(t)$ on the paramagnetic ground state is akin to that of the operator $\tilde{n}^{z}$ in the O(3) non-linear sigma model, that is, it either creates an excitation or it destroys one already present. For this reason the semi-classical approach developed in the previous section is easily generalizable.
For this model the $K$ matrix in the definition of $|\psi\rangle$ is known and equal to $K(k)=-\frac{i}{2}\tan\left[\frac{\theta_k-\theta_{k'}}{2}\right]$, where the $\theta$'s are the Bogoliubov angles defined in (\ref{eq:77bis}) before and after the quench respectively. 
In order to obtain (\ref{eq:ising_cor}), without rederiving everything from the beginning, one should formally replace $P^{a,b}=1/(q-1)^{2}$ and then take the limit $q \to 2$. 
Otherwise we can start again from (\ref{eq:56bis}), and take into account that for the TFIC excitations do not have internal quantum numbers (this is true both in the ordered and disordered phases). This feature greatly simplifies the algebra of the derivation, and we end up with the following general expression for the correlator of equation (\ref{eq:ising_cor}), valid in the thermodynamic limit:\\
\begin{equation}
\label{eq:75}
\begin{array}{l}
 C^{\rm Ising}(x, T, t) = C^{\rm Ising}_{T=0}(x, t)\,\exp{\left(-2t{\displaystyle\int_{0}^{\pi}\frac{dk}{\pi}f_k v_k\Theta[v_k t-x]}\right)}\\
\times  \exp{\left(-2T{\displaystyle\int_{0}^{\pi}\frac{dk}{\pi}f_k v_k \Theta[x-v_k T]}\right)} \exp{\left(-2x{\displaystyle\int_{0}^{\pi}\frac{dk}{\pi}f_k\Theta[x-v_k t]\Theta[v_k T-x]}\right)},\\
\end{array}
\end{equation}
where $C^{\rm Ising}_{T=0}(x, t)$ is given by:
\begin{equation}
\label{eq:76bis}
 C^{\rm Ising}_{T=0}(x, t)  \propto {\displaystyle\int_{-\pi}^{\pi}\frac{dk}{2\pi}\,\frac{{\rm e}^{-i\epsilon(k)t+ikx}}{\epsilon(k)}},
\end{equation}
where again as long as we work on the lattice the energy-momentum relation is $\epsilon(k)=[(h-\cos(k))^{2}+\sin^{2}(k)]^{1/2}$. 
Once you take the proper scaling limit $a \to 0$, then $\epsilon(k)$ becomes relativistic and you can replace the general prefactor with a Bessel function $K_0$.
Comparisons between theory and numerical data are shown in figures (\ref{Fig:numerics1},\ref{Fig:numerics1bis}) and (\ref{Fig:numerics2}, \ref{Fig:numerics2bis}). We note that in expression (\ref{eq:75}) $T$ is arbitrary (with the constraint $T>t>0$), and not necessarily much longer than the Fermi time $t_f=x/v_{max}$, as was in (\ref{eq:74tris}).

For larger quenches the semiclassical prediction, with  $f_k$ equal to $|K(k)|^{2}/(1+|K(k)|^{2})$ is not accurate. Indeed one has to take into account that for arbitrary values of the magnetic field $h>1$ generic excitations are no longer single spin-flips (even if this is not the only source of error, see below).

In general their shape is given by a superposition of states with an arbitrary number of spin-flips, with coefficients depending on the value of $h$ itself. 
Only when $h\gg 1$ are single excitations well-approximated by single-spin flips, as you treat them in the semi-classical approach. 
A first correction to the semiclassical results corresponds to substituting $f_k \to -1/2 \log|\cos(\Delta_k)|$, where $\Delta_k$ is the Bogoliubov angle given by: 
\begin{equation}
\label{eq:77bis}
\cos(\Delta_k)={\displaystyle\frac{h_0 h-(h_0 +h)\cos(k)+1}{\epsilon_{h_0}(k)\epsilon_{h}(k)}}.
\end{equation}
This substitution, which follows from asymptotically exact techniques (see (\cite{Calabrese5})), increases the agreement between the theoretical predictions and numerical data, expecially close to the quantum critical point $h=1$ where excitations are no longer localized objects. 
However in some plots mismatch between theory and numerics is still present, and we believe that it is due to the hypothesis of 
particle-number conservation of the {\em pure} semi-classical approach (i.e., the number of quasiparticles is conserved during the dynamical evolution of any configuration). If we considered the possibility of creating or destroying particles during the time evolution (which is realized by an operator insertion), we would get time-dependent corrections to the prefactor (\ref{eq:76bis}). Roughly speaking this would mean going beyond the leading order in the semi-classical approximation, a topic that is not addressed in detail in this paper. 
In the recent article (\cite{Evangelisti2}) the dynamic correlators after a quantum quench were studied by using the {\em form-factor} technique, finding that the prefactor of expression (\ref{eq:75}) can be written as:   
\be
\label{eq:exact1}
C^{\rm Ising}_{T=0}(x, t) \propto {\displaystyle\int_{-\pi}^{\pi}\frac{dk}{2\pi}\,\frac{{\rm e}^{ikx}}{\epsilon(k)}\left[{\rm e}^{-i\epsilon_{k}t}+2i\tan(\frac{\Delta_k}{2})\cos(\epsilon_k(T)){\rm sgn}(x-\epsilon'_k t)\right]}.
\ee
{\em Outside} the light-cone (when $x>v_{max}t$) the first contribution in (\ref{eq:exact1}) is exponentially small, whereas the second one behaves as a power law. This observation explains in turn why the {\em pure} semi-classical approximation typically fails when applied to this regime. In particular the semi-classics is not able to capture the behaviour of the equal-time correlator, that represents the extreme out-of-the light-cone case. \\

\subsection{The ferromagnetic phase}  
\label{sec:subIsing2}

\begin{figure}[t]
\begin{center}
\includegraphics[width=10cm]{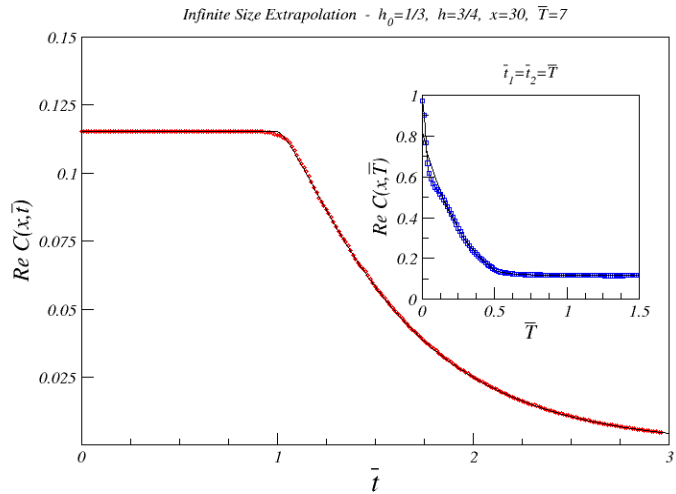}
\end{center}
\caption{Panel (inset): real part of the non-equal-time (equal-time) two point function in the ordered phase. Again the numerical data  are extrapolated in the thermodynamic limit ($L \to \infty$). In this phase, unlike in the paramagnetic one, the agreement between the semi-classics and the numerics works well also for the equal-time correlator}
\label{Fig:numerics3}
\end{figure}

\begin{figure}[t]
\begin{center}
\includegraphics[width=10cm]{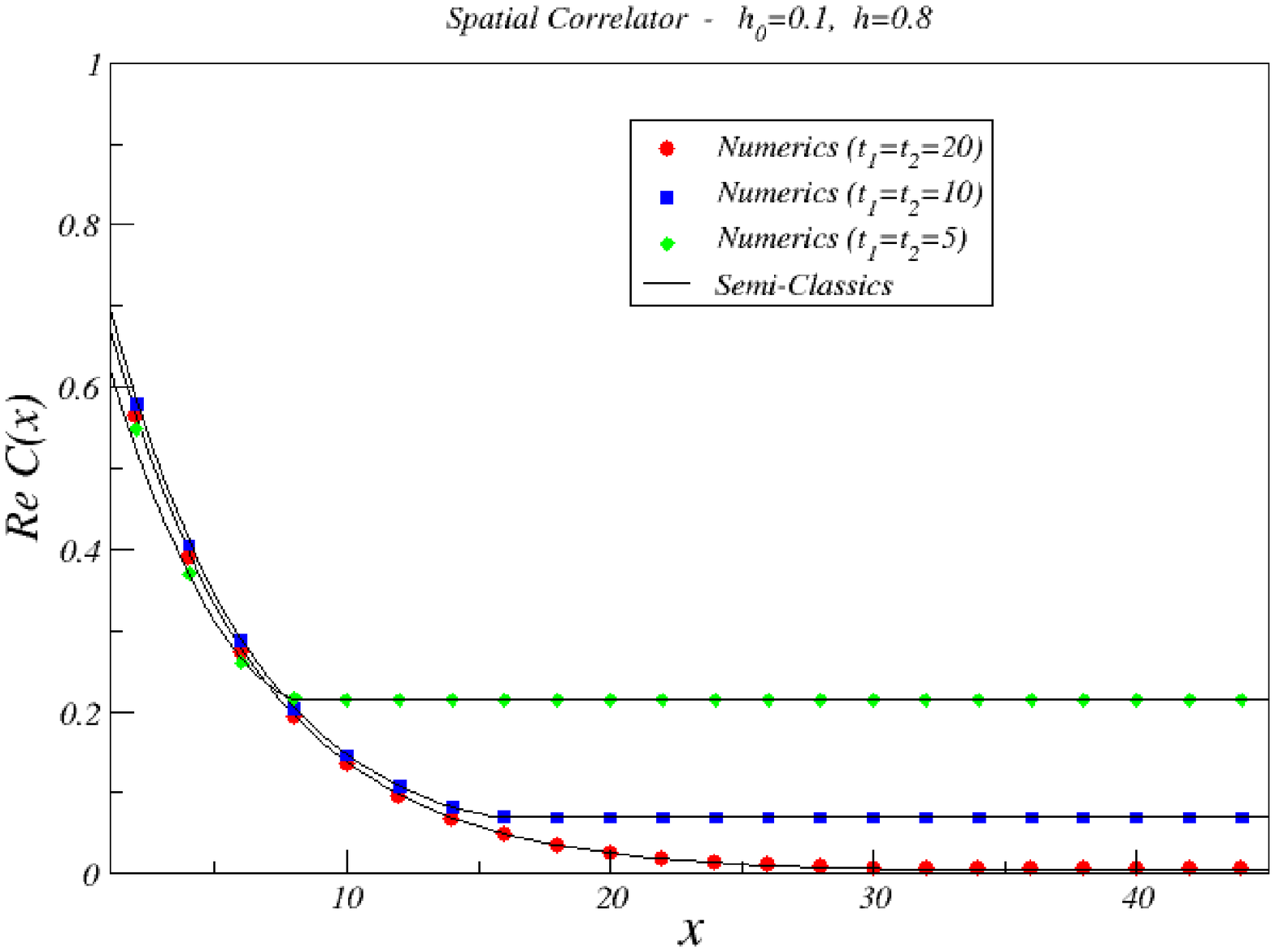}
\end{center}
\caption{Spatial correlation at equal times.}
\label{Fig:numerics3bis}
\end{figure}

Formula (\ref{eq:74tris}) cannot be directly used to derive the correlator of the Ising model in the ferromagnetic phase. In this case excitations are no longer spin-flips, in fact when $h<1$ these are domain walls (kinks). Following the general approach of F. Igloi and H. Reiger (\cite{Igloi1}, \cite{Igloi2}) we can easily obtain the two-point correlation function for arbitrary value of times and distances. The key ideas of this method have already been extensively discussed in their works (even if these authors did not consider explicitly the case of dynamical correlators), here we will just quickly review the main steps.
Let us start by recalling that when in Eq. (\ref{eq:Ising_ham}) $h=0$ the system is identical to the classical Ising spin chain. The ground state is two-fold degenerate and given by $|\psi_0\rangle = |+++\dots +\rangle$ and 
$|\psi_0\rangle = |---\dots -\rangle$, where $(+,-)$ refer to the alignment along the $x$-direction, and the first excited states are the $L-1$-fold degenerate given by the single kink states $|n\rangle=|+++\dots++--\dots---\rangle$ where $n$ denotes the kink position. If we switch on a small transverse field $h>0$, the low-lying excitations are, at first order in perturbation theory, superposition of these kink states. In practice, they are Fourier transforms of localized single kink states.
Analogously freely moving single kinks are therefore wave packets of the aformentioned low-lying excitations. Their energy agrees at first order in $h$ with the free fermion energies of Eq. (\ref{chap4_eq:36}), and their velocity is given by:
\begin{equation}
\label{eq:8}
v_{k}=\frac{\partial\epsilon_{k}}{\partial k}=\frac{h\sin(k)}{\epsilon_k}.
\end{equation}
Ballistically moving kinks are then the {\em fermionic} quasi-particle which we will use to formulate our semi-classical theory of the quantum quench dynamics of the transverse Ising model. Since by definition these quasi-particle are well-defined only at small fields in the ferromagnetic phase, we expect the theory to work for quenches only in the ordered phase. We will see that it actually works very well in the whole ferromagnetic phase not too close to the transition point $h=1$. \\
Immediatiately after the quench, for small $h$ and small $t$, the time-evolution of the system is given by:
\begin{equation}
\label{eq:9}
\begin{array}{lcl}
|\psi(t)\rangle & \approx &  \exp\left(ith{\displaystyle\sum_{i}\sigma_{i}^{z}}\right)|\psi_0\rangle\\
\, & = & {\displaystyle\prod_{i}[\cos(th)+i\sin(th)\sigma_{i}^{z}]}|\psi_0\rangle.
\end{array}
\end{equation}
This shows that by the action of the $\sigma_{i}^{z}$ operators initially single spins are flipped and therefore pairs of kinks are created at each lattice point, which then move ballistically with a speed proportional to the transverse magnetic field $h$. In this phase the maximal velocity is $v_{\text max}\approx h$.\\
In translationally invariant systems the creation probability of quasi-particles is uniformand will be denoted by $f_{k}(h_0, h)$. In the case of open boundary conditions there are corrections to a uniform creation probability close to the boundaries, which we will consider negligible for sufficiently large system sizes. \\
In a system at equilibrium at temperature $T$, this would be $f_{k}^{\text eq}(h_0, h)={\rm e}^{-\epsilon_{k}/T}$, while for relaxation at zero temperature $f_{k}(h_0, h)$
is the probability with which the modes with momentum number $k$ are occupied in the initial state $|\psi_0\rangle$, that is:
\begin{equation}
\label{eq:10}
f_{k}(h_0, h)= \langle\psi_0|\eta_{k}^{\dag}\eta_{k}|\psi_0\rangle,\quad\quad\quad T=0.
\end{equation}
Due to conservation of momenta after a global quench quasi-particles emerge pairwise at random position with velocities $+v_k$ and $-v_k$, as indicated in Figure (\ref{fig:1}).
 \begin{figure}[t]
   \begin{center}
   \includegraphics[width=7cm]{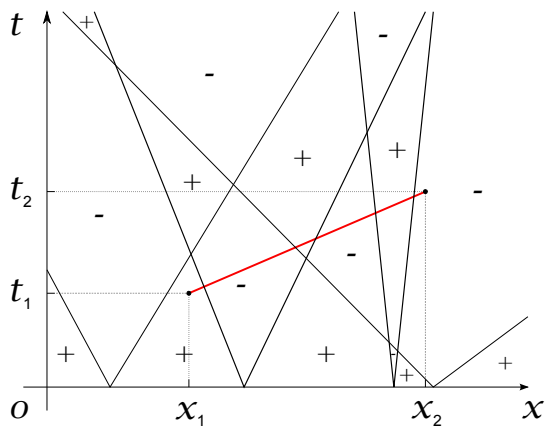}
   \end{center}
   \caption{Typical semi-classical contribution to the two-point  correlator function $C(r_1, t_1; r_2, t_2)$. It is worth noting that the eight trajectories of the four quasi-particles pairs intersect the line $(r_1, t_1; r_2, t_2)$ five times, i.e. an odd number of times, which implies that $\sigma_{r_1}^{x}(t_1)$ and $\sigma_{r_2}^{x}(t_2)$ have opposite orientation.}
\label{fig:1}
\end{figure}
Since quasi-particles represent kinks or domain walls, $\sigma^{x}$ changes sign each time a quasi-particle crosses the line connecting the points $(r_1,t_1)$ and $(r_2,t_2)$. Hence the correlation function in Eq. (\ref{eq:ising_cor}) can be evaluated in terms of classical particles moving ballistically by using a similar reasoning as in equilibrium, the difference being that here:
\begin{itemize}
\item Quasi-particles come {\em always} in pairs from a common off-spring at $t=0$;
\item The occupation number of quasi-particle states is not thermal.
\end{itemize}
In these note we consider only the case when $L \to \infty$, when we can avoid considering reflections by the boundaries.\\
At leading order in the semi-classical approach the number of quasi-particles is conserved during the time evolution of the system. In order to go beyond this order of approximation one has to consider quasi-particle creation and annihilation, phenomenon that can happen along the line $(r_1, t_1; r_2, t_2)$ with a probability given by a {\em form-factor}. In these notes we will not consider these corrections, that are negligible if the gas of excitations is enough dilute.\\
The idea behind the computation of the two-point function in the semi-classical approach is the following: if the quasi-particle trajectories intersect the line $(r_1, t_1; r_2, t_2)$ an odd number of times, the spins at $(r_1, t_1)$ and $(r_2, t_2)$ have opposite orientations (that is, $\sigma_{r_1}^{x}(t_1)=-\sigma_{r_2}^{x}(t_2)$), which contributes to the decay of the correlation between the two spins themselves. On the contrary, if the quasi-particle trajectories pass an even number of times, the spin have the same orientation, as if the trajectories did not pass the line $(r_1, t_1)$ and $(r_2, t_2)$ at all. \\
Let $Q(r_1, t_1; r_2, t_2)$ be the probability that a pair of quasi-particles that started from the same site at $t=0$ have passed the line $(r_1, t_1; r_2, t_2)$ a total odd number of times. Therefore the probability that for a given set of $n$ sites the kinks have passed (for each site a total odd number of times) this line is exactly: $Q^{n}(1-Q)^{L-n}$, where $L$ is the total number of sites. Summing over all possible cases we get:
\begin{equation}
\label{eq:11}
\begin{array}{lcl}
{\displaystyle\frac{C(r_1, t_1; r_2, t_2)}{C_{\rm eq}(r_1, r_2)}}  & = &    {\displaystyle\sum_{n=0}^{L}(-1)^{n}Q^{n}(1-Q)^{L-n}\frac{L!}{n!(L-n)!}}\\
\, & &\\
\, & = &  (1-2Q)^{L}\approx {\rm e}^{-2Q(r_1, t_1; r_2, t_2)L},
\end{array}
\end{equation}
where $C_{\rm eq}(r_1, r_2)$ is the equilibrium correlation function in the initial state and in the last step we have assumed the the probability $Q$ is small.
In order to calculate $Q$ one should average over the quasi-particles with momenta $k \in (-\pi, \pi)$, or equivalently over quasi-particle pairs which means a restriction to $k \in (0, \pi)$. By using the second method we have the expression:
\begin{equation}
\label{eq:11}
Q(r_1, t_1; r_2, t_2)=\frac{1}{2\pi}{\displaystyle\int_{0}^{\pi}}dk\,f_{k}(h_0,h)\cdot q_{k}(r_1, t_1; r_2, t_2)
\end{equation}
in terms of the occupation probability of equation (\ref{eq:10}) and the passing probability $q_{k}(r_1, t_1; r_2, t_2)$. This latter quantity measures the probability that the two trajectories of any quasi-particle pair  with momentum $k$ intersect the line $(r_1, t_1; r_2, t_2)$ together an odd number of times. 
Let us evaluate the passing probability $q_{k}(r_1, t_1; r_2, t_2)$ for $t_1 \neq t_2$, when $L \gg |r_2-r_1|$ (we do not want to take into account reflections against the boundaries, even if, for the time being, we consider the system large but finite). We call $t \equiv t_2-t_1$, $T \equiv t_2 + t_1$ and $x=|r_2-r_1|$. 
We have to consider three different cases, depending on the actual values of $(t,T)$ and $x$. 
\begin{figure}[htbp]
\centering
\subfigure[\protect\url{}\label{fig:scheme1}]%
{\includegraphics[scale=0.25]{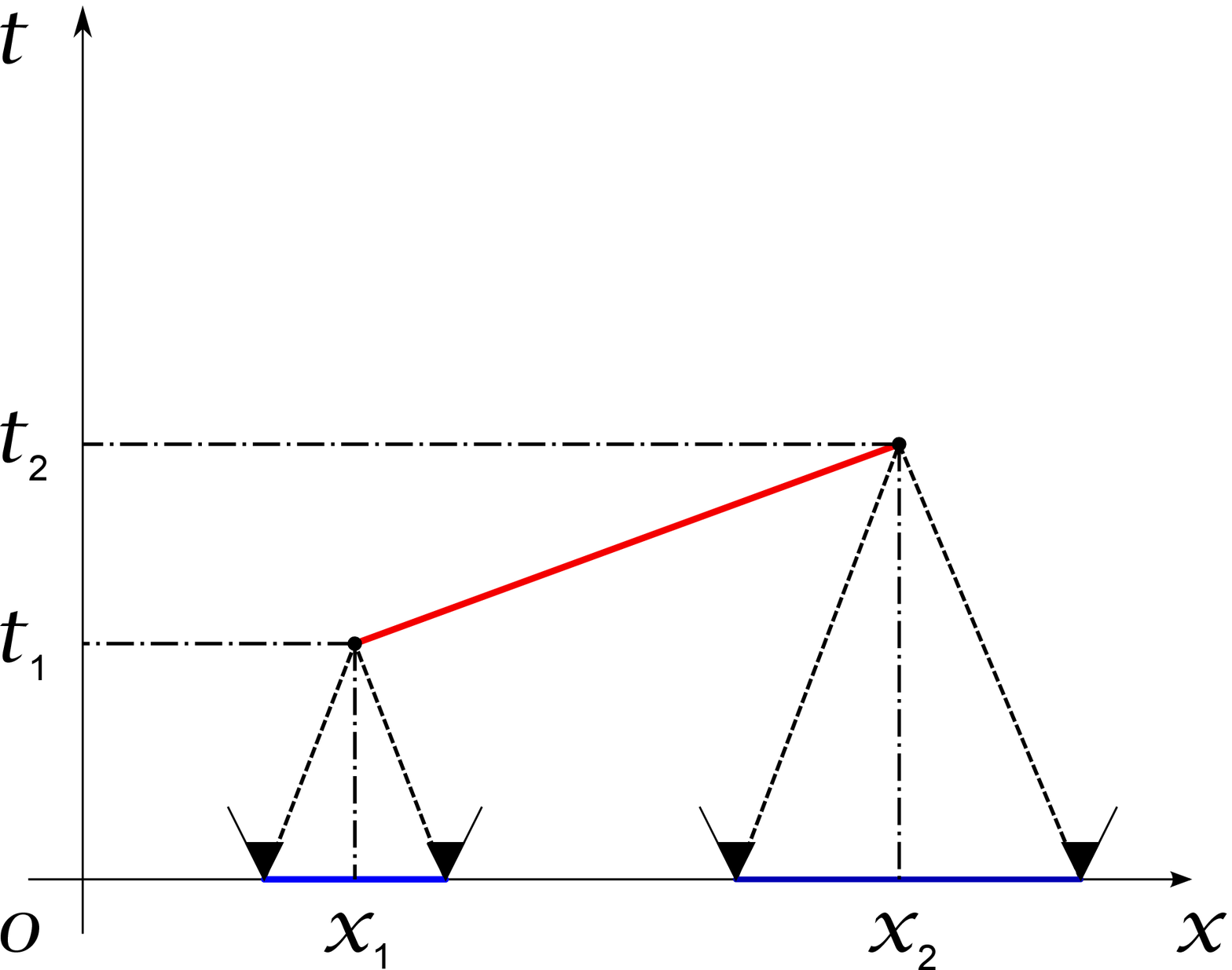}}\qquad
\subfigure[\protect\url{}\label{fig:scheme2}]%
{\includegraphics[scale=0.25]{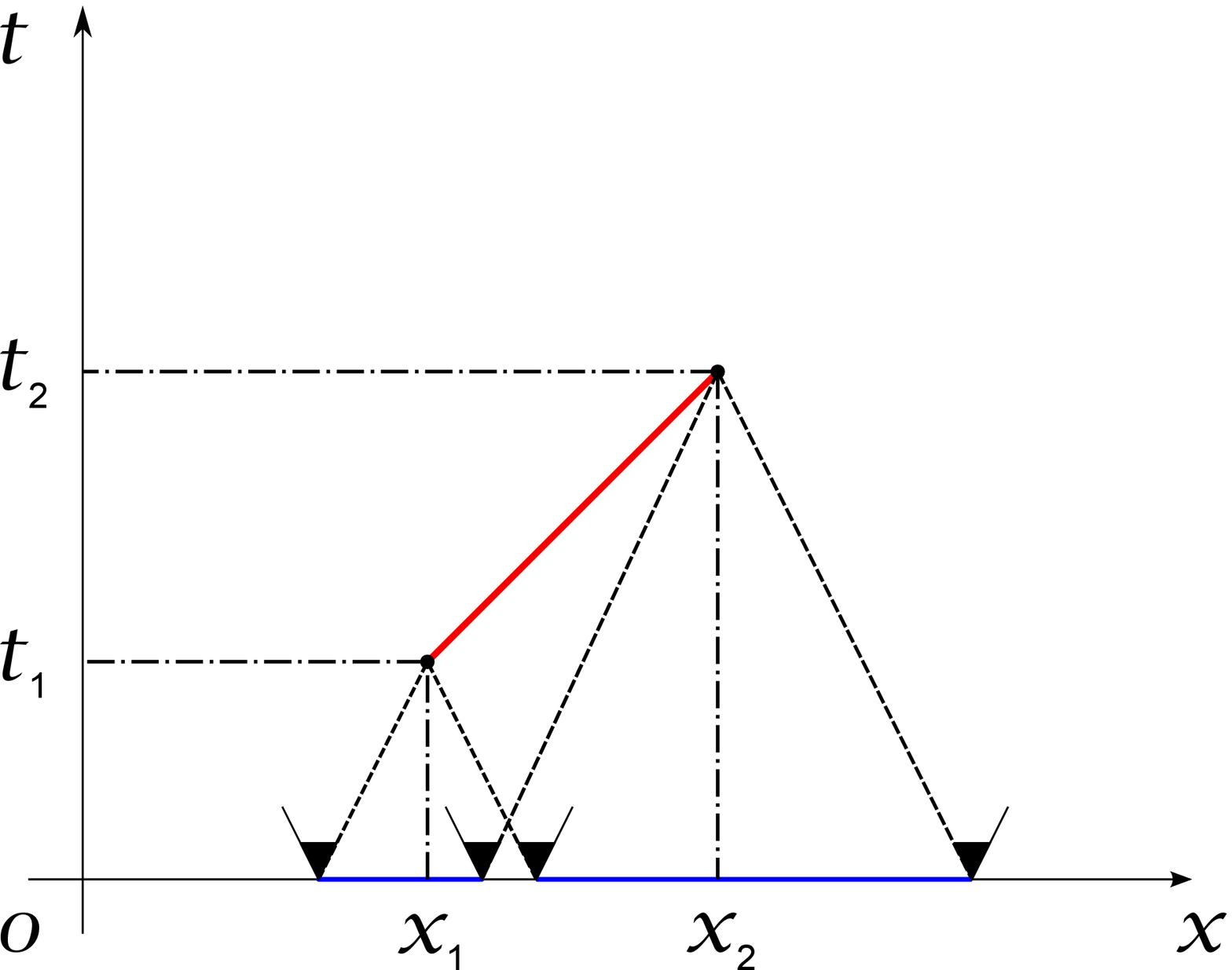}}\qquad
\subfigure[\protect\url{}\label{fig:scheme3}]%
{\includegraphics[scale=0.25]{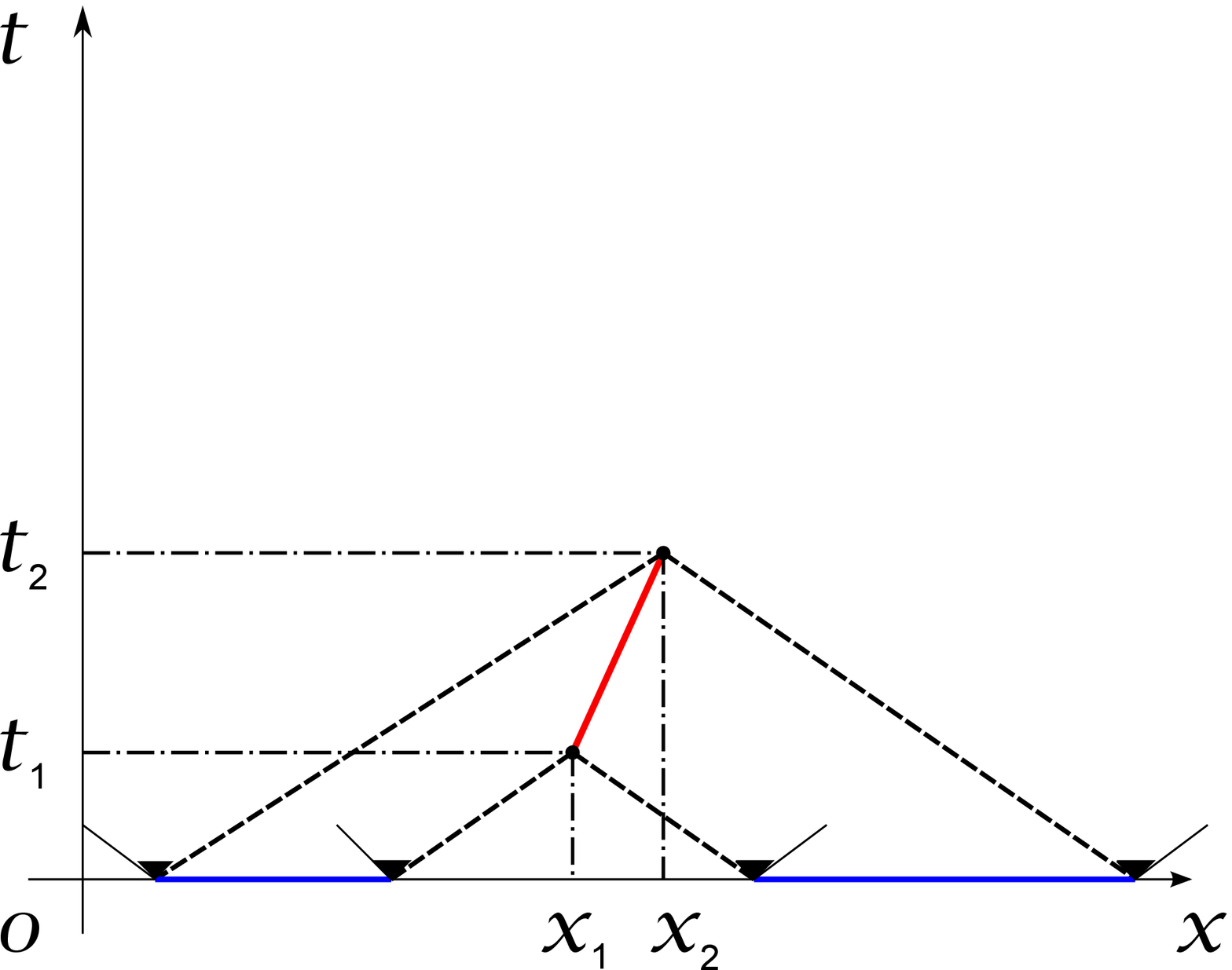}}\qquad
\caption{The passing probability $q_{k}(r_1, t_1; r_2, t_2)$ is given by the ratio between the blue area in figure and the total size of the system $L$.\label{fig:scheme}}
\end{figure}
We can summarize the general case as follows:
\begin{equation}
\label{eq:12}
q_{k}(r_1, t_1; r_2, t_2)= \begin{cases}2v_k T/L\quad\quad\quad x \geq v_k T \\
2x/L \quad\quad x < v_k T, \quad x \geq v_k t\\
2v_k t/L\quad\quad\quad\,\, x < v_k t.
\end{cases}
\end{equation}
Now we have all ingredients to write down the expression for the two-point correlator, and in particular the relaxation function in the ferromegnetic phase can eventually be written as: 
\begin{equation}
\label{eq:13}
\begin{array}{l}
{\displaystyle\frac{ C(x, T, t)}{C(h_0,h)}} \equiv  R(x, T, t) \\
\,\\
\quad=\exp{\left(-2t{\displaystyle\int_{0}^{\pi}\frac{dk}{\pi}f_k v_k\Theta[v_k t-x]}\right)}\times \exp{\left(-2T{\displaystyle\int_{0}^{\pi}\frac{dk}{\pi}f_k v_k \Theta[x - v_k T]}\right)}\\
  \\
\qquad\times\exp{\left(-2x{\displaystyle\int_{0}^{\pi}\frac{dk}{\pi}f_k\Theta[x-v_k t]\Theta[v_k T-x]}\right)}.\\
\end{array}
\end{equation}
where we se that the relaxation function in the ferromegnetic phase is equal to that in the paramagnetic phase
while the multiplicative constant, as introduced in (\cite{Calabrese5}, \cite{Calabrese6}) reads as:
\be
\label{eq:constant1}
C(h_0,h)={\displaystyle\frac{1-hh_0+\sqrt{(1-h^2)(1-h_0^2)}}{2\sqrt{1-hh_0}\sqrt[4]{1-h_0^2}}}.
\ee
In addition it is worth noticing that the semi-classical two-point correlator in the ordered phase of the Ising model is always a real function, as was the dominant contribution of the form-factor result at large ($x,T$) in (\cite{Evangelisti2}). 
In the ferromagnetic phase  the equal-time correlators can also be described by the semi-classic approach, in contrast to the paramagnetic phase, as is shown in the inset of figure (\ref{Fig:numerics3}, \ref{Fig:numerics3bis}). 

\section{Conclusions}
\label{s:conclusions}
In this chapter we have developed a semi-classical theory for the out-of-equilibrium quantum relaxation of the O(3) non-linear sigma model, after having prepared the system in a coherent superposition of Cooper pairs, a structure that is in agreement with the integrability of the theory in the bulk. For such quenches we analyzed the two-point function of the order parameter $\tilde{n}_{z}$ and argued that, in the long time limit its expression is given by formula (\ref{eq:74tris}), while for arbitrary times by equation (\ref{eq:74quatris}). The method employed here is a generalization of that used for studying the finite-temperature behaviour of a series of one-dimensional chains (\cite{Rapp2}, \cite{Sachdev1}, \cite{Rapp1}). As was already observed in other integrable models, the long-time behaviour of this two-point function after a quantum quench is {\em not} thermal. 
For equal times ($t_2=t_1\gg x/v_{max}$) the relaxation (\ref{eq:74tris}) reaches a stationary state, while for non-equal times ($t_2\neq t_1$) it is expressed as a function of the time difference $t$. These features are strong indications that the long-time limit of the two-point function (\ref{eq:5}) can be described by a statistical ensemble, and we expect it to be the GGE, being the model integrable.
It would be interesting to have an independent and explicit result in the GGE framework, in order to check the general belief also for the O(3) non-linear sigma model. The main difficulty here consists in finding an analytical expression for the lagrangian parameters of the GGE Hamiltonian, in order to explicitly carry out the statistical average.\\
The semi-classical approach has also been applied to predict the dynamics of the order-parameter two-point function of the transverse field Ising chain, in both phases.
In this case results for the auto-correlation and for the equal-time correlation are already present in literature (\cite{Igloi4}, \cite{Igloi1}), the novelty here is the generalization of the semi-classical method to the case of different-times correlators. Yet, the exact results recently obtained in \cite{Evangelisti2}, have allowed us to explore the limit of the {\em pure} semi-classics. This technique works very well within the ferromagnetic phase (also for pretty large quenches), whilst it is in the paramagnetic phase where it shows its real limits: it fails in predicting the equal-time two-point correlation functions, and in general the agreement in the out-of-the-light-cone case (see the insets of figures (\ref{Fig:numerics1}) and  (\ref{Fig:numerics2})) is only qualitative. Furthermore in the proximity of the critical point, the method - which in general is supposed to work well only for small quenches - needs to be improved with the substitution $f_k \to -1/2 \log|\cos(\Delta_k)|$, which takes into account the real shape of the excitations.\\
Nevertheless, our semi-classical theory can be applied straightforwardly to several other models, integrable and non-integrable, including the q-Potts model and the sine-Gordon theory, in order to compute correlators. These models will generate different combinatorial problems (the nature of the excitations is, of course, model-dependent), but the main ideas remain the same.



\chapter{Dynamical correlation after a quantum quench in integrable \newline systems}
\label{chap_final}
\ifpdf
    \graphicspath{{Chapter5/Chapter5Figs/PNG/}{Chapter5/Chapter5Figs/PDF/}{Chapter5/Chapter5Figs/}}
\else
    \graphicspath{{Chapter5/Chapter5Figs/EPS/}{Chapter5/Chapter5Figs/}}
\fi

In this chapter we consider dynamic (non equal time) correlation functions of local
observables after a quantum quench. We show that in the absence of
long-range interactions in the final Hamiltonian, the dynamics is
determined by the same ensemble that describes static (equal time)
correlations. For many integrable models static correlation functions
of local observables after a quantum quench relax to stationary values,
which are described by a \emph{generalized Gibbs ensemble} (GGE). The
same GGE then determines dynamic correlation functions and the basic
form of the fluctuation dissipation theorem holds, although the
absorption and emission spectra are not simply related as in the
thermal case. For quenches in the transverse field Ising chain (TFIC)
we derive explicit expressions for the time evolution of dynamic order
parameter correlators after a quench.\\
This chapter is essentially an expanded version of the article ``{\em Dynamical correlation after a quantum quench}'' written by F. Essler, S. Evangelisti and M. Fagotti, arXiv:1208.1961, published on Physical Review Letter, volume 109, (2012).

\section{Introduction to the problem}
\markboth{\MakeUppercase{\thechapter. Dynamical correlation after a quantum quench}}{\thechapter. Dynamical correlation after a quantum quench}

By virtue of their weak coupling to the environment ultra-cold atomic
gases provide ideal testing grounds for studying nonequilibrium
dynamics in isolated many-particle quantum systems. Recent
experiments~(\cite{Greiner},\cite{Kinoshita}, \cite{Hofferberth},\cite{Trotzky}, \cite{Cheneau}, \cite{Gring})
have observed essentially unitary time evolution on long
time scales. This has stimulated much theoretical research on
fundamental questions such as whether observables generically relax to
time independent values, and if they do, what principles determine their
stationary properties. Relaxational behaviour at first may appear
surprising, because unitary time evolution maintains the system in a
pure state at all times. However, it can be understood intuitively as
a property of a given finite subsystem in the thermodynamic limit,
with the role of the bath being played by the rest of the system.

Dimensionality and conservation laws strongly affect the
out-of-equilibrium dynamics. Ground breaking experiments by Kinoshita, 
Wenger and Weiss~(\cite{Kinoshita}) on trapped ${}^{87}{\rm Rb}$ atoms
established that three  dimensional condensates ``thermalize''
rapidly, i.e. relax quickly to a stationary state characterized by an
effective temperature, whereas the relaxation of quasi one-dimensional
systems is slow and towards an unusual non-thermal distribution. 
This difference has been attributed to the presence of
approximate conservation laws in the quasi-1D case, which are argued
to constrain the dynamics. The findings of Ref.~(\cite{Kinoshita})
sparked a tremendous theoretical effort aimed at clarifying the
effects of quantum integrability on the non-equilibrium evolution in
many-particle quantum systems, see
e.g. Refs~(\cite{Polkovnikov},\cite{Rigol}, \cite{Rigol2}, \cite{Calabrese3}, \cite{Cazalilla3}, \cite{Barthel}, \cite{Rossini1}, \cite{Rossini2}, \cite{Fioretto1}, \cite{Biroli}, \cite{Banulus}, \cite{Gogolin}, \cite{Rigol3}, \cite{Cazalilla4}, \cite{Mossel}, \cite{Caux})
and references therein. A widely held view, that has emerged from
these studies, is that the reduced density matrix of any
\emph{subsystem} (which determines correlation functions of all local
observables within the subsystem) is described in terms of either an
effective thermal (Gibbs) distribution or a so-called generalized
Gibbs ensemble (GGE)~(\cite{Rigol}). The former is believed to represent
the generic case, while substantial evidence suggests that the latter
arises for integrable models. 

Theoretical research so far has focussed on static properties in the
stationary state. A question of both great experimental relevance and
theoretical interest is what characterizes the \emph{dynamical}
properties at late times after a quench. These can be accessed by
experimental probes at finite energies, such as photoemission
spectroscopy~(\cite{ARPES}). In the first part of this chapter we prove
quite generally, that dynamical correlations of local
operators acting within a given subsystem in the stationary state
after a quantum quench are determined by the same distribution
function as static correlations. In particular this
means that whenever the GGE describes static correlations in the 
stationary state, it also applies to the dynamics. 

\section{Stationary state dynamics after a quantum quench}
We consider the following quench protocol. The system is prepared in
the ground state $|\Psi_0\rangle$ of a lattice Hamiltonian $H(h_0)$ with local
interactions, where $h_0$ is a system parameter such as a magnetic
field. At time $t=0$ we suddenly change $h_0$ to $h$ and the system
time evolves unitarily with Hamiltonian $H(h)$ thereafter. We are
interested in expectation values of the form ($t_1,\dots,t_n>0$)
\be
\langle \Psi_0(t)|{\cal O}_1(t_1)\ldots{\cal
  O}_n(t_n)|\Psi_0(t)\rangle,
\ee
where ${\cal O}_j$ are local observables.
We wish to demonstrate the following. If the stationary state of a
quantum many-body system after a quantum quench is described by a
density matrix $\rho_{\rm stat}$ such that for observables ${\cal
  O}_j$ acting only within a subsystem $S$ one has
\be
\lim_{t\to\infty}\langle\Psi_0(t)|
{\cal  O}_1\ldots{\cal O}_n|\Psi_0(t)\rangle
={\rm Tr}\big(\rho_{\rm stat}{\cal
  O}_1\ldots{\cal O}_n\big),
\label{static}
\ee
then dynamical correlations are described by the same density matrix,
i.e. for $t_1,\dots,t_n$ fixed we have
\bea
\lim_{t\to\infty}\langle\Psi_0(t)|
{\cal O}_1(t_1)\dots{\cal O}_n(t_n)|\Psi_0(t)\rangle\nonumber\\
={\rm Tr}\left(\rho_{\rm stat}{\cal
  O}_1(t_1)\dots{\cal O}_n(t_n)\right).
\label{dynamic}
\eea
The proof of this statement is based on the Lieb-Robinson
bound~(\cite{LR:1972}) and more specifically the following 
theorem by Bravyi, Hastings and Verstraete~(\cite{BHV:2006}): 
let $O_A$ be an operator that differs from the identity only
within a local region $A$. Now define the projection of the (non-local)
operator $O_A(t)$ to the subsystem $S\supset A$ by
\be
O^{(S)}_A(t)\equiv \frac{\mathrm{tr}_{\bar{S}}[
    O_A(t)]\otimes \mathrm I_{\bar{S}}}{\mathrm{tr}_{\bar{S}}[\mathrm I_{\bar{S}}]},
\ee
where $\bar{S}$ is the complement of $S$. If the time evolution is
induced by a short-range lattice Hamiltonian, then
\be\label{eq:upbound}
\lVert O_A(t)- O_A^{(S)}(t)\rVert\leq c|A|
e^{-\frac{d-v|t|}{\xi}}\, ,
\ee
where $\lVert.\rVert$ is the operator norm, $v$ is the maximal velocity at
which information propagates~(\cite{LR:1972}), $d$ is the (smallest)
distance between $\bar{S}$ and A, $|A|$ is the number of vertices in
set $A$, and $\xi$, $c$ positive constants. Assuming the operator
$\mathcal O_2$ to be bounded, $||\mathcal O_2||\leq \kappa$, we therefore have
\bea
\label{eq:ineq0}
&&|\langle \delta\mathcal O_1(t_1)
\mathcal O_2(t_2)\rangle_t|\leq
\lVert\delta\mathcal O_1(t_1)
\mathcal O_2(t_2)\rVert\nonumber\\
&&\, \leq\lVert\delta\mathcal O_1(t_1)\rVert\,
\lVert \mathcal O_2(t_2)\rVert
\leq c_1|A_1| \kappa\ e^{-\frac{d_1-v|t_1|}{\xi}},
\eea
where $\braket{.}_t$ denotes expectation value with respect to
$\ket{\Psi_0(t)}$ and $\delta \mathcal O_1(t)=\mathcal O_1(t)-\mathcal
O_1^S(t)$. 
The first inequality holds because the operator norm is an upper bound
for the expectation value on any state, while in the last step
we used (\ref{eq:upbound}). Eqn~\eqref{eq:ineq0} implies that
\be
\label{eq:ineq01}
\langle\prod_{j=1}^2 \mathcal O_j(t_j)\rangle_t=
\langle\mathcal O_1^S(t_1)\mathcal O_2(t_2)\rangle_t+
a_1(t_1,t_2,t)e^{-\frac{d_1-v|t_1|}{\xi}},
\ee
where $a_1(t_1,t_2,t)$ is a bounded function. By repeating the steps
leading to (\ref{eq:ineq01}) for the operators $\mathcal O_2(t_2)$ we arrive at
$\braket{\mathcal O_1(t_1)\mathcal O_2(t_2)}_t=\braket{\mathcal
  O_1^S(t_1)\mathcal O_2^S(t_2)}_t
+\sum_{i=1}^2 a_i(t_1,t_2,t)\exp\big(-\frac{d_i-v|t_i|}{\xi}\big)$,
where $a_2(t_1,t_2,t)$ is another bounded function. We may now use
the assumption~\eqref{static} for the expectation value on the right
hand side since all operators act within subsystem $S$
\be
\lim_{t\rightarrow\infty}\braket{\mathcal O_1(t_1)\mathcal O_2(t_2)}_t=\mathrm{Tr}(\rho_{\rm stat}\mathcal O_1^S(t_1)\mathcal O_2^S(t_2) )
+\sum_{i=1}^2 a_i(t_1,t_2)e^{-\frac{d_i-v|t_i|}{\xi}},
\label{eq:ineq1a}
\ee
where $\lim_{t\to\infty}a_i(t_1,t_2,t)=a_i(t_1,t_2)$ is assumed to
exists for simplicity\footnote{We can drop this assumption and bound
the sum on the r.h.s. in a $t$-independent way instead. The
corresponding contributions then vanish in the limit of an infinitely
large subsystem.}.
The chain of inequalities~\eqref{eq:ineq0} also holds
for the average with respect to the density matrix
$\rho_{\rm  stat}$, i.e.
\be
\label{eq:ineq2}
\mathrm{Tr}\big(\rho_{\rm stat}\mathcal O_1(t_1)\mathcal O_2(t_2)\big)=
\mathrm{Tr}\big(\rho_{\rm stat}\mathcal O_1^S(t_1)\mathcal O_2^S(t_2)\big)+
\sum_{i=1}^2 b_i(t_1,t_2)e^{-\frac{d_i-v|t_i|}{\xi}},
\ee
where $b_i(t_1,t_2)$ are bounded functions of $t_{1,2}$. Finally,
combining (\ref{eq:ineq1a}) and (\ref{eq:ineq2}) and then taking the size of
the subsystem $S$ to be infinite we obtain (\ref{dynamic}) in the case
$n=2$. The generalization to arbitrary $n$ is straightforward.

\section{Generalized Gibbs ensemble.}
We now concentrate on a quantum quench in an integrable model
in one dimension with Hamiltonian $H(h)\equiv I_1$ and
\emph{local} conservation laws $I_{n\geq 1}$, i.e. $[I_m,I_n]=0$.
The full (reduced) density matrix of the system (of a subsystem $A$)
at time $t$ after the quench is
\be
\rho(t)=|\Psi_0(t)\rangle\langle\Psi_0(t)|,\quad
\rho_A(t)={\rm Tr}_{\bar{A}}\big(\rho(t)\big),
\ee
where $\bar{A}$ is the complement of $A$. It is widely believed, and was
shown for quenches of the transverse field in the TFIC in
Refs~(\cite{Calabrese4}, \cite{Calabrese6}), that
\be
\lim_{t\to\infty}\rho_A(t)={\rm Tr}_{\bar{A}}\big(\rho_{\rm GGE}\big),
\label{GGE_static}
\ee
where
\be
\rho_{\rm GGE}=\frac{1}{Z_{\rm GGE}}e^{-\sum_m \lambda_m I_m},
\label{GGE}
\ee
is the density matrix of the GGE and ${Z_{\rm GGE}}$ 
ensures the normalization $\rm{tr}\big(\rho_{\rm GGE}\big)=1$.
Eqn~(\ref{GGE_static}) establishes that all \emph{local, equal time}
correlation functions of a given subsystem in the stationary 
state are determined by the GGE~(\ref{GGE}). Applying our result~(\ref{dynamic})
 to the case at hand, we conclude that dynamic correlation
functions are also given by the GGE, i.e.
\be
\lim_{t\to\infty}\langle\Psi_0(t)|
{\cal  O}_1(t_1)\ldots{\cal O}_n(t_n)|\Psi_0(t)\rangle
={\rm Tr}\big(\rho_{\rm GGE}{\cal
  O}_1(t_1)\ldots{\cal O}_n(t_n)\big).
\label{dynamicsGGE}
\ee
\section{Fluctuation Dissipation relation (FDR)}
A key question regarding dynamical properties in the stationary
state after a quench is whether a FDR holds~(\cite{Foini}). Given the
result~(\ref{dynamicsGGE}), we can answer this question for cases where
the stationary state is either described by a thermal distribution
with effective temperature $T_{\rm eff}$ or by a GGE. In the former
case, the standard thermal FDR with temperature $T_{\rm eff}$ applies.
 In this section we will set up the problem in a quite general way, in order to obtain a {\em weak} version of the fluctuation-dissipation theorem which is still valid when the steady state of the system is described by a GGE.\\
Let us consider the system described by a time-independent Hamiltonian $\hat{H}$, to which we add an external time-dependent field $F(t)$ that couples linearly to an observable $\hat{B}$ of the system. In the case of the transverse field Ising chain the role of $F(t)$ is played for instance by the magnetic field,  and $\hat{B}$ can be both $\sigma_{i}^{x}$ or $\sigma_{i}^{z}$. The complete time-dependent Hamiltonian is therefore 
\begin{equation}
\label{eq:8.1}
\hat{H}_{F}(t)=\hat{H}+F(t)\hat{B},
\end{equation}
when we assume that the external field $F(t)$ vanishes for $t$ earlier than a certain initial time $t_{0}$\footnote{This time must be very late if compared to the time scale of the previous section, because all this reasoning is valid when the system has already reached its stationary state for long times. In all this section we assume the unperturbed system to be stationary and characterized by a GGE.}. In general we would be interested in analysing the response to a superposition of external fields, but since in the linear response approximation the response to different perturbing terms add up independently, there is no loss of generality in considering the response to only one of them, as we did in (\ref{eq:8.1}). For $t\leq t_0$ the system is assumed to be in a stationary state which is described by a {\em generalized} Gibbs ensemble. For this reason it is now convenient to introduce a set of eigenstates of $\hat{H}_{GGE}\equiv \sum\lambda_k I_k$ , $\{|\psi_{n}^{GGE}\rangle\}$, with eigenvalues $E_{n}^{GGE}$. This {\em generalized} Hamiltonian comes from the definition of the GGE density matrix
\begin{equation}
\label{eq:8.1bis}
\rho_{GGE}\equiv \exp(-\sum\lambda_k I_k)/Z_{GGE} = \exp(-\hat{H}_{GGE})/Z_{GGE},
\end{equation}
and commutes with the real Hamiltonian $\hat{H}$, therefore there exists a set of eigenstates diagonalizing both Hamiltonians. 
We will indicate with $E_{n}$ the eigenvalues of the latter, and $P_n$ are the probabilities defined by
\begin{equation}
\label{eq:8.2}
P_n\equiv {\displaystyle\frac{{\rm e}^{-E_{n}^{GGE}}}{Z_{GGE}}},
\end{equation}
where $Z_{GGE}$ is the GGE partition function. The time-evolution of the system, in the Schrodinger picture is determined by the usual equation
\begin{equation}
\label{eq:8.3}
i\hbar{\displaystyle\frac{\partial}{\partial t}}|\psi_{n}^{GGE}(t)\rangle = \hat{H}_{F}(t)|\psi_{n}^{GGE}(t)\rangle,
\end{equation}
with the initial condition $|\psi_{n}^{GGE}(t_0)\rangle=|\psi_{n}^{GGE}\rangle$. The solution of this linear equation can be written in the following form
\begin{equation}
\label{eq:8.4}
|\psi_{n}^{GGE}(t)\rangle = \hat{U}(t, t_0)|\psi_{n}^{GGE}(t_0)\rangle,
\end{equation}
where $\hat{U}(t, t_0)$ is the unitary time-evolution operator, which gives the dynamical evolution from time $t_0$ to time $t$. In the absence of any perturbation the time-evolution of the system would be given by
\begin{equation}
\label{eq:8.5}
\hat{U}(t, t_0)= {\rm e}^{-\frac{i}{\hbar} \hat{H}(t-t_0)}.
\end{equation}
In order to set up a perturbative expansion of $\hat{U}$ in powers of $F(t)$ it is useful to re-write the previous equation as
\begin{equation}
\label{eq:8.6}
\hat{U}(t, t_0)= {\rm e}^{-\frac{i}{\hbar} \hat{H}(t-t_0)}\hat{U}_{F}(t, t_0)
\end{equation}
where $\hat{U}_{F}(t, t_0)$ is the part of the time-evolution that is due to the external field $F(t)$ and it obeys the equation of motion
\begin{equation}
\label{eq:8.7}
i\hbar{\displaystyle\frac{\partial}{\partial t}}\hat{U}_{F}(t, t_0)=F(t)\hat{B}(t-t_0)\hat{U}_{F}(t, t_0)
\end{equation}
with initial condition $\hat{U}_{F}(t, t_0)=\hat{1}$.  The time-dependent operator 
\begin{equation}
\label{eq:8.8}
\hat{B}(t)={\rm e}^{\frac{i}{\hbar}\hat{H}t}\hat{B}{\rm e}^{-\frac{i}{\hbar}\hat{H}t}, 
\end{equation}
is the Heisenberg representation of the operator $\hat{B}$. Equation (\ref{eq:8.6}) is useful because the time dependence of $\hat{U}_{F}$ is entirely due to the perturbation. Furthermore, substituting the zero-order approximation $\hat{U}_{F}(t, t_0)=\hat{1}$ on the right-hand side of equation (\ref{eq:8.7}) and integrating with respect to the time we get the first-order approximation to $\hat{U}_{F}(t, t_0)$ in the following form
\begin{equation}
\label{eq:8.9}
\hat{U}_{F,1}(t, t_0)=\left[\hat{1}-\frac{i}{\hbar}\int_{t_0}^{t}\hat{B}(t'-t_0)F(t')dt' \right].
\end{equation}
Hence, the complete time-evolution operator, {\em to first order in} $F$, takes the form
\begin{equation}
\label{eq:8.10}
\hat{U}_{1}(t, t_0)=  {\rm e}^{-i/\hbar \hat{H}(t-t_0)}  \left[\hat{1}-\frac{i}{\hbar}\int_{t_0}^{t}\hat{B}(t'-t_0)F(t')dt' \right].
\end{equation}
Let us consider a second observable $\hat{A}$ which, up to the time $t_0$, had the average equilibrium value
\begin{equation}
\label{eq:8.11}
\langle\hat{A}\rangle_{GGE}={\displaystyle\sum_{n}}P_n \langle\psi_{n}^{GGE}|\hat{A}|\psi_{n}^{GGE}\rangle,\qquad\qquad t\leq t_0.
\end{equation}
Our next goal is to compute the expectation value of $\hat{A}$ at time s later than $t_0$, under the influence of the perturbation.  Formally it is given by
\begin{equation}
\label{eq:8.11}
\langle\hat{A}\rangle_{GGE}(t)={\displaystyle\sum_{n}}P_n \langle\psi_{n}^{GGE}|\hat{A}(t)|\psi_{n}^{GGE}\rangle,
\end{equation}
where $\hat{A}(t)$ is the time-dependent operator which evolves under the action of the full Hamiltonian $\hat{H}_{F}$. Since we are interested only in the linear response to the external perturbation $F(t)$ we can make use of the linearized form of the time-evolution operator (\ref{eq:8.10}) and its hermitian conjugate. After some algebra we get the important result
\begin{equation}
\label{eq:8.12}
\langle\hat{A}\rangle_{GGE}(t)-\langle\hat{A}\rangle_{GGE}=-\frac{i}{\hbar}\int_{t_0}^{t}\langle[\hat{A}(t), \hat{B}(t')]\rangle_{GGE}F(t')dt',
\end{equation}
where both $\hat{A}(t)$ and $\hat{B}(t)$ are calculated via equation (\ref{eq:8.8}),  $[\hat{A}, \hat{B}]$ is the commutator of the two operators, and $\langle\dots\rangle_{GGE}$ denotes the average in the GGE. 
 Using again the time-independence of the unperturbed $\hat{H}$ we can write
\begin{equation}
\label{eq:8.13}
\langle[\hat{A}(t), \hat{B}(t')]\rangle_{GGE}=\langle[\hat{A}(\tau), \hat{B}]\rangle_{GGE}
\end{equation}
where $\tau\equiv t-t'>0$. Now let us consider the {\em retarded} response function $\chi_{AB}(\tau)$ which is defined as follows:
\begin{equation}
\label{eq:8.14}
\chi_{AB}(t_1, t_0)\equiv {\displaystyle\frac{\delta\langle A_{\hat{H}_{F}(t)}(t_1)\rangle}{\delta F(t_0)}}|_{F=0},
\end{equation} 
where now $\tau=t_1-t_0$, function that can also  be re-written in the following way by using the Kubo formula (see for instance \cite{Kubo} and references therein):
\begin{equation}
\label{eq:8.15}
\chi_{AB}(\tau)=-\frac{i}{\hbar}\Theta(\tau)\langle[\hat{A}(\tau), \hat{B}]\rangle_{GGE},
\end{equation} 
where $\Theta$ is the step function. Let us now make the change of variable $t'=t-\tau$ in equation (\ref{eq:8.12}), thus we can finally write the linear response for the observable $\hat{A}$ in the form
\begin{equation}
\label{eq:8.16}
\begin{array}{lcl}
\langle\hat{A}\rangle_{1}(t) & \equiv & \langle\hat{A}\rangle_{GGE}(t)-\langle\hat{A}\rangle_{GGE}\\
\, &  & \\
 & = &{\displaystyle \int_{0}^{t-t_0}}\chi_{AB}(\tau)F(t-\tau)d\tau.
\end{array}
\end{equation} 
From this equation it is evident that $\chi_{AB}(\tau)$ describes  the response function of the observable $\hat{A}$ at time $t$ to an impulse that couples to the observable $\hat{B}$ at an {\em earlier} time $t-\tau$. Because it describes the {\em after-effect} of a perturbation, it is appropriately called {\em retarded}, or even {\em casual} response function. Since the formula for the response function makes no reference to the initial time $t_0$, we can therefore let $t_0$ tend to $-\infty$, and all formulas still holds provided that the perturbing field approches zero for $t\to -\infty$ in such a way that the system can be assumed to have been in the unperturbed stationary state in the far past\footnote{State which is described by a GGE}:
\begin{equation}
\label{eq:8.17}
\langle\hat{A}\rangle_{1}(t)  = {\displaystyle \int_{0}^{\infty}}\chi_{AB}(\tau)F(t-\tau)d\tau.
\end{equation} 
It is worth noticing that, if $\hat{A}$ and $\hat{B}$ are hermitian operators, then $\chi_{AB}(\tau)$ is also real and it connects two real quantities.
In many pratical applications of the linear response formalism a central role is played by the response to a periodic perturbation of the type
\begin{equation}
\label{eq:8.19}
F(t) = F_{w}{\rm e}^{-iwt}+c.c. ,
\end{equation} 
where $F_{w}$ stands for a complex amplitude and $c.c.$ is the complex conjugate. We consider such perturbations because it is well known that every "good" function of time can be written as a superposition of periodic functions according to 
\begin{equation}
\label{eq:8.20}
F(t) = {\displaystyle\int_{-\infty}^{\infty}}\tilde{F}(w){\rm e}^{-iwt}\frac{dw}{2\pi},
\end{equation} 
where
\begin{equation}
\label{eq:8.21}
\tilde{F}(w) = {\displaystyle\int_{-\infty}^{\infty}}F(t){\rm e}^{iwt}dw.
\end{equation} 
Knowledge of the linear response to the first term of equation (\ref{eq:8.19}) will therefore suffice to determine the response to a general well behaved function $F(t)$. There is, however, a subtle point: the periodic potential we defined before does not vanish for $t\to-\infty$, and, for this reason, it seems that we are {\em not } entitled to assume that the system was in a stationary state in the far past. The standard trick  by which we can skip this problem is the so called "{\em adiabatic switching-on}" of the perturbation: we assume that the amplitude of the periodic potential is slowly turned on according the the law ${\rm e}^{\eta t}$, where $\eta$ is positive, and $\eta^{-1}$ represents a time scale much longer than the period of the perturbation. as long as the convergence factor is present, the whole formalism is applicable in the form already described. We will take the limit $\eta \to 0^{+}$ at the end of the calculation. If this limit exists, we are confident to describe the physical response of the system to a steady  periodic perturbation field, that is, a periodic field that has been going on long enough to erase any memory of the artificial switching-on process.  Mathematically we write $F(t)$ as 
\begin{equation}
\label{eq:8.22}
F(t) = {\displaystyle\int_{-\infty}^{\infty}}\tilde{F}(w){\rm e}^{-i(w+i\eta)t}\frac{dw}{2\pi},
\end{equation} 
and inserting (\ref{eq:8.19}) into equation (\ref{eq:8.17}) yields:
\begin{equation}
\label{eq:8.23}
\langle\hat{A}\rangle_{1}(t) = \langle\hat{A}\rangle_{1}(w){\rm e}^{-iwt}+c.c. ,
\end{equation} 
where 
\begin{equation}
\label{eq:8.24}
\langle\hat{A}\rangle_{1}(w) = \chi_{AB}(w)F_{w}, 
\end{equation}
and 
\begin{equation}
\label{eq:8.25}
\chi_{AB}(w)=-\frac{i}{\hbar}\,{\displaystyle\lim_{\eta\to 0}\int_{0}^{\infty}}\langle[\hat{A}(\tau), \hat{B}]\rangle_{GGE}\,{\rm e}^{i(w+i\eta)\tau}d\tau
\end{equation}
is the Fourier transform of the response function. Equation (\ref{eq:8.25}) defines a frequency dependent response function for any pair of operators $\hat{A}$ and $\hat{B}$. \\
Let us now expand the commutator in equation (\ref{eq:8.25}) in a complete set of exact eigenstates $|\psi_{n}^{GGE}\rangle$ of $\hat{H}_{GGE}$ (which are eigenstates of $\hat{H}$ as well):
\begin{equation}
\label{eq:8.26}
\begin{array}{lcl}
\langle[\hat{A}(\tau), \hat{B}]\rangle_{GGE} & = &  {\displaystyle\sum_{mn}}P_m\, ({\rm e}^{iw_{mn}\tau}A_{mn}B_{nm}-{\rm e}^{iw_{nm}\tau}B_{mn}A_{nm}) \\
& = &   {\displaystyle\sum_{mn}}(P_m - P_n){\rm e}^{iw_{mn}\tau}A_{mn}B_{nm}
\end{array}
\end{equation}
where the notation $O_{nm}\equiv\langle\psi_{n}^{GGE}|\hat{O}|\psi_{m}^{GGE}\rangle$ has been introduced to denote the matrix elements of an operator $\hat{O}$, and $w_{nm}=\frac{E_{n}-E_{m}}{\hbar}=-w_{mn}$ are the excitation frequancies of the system, while $E_{n}$'s are the eigenvalues of $|\psi_{n}^{GGE}\rangle$ with respect to $\hat{H}$. If we now insert equation (\ref{eq:8.26}) into equation (\ref{eq:8.25}) and perform the time integration, we obtain the {\em exact eigenstate representation} - also known as {\em Lehman representation}- of the response function:
\be
\label{eq:8.27}
\chi_{AB}(w)=\frac{1}{\hbar}  {\displaystyle\sum_{mn}\frac{P_m-P_n}{w-w_{nm}+i\eta}}A_{mn}B_{nm}
\ee
where it is implicit that $\eta \to 0^{+}$. Notice that $\chi_{AB}(w)$ is analytic in the upper half of the complex plane and has only simpe poles in the lower half. Let us now separate the real and imaginary part of the response function, with the help of the following formula:
\begin{equation}
\label{eq:8.28}
{\displaystyle\lim_{\eta\to 0^{+}}}\frac{1}{w-x+i\eta}=\mathcal{P}\frac{1}{w-x}-i\pi\delta(w-x),
\end{equation}
where the first term is the Cauchy-Hadamard {\em principal value} distribution. By using this formula we can immediately get the following two results:
\begin{equation}
\label{eq:8.29}
Re\,\chi_{AB}(w)=\frac{1}{\hbar}\mathcal{P}{\displaystyle\sum_{mn}\frac{P_m-P_n}{w-w_{nm}}}A_{mn}B_{nm},
\end{equation}
and
\begin{equation}
\label{eq:8.30}
\begin{array}{lcl}
Im\,\chi_{AB}(w)  & = &  - \frac{\pi}{\hbar}{\displaystyle\sum_{mn}(P_m-P_n)A_{mn}B_{nm}\,\delta(w-w_{nm})}\\
& = &  - \frac{\pi}{\hbar}{\displaystyle\sum_{mn} P_m [A_{mn}B_{nm}\,\delta(w-w_{nm})-B_{mn}A_{nm}\,\delta(w+w_{nm})]}\\
& \equiv &  - \frac{\pi}{\hbar}[S_{AB}(w)-S_{BA}(-w)],
\end{array}
\end{equation}
where in the last line we have introduced the so-called {\em dynamical structure factor}
\begin{equation}
\label{eq:8.31}
S_{AB}(w)=\frac{1}{2\pi}{\displaystyle\int_{-\infty}^{\infty}}\langle\hat{A}(t)\hat{B}\rangle_{GGE}\,{\rm e}^{iwt}dt,
\end{equation}
which can also be written as:
\begin{equation}
\label{eq:8.32}
S_{AB}(w)={\displaystyle\sum_{mn}} P_m A_{mn}B_{nm}\,\delta(w-w_{nm}).
\end{equation}
Equation (\ref{eq:8.30}) can be considered as a {\em weak} version of the well-known {\em fluctuation-dissipation theorem}, which remains valid for a generic choice of the probabilities $P_m$, therefore it is also applicable to the GGE. The only important assumption we have made about the probabilities $P_n$ throughout these notes is that the operator that appears at the exponent of the density matrix must commute with the Hamiltonian $\hat{H}$ (this is surely the case for the TFIC). We also notice that the imaginary part of the response function is always negative for positive $w$ and positive for negative $w$.\\
If we wanted to explicitly show the momentum dependence of the previous quantities we would start by writing the linear response function of observables $A_j$ and $B_l$ acting on sites $j$ and $l$ as  
\bea
\chi_{AB}(\omega,{\bf q})=-\frac{i}{\hbar L}\sum_{j,l}\int_0^\infty d\tau
e^{i\omega \tau-i{\bf q}({\bf r}_j-{\bf r}_l)}\nonumber\\
\times\ \mathrm{tr}[\rho_{\rm GGE}[A_j(\tau),B_l]].
\eea
On the other hand, the spectral function (dynamic structure factor) of the same two
observables in the stationary state would be given by 
\bea
S_{A B}(\omega,{\bf q})=\frac{1}{L}\sum_{j,l}
\int_{-\infty}^\infty\frac{\mathrm d \tau}{2\pi}
  e^{i\omega \tau-i{\bf q}\cdot({\bf r}_l-{\bf r}_j)}\nonumber\\
\times\ \mathrm{tr}[\rho_{\rm GGE}A_{l}(\tau) B_j].
\eea
Again using a Lehmann representation in terms of Hamiltonian eigenstates it
is straightforward to show that
\be
-\frac{\hbar}{\pi}\mathrm{Im}\ \chi_{AB}(\omega,{\bf q})=
S_{AB}(\omega,{\bf q})-S_{BA}(-\omega,-{\bf q}).
\ee
However, as was already noted in
Ref.~(\cite{Foini}) for the TFIC, unlike in the thermal
(Gibbs) case, the negative frequency part $S_{BA}(-\omega,-{\bf q})$ is
not related to the positive frequency part by a simple relation of the
form $S_{AB}(-\omega,-{\bf q})=f(\omega)S_{BA}(\omega,{\bf q})$,
where $f(\omega)$ is independent of $A$ and $B$.

\section{Transverse field Ising chain}
We now focus on the dynamics after a quantum quench in a particular
example, the TFIC described by the Hamiltonian
\be
H(h)=-J\sum_{j=1}^L\Bigl[\sigma_j^x\sigma_{j+1}^x+h\sigma_j^z\Bigr],
\label{Hamiltonian}
\ee
where $\sigma_j^\alpha$ are the Pauli matrices at site $j$, $J>0$ and we
impose periodic boundary conditions $\sigma_{L+1}^\alpha=\sigma^\alpha_1$. 
The model~(\ref{Hamiltonian}) is a crucial paradigm of quantum critical
behaviour and quantum phase transitions~(\cite{Sachdev1}). 
At zero temperature and in the thermodynamic limit it exhibits
ferromagnetic ($h < 1$) and paramagnetic ($h > 1$) phases, separated
by a quantum critical point at $h_c=1$. 
For $h<1$ and $L\rightarrow\infty$ there are two degenerate ground
states. Spontaneous symmetry breaking selects a unique ground state,
in which spins align along the $x$-direction. On the other hand, for
magnetic fields $h>1$ the ground state is non-degenerate and, as the
magnetic field $h$ is increased, spins align more and more along the
$z$-direction.  The order parameter for the quantum phase transition
is the ground state expectation value $\braket{\sigma^x_j}$.
We note that the model~(\ref{Hamiltonian}) is (approximately) realized
in systems of cold Rb atoms confined in an optical lattice~(\cite{ising-ca}).\\

Two point dynamical correlation functions are of particular importance
due to their relationships to response functions measured in
photoemission and scattering experiments. The two-point function of
transverse spins
$\langle\Psi_0(t)|\sigma^z_{j+\ell}(\tau_1)\sigma^z_j(\tau_2)|\Psi_0(t)\rangle$
in the TFIC can be calculated by elementary means~(\cite{McCoy1})
as it is local in terms of Jordan-Wigner fermions. Our goal is to
determine the dynamical order-parameter two-point function 
\be
\rho^{xx}(\ell,t+\tau_1,t+\tau_2)=\langle\Psi_0(t)|\sigma^x_{1+\ell}(\tau_1)
\sigma^x_{1}(\tau_2)|\Psi_0(t)\rangle,
\ee
after quenching the transverse field at time $t=0$ from $h_0$ to $h$
for times $\tau_{1,2}\geq 0$. This can be achieved by employing a
generalization of the form factor methods recently developed in
Ref.~(\cite{Calabrese5}) to the non-equal-time case, and augmenting the
results obtained in this way by exploiting the knowledge of exact
limiting behaviours derived in Refs~(\cite{Calabrese5}, \cite{Calabrese6}). 
Our approach is outlined in (\ref{sec:formfactor}).
For quenches \emph{within the ordered phase} ($h_0,h<1$) we
obtain for large positive $\ell$, $t$ 
\be
\rho^{xx}(\ell,t+\tau,t)\simeq C^x_{\rm FF}(h_0,h)\ R(\ell,\tau,t),
\label{rhoxx}
\ee
where
\be
R(\ell,\tau,t)=\exp\big[\int_0^\pi\frac{\mathrm dk}{\pi}\log\big(\cos\Delta_k\big)\times\min\left\{\max\{\varepsilon^\prime_h(k)\tau,\ell\}, \varepsilon^\prime_h(k)(2t+\tau)\right\}\big],
\label{R}
\ee
and
\be
C^x_{\rm FF}(h_0,h)=\frac{1-h
  h_0+\sqrt{(1-h^2)(1-h_0^2)}}{2\sqrt{1-h h_0}\sqrt[4]{1-h_0^2}}.
\label{Rbis}
\ee
Here \mbox{$\varepsilon_h(k)=2J\sqrt{1+h^2-2h\cos k}$} is the dispersion
relation of elementary
excitations of the Hamiltonian $H(h)$, \mbox{$\cos\Delta_k=
4J^2(1+hh_0-(h+h_0)\cos k)/\varepsilon_h(k)\varepsilon_{h_0}(k)$}
and \mbox{$\varepsilon'_{h}(k)=d\varepsilon_h(k)/dk$}. An important scale in the problem is
given by the ``Fermi-time''
\be
t_F=\frac{\ell}{2v_{\rm max}}\, ,\quad v_{\rm max}={\rm max}_k\ \varepsilon'_h(k),
\ee
where $v_{\rm max}$ is the maximal propagation velocity of the elementary
excitations of the post-quench Hamiltonian $H(h)$.
We note that the dominant contribution
at large $\ell, t$ (\ref{R}) has a vanishing imaginary part. This is
similar to the corresponding correlator at finite temperature in
equilibrium~(\cite{Sachdev1}).
In order to assess the accuracy of the asymptotic result~(\ref{R}) at short and
intermediate times and distances we have computed the correlator~(\ref{rhoxx})
numerically on large, open chains by means of a determinant
representation and then extrapolated the results to the thermodynamic
limit. 
\begin{figure}[t]
\begin{center}
\includegraphics[width=10cm]{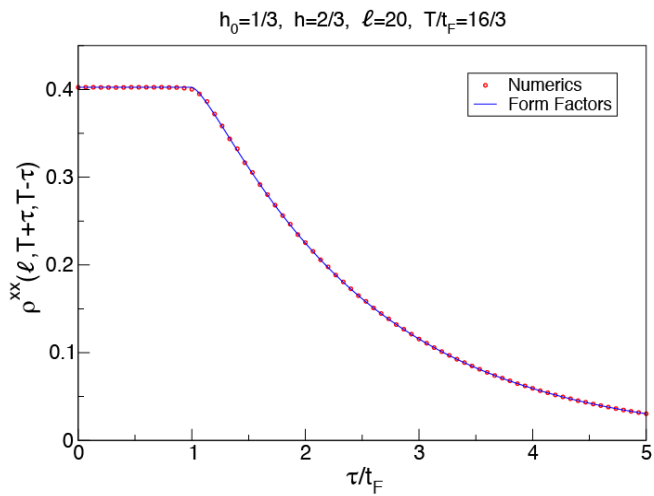}
\caption{Non-equal-time two point function after a quench in the
ordered phase from $h_0=1/3$ to $h=2/3$. The distance and time $T$ are
fixed at $\ell=20$ and $T/t_F=16/3$ respectively.}
\end{center}
\label{fig:ordered}
\end{figure}
A comparison between (\ref{rhoxx}) and the numerical results for a quench
from $h_0=1/3$ to $h=2/3$ and distance $\ell=20$ is shown in
Fig.~\ref{fig:ordered}. The agreement is clearly excellent. The
qualitative behaviour of $\rho^{xx}(\ell,T+\tau,T-\tau)$ is as
follows: $\tau=0$ corresponds to the known~(\cite{Calabrese5}) equal-time
correlator at time $T$ after the quench. The correlator remains
essentially unchanged until $\tau=t_F$ (corresponding to
$\tau_1-\tau_2=2t_F$ in \ref{rhoxx}), where a horizon effect occurs. At
later times $\tau>t_F$ the correlator decays exponentially.

For quenches \emph{within the disordered phase} ($h_0,h>1$), we obtain
for \\ 
$v_{\rm max}(2t+\tau)>\ell$
\be
\rho^{xx}(\ell,t+\tau,t)\simeq
h C^x_{\rm FF}(h_0^{-1},h^{-1}) F(\ell,\tau,t) R(\ell,\tau,t),
\label{rhoxxdiso}
\ee
where 
$R(\ell,\tau,t)$ and
$C^{x}_{\rm FF}$ are given by (\ref{R}) 
and 
\be
F(\ell,\tau,t)=\int_{-\pi}^\pi \frac{{\mathrm dk}J e^{i\ell k}}{\pi\varepsilon_h(k)}
\Big[e^{-i\varepsilon_k\tau}
+2i\tan\big(\frac{\Delta_k}{2}\big)\cos\big(\varepsilon_k(2t+\tau)\big)\mathrm{sgn}\big(\ell-\varepsilon_k^\prime\tau\big)\Big].
\label{F}
\ee
In the complementary regime $v_{\rm max}(2t+\tau)<\ell$ the correlator
is exponentially small and the expressions~(\ref{rhoxxdiso},~\ref{F}) no
longer apply. Outside the ``light-cone'' \\ $v_{\rm max}\tau<\ell$ the
first contribution in \eqref{F} is exponentially small, whereas the
second one decays as a power-law. The result~(\ref{rhoxxdiso}) is
obtained by a generalization of the form factor~(\cite{FF1}, \cite{FF2})
approach developed in
Ref.~(\cite{Calabrese6}) and is based on an expansion in the density of
excitations of $H(h)$ in the initial state after the quench. Hence it
is most accurate for quenches where this density is low and breaks
down for quenches from/to the quantum critical point. In
Figs~\ref{fig:1} and \ref{fig:2} we compare the asymptotic result~(\ref{rhoxx})
 to numerics obtained in the way described above. The
agreement for the chosen set of parameters ($\ell=30$, $h_0=2$, $h=3$
and $T/t_F=16/3$) is seen to be excellent.
\begin{figure}[t]
\begin{center}
\includegraphics[width=10cm]{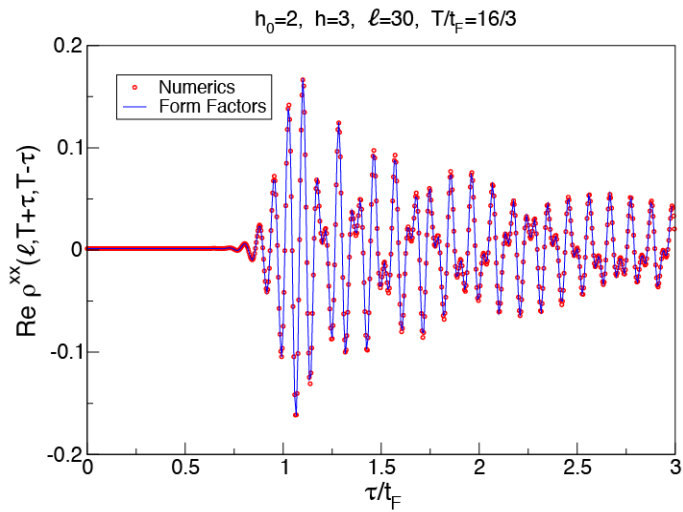}
\caption{Real part of the non-equal-time two point function after a
quench in the disordered phase from $h_0=2$ to $h=3$. The distance and
time $T$ are fixed at $\ell=30$ and $T/t_F=16/3$ respectively. Data
points are numerical results (see the text for details) and the solid
line is eqn~(\ref{rhoxxdiso}).}
\end{center}
\label{fig:1}
\end{figure}
\begin{figure}[t]
\begin{center}
\includegraphics[width=10cm]{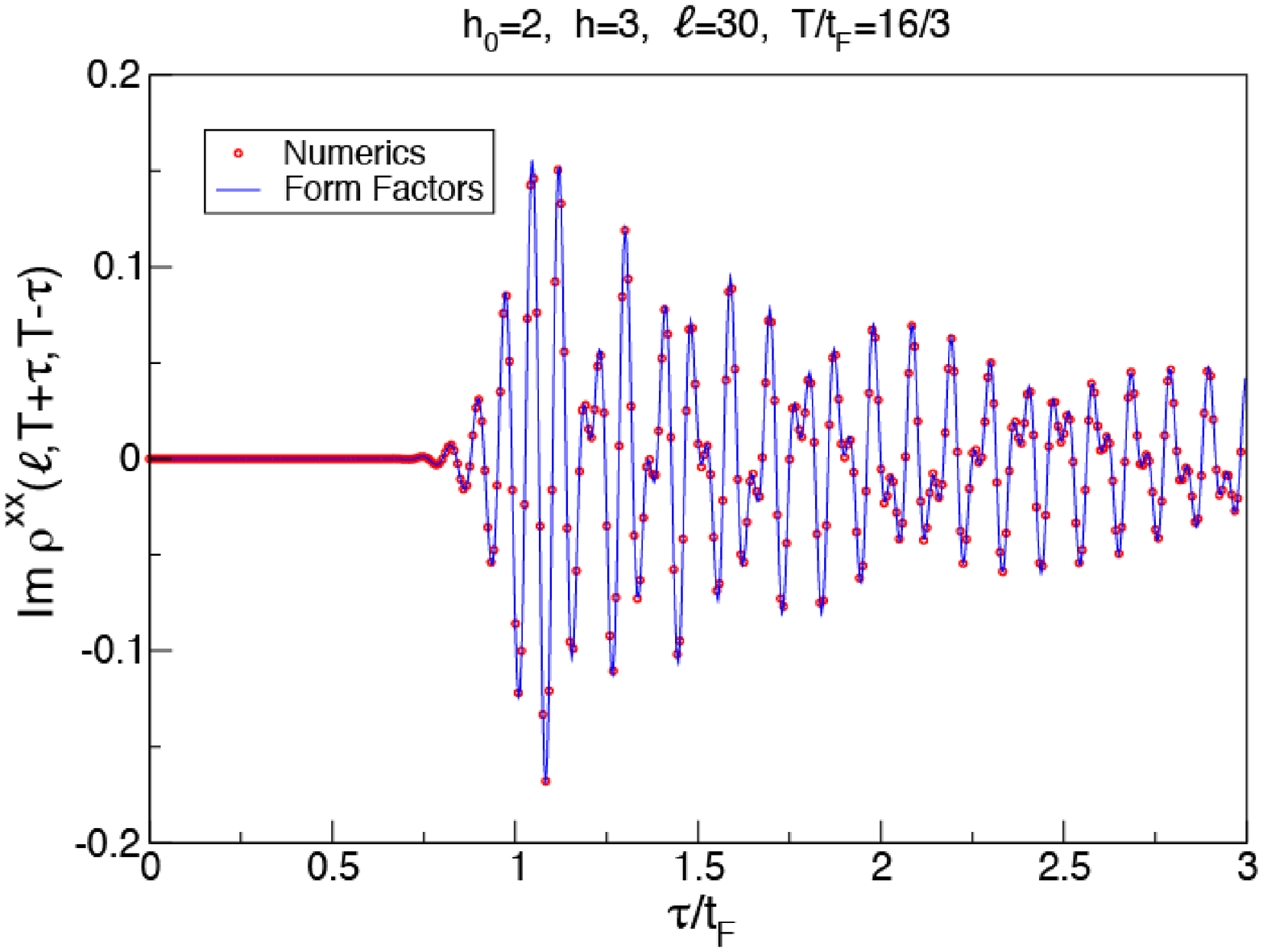}
\caption{Imaginary part of the non-equal-time two point function after a
quench in the disordered phase from $h_0=2$ to $h=3$. The distance and
time $T$ are fixed at $\ell=30$ and $T/t_F=16/3$ respectively. Data
points are numerical results (see the text for details) and the solid
line is eqn~(\ref{rhoxxdiso}).}
\end{center}
\label{fig:2}
\end{figure}
The value of $\rho^{xx}(\ell,T+\tau,T-\tau)$ at $\tau=0$ equals 
the known equal-time correlator at time $T$ after the quench~(\cite{Calabrese6}), 
which is small in the case considered. The correlator remains
largely unchanged up to a horizon at $\tau=t_F$ (corresponding to
$t=t_F/2$ in (\ref{rhoxx})), and for times $\tau>t_F$ exhibits an
oscillatory $\tau^{-1/2}$ power-law decay. 
We note that the result~(\ref{rhoxx},~\ref{R}) can be obtained in an
alternative way by generalizing the semiclassical approach of
Ref.~(\cite{IR:2011b}) (see also \cite{Sachdev1}, \cite{Rossini1}, \cite{Rossini2}) to the
non-equal time case, and then elevating it using exact limiting 
results of Refs~(\cite{Calabrese5}, \cite{Calabrese6}). While this method fails
to reproduce the result for quenches in the disordered phase outside
the light-cone, i.e. $v_{\rm max}\tau<\ell$, it provides a 
physical picture. The behaviour is similar to the finite temperature
case \cite{Sachdev1} and for $h_0,h<1$ can be understood in
terms of classical motion of domain walls. For $h_0,h>1$
(and within the light-cone), quantum fluctuations (associated with the
function $F$ in (\ref{F})) give rise to the oscillatory behaviour seen in 
Fig.~\ref{fig:2}, while relaxation occurs at longer scales and
is again driven by classical motion of particles (spin flips) (\cite{Sachdev1}). 
A simple picture emerges when we Fourier transform
$\rho^{xx}(\ell,t+\tau,t)$ at $t\to\infty$ for small quenches. As a
function of $\omega$ for fixed $q$ the resulting ``dynamical
structure factor'' for quenches within the disodered phase 
is dominated by a narrow, asymmetric peak around $\omega=\varepsilon_h(q)$,
while for $h_0,h<1$ we observe a broadening of the
$\delta$-function peak associated with the ferromagnetic order
in the initial state. Both of these are qualitatively similar to the finite-T
equilibrium response (\cite{Sachdev1},\cite{finiteT}). 
Having established in the first part of this work that the
$t\to\infty$ limit of $\rho^{xx}(\ell,t+\tau_1,t+\tau_2)$ is described
by the GGE, an important question is how quickly this limiting
behaviour is approached. It follows from (\ref{rhoxx},~\ref{rhoxxdiso}) that
for quenches within the ordered (disordered) phase the limiting value
for fixed $\tau_{1,2}$ and $\ell$ is approached as a $t^{-3}$
($t^{-3/2}$) power law.

\section{Form Factor approach}
\label{sec:formfactor}
The dynamical two-point functions~(\ref{rhoxx},~\ref{rhoxxdiso}) are
determined by a generalization of the form factor
approach recently developed in Ref.~(\cite{Calabrese5}). The latter is
based on a Lehmann representation of two-point functions in terms of
simultaneous eigenstates of the momentum operator $P$ and the
post-quench Hamiltonian $H(h)$
\bea
H(h)|k_1,\ldots,k_n\rangle_{\tt a}&=&
\Big[\sum_{j=1}^n\varepsilon_h(k_j)\Big]
|k_1,\ldots,k_n\rangle_{\tt a}\, ,\nonumber\\
P|k_1,\ldots,k_n\rangle_{\tt a}&=&
\Big[\sum_{j=1}^nk_j\Big]
|k_1,\ldots,k_n\rangle_{\tt a},
\eea
where ${\tt a}=\rm R,NS$ correspond to periodic/antiperiodic boundary
conditions on the Jordan-Wigner fermions~(\cite{Calabrese5}). For a quench
within the disordered phase the state $|\Psi_0(t)\rangle$ has the
following representation in a large, but finite volume $L$
\be
|\Psi_0(t)\rangle=\frac{|B(t)\rangle_{\rm
    NS}}{\sqrt{{}_{\rm NS}\langle B(t)|B(t)\rangle_{\rm NS}}},
\ee
where
\begin{multline}
|B(t)\rangle_{\rm
  NS}=\sum_{n=0}^\infty\frac{i^n}{n!}\sum_{0<p_1,\ldots,p_n\in\rm NS} 
\prod_{j=1}^nK(p_j)
e^{-2i\varepsilon_{h}(p_j)t}\\
\times\ |-p_1,p_1,\ldots,
-p_n,p_n\rangle_{\rm NS}\, ,
\end{multline}
\be
K(k)=\frac{\sin(k)\ (h_0-h)}{\frac{\varepsilon_{h_0}(k)
\varepsilon_{h}(k)}{(2J)^{2}}+1+hh_0-(h+h_0)\cos(k)}\, . 
\ee
The function $K(p)$ is related to the quantity $\cos(\Delta_p)$
defined in the main text by $K(p)=\tan(\Delta_p/2)$.
The dynamical order parameter two-point function 
\be
\rho^{xx}(\ell,t+\tau_1,t+\tau_2)=
\frac{{}_{\rm NS}\langle
B(t)|\sigma^x_{m+\ell}(\tau_1)\sigma^x_m(\tau_2)|B(t)\rangle_{\rm NS}}
{{}_{\rm NS}\langle B|B\rangle_{\rm NS}}
\ee
has the following Lehmann representation
\begin{multline}
{}_{\rm NS}\langle
B(t)|\sigma^x_{\ell+m}(\tau_1)\sigma^x_{m}(\tau_2)|B(t)\rangle_{\rm NS}\!=\\
\sum_{m,n=0}^\infty\frac{i^{n-m}}{n!m!}
\sum_{\genfrac{}{}{0pt}{}{0<p_1,\ldots,p_n\in {\rm
      NS}}{0<k_1,\ldots,k_m\in{\rm NS}}} 
\left[\prod_{j=1}^nK(p_j)e^{-2i(t+\tau_2)\varepsilon_h(p_j)}
\right]\left[\prod_{l=1}^mK(k_l)e^{2i(t+\tau_1)\varepsilon_h(k_l)}\right]\\
\times\sum_{s=0}^\infty\sum_{q_1,\ldots,q_s\in {\rm R}}
\frac{\prod_{r=1}^se^{i(\tau_2-\tau_1)\varepsilon_h(q_r)+iq_r\ell}}{s!}\
\langle k_m,-k_m,\ldots,k_1-k_1|\sigma^x_m
|q_1,\ldots,q_s\rangle\\
\times \langle q_s,\ldots,q_1|\sigma^x_m
|-p_1,p_1,\ldots, -p_n,p_n\rangle,
\label{lehmannquench2}
\end{multline}
\be
{}_{\rm NS}\langle B|B\rangle_{\rm NS}=
\exp\Bigl[\sum_{0<q\in {\rm NS}}\log\bigl(1+K^2(q)\bigr)\Bigr].
\ee
The form factors
\be
\langle k_m,-k_m,\ldots,k_1-k_1|\sigma^x_m |q_1,\ldots,q_s\rangle
\ee
are known exactly~(\cite{FF1}, \cite{FF2}) see eqns (109)-(111) of Ref.~(\cite{Calabrese5}). 
The leading behaviour of (\ref{lehmannquench2}) is
evaluated by considering it as a formal expansion in powers of the
function $K(p)$. As shown in Ref.~(\cite{Calabrese5}) this corresponds
to an expansion, where the small parameter is the density
of excitations of the post-quench Hamiltonian $H(h)$ in the initial
state $|\Psi_0(0)\rangle$. We determine the dominant contributions at
large $\ell$, $t$ and $|\tau_1-\tau_2|$ to (\ref{lehmannquench2}) for a
given order in the formal expansion in powers of $K(p)$, and then
sum these to all orders. The structure of this calculation is similar
to the equal time case ($\tau_1=\tau_2$) considered in
Ref.~(\cite{Calabrese5}), but the details differ substantially 
and will be reported elsewhere. The result of the form
factor calculation for quenches within the disordered phase is
\be
\rho^{xx}(\ell,t+\tau,t)\simeq
2\sqrt{h}(h^2-1)^\frac{1}{4} F(\ell,\tau,t) R_0(\ell,\tau,t)\, ,
\ee
where the function $F$ is given in (\ref{F}) and
\begin{multline}
R_0(\ell,\tau,t)=\exp\Big[-2\int_0^\pi\frac{\mathrm d
    k}{\pi}K^2(k)\\
\times \min\Big\{\max\{\varepsilon^\prime_h(k)\tau,\ell\}, 
\varepsilon^\prime_h(k)(2t+\tau)\Big\}\Big].
\end{multline}
We now use that the general structure of the resummation for
$\tau_{1,2}\neq 0$ is the same as for $\tau_{1,2}=0$. This allows us
to go beyond the low-density expansion by exploiting results obtained
in Refs~(\cite{Calabrese5}, \cite{Calabrese6}) for $\tau_1=\tau_2=0$ by means of
determinant techniques. In this way we arrive at eqn~(\ref{rhoxxdiso}).
Quenches within the ordered phase are analyzed in the same way.

\section{Conclusions}  %
In this chapter we have considered dynamical correlation functions of local
observables after a quantum quench. We have shown, that dynamical
correlators of local observables in the stationary state are governed
by the same ensemble that describes static correlations. 
For quenches in the TFIC this implies that they
are given by a GGE, for which the basic form of
the fluctuation dissipation theorem holds. We have obtained explicit
expressions for the time evolution of dynamic order parameter
correlators after a quench in the TFIC.




\backmatter 
\appendix
\chapter{Appendix A}
\label{app:elliptic}

Elliptic functions are the extensions of the trigonometric functions to work with doubly periodic expressions, i.e. such that
\be
   f \left( z + 2 n \omega_1 + 2 m \omega_3 \right) = f \left( z + 2 n \omega_1 \right) = f (z) \; ,
\ee
where $\omega_1, \omega_3$ are two different complex numbers and $n,m$ are integers (following \cite{lawden}, we have
$ \omega_1 + \omega_2 + \omega_3 =0$).

The Jacobi elliptic functions are usually defined through the pseudo-periodic {\it theta functions}:
\bea
\label{thetadef}
 \theta_1(z;q) &=& 2 \sum_{n = 0}^\infty (-1)^n q^{(n + 1/2)^2} \sin [(2n + 1) z] \; ,\\ \label{theta1}
 \theta_2(z;q) &=& 2 \sum_{n = 0}^\infty q^{(n + 1/2)^2} \cos[(2n+1) z] \; ,\\
 \theta_3(z;q) &=&  1 + 2 \sum_{n=1}^\infty q^{n^2} \cos (2 n z) \; , \\
 \theta_4(z;q) &= & 1 + 2 \sum_{n=1}^\infty (-1)^n q^{n^2} \cos (2 n z) \; ,
\eea

where the {\it elliptic nome} $q$ is usually written as $q \equiv \eu^{\ii \pi \tau}$ in terms of the elliptic parameter $\tau$ and accordingly
\be
  \theta_j (z | \tau) \equiv \theta_j (z;q) \; .
\ee
The periodicity properties of the theta functions are
\bea
\label{thetaperiod}
 \hskip -1cm \theta_1(z | \tau) &=&  - \theta_1 (z +  \pi | \tau) = - \lambda \theta_1 (z + \pi \tau | \tau) =
 \: \: \: \lambda \theta_1 (z + \pi + \pi \tau | \tau ) \; ,\\ 
 \hskip -1cm \theta_2(z | \tau) &=&  - \theta_2 (z +  \pi | \tau) = \: \: \: \lambda \theta_2 (z + \pi \tau | \tau) =
 - \lambda \theta_2 (z + \pi + \pi \tau | \tau) \; ,\\ 
 \hskip -1cm \theta_3(z | \tau) &= & \: \: \: \theta_3 (z +  \pi | \tau) = \: \: \: \lambda \theta_3 (z + \pi \tau | \tau) =
 \: \: \: \: \lambda \theta_3 (z + \pi + \pi \tau | \tau) \; ,\\ 
\hskip -1cm  \theta_4(z | \tau) &=&  \: \: \: \theta_4 (z +  \pi | \tau) = - \lambda \theta_4 (z + \pi \tau | \tau) =
 - \lambda \theta_4 (z + \pi + \pi \tau | \tau) \; ,
\eea

where $\lambda \equiv q \eu^{2 \ii z}$.

Moreover, incrementation of $z$ by the half periods ${1 \over 2} \pi, {1 \over 2} \pi \tau$, and ${1 \over 2} \pi (1 + \tau)$ leads to 
\bea
\label{thetaphalfperiod}
 \hskip -1.5cm \theta_1(z | \tau) &=&  - \theta_2 (z +  {\textstyle{1 \over 2}} \pi | \tau) = - \ii \mu \theta_4 (z + {\textstyle{1 \over 2}} \pi \tau | \tau) =
 -\ii \mu \theta_3 (z + {\textstyle{1 \over 2} \pi + {1 \over 2}}  \pi \tau | \tau ) \; ,\\ 
 \hskip -1.5cm \theta_2(z | \tau) &=&  \: \: \:  \theta_1 (z +  {\textstyle{1 \over 2}} \pi | \tau) = \: \: \: \: \mu \theta_3 (z + {\textstyle{1 \over 2}} \pi \tau | \tau) =
 \: \: \: \: \mu \theta_4 (z + {\textstyle{1 \over 2} \pi + {1 \over 2}}  \pi \tau | \tau ) \; ,\\ 
 \hskip -1.5cm \theta_3(z | \tau) &=&  \: \: \:  \theta_4 (z +  {\textstyle{1 \over 2}} \pi | \tau) = \: \: \:  \: \mu \theta_2 (z + {\textstyle{1 \over 2}} \pi \tau | \tau) =
 \: \: \: \: \mu \theta_1 (z + {\textstyle{1 \over 2} \pi + {1 \over 2}}  \pi \tau | \tau ) \; ,\\ 
 \hskip -1.5cm \theta_4(z | \tau) &=&  \: \: \: \theta_3 (z +  {\textstyle{1 \over 2}} \pi | \tau) = - \ii \mu \theta_1 (z + {\textstyle{1 \over 2}} \pi \tau | \tau) =
 \: \: \:  \ii \mu \theta_2 (z + {\textstyle{1 \over 2} \pi + {1 \over 2}}  \pi \tau | \tau ) \; ,
\eea
with $\mu \equiv q^{1/4} \eu^{\ii z}$.

The Jacobi elliptic functions are then defined as
\bea
\label{jacobidef}
 \sn (u; k) & = &  {\theta_3 (0 | \tau) \over \theta_2 (0 | \tau) } \; {\theta_1 (z | \tau) \over \theta_4 (z | \tau) } \; ,\\ 
 \cn (u; k) & =  & {\theta_4 (0 | \tau) \over \theta_2 (0 | \tau) } \; {\theta_2 (z | \tau) \over \theta_4 (z | \tau) } \; ,\\ 
 \dn (u; k) & =  & {\theta_4 (0 | \tau) \over \theta_3 (0 | \tau) } \; {\theta_3 (z | \tau) \over \theta_4 (z | \tau) } \; ,
\eea
where $u = \theta_3^2 (0 | \tau) \, z = {2 \over \pi} \, K (k) \, z$, $\tau \equiv \ii {K (k') \over K (k)} $, $k' \equiv \sqrt{ 1 -k^2}$, and 
\be
   K (k) \equiv \int_0^1 { \de t \over \sqrt{ (1 - t^2) (1-k^2 t^2)}}
   = \int_0^{\pi \over 2} { \de \theta \over \sqrt{1 - k^2 \sin^2 \theta}}
   \label{Kdef}
\ee
is the elliptic integral of the first type. 
Additional elliptic functions can be generated with the following two rules: denote with $p,q$ the letters $s, n, c, d$, then
\be
  {\rm pq} (u;k) \equiv {1 \over {\rm qp} (u;k) } \; , \qquad \qquad
  {\rm pq} (u;k) \equiv {{\rm pn} (u;k) \over {\rm qn} (u;k)} \; .
\ee

For $0 \le k \le 1$, $K(k)$ is real, $\ii K^\prime (k) \equiv \ii K (k')$ is purely imaginary and together they are called 
{\it quater-periods}. The Jacobi elliptic functions inherit their periodic properties from the theta functions, thus:
\bea
\label{jacobiperiod}
 \hskip-2cm \sn (u;k) &=&  - \sn (u + 2K;k) = \: \: \: \sn (u + 2 \ii K^\prime;k) = - \sn (u +2K +2\ii K^\prime; k ) \; ,\\ 
 \hskip-2cm \cn (u;k) &=&  - \cn (u + 2K;k) = - \cn (u + 2 \ii K^\prime;k) = \: \: \: \cn (u +2K +2\ii K^\prime; k ) \; ,\\ 
 \hskip-2cm \dn (u;k) &=&  \: \: \: \dn (u + 2K;k) = - \dn (u + 2 \ii K^\prime;k) = - \dn (u +2K +2\ii K^\prime; k ) \; .
\eea

The inverse of an elliptic function is an elliptic integral. The {\it incomplete elliptic integral of the first type} is usually written as
\be
   F (\phi;k) = \int_0^\phi {\de \theta \over \sqrt{1 - k^2 \sin^2 \theta} } 
   = \sn^{-1} \left( \sin \phi ; k \right) \; ,
   \label{Fdef}
\ee
and it is the inverse of the elliptic $\sn$. Clearly, $F \left( {\pi \over 2}; k \right) = K (k)$. Further inversion formulae for the elliptic functions can be found in \cite{lawden}.

Additional important identities include
\be
   k = k ( \tau ) =  {\theta_2^2 (0 | \tau) \over \theta_3^2 (0 | \tau) } \; , \qquad \qquad
   k' = k' ( \tau ) =  {\theta_4^2 (0 | \tau) \over \theta_3^2 (0 | \tau) } \; , 
\ee
and
\bea
    \sn^2 (u;k) + \cn^2 (u;k) & = & 1 \; , \label{ellid1}  \\
    \dn^2 (u;k) + k^2 \sn^2 (u;k) & = & 1 \; , \label{ellid2}  \\
    \dn^2 (u;k) - k^2 \cn^2 (u;k) & = & k^{\prime 2} \label{ellid3} \; .
\eea    

Taken together, the elliptic functions have common periods $2 \omega_1 = 4 K(k)$ and $2\omega_3 =  4 \ii K^\prime (k)$. 
These periods draw a lattice in $\mathbb{C}$, which has the topology of a torus. While the natural domain is given 
by the rectangle with corners $(0,0),(4K,0),(4K,4 \ii K^\prime), (0,4 \ii K^\prime)$,  any other choice would be exactly equivalent. 
Thus, the parallelogram defined by the half-periods
\be
   \omega_1^\prime = a \, \omega_1 + b \, \omega_3 \; , \qquad \quad
   \omega_3^\prime = c \, \omega_1 + d \, \omega_3 \; ,
\ee
with $a, b, c, d$ integers such that $ad - bc = 1$, can suit as a fundamental domain. This transformation, which also changes the 
elliptic parameter as
\be
   \tau^\prime = {\omega_3^\prime \over \omega_1^\prime } = { c + d \, \tau \over a + b \, \tau } \; ,
\ee
can be casted in a matrix form as
\be
   \left( \begin{array}{c} \omega_1^\prime \\ \omega_3^\prime \end{array} \right) =
   \left( \begin{array}{c c} a & b \\ c & d \end{array} \right)
   \left( \begin{array}{c} \omega_1 \\ \omega_3 \end{array} \right)
\ee
and defines a {\it modular transformation}. The modular transformations generate the {\it modular group} ${\rm PSL}(2,\mathbb{Z})$.
Each element of this group can be represented (in a {\bf not} unique way) by a combination of the two transformations
\bea
   {\bf S} = \left( \begin{array}{c c} 0 & 1 \\\ -1 & 0 \end{array} \right) & \quad \longrightarrow \quad & \tau^\prime = - {1 \over \tau} \; , \\
   {\bf T} = \left( \begin{array}{c c} 1 & 0 \\\ 1 & 1 \end{array} \right) & \quad \longrightarrow \quad & \tau^\prime =  \tau + 1 \; ,
\eea
which satisfy the defining relations
\be
    {\bf S}^2 = {\bf 1} \; , \qquad \qquad \left( {\bf ST} \right)^3 = {\bf 1} \; .
\ee

The transformation properties for the ${\bf S}$ transformation are
\bea
\label{SthetaTrans}
 \theta_1 \left( z | - {1 \over \tau} \right) &=& -\ii(\ii \tau)^{\frac 1 2} \eu^{\frac{\ii \tau z^2}{\pi}}
 \theta_1( \tau z | \tau) \; , \\
 \theta_2 \left( z |  - {1 \over \tau} \right) &=& (-\ii \tau)^{\frac 1 2} \eu^{\ii \tau z^2 \over \pi}
 \theta_4 (\tau z | \tau) \; , \\
 \theta_3 \left( z  |  - {1 \over \tau} \right) &=& (-\ii \tau)^{\frac 1 2} \eu^{\ii \tau z^2 \over \pi}
 \theta_3 (\tau z | \tau) \; , \\
 \theta_4 \left (z |  - {1 \over \tau} \right) &=& (-\ii \tau)^{\frac 1 2} \eu^{\ii \tau z^2 \over \pi} \theta_2
(\tau z | \tau) \; .
\eea
From which follows
\be
   k \left( - {1 \over \tau} \right) = k^\prime ( \tau ) \; , \qquad \qquad \qquad
   k^\prime  \left( - {1 \over \tau} \right) = k ( \tau ) \; ,
\ee
and
\be
  K \left( - {1 \over \tau} \right) = K' ( \tau) = K (k') \; , \qquad 
  \ii K^\prime \left( - {1 \over \tau} \right) = \ii K (\tau) = \ii K (k) \; .
\ee
Moreover,
\bea
  \sn (u; k') = - \ii { {\rm sn} (\ii u ; k) \over {\rm cn} ( \ii u ; k ) } \; , & \qquad &
  \cn (u; k') =  { 1 \over {\rm cn} ( \ii u ; k ) } \; , 
  \nonumber \\
  \dn (u; k') =  { {\rm dn} (\ii u ; k) \over {\rm cn} ( \ii u ; k ) } \; , & \qquad & \ldots
\eea
   
For the ${\bf T}$ transformation we have
\bea
\label{TthetaTrans}
 \theta_1 \left( z |  \tau + 1 \right) &= & \eu^{\ii \pi/4} \theta_1 ( z | \tau) \; , \\
 \theta_2 \left( z |  \tau + 1 \right) &=& \eu^{\ii \pi/4} \theta_2 ( z | \tau) \; ,\\
 \theta_3 \left( z |  \tau + 1 \right) &=&  \theta_4  ( z | \tau) \; , \\
\theta_4 \left (z |  \tau + 1 \right) &=&  \theta_3 ( z | \tau) \; ,
\eea
and thus
\be 
  k ( \tau +1 ) = \ii {k (\tau) \over k^\prime (\tau) } \; , \qquad \qquad \quad
  k^\prime (\tau + 1 ) =  {1 \over k^\prime (\tau) } \; ,
\ee
and
\be
  K (\tau +1 ) = k^\prime (\tau) K (\tau) \; , \qquad \quad
  \ii K^\prime (\tau + 1) =  k^\prime \left[ K (\tau) + \ii K^\prime (\tau) \right] \; .
\ee
Finally
\bea
  \sn \left( u; \ii {k \over k'} \right) =  k- { {\rm sn} (u/k' ; k) \over {\rm dn} ( u/k' ; k ) } \; , & \quad &
  \cn  \left( u; \ii {k \over k'} \right)  =  { {\rm cn} (u/k' ; k) \over {\rm dn} ( u/k' ; k ) }  \; ,
  \nonumber \\
  \dn  \left( u; \ii {k \over k'} \right)  =  { 1 \over {\rm dn} ( u/k' ; k ) } \; , & \quad & \ldots
\eea

While it is true that by composing these two transformations we can generate the whole group, for convenience we 
collect here the formulae for another transformation we use in the body of the paper: $\tau \to { \tau \over 1 - \tau}$, 
corresponding to ${\bf STS}$
\bea
 \theta_1 \left( z |  {\tau \over 1 - \tau} \right) &=&  \ii^{\frac 1 2} F \: \theta_1 \Big ( (1 - \tau) z | \tau \Big) \; , \\
 \theta_2 \left( z |  {\tau \over 1 - \tau} \right) &=&  F \: \theta_3 \Big ( (1 - \tau) z | \tau \Big) \; , \\
 \theta_3 \left( z  | {\tau \over 1 - \tau} \right) &=&  F \: \theta_2 \Big ( (1 - \tau) z | \tau \Big) \; , \\
 \theta_4 \left (z |  {\tau \over 1 - \tau} \right) &=&  \ii^{\frac 1 2} F \: \theta_4 \Big ( (1 - \tau) z | \tau \Big) \; ,
\eea
\label{SthetaTrans}
where $F = (1 - \tau)^{1 \over 2} \eu^{\ii {(\tau -1) z^2 \over \pi}}$.
We have
\be
   k \left( {\tau \over 1- \tau} \right) = {1 \over k ( \tau )} = {1 \over k} \; , \qquad \qquad \quad
   k^\prime  \left( {\tau \over 1 - \tau} \right) = \ii {k^\prime ( \tau ) \over k (\tau)} = \ii {k' \over k} \; ,
\ee
and
\be
  K \left( {\tau \over 1 - \tau} \right) = k \left[ K ( \tau) - \ii K^\prime (\tau) \right]  \; , \qquad 
  \ii K^\prime \left( {\tau \over 1 - \tau} \right) = \ii k K^\prime (\tau) \; .
\ee
Moreover,
\bea
  \sn \left( u; {1 \over k} \right) = k \: {\rm sn} \left ( {u \over k} ; k \right) \; , & \qquad &
  \cn \left( u; {1 \over k} \right) =  {\rm dn}  \left ( {u \over k} ; k \right) \; , 
  \nonumber \\
  \dn \left( u; {1 \over k} \right) =  {\rm cn}  \left ( {u \over k} ; k \right) \ \; . & \qquad & 
  \label{jacobiSTS}
\eea



\chapter{Appendix B}
\label{appB}
In this appendix we list the code we have used to compute the two-point function of the order parameter in the TFIC, in both the ferromagnetic and paramagnetic phase, with open boundary conditions (OBC). In particular this version of the code computes the equal-time correlator, but its extension to the non-equal time is straightforward. 

\begin{lstlisting}

(*Beginning of the programme*)


s11 = {{1, 0}, {0, 0}};
s12 = {{0, 1}, {0, 0}};
s21 = {{0, 0}, {1, 0}};
s22 = {{0, 0}, {0, 1}};

L = 512;        (* Size of the system *)

h = 3/4;        (* Initial Magnetic field *)

h0 = 1/3;       (* Final magnetic field *)

l = 30;         (*Spatial distance between operators*)

Tmax = 100;     (*Maximal simulated time*)


Solve[h0 == (h - 2 eps + h eps^2)/(1 - 2 h eps + eps^2 ), eps] 

(*The previous ``solve'' gives us a measure*)
      (*of the size of the quench*)

f = {};

If[h > 1, Monitor[For[j = 1, j <= L, j++,
   AppendTo[f, k /. FindRoot[(Sin[L k] == h Sin[(L + 1) k]), 
	 {k, Pi*j/(L + 1)}]];], j];, Monitor[AppendTo[f, I k /. 
	 FindRoot[(Sinh[L k] == h Sinh[(L + 1) k]), {k, -Log[h]}]];
   For[j = 1, j <= L - 1, j++, AppendTo[f, k /. FindRoot[
	 (Sin[L k] == h Sin[(L + 1) k]),{k, Pi*j/(L)}]];],j];
  ]

If[Length[Union[f]] != L, Print["Warning: degenerate roots!"]];

f0 = {};

If[h0 > 1, Monitor[ For[j=1, j <= L, j++, 
   AppendTo[f0, k /. FindRoot[(Sin[L k] == h0 Sin[(L+1) k]), 
	 {k, Pi*j/(L+1)}]];], j];, Monitor[ AppendTo[f0, I k /. 
	 FindRoot[(Sinh[L k]==h0 Sinh[(L+1) k]), {k, -Log[h0]}]];
   For[j=1,j<= L-1, j++, AppendTo[f0, k /. FindRoot[
	 (Sin[L k]==h0 Sin[(L+1) k]), {k, Pi*j/(L)}]];],j];
  ] 

If[Length[Union[f0]] != L, Print["Warning: degenerate roots!"]];

[epsilon]=Sqrt[1 + h^2 - 2 h Cos[f]];

Nk = Sqrt[2/(L+h^2 (L+1)-h (1+2 L) Cos[f])];
Nk0 =  Sqrt[2/(L+h0^2 (L+1)-h0 (1+2 L) Cos[f0])];

ClearAll[phi, phi0, psi, psi0, [Phi], [Phi]0, [Psi], [Psi]0];

phi[j_, i_] = "UnDefined";

[Phi]= 
Function[If[phi[#1,#2]=="UnDefined",If[h>1,
phi[#1, #2]=(-1)^#2 Nk[[#1]](Sin[(#2-1) 
f[[#1]]]-h Sin[#2*f[[#1]]]),phi[#1,#2]=
If[(#1==1)&&(Abs[Cos[f[[#1]]]-(1+h^2)/2/h]<
10^(-7)),-Sqrt[1-h^2]/h*(-h)^#2,(-1)^#2 
Nk[[#1]]*(-h*Sin[#2*f[[#1]]]+Sin[(#2-1)
*f[[#1]]])]]];phi[#1, #2]];
								
phi0[j_, i_] = "UnDefined";

[Phi]0= 
Function[If[phi0[#1, #2]=="UnDefined",If[h0>1,
phi0[#1, #2]=(-1)^#2 Nk0[[#1]] (Sin[(#2-1) 
f0[[#1]]]-h0 Sin[#2*f0[[#1]]]),phi0[#1,#2]=
If[(#1==1)&&(Abs[Cos[f0[[#1]]]-(1+h0^2)/2/h0]<
10^(-7)),-Sqrt[1-h0^2]/h0*(-h0)^#2,(-1)^#2 
Nk0[[#1]]*(-h0*Sin[#2*f0[[#1]]]+Sin[(#2-1)*
f0[[#1]]])]]];phi0[#1, #2]];
								
psi[j_, i_]= "UnDefined";

[Psi] = Function[If[psi[#1,#2]=="UnDefined", psi[#1,#2]=
        (-1)^(L-#1)[Phi][#1,L+1-#2]]; psi[#1,#2];

psi0[j_, i_] = "UnDefined";

[Psi]0 = Function[If[psi0[#1,#2]=="UnDefined", psi0[#1,#2]= 
         (-1)^(L-#1)[Phi]0[#1,L+1-#2]]; psi0[#1,#2]];

Phi = Array[[Phi], {L, L}];
Phi0 = Array[[Phi]0, {L, L}];
Psi = Array[[Psi], {L, L}];
Psi0 = Array[[Psi]0, {L, L}];

G = -Transpose[Phi0].Psi0;
Gt = Transpose[G];

SetSharedVariable[correl];
correl = {};
Monitor[

ParallelDo [
t2 = t;    
t1 = t;
	
C1 = DiagonalMatrix[Cos[[Epsilon]*t1]];
C2 = DiagonalMatrix[Cos[[Epsilon]*t2]];
S1 = DiagonalMatrix[Sin[[Epsilon]*t1]];
S2 = DiagonalMatrix[Sin[[Epsilon]*t2]];

Q11 = Transpose[Phi].C1.Phi;
Q12 = Transpose[Phi].C2.Phi;
Q21 = Transpose[Psi].S1.Phi;
Q21t = Transpose[Q21];
Q22 = Transpose[Psi].S2.Phi;
Q22t = Transpose[Q22];
Q31 = Transpose[Psi].C1.Psi;
Q32 = Transpose[Psi].C2.Psi;

A2A2 = Q12.Q12 + Q22t.Q22 + I Q12.G.Q22 - I Q22t.Gt.Q12;
A1A1 = Q11.Q11 + Q21t.Q21 + I Q11.G.Q21 - I Q21t.Gt.Q11;
A2A1 = Q12.Q11 + Q22t.Q21 + I Q12.G.Q21 - I Q22t.Gt.Q11;
B2B2 = - Q22.Q22t - Q32.Q32 + I Q22.G.Q32 - I Q32.Gt.Q22t;
B1B1 = - Q21.Q21t - Q31.Q31 + I Q21.G.Q31 - I Q31.Gt.Q21t;
B2B1 = - Q22.Q21t - Q32.Q31 + I Q22.G.Q31 - I Q32.Gt.Q21t;
A2B2 = I  Q12.Q22t - I Q22t.Q32 +  Q12.G.Q32 + Q22t.Gt.Q22t;
A1B1 = I  Q11.Q21t - I Q21t.Q31 +  Q11.G.Q31 + Q21t.Gt.Q21t;
A2B1 = I  Q12.Q21t - I Q22t.Q31 +  Q12.G.Q31 + Q22t.Gt.Q21t;
B2A2 = I  Q22.Q12 - I Q32.Q22 -  Q22.G.Q22 - Q32.Gt.Q12;
B1A1 = I  Q21.Q11 - I Q31.Q21 -  Q21.G.Q21 - Q31.Gt.Q11;
B2A1 = I  Q22.Q11 - I Q32.Q21 -  Q22.G.Q21 - Q32.Gt.Q11;

n = L/2 - l/2;         (*Position of the first operator*)
	
Gamma1 = UpperTriangularize[
Join[Join[

KroneckerProduct[A2A2[[;; n, ;; n]], s11][[;; 2 n - 1, ;; 
2 n - 1]] + 
KroneckerProduct[A2B2[[;; n, ;; n]], s12][[;; 2 n - 1, ;; 
2 n - 1]] + 
KroneckerProduct[B2A2[[;; n, ;; n]], s21][[;; 2 n - 1, ;; 
2 n - 1]] + 
KroneckerProduct[B2B2[[;; n, ;; n]], s22][[;; 2 n - 1, ;; 
2 n - 1]],
KroneckerProduct[A2A1[[;; n, ;; n + l]], s11][[;; 2 n - 1, ;; 
2 n + 2 l - 1]] + 
KroneckerProduct[A2B1[[;; n, ;; n + l]], s12][[;; 2 n - 1, ;; 
2 n + 2 l - 1]] + 
KroneckerProduct[B2A1[[;; n, ;; n + l]], s21][[;; 2 n - 1, ;; 
2 n + 2 l - 1]] + 
KroneckerProduct[B2B1[[;; n, ;; n + l]], s22][[;; 2 n - 1, ;; 
2 n + 2 l - 1]], 2],

Join[ConstantArray[0, {2 n + 2 l - 1, 2 n - 1}],

KroneckerProduct[A1A1[[;; n + l, ;; n + l]], s11][[;; 
2 n + 2 l - 1, ;; 2 n + 2 l - 1]] + 
KroneckerProduct[A1B1[[;; n + l, ;; n + l]], s12][[;; 
2 n + 2 l - 1, ;; 2 n + 2 l - 1]] + 
KroneckerProduct[B1A1[[;; n + l, ;; n + l]], s21][[;; 
2 n + 2 l - 1, ;; 2 n + 2 l - 1]] + 
KroneckerProduct[B1B1[[;; n + l, ;; n + l]], s22][[;; 
2 n + 2 l - 1, ;; 2 n + 2 l - 1]], 2]], 1];

Gamma2 = Transpose[Gamma1] ;
GGamma = Gamma1 - Gamma2; 

AppendTo[correl, {t, Sqrt[Det[GGamma]]} ]; 

 (*Here we compute the square of the Det of Gamma*)
(*because it is much faster than the Pfaffian code*)

, {t, 0, Tmax, 0.5}
   
];,
{ListPlot[{
Table[{Sort[correl][[l, 1]], Re[Sort[correl][[l, 2]]]}, 
{l, Length[correl]}], 
Table[{Sort[correl][[l, 1]], -Re[Sort[correl][[l, 2]]]}, 
{l, Length[correl]}]
}
, Joined -> True], 
ListPlot[{Table[{Sort[correl][[l, 1]], 
Im[Sort[correl][[l, 2]]]}, {l, Length[correl]}], 
Table[{Sort[correl][[l, 1]], -Im[Sort[correl][[l, 2]]]}, 
{l, Length[correl]}]}, Joined -> True]}]
						
correl = Sort[correl];		
	
(*End of the programme*)


\end{lstlisting}

 Notice that [Phi] as a variable is different from phi, and also [Phi]0 is another variables with respect to phi0 (the same for all the variable psi-like).



\bibliographystyle{plainnat}
\renewcommand{\bibname}{References} 
\bibliography{References/references} 

\end{document}